%% file: Thesis.tex
\documentclass[sotoncolour]{uosthesis}      
\graphicspath{{Figures/}}   
\usepackage{bibentry}          
\nobibliography*               
\usepackage{notoccite}
\usepackage{attrib}            
\hypersetup{colorlinks=true}   
\input{Definitions}            
\usepackage{color}
\usepackage{hyperref}
\usepackage{tcolorbox}
\usepackage{mathrsfs}
\usepackage{upgreek}
\usepackage{float}
\usepackage{float}
\usepackage{tensor}
\usepackage[justification=justified]{caption}
\usepackage{cite}
\usepackage{bm}
\usepackage[intoc]{nomencl}
\makenomenclature

\counterwithout*{footnote}{chapter}

\department  {School of Mathematical Sciences}
\DEPARTMENT  {\MakeUppercase{\deptname}}
\group       {Gravity Group}
\GROUP       {\MakeUppercase{\groupname}}
\faculty     {Faculty of Social Sciences}
\FACULTY     {\MakeUppercase{\facname}}
\title      {Self-force in hyperbolic black hole encounters}
\authors    {Oliver Francis Long} 
\addresses  {\groupname\\\deptname\\\univname}
\date       {\today}

\orcidid{0000-0002-3897-9272}
\subject    {}

\begin{document}
\pagenumbering{gobble} 
\copyrightDeclaration{} 
                              
\frontmatter
\maketitle
\include{Abstract}
\tableofcontents
\listoffigures
\listoftables
\authorshipdeclarationdigital{\\ Oliver Long and Leor Barack. Time-domain metric reconstruction for hyperbolic scattering. {\it Phys.\ Rev.\ D}, 104(024014), Jul 2021.}
\input{Acknowledgements}
\include{Nomenclature}

\mainmatter

\input{IntroGW}

\input{IntroEMRI}

\input{IntroSF}
\input{IntroScatter}
\input{IntroThesis}

\chapter{Review of scattering geodesics in Schwarzschild spacetime}
\label{chapter:ScatGeo}
\input{Geodesics}

\chapter{Conservative self-force correction to the scatter angle}
\label{chapter:SFScatAngle}
\input{Correction}

\chapter{Time-domain code development: Scalar field}
\label{chapter:ScalarField}
\input{ScalarVacuum}
\input{ScalarSource}
\input{ScalarCircular}
\input{ScalarScatter}

\chapter{Scalar self-force on hyperbolic orbits}
\label{chapter:ScalarSF}
\input{ScalarSF}

\chapter{Time-domain reconstruction of the metric perturbation: Vacuum}
\label{chapter:Vacuum}
\input{Reconstruction}
\input{HertzFormulation}
\input{Vacuum}
\input{RWFormulation}

\chapter{Time-domain reconstruction of the metric perturbation: Particle formulation}
\label{chapter:ParticleForm}
\input{NonVacuumReconstruction}
\input{HertzNonVacFormulation}
\input{HertzJumps}
\input{RWJumps}

\addtocontents{toc}{\protect\newpage}
\chapter{Time-domain reconstruction of the metric perturbation: Particle implementation}
\label{chapter:Implementation}
\input{RWCircular}
\input{RWScatter}

\input{Conclusions}

\appendix
\renewcommand*{\chaptername}{\appendixname}
\input{BPTReconstruction}
\input{WeylJumps}
\input{FDS}
\backmatter
\bibliographystyle{unsrt}
\bibliography{Thesis}
\end{document}

%% file: Definitions.tex
\newcommand{\BibTeX}{{\rm B\kern-.05em{\sc i\kern-.025em b}\kern-.08em T\kern-.1667em\lower.7ex\hbox{E}\kern-.125emX}}

\newcounter{address}
\newcommand{\vinf}{v_{\infty}}
\newcommand{\uinf}{u_{\infty}}
\newcommand{\Einf}{E_{\infty}}
\newcommand{\El}{{\sf El}}
\newcommand{\Linf}{L_{\infty}}
\newcommand{\chiinf}{\chi_{\infty}}
\newcommand{\rmin}{R_{\rm min}}
\newcommand{\Rdot}{\dot{R}}
\newcommand{\Rdotinf}{\dot{R}_\infty}
\newcommand{\J}[1]{\left[#1\right]}

\newcommand{\vf}{\varphi}

\newcommand{\SP}[1]{\left(#1\right)}
\newcommand{\bP}[1]{\big(#1\big)}

\newcommand{\BB}[1]{\Big[#1\Big]}

\newcommand{\tet}[2]{e_{\bm #1}^{#2}}				

\newcommand{\RomanNumeralCaps}[1]{\uppercase\expandafter{\romannumeral #1}}

\newcounter{alg}

%% file: Abstract.tex

\begin{abstract}
Self-force methods can be applied in calculations of the scatter angle in two-body hyperbolic encounters, working order by order in the mass ratio (assumed small) but with no recourse to a weak-field approximation. This, in turn, can inform ongoing efforts to construct an accurate description of the general-relativistic binary dynamics via an effective-one-body description or other approaches. Existing self-force methods are to a large extent specialised to bound, inspiral orbits. Here we derive the first-order conservative self-force correction to the scattering angle, show its agreement with recent post-Minkowsian results, and develop a technique for (numerical) self-force calculations that can efficiently tackle scatter orbits. In the method, the metric perturbation is reconstructed from a Hertz potential that satisfies (mode-by-mode) a certain inhomogeneous version of the Teukolsky equation. The crucial ingredient in this formulation are certain jump conditions that the (multipole modes of the) Hertz potential must satisfy along the worldline of the small body's orbit. We present a closed-form expression for these jumps, for an arbitrary geodesic orbit in Schwarzschild spacetime. To begin developing the numerical infrastructure, a scalar-field evolution code on a Schwarzschild background (in 1+1D) is developed. Following this, results for the conservative scalar self-force corrections to the scatter angle are calculated. We continue by constructing a Teukolsky evolution code on a Schwarzschild background. This produces numerically unstable solutions due to unphysical homogeneous solutions of the Teukolsky equation at the horizon and null infinity being seeded by numerical error. This can be resolved by a change of variables to a Regge-Wheeler-like field. We then present a full numerical implementation of this method for circular and scatter orbits in Schwarzschild. We conclude with a discussion of the outlook for self-force calculations on scatter orbits.
\end{abstract}

%% file: Acknowledgements.tex

\acknowledgements{First and foremost I would like to thank my supervisor Leor Barack without whom none of this would have been possible. Leor's welcoming demeanour, patience, and expertise has continued to make him a joy to work with. He has always been willing to carve out time including meetings several miles above the Dorset countryside. 

I would also like to thank Adam Pound, Andrew Spiers, Sam Upton, Mekhi Dhesi, and the rest of the self force group in Southampton for supporting me throughout the start of my research career. I have been warmly welcomed by the Capra community as well as the wider gravitational wave community and I hope to continue conducting research in this revolutionary field.

A further thank you to Tom Hutchins, Emma Osbourne, Alex Wright, Garvin Yim, and the rest of office 54/2019 for providing ample distractions and without whom I would have finished this degree in a much more timely manner.

In my time at Southampton I have lived with some wonderful people so I would like to thank Nova, Joe Litchfield, Justine McAvoy, Dominic So, Hazel Mills, Reggie Roy, Sud Sivaneswaran, Robert Elkington, and Ella Benson for all of the antics.

Southampton University Mountaineering Club (SUMC) has been a wonderful outlet during my time here by taking me to many places and giving me some of my most fond memories. I would like to give a special shoutout to Allie Soave, Ludo Bello, Luke Colbeck-Tate, Milly Thurgood, Joh Kirby, Fraser Grandfield, and Laurence Rawlings who have become some of my closest friends and always been there for me. SUMC has introduced me to many more of my friends including Fred, Robin, Harry, Rob, Russ, Tom, Verity, Ben, Cara, Phoebe, Rob, Iain, Abe, Ben, Abi, Will, Tom, Jack, Cara, Adam, Christian, Felix, Ellie, Josie, Tom, Nicki, Becca, Amber, Eilean {\em et al}. Additionally, I would like to thank Gill for my amazing pair of patchwork trousers.

I would also like to acknowledge the continued support from friends that I made before my time at Southampton including Elliot Thomas, Patrick Buck, Tom Hutchins, Charlie Lyth, Jack Taylor, Jack Helliwell, Ben Leather, Ellie Brown, and Bea Bottura. 

Lastly I would like to thank my parents Gaynor and Jon, as well as my wider family, for supporting me through all my endeavours.}

%% file: Nomenclature.tex

\nomenclature[A]{SF}{Self-force}
\nomenclature[A]{NR}{Numerical Relativity}
\nomenclature[A]{PM}{Post-Minkowskian}
\nomenclature[A]{PN}{Post-Newtonian}
\nomenclature[A]{EOB}{Effective one-body}
\nomenclature[A]{QFT}{Quantum field theory}
\nomenclature[A]{EFT}{Effective field theory}

\nomenclature[A]{IRG}{Ingoing radiation gauge}
\nomenclature[A]{ORG}{Outgoing radiation gauge}

\nomenclature[A]{BPT}{Bardeen-Press-Teukolsky}
\nomenclature[A]{RW}{Regge-Wheeler}
\nomenclature[A]{EFE}{Einstein's field equations}
\nomenclature[A]{TD}{Time-domain}

\nomenclature[A]{ODE}{Ordinary differential equation} 
\nomenclature{PDE}{Partial differential equation}

\nomenclature{GR}{General Relativity}
\nomenclature{GW}{Gravitational wave}
\nomenclature{EMRI}{Extreme-mass ratio inspiral}
\nomenclature{BBH}{Binary-black hole}

\printnomenclature

%% file: IntroGW.tex

\chapter{Introduction}
\label{Chapter:Introduction}

\section{Gravitational waves}

The early belief in the existence of gravitational waves (GWs) varied, much like the waves themselves. They were discussed by Heaviside \cite{Heaviside1893} in the late 19th century and later proposed by Poincare in 1905 through analogies of electromagnetic and gravitational fields \cite{Poincare1905}. The first mathematical proof within General Relativity (GR) came from Einstein himself who discovered three separate types of waves which were later categorised by Hermann in 1922 \cite{Hermann1922}. In the same year, Eddington found that two of these were artefacts of the coordinate system used by Einstein but did not rule out the existence of the third \cite{Eddington1922}. Gravitational waves were again put into doubt when Einstein, with Rosen, concluded that they could not exist in GR due to the presence of a singularity. Robertson realised that this was only a coordinate singularity which lead to Einstein concluding that gravitational waves do in fact exist \cite{Einstein1937}. The confusion due to coordinate systems was finally settled several decades later when Pirani used the coordinate independent Riemann curvature tensor to reach the same conclusion \cite{Pirani1956}.

With the scientific community agreed on their existence, there became an invested interest in attempting to observe the phenomenon. In 1969, Weber announced the first detection of gravitational waves \cite{Weber1969}, using resonant mass detectors which became known as Weber bars. However, doubt was cast on the validity of the discovery due to the frequency of reported detections originating from the galactic centre. The timescale for the Milky Way to emit all its energy via GW emission was significantly smaller than the widely accepted age of the galaxy. Further uncertainty came due to other independent groups failing to detect the same signal with their own Weber bars. This led the scientific community to come to the consensus that Weber's results were spurious.

\begin{figure}[h!]
\centering
\includegraphics[width=0.65\linewidth]{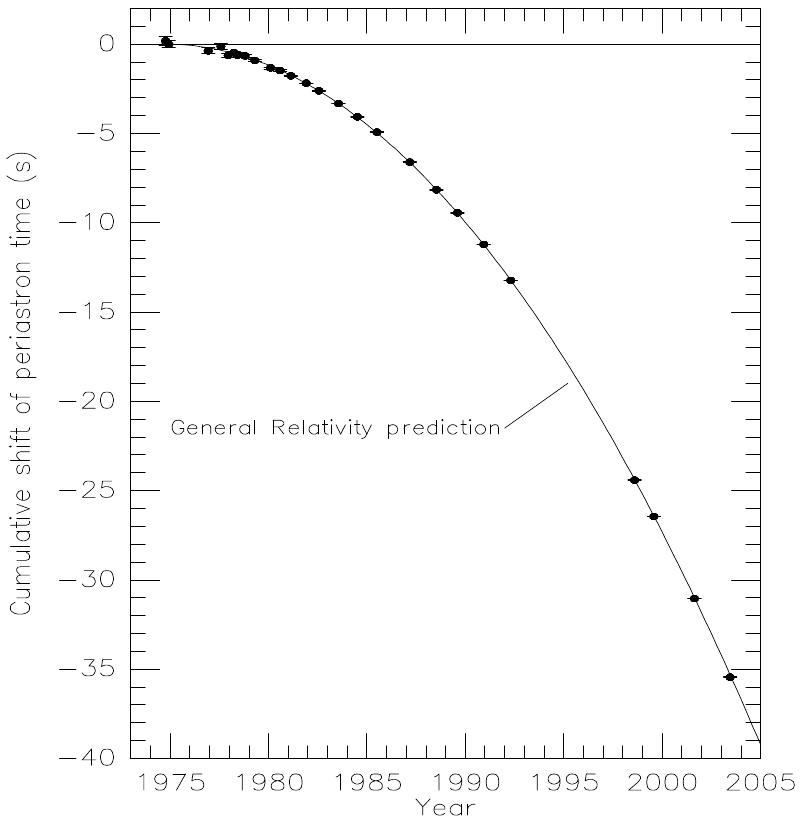}
\caption[Orbital decay of the Hulse-Taylor pulsar (PSR B1913+16)]{Orbital decay of the Hulse-Taylor pulsar (PSR B1913+16). The data points indicate the observed change in the epoch of periastron with date while the parabola illustrates the theoretically expected change in epoch for a system emitting gravitational radiation, according to GR. Image courtesy: \cite{WeisbergTaylor2004}.}
\label{HulseTaylorPulsar}
\end{figure}

The detection of the first binary pulsar PSR B1913+16 by Hulse and Taylor \cite{HulseTaylor1975}, which won them the 1993 Nobel Prize in Physics, allowed a unique viewpoint into the weak-field regime of the two-body problem in GR using electromagnetic observations. The loss of energy and angular momentum of the system, observed using pulsar timing over several years, precisely matched the predicted losses in GR due to GW emission \cite{Taylor1979,Weisberg1982}. This provided the first indirect evidence of the existence of gravitational waves. Observations of the pulsar have continued for many years and observations continue to closely match the theory as shown in Figure \ref{HulseTaylorPulsar}.

The indirect detection only fuelled the race to directly detect a gravitational wave. In spite of Weber's discredit, some groups continued to improve on his design while others toyed with laser interferometers. These measure the relative changes in the length of two perpendicular arms due to the propagation of a GW using the interference of light beams which travel down each arm. Several laser interferometers were constructed including GEO600 (British-German collaboration), Laser Interferometer Gravitational-Wave Observatory (LIGO; USA), and Virgo (Italy) but these lacked the sensitivity required to detect the minuscule changes in length required for the detection of GWs. Undeterred, LIGO and Virgo sort to remedy this with upgrades to ``advanced'' detectors aimed to increase the sensitivity of the detectors by a factor 10 relative to the original designs. 

\begin{figure}[h!]
\centering
\includegraphics[width=\linewidth]{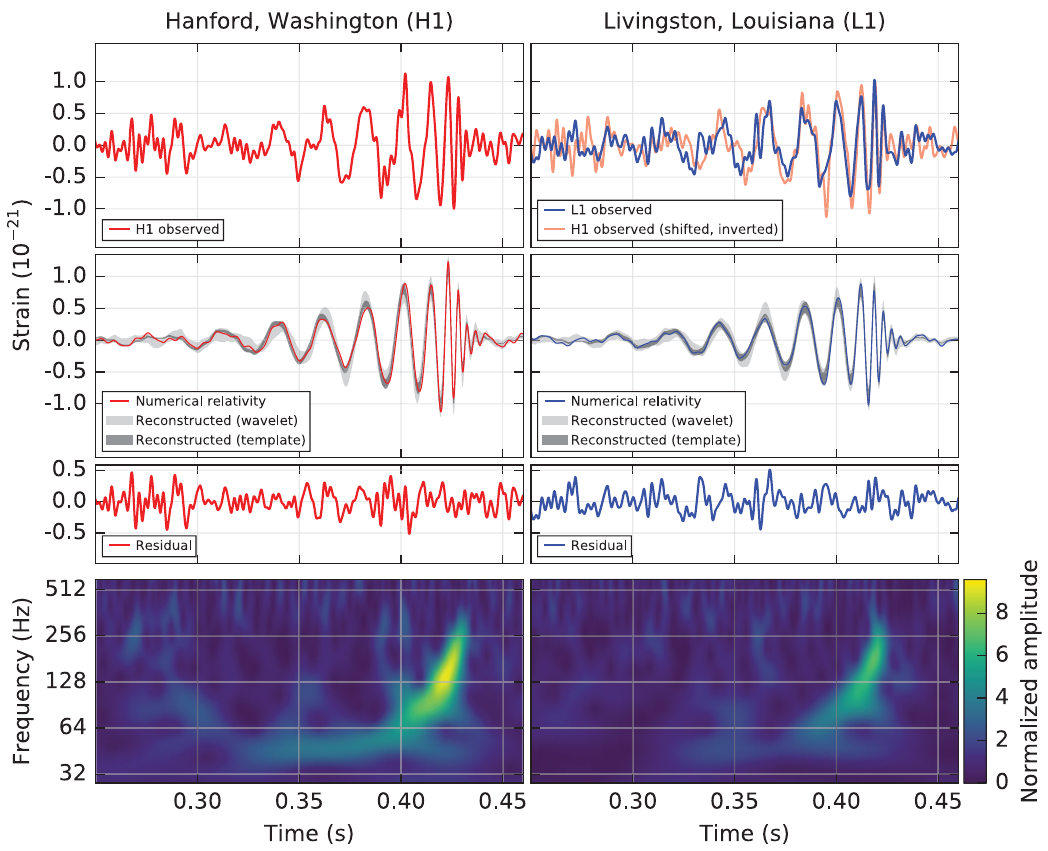}
\caption[The gravitational-wave event GW150914 as observed by LIGO]{The gravitational-wave event GW150914 as observed by LIGO Hanford (H1, left column panels) and LIGO Livingston (L1, right column panels). Times are shown relative to September 14, 2015 at 09:50:45 UTC. Image courtesy: \cite{GW150914}.}
\label{GW150914Signal}
\end{figure}

On 15th September 2015, the newly upgraded advanced LIGO (aLIGO) detected the first gravitational wave signal, GW150914, at both Livingston and Hanford sites \cite{GW150914}. The detected signal, shown in Figure \ref{GW150914Signal}, lasted 0.2 seconds and had a peak strain of $10^{-21}$, which corresponds to the length of the $4\; {\rm km}$ LIGO arms differing by $10^{-18}\; {\rm m}$. Match filtering the signal against the library bank of theoretical waveforms determined that the collision consisted of two black holes with masses $35^{+5}_{-3}\; M_\odot$ and $30^{+3}_{-4}\; M_\odot$ which formed a single black hole of mass $62^{+4}_{-3}\; M_\odot$ with $3.0 \pm 0.5\; M_\odot$ worth of energy being emitted as gravitational waves. The event was revolutionary as it was not only the first gravitational wave signal detected but also the first direct detection of a black hole (BH), the first observed binary-black hole (BBH) collision, and one of the best tests of strong-field GR to date \cite{GW150914GR}. The LIGO Scientific Collaboration announced the discovery on 11th February 2016 which was widely celebrated  and commended, including with Barry Barish, Kip Thorne, and Rainer Weiss receiving the 2017 Nobel Prize in Physics ``for decisive contributions to the LIGO detector and the observation of gravitational waves." The detection of GW150914 began a new age of gravitational wave astronomy.

A second type of GW source, a binary-neutron star (BNS) collision, was detected by both LIGO sites and the newly operational advanced Virgo on 17th August 2017 \cite{GW170817}. The signal, dubbed GW170817, had a lower amplitude than previous BBH signals but lasted significantly longer such that the signal-to-noise ratio (SNR) integrated over the entire signal was within detectible levels. Mere seconds after the GW event, a short gamma ray burst (GRB), designated GRB 170917A, was detected. The probability of incidental near-simultaneous temporal and spatial observations was so minute that it was concluded that they originated from the same event and thus BNS mergers were confirmed as a progenitor of short GRBs \cite{GW170817GRB}. The improved sky localisation of the GW signal due to Virgo allowed other electromagnetic (EM) telescopes to detect the first EM counterpart to a GW event in the galaxy NGC 4993. Over the following days signals from radio to X-rays were detected from the galaxy. The combination of the GW and EM signal has allowed significant breakthroughs including the restrictions on neutron star equations of state \cite{GW170817EOS}.

\begin{figure}[h!]
\centering
\includegraphics[width=\linewidth]{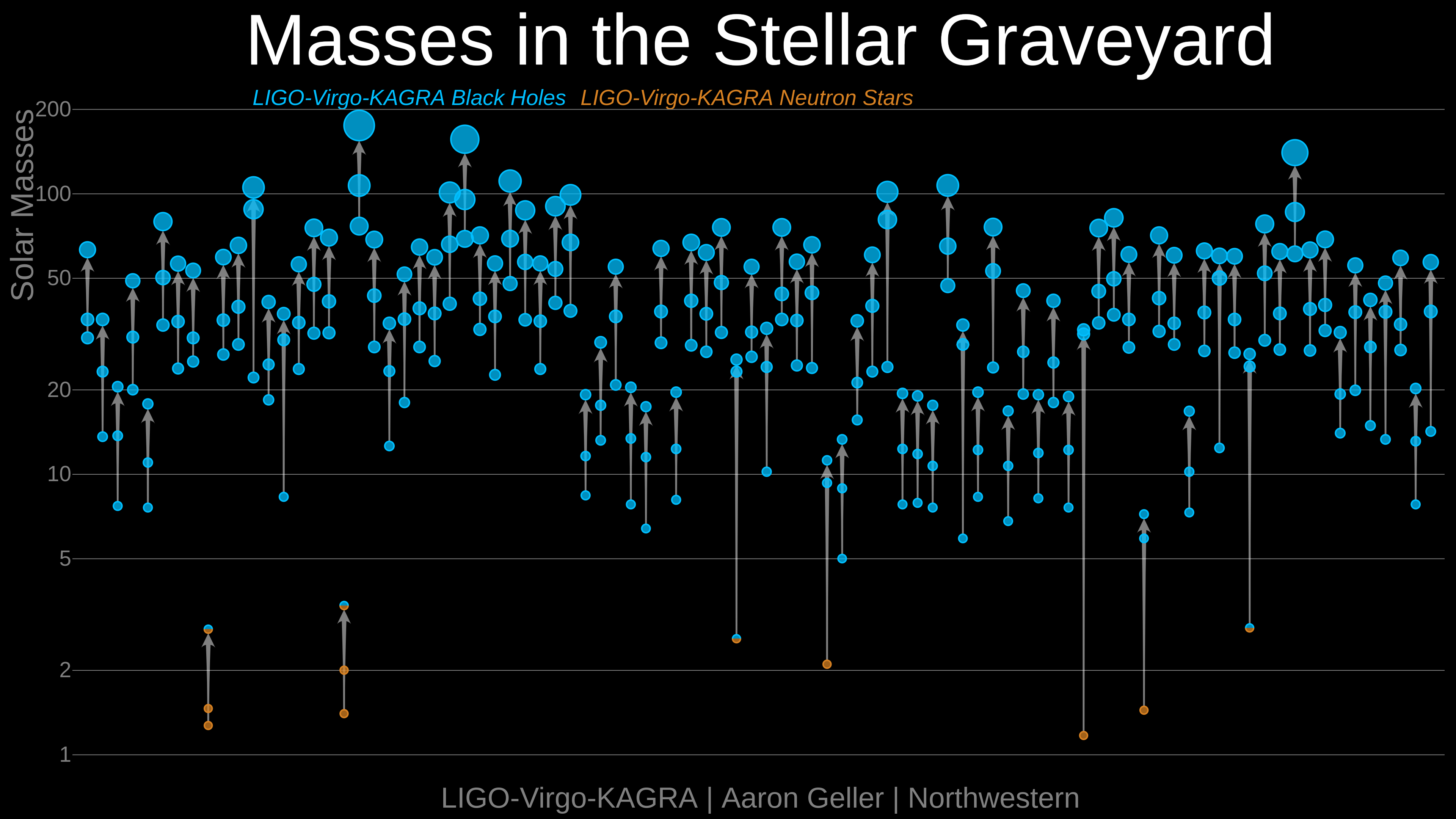}
\caption[The masses of all compact binaries detected by LIGO/Virgo/KAGRA]{The masses of all compact binaries detected by LIGO/Virgo/KAGRA up to the end of O3b, with black holes in blue and neutron stars in orange. The objects are arranged in order of discovery date. Image courtesy: LIGO-Virgo/Aaron Geller/Northwestern.}
\label{GWCatalog}
\end{figure}

With BBH and BNS mergers both observed it was only a matter of time before we observed a neutron star-black hole (NSBH) collision. The first confirmed NSBH signal was detected during the first part of observing run 3 (O3a) on 5th January 2020 with a second detection 10 days later \cite{NSBH}. Detections of the three types of binary merger are becoming much more frequent with the addition of more detectors, including KAGRA (Japan) in 2020, and continued upgrades to the existing observatories. By the end of observing run 3 there have been a total of 90 observed gravitational wave events as shown in Figure \ref{GWCatalog}.

\begin{figure}[h!]
\centering
\includegraphics[width=\linewidth]{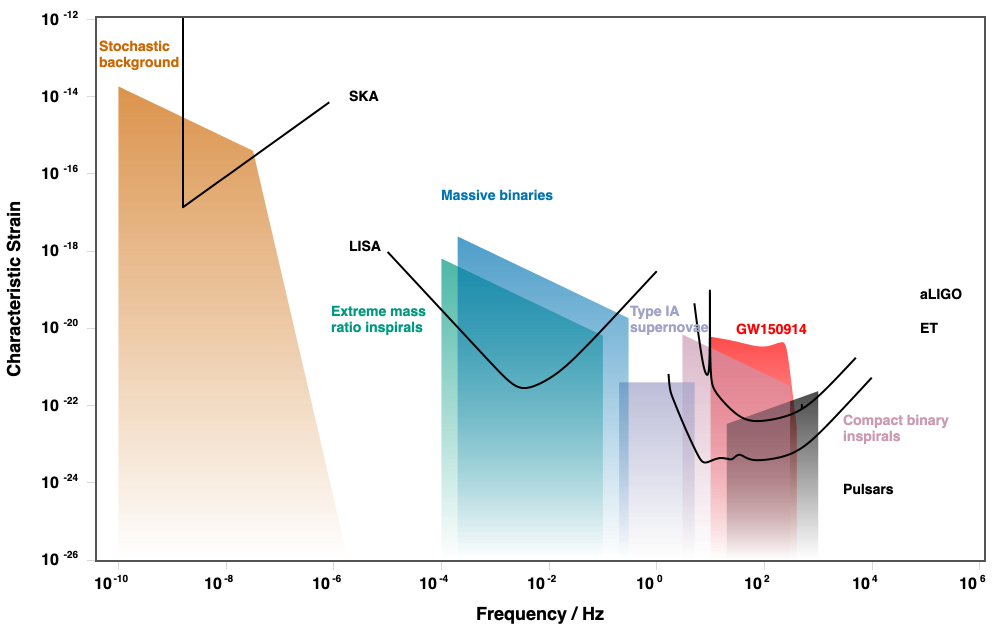}
\caption[Noise curves of current and future gravitational wave detectors]{Noise curves of current and future gravitational wave detectors including the characteristic strain of various GW sources. Image courtesy: \cite{gwplotter}.}
\label{GWNoiseCurves}
\end{figure}

Looking to the future, there are many exciting plans for gravitational wave astronomy. As well as the continued upgrade of the current detectors, there are a series of potential 3rd-generation ground-based detectors proposed including LIGO Voyager (at the current LIGO sites) as well as the Einstein Telescope (ET; Europe) and Cosmic Explorer (CE; USA). These improved designs will not only let us see some of the earliest mergers in the universe but also the potential to detect continuous wave sources including accreting neutron stars \cite{Glampedakis2018}. The future space-based Laser Interferometer Space Antenna (LISA) detector, due for launch in 2034, will allow us to detect GW signals within the millihertz range such as supermassive black hole (SMBH) binaries. One exciting prospect is multiband GW astronomy where loud ground-based sources are detected earlier in their inspiral by LISA. This allows a prediction of the time of merger for the ground-based observatories and sky localisation for the detection of EM counterparts \cite{Sesana2016}. The stochastic background of GW is expected to consist of a history of SMBH mergers with frequencies ranging from microhertz to nanohertz. We can detect in this range using millisecond pulsar timing arrays such as the future Square Kilometre Array (SKA). The propagation of a GW alters the distance between Earth and the pulsar causing the observed time of the pulsar signal to be offset by tens of nanoseconds. Figure \ref{GWNoiseCurves} shows the noise curves of several GW detector as well as the characteristic strains of sources that they intend to observe. The future of gravitational wave astronomy will continue to push our understanding of the universe around us. 

%% file: IntroEMRI.tex

\section{Extreme-mass ratio inspirals}

\begin{figure}[h!]
\centering
\includegraphics[width=50mm]{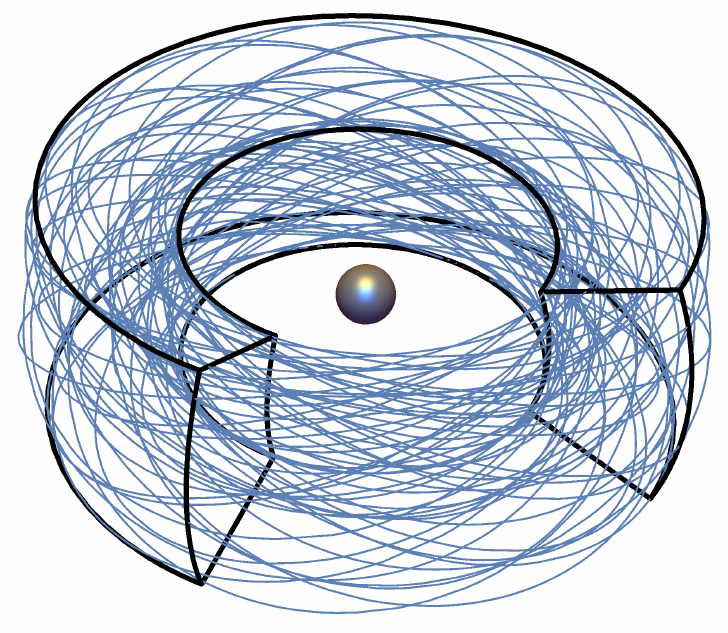}\qquad\qquad 
\includegraphics[width=55mm]{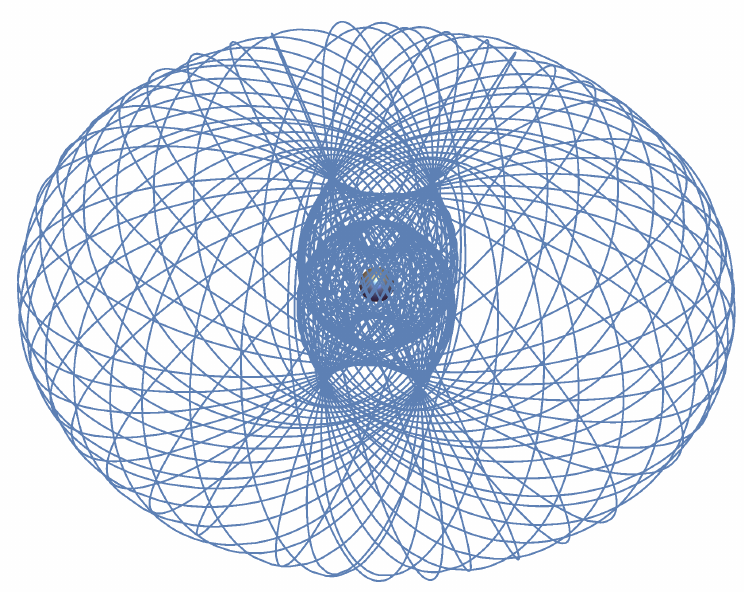}
\caption[Examples of typical and resonant geodesics in Kerr spacetime]{{\it Left}: A typical geodesic in Kerr spacetime. The orbit fills the space within the torus-shaped region. {\it Right}: A resonant orbit that has a 3:2 ratio between the radial and longitudinal periods of the orbit. Images courtesy: \cite{BarackPound2018}.}
\label{KerrGeoPlot}
\end{figure}

One interesting source of gravitational waves are {\it extreme-mass ratio inspirals} (EMRIs). These binary systems consist of a stellar mass black hole (or neutron star) of mass $\mu$ slowly inspiraling into a supermassive black hole with mass $M$ of order $10^5$ -- $10^7\; M_\odot$. On short timescales, the small compact object (CO) follows the standard geodesic motion of the large central black hole with constants of motion energy $E$, azimuthal angular momentum $L$, and Carter constant $Q$. This motion is {\it ergodic} such that the trajectory of the CO will trace out an entire region of space as shown in Figure \ref{KerrGeoPlot}. There exists special orbits where two of the three orbital frequencies (radial, azimuthal, and longitudinal) have an integer ratio. In these cases the motion is no longer ergodic and are instead confined to a 2-dimensional surface, also shown in Figure \ref{KerrGeoPlot}.

The evolution of the orbit is driven by the ``back-reaction'' from emitted gravitational waves which modifies the trajectory on timescales much larger than the orbital frequencies. This slow ``adiabatic'' effect causes the trajectory of the CO to slowly evolve through background geodesics. The timescale of the change of the constants of motion is given by $E/\dot E \sim M^2/\mu$, which is commonly referred to as the {\it radiation-reaction} time. The transition between geodesics is generally smooth except where the motion passes through an orbital resonance. In this case, the repeated regular motion can cause large changes in the orbital parameters on timescales much shorter than the radiation-reaction time.

Astrophysical EMRIs are expected to emit gravitational waves in millihertz frequencies which puts them at the prime sensitivity of the LISA detector (see Figure \ref{GWNoiseCurves}). A typical LISA EMRI will emit $10^5$ -- $10^6$ detectable gravitational-wave cycles over time periods of the order of years. During this time, the slowly evolving geodesic nature of the trajectory means that CO will map out the spacetime surrounding the central black hole in extreme detail. This will give us one of the most comprehensive tests of GR in the strong-field regime to date. Studies have shown that EMRI signals can be used to determine the central object's mass and spin with high accuracy as well as confirm if it is a Kerr black hole or tightly constrain any alternative theories to GR. A study \cite{BabakGair2017} estimates that we can expect to observe up to 100 EMRIs during LISA's lifetime. However, the amplitude of even the strongest EMRI signals will be drowned out by LISA's detector noise. We will require a template bank of highly accurate theoretical waveforms in order to extract the signal. 

\begin{figure}[h!]
\centering
\includegraphics[width=0.6\linewidth]{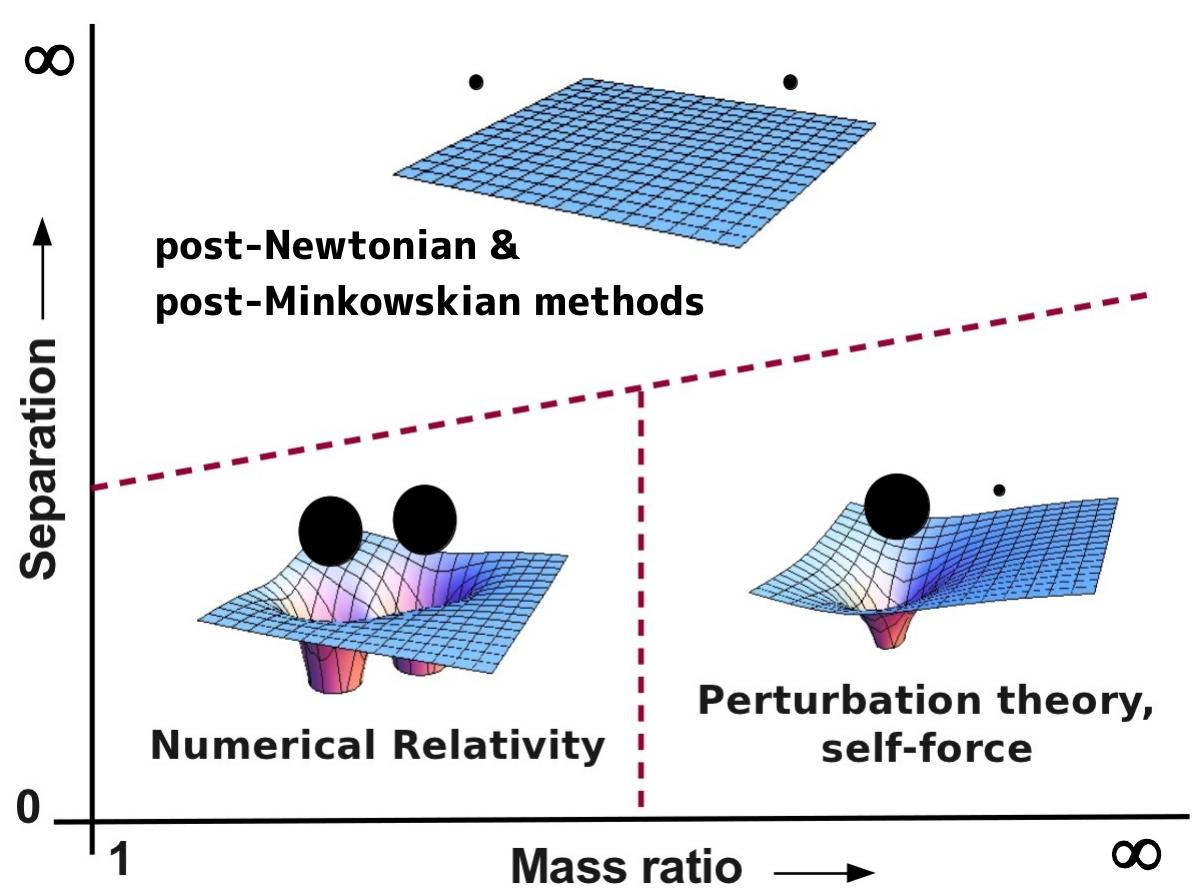}
\caption[Domains of the techniques for solving the two-body problem in GR]{Domains of the techniques for solving two-body problem in GR. Cartoons illustrate the principle behind each approach: PN expands about flat space, SF expands about the exact fixed geometry of a central black hole, and NR tackles the full nonlinear dynamics. Image and caption courtesy: \cite{BarackPound2018}.}
\label{GWModelling}
\end{figure}

In order to construct theoretical waveforms we need to solve the two-body problem of GR, where the spacetime metric is determined as a solution of Einstein’s field equations: a notoriously complicated set of 10 coupled nonlinear second-order partial differential equations (PDEs). The mainstay approach employs the methods of Numerical Relativity (NR), in which the field equations are solved numerically as an initial-value problem using finite-difference or spectral schemes. This is extremely computationally expensive, especially when there are multiple scales involved, as when the two objects are very far apart. In the large distance regime we can utilise weak-field approximations such as the post-Newtonian (PN) and post-Minkowskian (PM) theories which expands quantities order by order in $G/c^2$ and $G$ respectively. The combination of PN in the weak-field transitioning to NR in the strong-field is the currently favoured method for producing waveform template banks for ground-based observatories. However, this method is not suitable to construct EMRI waveforms. The large disparity in the two masses means that the scales required to resolve the areas surrounding each object are vastly different. This, combined with the increased number of orbits in the strong-field regime, vastly increases the computational requirements such that the run time of NR calculations scales as $(M/\mu)^2$. It is possible to still utilise PN (or PM) expansions in the weak field for EMRI calculations, however, we would lose the ability to probe the strong-field regime. Fortunately, there is a natural small parameter in EMRI systems: binary mass-ratio $\eta:=\mu/M$. Self-force (SF) theories use a perturbative approach via order by order expansions in the mass ratio to model EMRIs. A schematic plot of the relative domains of NR, PN, PM, and SF methods is shown in Figure \ref{GWModelling}.

%% file: IntroSF.tex

\section{Self-force theory}

In this section we present a brief overview of the self-force theory including formulation of the problem and a discussion of the various computational methods and current progress. When discussing the computational method we will focus on first-order SF and then detail the approach to second-order calculations in Section \ref{sec:2GSF}. More detailed discussions can be found in the review by Barack \& Pound \cite{BarackPound2018} or for a more technical description we refer to \cite{PoissonPoundVega2011, Pound2015Review, PoundWardell2021} for self-force formulation and \cite{Barack2009, Wardell2015} for computational methods.

\subsection{Expanding in the mass-ratio}

The fundamental principle in self-force theory is the expansion of quantities in terms of the small mass ratio $\eta=\mu/M$. The exact metric of the binary system ${\sf g}_{\mu\nu}$ can be expanded such that
\begin{equation}
{\sf g}_{\mu\nu} = g_{\mu\nu} + h^{(1)}_{\mu\nu} + h^{(2)}_{\mu\nu} + {\cal O} (\eta^3).
\label{eqn:MetricExpansion}
\end{equation}
The zeroth order term $g_{\mu\nu}$ is the metric of the central object, known as the background metric, and $h^{(n)}_{\mu\nu}(\propto \eta^n)$ are metric perturbations created by the CO. The full metric must obey Einstein's field equations (EFE)
\begin{equation}
G_{\mu\nu}[{\sf g}] =8 \pi T_{\mu\nu},
\label{eqn:EinsteinEquation}
\end{equation}
where $G_{\mu\nu}$ is the Einstein tensor and $T_{\mu\nu}$ is the stress-energy tensor. Substituting (\ref{eqn:MetricExpansion}) into EFE gives the left-hand side of Eq.\ (\ref{eqn:EinsteinEquation}) as
\begin{equation}
G_{\mu\nu}[{\sf g}] = G_{\mu\nu}[g] + \delta G_{\mu\nu}[h^{(1)}] + \left(\delta G_{\mu\nu}[h^{(2)}]+\delta^2 G_{\mu\nu}[h^{(1)}]\right)+ {\cal O}(\eta^3).
\end{equation}
Here $\delta G_{\mu\nu}[h^{(n)}]$ is linear in $h^{(n)}_{\mu\nu}$ and $\delta^2 G_{\mu\nu}[h^{(1)}]$ is quadratic with the (schematic) form $\partial h^{(1)}_{\mu\nu}\partial h^{(1)}_{\alpha\beta}+h^{(1)}_{\mu\nu}\partial^2 h^{(1)}_{\alpha\beta}$. Consider the case that the energy-momentum tensor is approximately that of a point-particle such that it takes the form
\begin{equation}
T_{\mu\nu} = T^{(1)}_{\mu\nu} + T^{(2)}_{\mu\nu} + {\cal O}(\eta^3),
\end{equation}
where $T^{(n)}\propto \eta^n$. 
We can form the {\it linearised} Einstein equation by taking only the terms proportional to $\eta^1$ in the expansions of each side of Eq.\ (\ref{eqn:EinsteinEquation}) to give
\begin{equation}
\delta G_{\mu\nu}[h^{(1)}] = 8\pi T^{(1)}_{\mu\nu}.
\label{eqn:LinearisedEinsteinEquation}
\end{equation}

\begin{figure}[h!]
\centering
\includegraphics[width=0.8\linewidth]{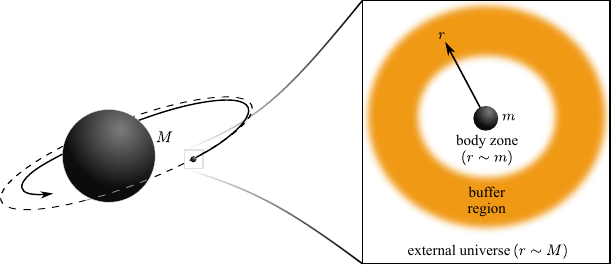}
\caption[Regions involved in matched asymptotic expansions]{Regions involved in matched asymptotic expansions specialised to an EMRI. The body zone and external universe correspond to distances $r\sim \mu (=m)$ and $r\sim M$ respectively. The buffer region corresponds to $\mu\ll r \ll M$ lying between the other two regions. Image courtesy: \cite{BarackPound2018}.}
\label{MatchedAsymptotcExpansion}
\end{figure}

The method of matched asymptotic expansions can be used to determine an effective equation of motion for the compact object. This is done by performing a perturbative analysis of solutions to the EFE in two asymptotic regimes: body zone (dominated by the effects of the compact object) and far zone or external universe (dominated by the effects of the central object). However, there is a so-called ``buffer region'' where both expansions are valid as shown in Figure \ref{MatchedAsymptotcExpansion}. Matching the two perturbative solutions in the buffer zone allows us to form an equation of motion for the compact object in the far zone limit.

The analysis shows that at first-order in $\eta$, the compact object is {\em exactly} described by a point-particle in the background spacetime $g_{\mu\nu}$ \cite{DEath1975}. It also shows that the metric perturbations can be split into two (non-physical) parts such that $h^{(n)}_{\mu\nu}=h^{{\rm S} (n)}_{\mu\nu} + h^{{\rm R} (n)}_{\mu\nu}$. The singular field $h^{{\rm S} (n)}_{\mu\nu}$ contains the divergent nature of the solution near the compact object whereas the regular field $h^{{\rm R} (n)}_{\mu\nu}$ is smooth and homogeneous. The (first-order) equation of motion for the compact object in the far zone is entirely encapsulated in a particular choice of regular field, known as the Detweiler-Whiting regular field, such that we can write the change in the particle's trajectory $x_p^\mu$ with proper time $\tau$ as
\begin{equation}
\mu\frac{D^2 x_p^\mu}{d\tau^2} = \lim_{x\rightarrow x_p}\frac{1}{2} \mu g^{\mu\nu}\left(h^{{\rm R}(1)}_{\rho\sigma;\nu}(x) - 2h^{{\rm R}(1)}_{\nu\rho;\sigma}(x)\right)u^\rho u^\sigma =: F^\mu(x_p),
\label{eqn:SFdefinition}
\end{equation}
where $u^\mu:= dx_p^\mu/d\tau$ is the particle's four-velocity, $D/d\tau := u^\mu \nabla_\mu$ is the covariant derivative along the worldline and a semi-colon represents a covariant derivative. This equation of motion defines the first-order self-force $F^\mu$. We can rewrite this equation of motion in terms of an effective metric defined by $\tilde g_{\mu\nu} := g_{\mu\nu} + \eta h^{{\rm R}(1)}_{\mu\nu}$ such that
\begin{equation}
\frac{\tilde D^2 x_p^\mu}{d \tilde\tau^2} = {\cal O}(\eta^2),
\end{equation}
where $\tilde \tau$ (etc.) are defined in the effective spacetime. This result is the geodesic equation of motion in the {\it effective} spacetime hence we can treat the first-order source as a point-particle test body moving in the effective spacetime.

\subsection{Conservative and dissipative self-force}

We can split the self-force into the (time-symmetric) conservative $F_\mu^{\rm cons}$ and (time-antisymmetric) dissipative $F_\mu^{\rm diss}$ pieces. The dissipative piece drives the evolution through the geodesics and removes energy and angular momentum from the binary system. We can view this as the ``work done'' on the particle is balanced by the emitted radiation. However, the lack of a notion of local energy within GR prevents us from comparing these in a momentary sense. We can instead form ``time-averaged'' balance laws in terms of the asymptotic fluxes of energy $\cal \dot E$ and angular momentum $\cal \dot L$ which take the form
\begin{equation}
\langle F_t^{\rm diss}/u^t \rangle = \langle {\cal \dot E_\infty} \rangle + \langle {\cal \dot E_H} \rangle, \qquad \qquad -\langle F_\varphi^{\rm diss}/u^t \rangle = \langle {\cal \dot L_\infty} \rangle + \langle {\cal \dot L_H} \rangle,
\label{eqn:BalanceLaws}
\end{equation}
where subscripts $\infty$ and $\cal H$ represent the fluxes at infinity and the horizon respectively, $\langle \cdot \rangle$ represents time averaging and an overdot denotes a derivative with respect to Boyer-Lindquist time $d/dt$. It is important to consider how long we need to average over for the balance laws to hold. For generic orbits in Kerr we have to average over infinite time but for intrinsically periodic orbits (i.e.\ any orbit in Schwarzschild or circular, equatorial or resonant orbits in Kerr) we can average over a single orbit \cite{QuinnWald1999, SagoTanaka2006, GanzHikida2007, FlanaganHughes2014}. An interesting observation comes from observing the signs of the various terms of the balance laws. The left-hand sides of Eqs.\ (\ref{eqn:BalanceLaws}) and fluxes at infinity are always positive but in certain cases in Kerr $\langle {\cal \dot E_H} \rangle$ and $\langle {\cal \dot L_H} \rangle$ terms can be negative. This corresponds to {\it superradiance} where some of the central black hole's rotational energy and angular momentum are transferred to the orbit \cite{TeukolskyPress1974, Hughes2000Motion, GlampedakisKennefick2002}. So far we have ignored the third constant of motion in Kerr spacetimes, the Carter constant $Q$. There exists a simple formula for $\langle \dot Q \rangle$ which requires information not only from the asymptotic radiation but also quantities that are locally defined as integrals along the orbit \cite{Mino:2003yg, Mino2005, SagoTanaka2006, GanzHikida2007}. Formulae for the three time-averaged rate of change of the three constants of geodesic motion allows us to calculate full generic EMRI inspirals in Kerr, to leading order, without having to directly evaluate the SF.

The conservative piece of the self-force does not drive the inspiral but it does have measurable effects, such as altering the rate of periastron advance \cite{OsburnWarburton2016}. These conservative effects can be cumulative and effect the long-term evolution of the system. Hinderer and Flanagan \cite{HindererFlanagan2008} found that the generalised angle variables $q^\mu = \{q^t, q^r, q^\theta, q^\varphi\}$ (associated with the Boyer-Lindquist $t$, $r$, $\theta$ and $\varphi$ motions) can be expanded in the mass-ratio such that
\begin{equation}
q^\mu = \frac{1}{\eta} \left( q^\mu_{(0)} + \eta q^\mu_{(1)} + {\cal O}(\eta^2) \right).
\label{eqn:qExpansion}
\end{equation}
For LISA waveforms we will require the phase evolution accurate to the precision of radians across the entire inspiral hence we require the sub-leading term in the expansion. Ref.\ \cite{HindererFlanagan2008} also showed that while the leading ``adiabatic'' order term (0PA) can be determined purely from the time-averaged {\it dissipative} piece of the first-order SF, we need the {\it full} first-order force plus the time-averaged {\it dissipative} piece of the second-order SF to include the ``post-adiabatic'' (1PA) sub-leading term. This has been the main motivation for the formulation of second-order self-force.

\subsection{Practical computation schemes: mode-sum and puncture methods}
\label{sec:SFComp}

Practical calculations of the self-force rely on taking the singular field away from the full physical field during formation of the problem, as done in {\it puncture} methods, or post-calculation of the metric as done in {\it mode-sum regularisation}. The latter, introduced by Barack and Ori \cite{BarackOri2000}, is done at the level of the self-force by introducing the full and singular parts of the first-order self-force defined by
\begin{equation}
F_{\rm full}^\mu(x_p) := \lim_{x\to x_p} \mu \nabla^{\mu\nu\rho} h^{(1)}_{\nu\rho}(x), \qquad
F_{\rm S}^\mu(x_p) := \lim_{x\to x_p} \mu \nabla^{\mu\nu\rho} h^{{\rm S} (1)}_{\nu\rho}(x),
\end{equation}
respectively, where $\nabla^{\mu\nu\rho}$ represents the operator appearing in Eq.\ (\ref{eqn:SFdefinition}). Mode-sum regularisation involves decomposing each of the forces into spherical harmonic $Y_{\ell m}$ modes using Boyer-Lindquist coordinates $(t,r,\theta,\varphi)$ such that $F^{\mu} =\sum_{\ell=0}^\infty F^{\mu\ell}$ where $F^{\mu\ell} := \sum_{m=-\ell}^\ell F_{\ell m}^\mu (t,r) Y_{\ell m} (\theta, \varphi)$, and similarly for $F_{\rm full}^\mu$ and $F_{\rm S}^\mu$. We can now write the self-force as
\begin{equation}
F^\mu(x_p) = \lim_{x\to x_p} \sum_{\ell=0}^\infty \left[ F_{\rm full}^{\mu\ell}(x) - F_{\rm S}^{\mu\ell}(x) \right].
\end{equation}
Ref.\ \cite{BarackOri2000} showed that the divergent parts of the fields can be removed using a simple $\ell$-mode expansion such that
\begin{equation}
F^\mu(x_p) = \lim_{x\to x_p} \sum_{\ell=0}^\infty \left[ F_{\rm full}^{\mu\ell}(x) - F^\mu_{[-1]} {\cal L}_{-1} - F^\mu_{[0]} \right].
\label{eqn:ModeSumReg}
\end{equation}
where $F^\mu_{[n]}$ are ``regularisation parameters'' and ${\cal L}_{-1}:=2\ell+1$. In practical calculations we can only compute a finite number of $\ell$-modes. For a summation up to finite $\ell=\ell_{\rm max}$, Eq. (\ref{eqn:ModeSumReg}) converges linearly with $1/\ell_{\rm max}$. We can increase the convergence through the use of higher-order regularisation parameters such that
\begin{equation}
F^\mu(x_p) \sim \lim_{x\to x_p} \sum_{\ell=0}^{\ell_{\rm max}} \left[ F_{\rm full}^{\mu\ell}(x) - F^\mu_{[-1]}{\cal L}_{-1} - F^\mu_{[0]} - F^\mu_{[2]} {\cal L}_2 - F^\mu_{[4]}{\cal L}_4 - {\cal O}\left(\ell^{-6}\right) \right],
\label{eqn:ModeSumRegFinite}
\end{equation}
where ${\cal L}_n$ are certain polynomials in $\ell$ which decay as $\ell^{-n}$ in the large-$\ell$ limit [explicit definitions are given in Eq.\ (\ref{eqn:ellExpansion})]. The higher-order terms are are absent from Eq.\ (\ref{eqn:ModeSumReg}) as the form of the expansion in $\ell$ is chosen such that $\sum_{\ell=0}^\infty {\cal L}_n = 0 \: \forall \: n>0$. 

The mode-sum regularisation scheme has several benefits. One of the remarkable parts of this scheme is that the regularisation parameters are {\it finite} quantities. This, along with the smoothness of the regular field, ensures that we only ever have to perform finite subtractions during our calculations even though we are subtracting a singular field. Another advantage is that many of the (often numerical) calculations of the metric perturbations are performed mode-by-mode hence the necessary inputs are readily available with no extra input. Mode-sum regularisation has been the primary method for self-force calculations and its various implementations have been used to obtain many breakthrough results. These include generic bound orbits in Kerr \cite{vandeMeent2018} and the work detailed in this thesis. 

The puncture (or effective source) method addresses the subtraction of the singular field by performing the regularisation at the level of the field equation. This involves solving directly for a (local) approximate form of $h^{{\rm R} (1)}_{\mu\nu}$ known as the residual field which is defined such that
\begin{equation}
h^{\cal R}_{\mu\nu} = h_{\mu\nu} - h^{\cal P}_{\mu\nu},
\label{eqn:ResidualFieldDefinition}
\end{equation}
where we have introduced the puncture field $h^{\cal P}_{\mu\nu}$ and omitted the `$(1)$' superscript for brevity. The puncture field is an analytic approximation to the singular field that satisfies
\begin{equation}
\lim_{x\to x_p}(h_{\mu\nu}^{\cal P}-h_{\mu\nu}^{\rm S})=0, \qquad
\lim_{x\to x_p}(\nabla^{\mu\nu\rho} h^{\cal P}_{\nu\rho}-\nabla^{\mu\nu\rho} h^{\rm S}_{\nu\rho})=0,
\end{equation}
near the particle. We can form an equation for the residual field by substituting Eq.\ (\ref{eqn:ResidualFieldDefinition}) into the linearised Einstein equation (\ref{eqn:LinearisedEinsteinEquation}) to give
\begin{equation}
\delta G_{\mu\nu}[h^{\cal R}] = 8\pi T_{\mu\nu} - \delta G_{\mu\nu}[h^{\cal P}] =: T^{\rm eff}_{\mu\nu}.
\end{equation}
This new ``effective source'' $T^{\rm eff}_{\mu\nu}$ lacks the delta function in the original source with its smoothness determined by how well the puncture field approximates the singular field. We can extract the SF directly from the residual field such that
\begin{equation}
F^\mu(x_p) = \lim_{x\to x_p} \mu \nabla^{\mu\nu\rho} h^{\cal R}_{\nu\rho}(x).
\end{equation}
In practical calculations the puncture scheme is usually only applied in a region local to the particle such that we can neglect the behaviour of the puncture and effective sources at large distances from the particle. We can solve for the residual field within a ``worldtube'' which surrounds the particle's worldline and for the full field elsewhere, with changes on the boundary of the worldtube being determined by the analytic puncture field.

\subsection{Choice of gauge}

When performing calculations in GR we have a choice of gauge. Self-force theory, including much of the above, was originally formulated in the Lorenz gauge where we define the gauge condition $\nabla^\mu \bar h_{\mu\nu}=0$ using a new ``trace-reversed'' variable $\bar h_{\mu\nu}:=h_{\mu\nu}-\frac{1}{2}g_{\mu\nu}g^{\alpha\beta}h_{\alpha\beta}$. This new variable allows us to write the linearised Einstein field equation (\ref{eqn:LinearisedEinsteinEquation}) in a simple hyperbolic form 
\begin{equation}
\delta G_{\mu\nu}[h] = -\frac{1}{2} \nabla^\rho \nabla_\rho \bar h_{\mu\nu} - R_\mu{}^\alpha{}_\nu{}^\beta \bar h_{\alpha\beta} = 8\pi T_{\mu\nu},
\label{eqn:LinearisedEinsteinLorenz}
\end{equation}
where $R_\mu{}^\alpha{}_\nu{}^\beta$ is the Riemann tensor of the background metric. One important thing to note is that the SF is not a gauge invariant quantity. The change in the self-force $\delta F^\mu$ relative to the Lorenz-gauge definition due to a continuous gauge transformation of the form $x^\mu \rightarrow x^\mu + \xi^\mu$ is
\begin{equation}
\delta F^\mu = -\mu \left(g^{\mu\nu} + u^\mu u^\nu\right) \left( \xi_{\nu;\alpha\beta} + \xi_\gamma R^\gamma{}_{\alpha\nu\beta} \right) u^\alpha u^\beta,
\end{equation}
where $\xi^\mu$ is continuous and a semi-colon represents a covariant derivative \cite{BarackOri2001}. In fact, it is possible to completely nullify the SF with a specific choice of gauge. However, in this case the information of the physical effects of the SF is encoded in the metric perturbation \cite{Detweiler2008}. It is the combination of the self-force and the metric perturbation which contain the physical gauge-invariant information.

Performing direct calculations in the Lorenz gauge is difficult. There are no known basis to perform decoupled mode decompositions in Kerr hence the full 10 coupled PDEs have to be solved in the time-domain (TD) as 3+1D or 2+1D calculations, which are extremely computationally expensive. In Schwarzschild, the problem can be decomposed using tensorial spherical harmonics but this system still involves solving the set of 10 equations which involve coupling between tensorial components \cite{BarackLousto2005}. Additionally, the Schwarzschild field can be further decomposed into Fourier modes allowing for calculations in the frequency domain (FD) \cite{AkcayWarburtonBarack2013, WardellWarburton2015}. Evolutions in the Lorenz gauge can excite pure-gauge modes which grow linearly with time \cite{DolanBarack2013} in the TD and there are difficulties associated with resolving low-frequency modes in the FD \cite{AkcayWarburtonBarack2013}. These have been overcome in select cases (see \cite{DolanBarack2013, BarackSago2007, BarackSago2010} for TD and \cite{AkcayWarburtonBarack2013} for FD) but finding general solutions continues to be an open problem.

\begin{figure}[h!]
\centering
\includegraphics[width=33mm]{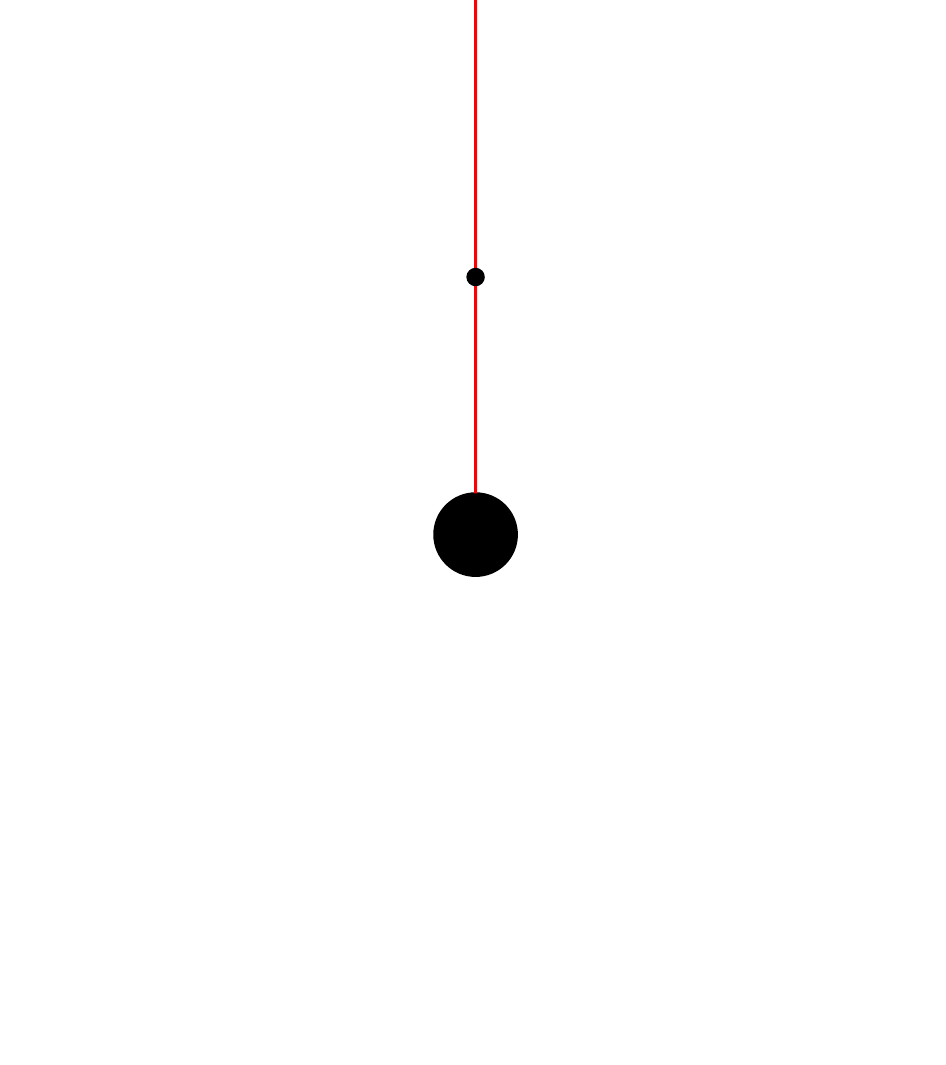}\quad
\includegraphics[width=33mm]{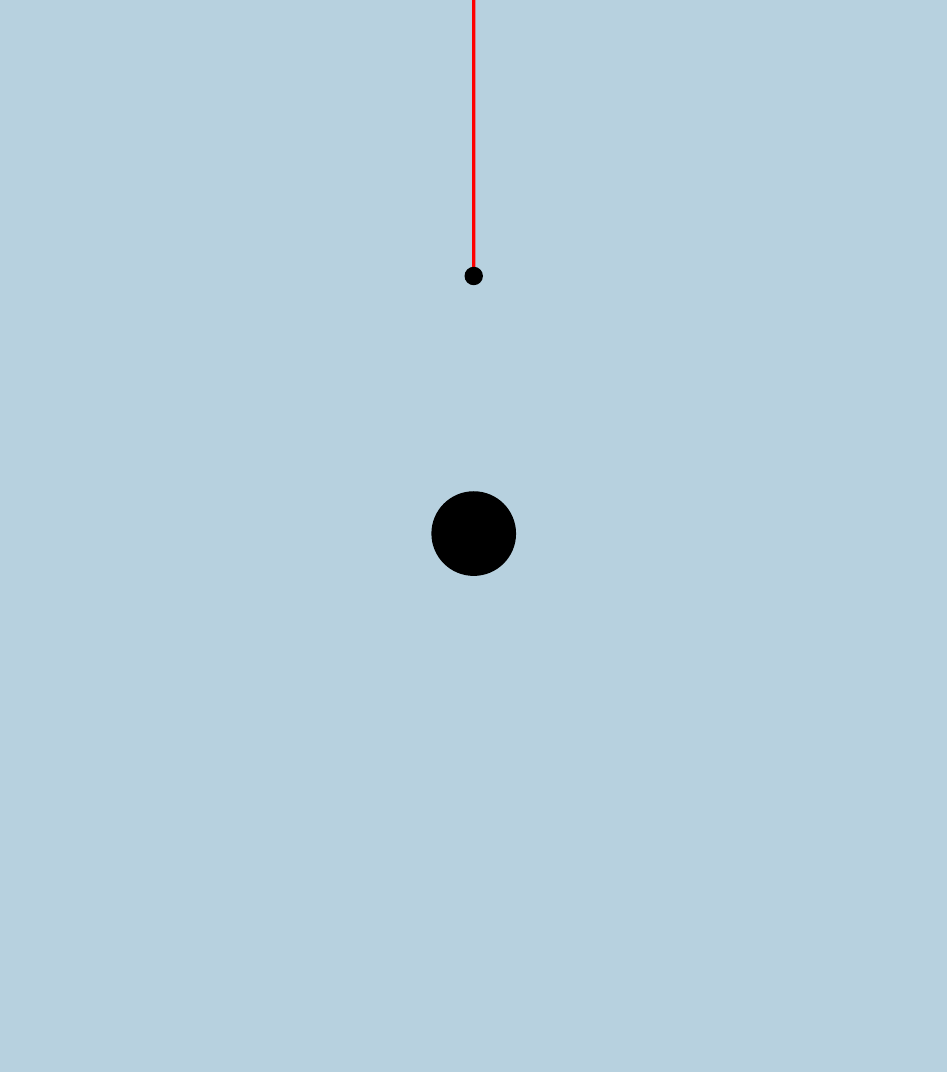}\quad
\includegraphics[width=33mm]{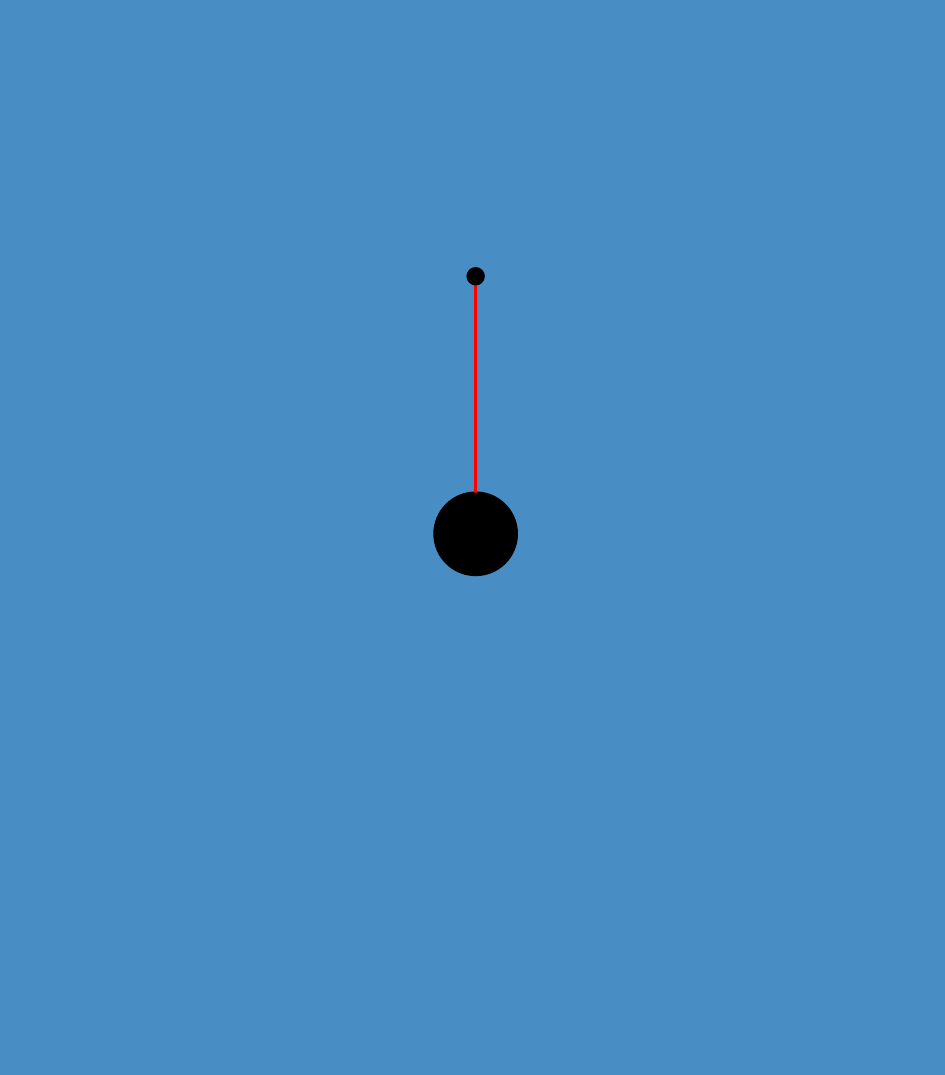}\quad
\includegraphics[width=33mm]{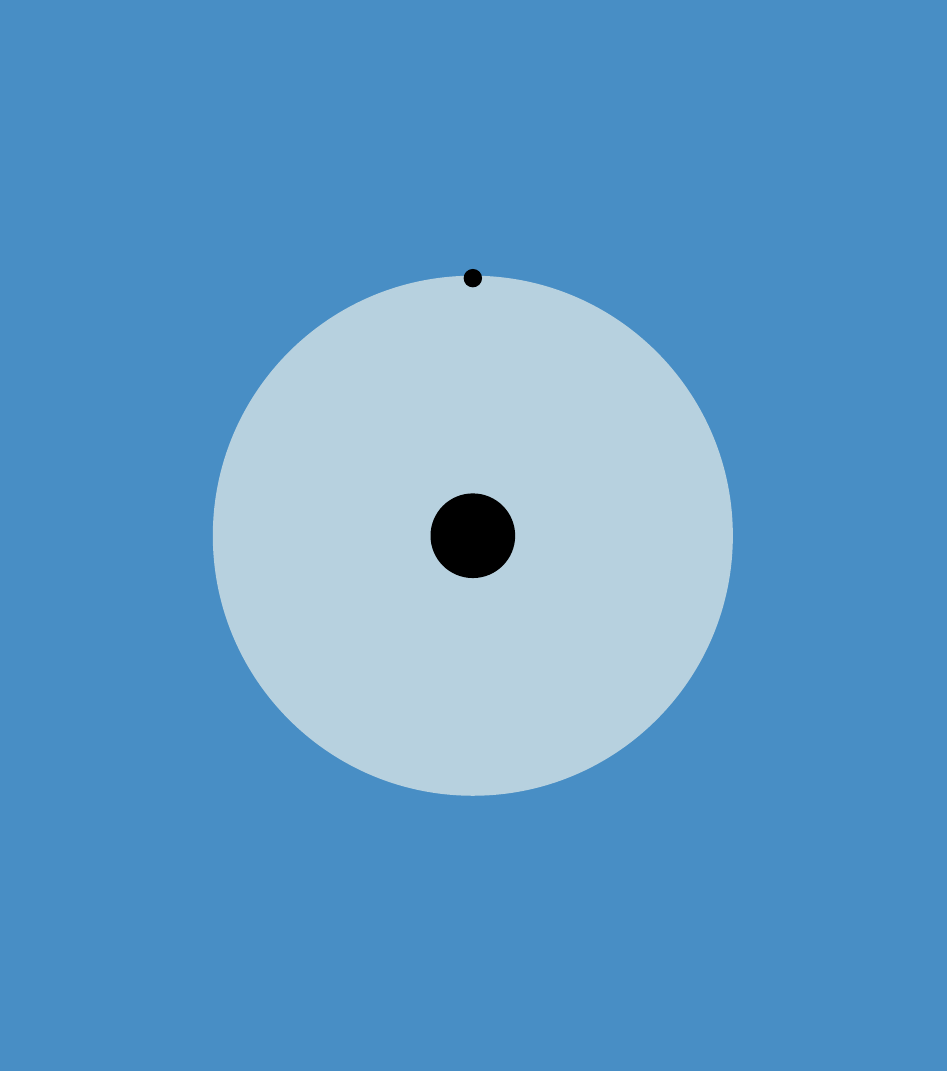}
\caption[Illustrations of the ``string'' solutions of the metric perturbation in a radiation gauge]{Illustrations of the ``string'' solutions of the metric perturbation in a radiation gauge with the string singularity shown in red. We show the full-string solution ({\it left}), two half-string solutions ({\it centre}) and the no-string solution ({\it right}). The no-string solution is constructed by combining the regular regions of each half-string solution.}
\label{StringSolutions}
\end{figure}

The complications associated with calculations in the Lorenz gauge prompted significant efforts to extend the SF formulation to alternative gauges \cite{BarackOri2001, GrallaWald2008, GrallaWald2011, Gralla2011, PoundMerlinBarack2014, Pound2015}. The Regge-Wheeler-Zerilli gauge \cite{ReggeWheeler1957, Zerilli1970} in Schwarzschild and the radiation gauges in Kerr \cite{Chrzanowski1975,CohenKegeles1975} have been traditionally favoured gauges for black hole perturbation theory. In these gauges calculations can be reduced to solving vastly simpler differential equations for certain scalar quantities, from which we can reconstruct the full metric perturbation. However, these alternative gauges are not naturally well-suited for a point-particle source. The radiation gauge solution has a 1-dimensional ``string-like'' singularity which plagues the reconstructed metric perturbation. In the most general case the singularity emanates from the particle in both radial directions. This is commonly referred to as the ``full-string'' solution. However, we can utilise a spare gauge freedom to reduce the impact of the singularity. It is possible to remove the singularity {\it either} interior or exterior to the particle (relative to the central object) \cite{BarackOri2001}. These ``half-string'' solutions both have regular and singular regions. Ref.\ \cite{PoundMerlinBarack2014} showed that it is possible to combine the regular regions of the half-string solutions to form a ``no-string'' solution which is smooth everywhere, except on the boundary between the two solutions. Figure \ref{StringSolutions} shows graphic representations of the three types of string solutions. We present a more technical explanation of these string solutions in Section \ref{sec:PointParticleSource}. The method of calculating the self-force in this no-string radiation gauge has been widely successful including van de Meent's calculations of the first-order SF for generic bound orbits in Kerr \cite{vandeMeent2018}.

\subsection{Time and frequency domain calculations}

As briefly mentioned earlier, when performing calculations we can choose to do so in the time or frequency domains. FD calculations use Fourier transforms to separate quantities by frequencies of the positional elements of the orbit, such as radial and azimuthal frequencies. This simplifies PDEs to ODEs as all time derivatives are replaced with factors of the frequencies forming a discrete spectrum. This method is useful for bound orbits which naturally have a periodic nature with well defined orbital frequencies but there has been interest in formulation for other type of orbits \cite{Hopper:2017qus, Hopper2018}. Additionally, FD calculations become increasingly computationally expensive for highly eccentric orbits where high-frequencies modes have larger contributions \cite{vandeMeent2016}. The reliance on orbital frequencies also makes it difficult to evolve the orbit (relative to a TD calculation) as is required for a full inspiral calculation. TD calculations involve explicitly evolving the equations in time using methods such as finite-differencing. These methods are often computationally more expensive than their FD counterparts but offer more freedom in the types of orbits that can be calculated.

\subsection{Second-order self-force}
\label{sec:2GSF}

Let us now focus on the second-order self-force. We can form the equation of motion up to second-order through matched asymptotic expansions to give
\begin{equation}
\frac{D^2 x_p^\mu}{d\tau^2} = \frac{1}{2} \left( g^{\mu\nu} -h_{\rm R}^{\mu\nu} \right)\left(h^{\rm R}_{\rho\sigma;\nu} - 2h^{\rm R}_{\nu\rho;\sigma}\right)u^\rho u^\sigma + {\cal O}(\eta^3),
\label{eqn:2GSFEoM}
\end{equation}
where $h^{\rm R}_{\mu\nu}:= \eta h^{{\rm R}(1)}_{\mu\nu} + \eta^2 h^{{\rm R}(2)}_{\mu\nu}$ \cite{Pound:2012nt}. As with the first-order case, we can define an effective metric $\tilde g_{\mu\nu}:= g_{\mu\nu} + h^{\rm R}_{\mu\nu}$ and rewrite the equation of motion in terms of the effective metric to give
\begin{equation}
\frac{\tilde D^2 x_p^\mu}{d \tilde\tau^2} = {\cal O}(\eta^3).
\end{equation}
Once again we find that the object moves as a test body in the effective metric.

Consider taking only the second-order terms in the expansion of EFE (\ref{eqn:EinsteinEquation}) which gives 
\begin{equation}
\delta G_{\mu\nu}[h^{(2)}] = 8\pi T^{(2)}_{\mu\nu} - \delta^2 G_{\mu\nu}[h^{(1)}].
\label{eqn:2GSFEFE}
\end{equation}
The nonlinearity of the field equations means that in a general gauge this form is not well defined. The second-order Einstein tensor has the schematic form $\delta^2 G(h)\sim h\partial^2h + \partial h\partial h$. This means that when it acts on the first-order metric perturbation (and its $1/r$ singularity near the worldline), the second-order Einstein tensor diverges as $\delta^2 G[h^{(1)}]\sim 1/r^4$. As a result $\delta^2 G[h^{(1)}]$ is not locally integrable at the worldline and does not have a unique definition of a distribution which intersects the worldline. The consequence of this is that it is impossible to define a unique stress-energy tensor at second-order and Eq.\ (\ref{eqn:2GSFEFE}) is not a well defined equation.

Fortunately, there exists a class of ``highly-regular'' gauges which offers a fine-tuning solution that removes the most singular term of $h^{(2)}_{\mu\nu}$ and reduces the divergence of the source such that $\delta^2 G[h^{(1)}] \sim 1/r^2$ \cite{Pound:2017psq}. This reduced divergence allows us to construct a unique source as a distribution and thus construct a field equation for $h^{(2)}_{\mu\nu}$ valid for all $r\geq 0$ \cite{UptonPound2021}.

The no-string construction that has been highly successful at first-order is no longer valid at second-order. The spatially extended source means that it is not possible to perform calculations in vacuum regions as they do not exist. Green, Hollands and Zimmerman (GHZ) \cite{GreenHollandsZimmerman2020} proposed an alternative formulation to the no-string gauge where the singularities associated with the metric perturbation are removed by a certain ``corrector tensor'' during reconstruction of the metric (from curvature scalars) rather than at the formulation level. Recent work \cite{ToomaniZimmerman2021} has shown how the GHZ formalism can be applied to self-force calculations and provides a promising method for performing second-order calculations.

Current calculations at second-order have been restricted to quasi-circular orbits in Schwarzschild. These have included the gravitational binding energy \cite{PoundWardell2020}, gravitational-wave energy flux \cite{WarburtonPound2021} and the inspiral waveform \cite{WardellPound2021}. 

\subsection{Interfaces with other two-body GR models}

Figure \ref{GWModelling} shows a schematic plot of the relative domains of various techniques for solving the two-body problem in GR. These boundaries are fairly fluid which allows interfaces, collaborations, and comparisons between the models including the combination of PN and NR data being combined to construct the waveforms for LIGO/Virgo/KAGRA data analysis. 

\begin{figure}[h!]
\centering
\includegraphics[width=\linewidth]{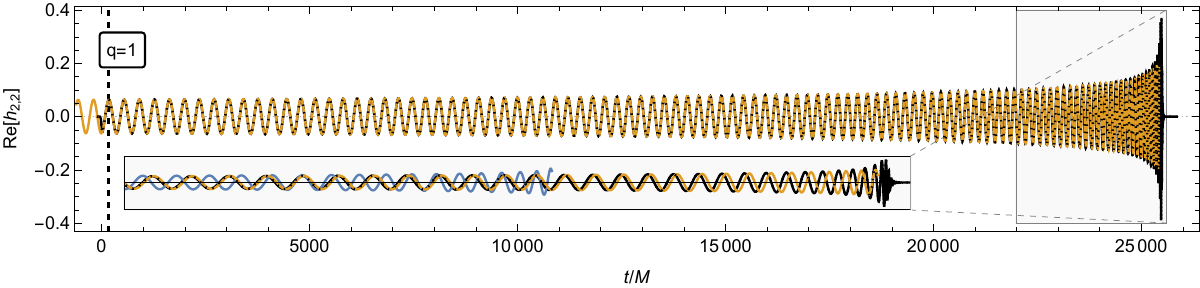}
\caption[Second-order waveform for a mass ratio $1:1$ nonspinning binary]{Second-order waveform for a mass ratio $1:1$ nonspinning binary (orange). Also included for comparison is the waveform for the same binary produced using an NR simulation (black) \cite{SXS:BBH:1132}. The inset shows a zoomed region near the merger and also shows the corresponding first-order waveform (blue). The waveforms are aligned in time and phase at $t = 170M$. Image courtesy: \cite{WardellPound2021}.}
\label{1PAWaveform}
\end{figure}

It is possible to determine where SF has sufficient accuracy for data analysis purposes by comparing with the ``exact'' results of NR. This was first done in Ref.\ \cite{LeTiec:2011bk} who showed that one can get remarkable agreement, even up to mass ratios of $1:1$, if you replace $\eta$ with the {\it symmetric} mass ratio $\nu:=\mu M/(\mu+M)^2 = \eta + {\cal O}(\eta^2)$. The accuracy of results calculated using SF methods with expansions in $\nu$ in the comparable mass regime is further increased at second-order, as shown in Figure \ref{1PAWaveform}, where the post-adiabatic waveform closely matches the NR result for much of the inspiral. This agreement with NR results at comparable mass ratios when expanding in the symmetric mass-ratio is yet to be theoretically explained.

Recent work \cite{DhesiRuter2021} has shown how SF and NR could be used in harmony to tackle the problem of intermediate-mass-ratio inspirals (IMRIs) which have mass ratios $10^{-2} \lesssim \eta \lesssim 10^{-4}$. The method excises a region around the smaller black hole such that the disparity in the length scales is decreased alleviating some of the computational burden. Boundary conditions calculated using SF methods can be applied to the excised region and used within the NR calculations.

Interfaces between different perturbative methods of the two-body problem have also proved extremely useful. Expansions valid in different regimes can be used to check the convergence of other methods such as using an analysis of how the PN terms depend on $\eta$ to test the convergence properties of the SF expansion. Similarly, it is also possible to predict higher-order terms using the other expansion. SF information has been used in such a way to determine the 5PN equation of motion through 4PM order \cite{BiniDamour2019} as well as the resolution of ambiguities in, and inconsistencies amongst the first derivations of the 4PN equation of motion \cite{vandeMeent:2016hel}.

%% file: IntroScatter.tex

\section{Black-hole scattering}

The modelling of binary-black hole scattering has been largely neglected (relative to bound orbits) due to the fact that the brief encounters are unlikely strong enough to be directly detected by LISA. However, there is still strong motivation to study the unbound systems.

\begin{figure}[h!]
\centering
\includegraphics[width=0.8\linewidth]{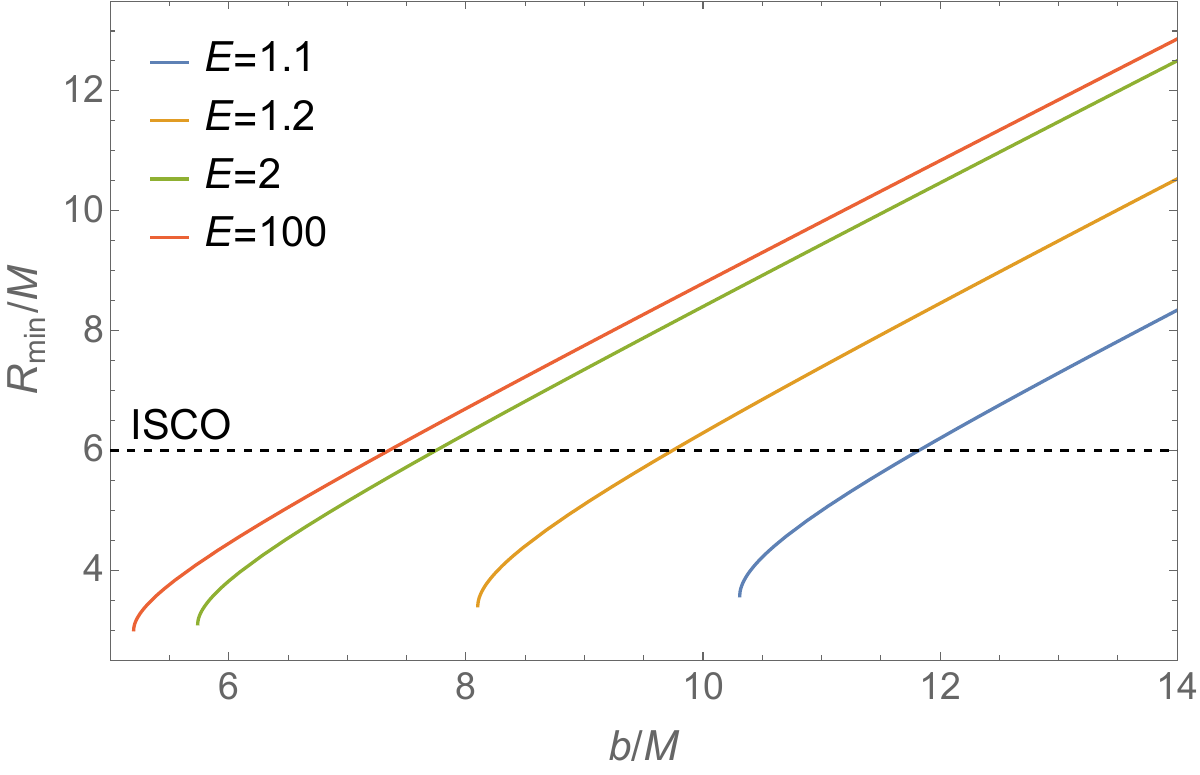} 
\caption[The periastron distance as a function of impact parameter]{The periastron distance $R_{\rm min}$ of a geodesic around a Schwarzschild black hole as a function of impact parameter $b$ for 4 different values of orbital energy $E$. Also shown is the Innermost Stable Circular Orbit (ISCO). The minimal value of impact parameter $b$ (obtained for $E\to\infty$) is $3\sqrt{3}M \simeq 5.196 M$. The minimal value of $R_{\rm min}$ (also obtained for $E\to\infty$) is $3M$. Note how efficient scatter orbits are in probing the sub-ISCO part of spacetime, even at relatively low energy. Explicit definitions of $R_{\rm min}$, $b$ and $E$ are given in Chapter \ref{chapter:geodesics}.}  
\label{periastron}
\end{figure}

Let us consider timelike geodesic motion around a Kerr black hole. Scatter orbits can have much higher energies and angular momenta than bound orbits. This means that scatter geodesics can have smaller points of closest approach without transitioning to a plunge orbit and thus can probe much deeper into the gravitational potential.
Figure \ref{periastron} highlights this fact for a geodesics around a Schwarzschild black hole. These strong-field self-force calculations of scatter orbits can provide a benchmark for (semi-)analytic models such {\it effective one-body} (EOB).

\begin{figure}[h!]
\centering
\includegraphics[width=\linewidth]{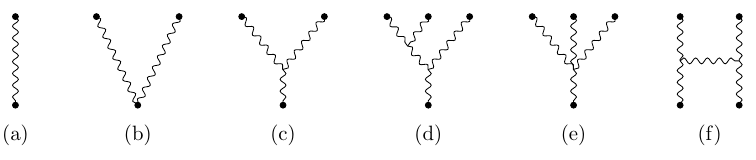} 
\caption[Sample of Feynman diagrams needed for the computation of the scattering angle]{Sample of Feynman diagrams needed for the computation of the scattering angle in the {\it effective field theory} (EFT) approach to 3PM order. Only the diagrams to one-loop order in (a)-(c) are needed to calculate 2PM order. Image courtesy: \cite{Kalin:2019rwq}.} 
\label{Feynman}
\end{figure}

There has been a new approach for solving the binary-black hole problem using methods from high-energy physics. This ``amplitudes'' approach is based on the study of (classical or quantum) scattering amplitudes in unbound gravitationally interacting two-body systems such as those involving the Feynman diagrams in Figure \ref{Feynman}. The idea is to map information about scattering observables (like the scattering angle) into some form of a two-body gravitational potential \cite{ChoKalin2021}. This provides a strikingly fast route to the formulation of an equation of motion for inspiralling binaries in GR. These efforts culminated with the recent breakthrough work of deriving the conservative part of the two-body equation of motion at 4PM order \cite{Bern4PM,BernParraMartinez2021}. Additionally, there have been results which reverse the process where the scattering of waves within a GR context have been used to calculate quantities from {\it quantum field theory} (QFT), such as scattering cross sections \cite{BautistaGuevara2021}. This extraordinary connection between high-energy and gravitational physics provides unique opportunities for progress in both fields and provides strong motivation to study the self-force scatter problem. Direct calculations of the self-force corrections to the scatter angle can provide comparison data for the amplitude calculations as well as the potential to extract QFT results from the gravitational calculation.

Previous results which have utilised SF information to inform other methods (such as those presented at the end of the previous section) were all obtained as a byproduct of EMRI-motivated calculations. It has been advocated by several authors that direct SF calculations of scattering observables would provide a most powerful handle on the PM dynamics and lead to transformative improvement in EOB models. Damour recently showed that a first-order (second-order) SF calculation of the scattering angle determines the {\it complete} two-body Hamiltonian through 4PM (6PM) order to all orders in the mass ratio \cite{Damour2020}. This opportunity to extract information on bound motion across {\it all} mass-ratios provides the strongest motivation yet to tackle the binary-black hole scattering problem with self-force methods.

The standard FD domain methods for bound EMRI calculations are a priori unsuitable for scatter calculations due to their reliance on a discrete spectrum of frequencies. While there has been some work to formulate the scatter problem in the FD \cite{Hopper:2017qus, Hopper2018}, the majority of progress has been in the TD. The first unbound problem considered was the case of a {\it marginally} bound orbit where the particle follows a parabolic-like trajectory starting at rest at infinity \cite{Baracketal2019}. This TD calculation gave insight on how to formulate gauge-invariant quantities for unbound orbits and showed that the Lorenz-gauge implementation is too computationally inefficient for precision scatter orbit calculations. My numerical calculations, also presented in this thesis, tackled hyperbolic orbits in Schwarzschild taking the alternative route of using a TD implementation in the radiation gauge \cite{LongBarack2021}. More recently, Gralla \& Lobo presented an analytical calculation of SF corrections to the scatter angle at leading PM order \cite{GrallaLobo2021}.

%% file: IntroThesis.tex

\section{This thesis}

In this thesis we focus on the largely unexplored case of {\em unbound} self-force calculations. We develop an alternative TD method designed to tackle the scatter case and present implementations for both scalar and gravitational self-force calculations.

We begin in Chapter \ref{chapter:ScatGeo} with a review of scattering geodesics in Schwarzschild, including different possible parameterisations and a derivation of the geodesic scatter angle. Chapter \ref{chapter:SFScatAngle} analyses the self-forced equations of motion and derives the self-force corrections to the geodesic orbital parameters. The main result of this chapter are two independently derived formulae for the conservative self-force correction to the scatter angle. We expand these equations within the PM regime showing that the relations are equivalent to, and agree with, the recent PM results of Gralla \& Lobo \cite{GrallaLobo2021}. 

Chapter \ref{chapter:ScalarField} starts with the formulation of our 1+1D numerical method with calculations of a scalar field on a Schwarzschild background for the vacuum case. The code employs a finite--difference scheme on a characteristic grid based on Eddington--Finkelstein coordinates---a simple tried-and-tested architecture. We extend our implementation with the addition of a point-particle source for both circular and scatter orbits. In Chapter \ref{chapter:ScalarSF} we present scalar self-force results for a scatter orbit in the strong-field regime and extract new physics including post-periastron undulation, which we attribute to quasinormal-mode excitations. We use these SF results to perform first-of-their-kind (numerical) calculations of the correction to the scatter angle.

We continue in Chapter \ref{chapter:Vacuum} with a review of vacuum metric reconstruction and a formulation of an initial-value problem for the Hertz potential. We then present an extension of our 1+1D code for the numerical integration of the Bardeen-Press-Teukolsky (BPT) equation. We demonstrate, however, that a naïve implementation of this standard scheme fails when applied to the Teukolsky equation with spin parameter $s=\pm 2$, due to divergences that develop at late time (an exponential divergence for $s=+2$ and a $\sim t^4$ divergence for $s=-2$). We attribute these divergences to certain growing modes of the Teukolsky equation. These modes violate the physical boundary conditions, but since boundary conditions are not actively imposed in our characteristic scheme, they are allowed to grow. We explain why the issue is not encountered in existing time-domain Teukolsky codes based on hyperboloidal slicing with compactification \cite{Racz:2011qu,Zenginoglu:2012us,Harms2013}. Here, restricting to the Schwarzschild case, we opt for a simpler solution. We circumvent the problem of growing modes by transforming to a new field variable (using a time-domain version of the Chandrasekhar transformation), which, in the vacuum case, satisfies the Regge--Wheeler (RW) equation, for which the problem does not occur.

Chapter \ref{chapter:ParticleForm} begins by detailing how introducing a source causes the standard metric reconstruction procedure to fail. We present our attempts to circumvent this problem and summarise the non-vacuum metric reconstruction procedure proposed by Green {\em et al}.\ \cite{GreenHollandsZimmerman2020}. Continuing, we review metric reconstruction in a no-string gauge, specialising to a Schwarzschild background and casting the procedure in a form suitable for a time-domain implementation. In the no-string construction, the spacetime outside the central black hole is split into two vacuum domains, $r>R(t)$ and $r<R(t)$, where $r$ and $t$ are Schwarzschild coordinates and $r=R(t)$ along the particle's trajectory.  The crucial ingredient in our formulation are jump conditions that the Hertz potential, RW variable and their derivatives must satisfy on the (time-dependent) two-sphere $r=R(t)$. These conditions are derived for an arbitrary timelike geodesic trajectory. 

In Chapter \ref{chapter:Implementation} we finally present a full numerical implementation of our method, first for a circular orbit. We evolve the field equation for the RW variable, and from it compute (multipole mode by multipole mode) the no-string IRG Hertz potential. Our results show agreement with analytic solutions for static modes and those of Barack \& Giudice \cite{Barack:2017oir}. We extend our implementation to the scatter case and present the first calculations of the no-string Hertz potential for an unbound orbit, both at $\mathscr{I}^+$ and along the scatter trajectory. We thus numerically construct the necessary input for a calculation of the self-force along the orbit.

We conclude in Chapter \ref{chapter:conclusions} by summarising the main results presented in this thesis and reviewing the extra steps needed to carry our the calculation of the self-force from the Hertz potential. We also discuss potential interfaces with other two-body GR models and the prospects of extending our method to the case of a Kerr background.   

The conventions used in this thesis are a $(-+++)$ metric signature and geometrized units with $G=c=1$. Commas and semi-colons represent partial and covariant derivative respectively. Complex conjugation is denoted by an overbar. For quantities that arise in the Newman-Penrose formalism we follow the sign conventions of Ref.\ \cite{Merlin2016}, as summarised in Appendix A therein; for ease of reference we review the relevant details here, in Appendix \ref{App:convention}, specialised to the Schwarzschild case. 

%% file: Geodesics.tex

\label{chapter:geodesics}

In this chapter we will review scattering timelike geodesics of a Schwarzschild black hole. We present several parameterisations as well as calculate important quantities such as the periastron and the scatter angle. 

\section{Equation of motion and parametrisation}

With motion in the equatorial plane we have the first integrals
\begin{eqnarray}
\dot{t}&=& E/f, \label{tdot}\\
\dot{\varphi} &=& L/r^2, \label{phidot}\\
\dot{r} &=& \pm \sqrt{E^2-V(r;L)},  \label{rdot}
\end{eqnarray}
where an overdot is $d/d\tau$, $f:=1-2M/r$, $E$ and $L$ are specific energy and angular momentum, and
\begin{equation}\label{V}
V(r;L)=f(r)\left(1+L^2/r^2\right).
\end{equation}

For hyperbolic encounters we require $E>1$. For such orbits, $E$ is just the usual $\gamma$ factor of special relativity where the energy is $\gamma m c^2$, so the {\it specific} energy (taking $c=1$) is $\gamma$. This can also be confirmed with a direct calculation. Letting ${\bf v}_\infty$ be the 3-velocity at infinity, we have
\begin{equation}
v^r_{\infty} = \frac{\dot{r}_{\infty}}{\dot{t}_\infty}=\pm \frac{\sqrt{E^2-1}}{E},
\quad\quad
(r v^\varphi)_\infty = \frac{(r\dot{\varphi})_{\infty}}{\dot{t}_\infty}=\frac{rL/r^2}{E/f}\Big\vert_{r\to\infty} = 0,
\end{equation}
and so
\begin{equation}
\vinf:=|{\bf v}_\infty|= \sqrt{(v^r_{\infty})^2+(r v^\varphi)^2_\infty}
=\frac{\sqrt{E^2-1}}{E},
\end{equation}
giving
\begin{equation}
\gamma = (1-v_\infty^2)^{-1/2}= E. 
\end{equation}
We will use $\vinf$ as one of the two (pseudo-)invariant orbital parameters.

The particle actually scatters back to infinity (and does not fall into the black hole) only if $L>L_{\rm crit}(E)$, where the critical value of the angular momentum is the (relevant) simultaneous solution of $\partial_r V(r;L)=0$ and $E^2=V(r;L)$. This gives
\begin{equation}
L_{\rm crit}(E) = \frac{M}{\uinf}\sqrt{(27E^4+9\alpha E^3-36 E^2-8\alpha E+8)/2},
\end{equation}
where $\alpha:=\sqrt{9E^2-8}$ and 
$$\uinf:= |\dot{r}_{\infty}|=\sqrt{E^2-1}=\vinf E.$$

\begin{figure}[h!]
\centering
\includegraphics[width=0.85\linewidth]{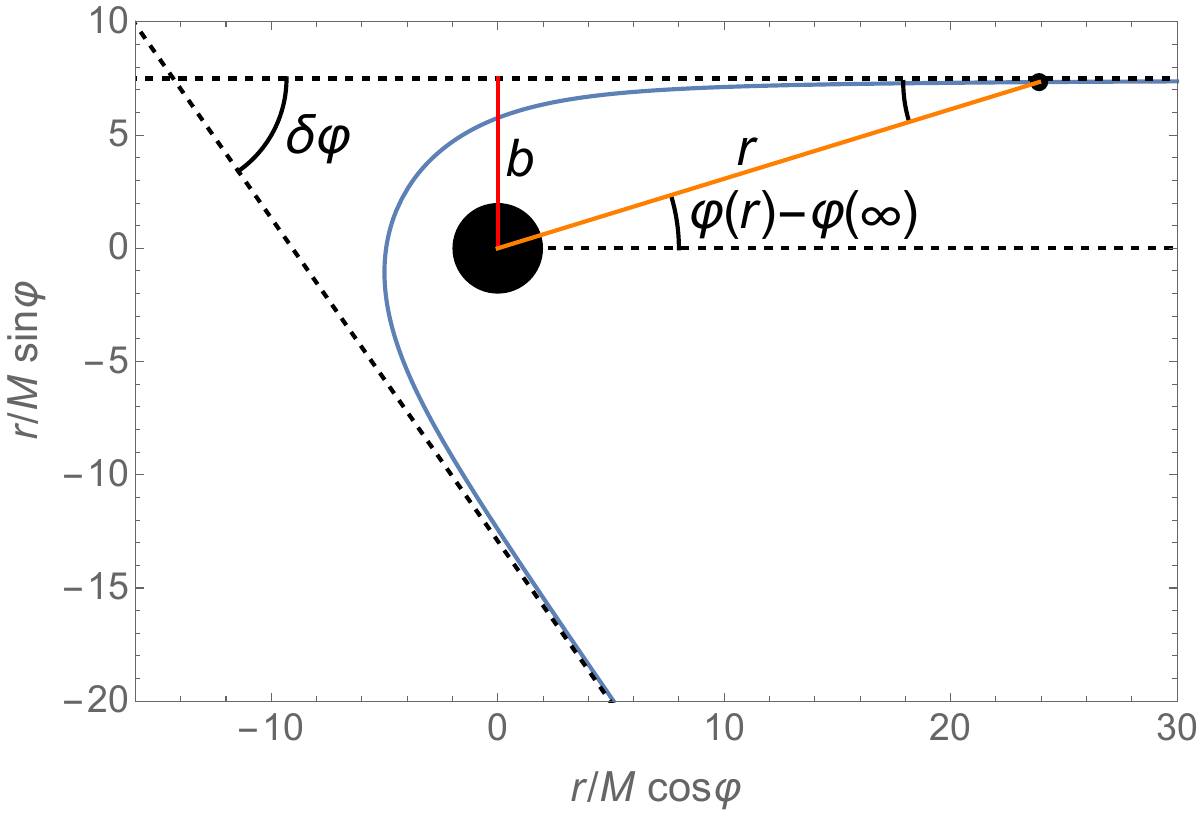} 
\caption[A visualisation of the definitions of the impact parameter and the scatter angle]{A visualisation of the definitions of the impact parameter $b$ and the scatter angle $\delta\varphi$ [to be defined in Eq.\ (\ref{eqn:ScatterAngleGeodesic})] for a geodesic scatter orbit. The orbit shown has parameters $(v_\infty, b) = (0.7,15M/2)$ with $r_p\simeq4.78M$ and $\delta\varphi\simeq125^\circ$.}  
\label{ImpactParScatterAngle}
\end{figure}

Instead of $L$, we can use as a second parameter the more geometrically motivated ``impact parameter'', $b$, defined as 
\begin{equation}
b:=\lim_{r\to\infty} r\sin\left|\varphi(r)-\varphi(\infty)\right|.
\end{equation}
A graphical illustration of this definition is shown in Figure \ref{ImpactParScatterAngle}. From Eqs.\ (\ref{phidot}) and (\ref{rdot}) we have at large $r$
\begin{equation}
\varphi(r)-\varphi(\infty)= \int_{\infty}^{r}\frac{\dot{\varphi}}{\dot{r}}dr'\simeq - \frac{L}{r \dot{r}_\infty},
\end{equation}
giving
\begin{equation}
b=\frac{L}{\uinf} = \frac{L}{\sqrt{E^2-1}} = \frac{L}{\vinf E}.
\end{equation}

For a scatter orbit we need $b>b_{\rm crit}$ where $b_{\rm crit}:=L_{\rm crit}(E)/\uinf$.
The minimal possible value of $b$ is 
\begin{equation}
b_{\rm min} = \lim_{E\to\infty}b_{\rm crit}(E) = 3\sqrt{3}M \simeq 5.196 M .
\end{equation}

We henceforth use the pair $\{\vinf,b\}$ as a (pseudo-)invariant parametrisation of the scatter orbit.

\section{Geodesic equation in terms of its roots}

For the geodesic orbit we can form an expression for radial geodesic motion utilising the turning points of the radial first integral (\ref{rdot}). We can write
\begin{equation}
E^2-V(r;L)= (E^2-1)(r-r_p)(r-r_1)(r-r_3)/r^3,
\label{eqn:RadialRdot}
\end{equation}  
where for scatter orbits the roots satisfy $r_1<0<r_3<r_p$ and are given explicitly by \cite{maarten}
\begin{equation}\label{r1r2r3}
r_p=\frac{6M}{1-2\zeta \sin\left(\frac{\pi}{6}-\xi\right)},\quad\quad
r_1=\frac{6M}{1-2\zeta \sin\left(\frac{\pi}{6}+\xi\right)},\quad\quad
r_3=\frac{6M}{1+2\zeta \cos \xi},
\end{equation}
with
\begin{equation}
\zeta:= \sqrt{1-12M^2/L^2}, \quad\quad
\xi:= \frac{1}{3}\arccos\left(\frac{1+(36-54E^2)M^2/L^2}{\zeta^3} \right).
\label{eqn:zetaxi}
\end{equation}
These roots are not independent. We can solve the first two equations in (\ref{r1r2r3}) for $\{\cos\xi,\sin\xi\}$ in terms of $\{r_p,r_1\}$, and subsequently express $r_3$ explicitly in terms of $\{r_p,r_1\}$ such that
\begin{equation}\label{r3}
r_3= \frac{2M r_p r_1}{r_p r_1-2M(r_p + r_1)}.
\end{equation}

\section{The $e,p$ parametrisation}

It is convenient to represent the radial motion in the form 
\begin{equation}\label{rofchi}
r=\frac{Mp}{1+e\cos\chi},
\end{equation}
where $e>1$, and $\chi$ is a new parameter along the orbit, taking the values $\chi\in (-\chi_\infty,\chi_\infty)$ with 
\begin{equation}\label{chiinf}
\chi_\infty=\arccos(-1/e).
\end{equation}
The roots $r_p$ and $r_1$ correspond to $\chi=0$ and $\chi=\pi/2$ which gives the results 
\begin{equation}\label{r1r2pe}
r_p= \frac{M p}{1+e}, \qquad r_1= \frac{M p}{1-e}.
\end{equation}
and thus using Eq.\ (\ref{r3}) we can write
\begin{equation}\label{r3pe}
r_3 = \frac{2Mp}{p-4}.
\end{equation}

In the bound-orbit case, the orbit oscillates between $r_p \leq r \leq r_1$ and the relation between $\{E,L\}$ and $\{e,p\}$ is uniquely determined from the two conditions $E^2=V(r_p,L)$ and $E^2=V(r_1,L)$. In the scatter case, $r_1$ is unphysical (negative) so $e(E,L)$ and $p(E,L)$ are a-priori ambiguous [this is equivalent to the ambiguity in $\chi(t)$ or $\chi(\tau)$]. However, it is convenient to still impose the extra condition imposed by $r_1$ which fixes the relations $e(E,L)$ and $p(E,L)$ to be as they are in the bound-orbit case:
\begin{equation}
E^2 =
\frac{(p-2)^2-4e^2}{p(p-3-e^2)}, \quad\quad
L^2 =
\frac{p^2M^2}{p-3-e^2}.
\label{ELep}
\end{equation} 
To invert these relations we need to solve cubic equations, and the result is cumbersome. However, it is relatively simple to express $\{e,p\}$ in terms of $\{L,r_p\}$ or $\{E,r_p\}$ [where $r_p$ itself can be obtained from $(E,L)$ using the expression in Eqs.\ (\ref{r1r2r3})]:
\begin{equation}\label{eofL}
e=\frac{L^2 r_p-2Mr_p^2+\sqrt{L^4(r_p^2+4Mr_p-12M^2)-16M^2 L^2r_p^2}}{2M(L^2+r_p^2)},
\end{equation}
or
\begin{align}\label{eofE}
e=\:&\frac{\sqrt{[r_p(E^2-1)+2M][(r_p+2M)^2(r_p(E^2-1)+2M)-16E^2 M^2r_p]}}{2E^2 Mr_p} \nonumber\\
&+\frac{r_p^2(E^2-1)+4M^2}{2E^2 Mr_p},
\end{align}
with $p=(r_p/M)(1+e)$.

Once $\{e,p\}$ are fixed in terms of $\{E,L\}$, the relation between $\chi$ and $t$ is also fixed, and it as in the bound-orbit case:
\begin{equation}
\frac{dt}{d\chi} = \frac{\dot{t}}{\dot{r}}\frac{dr}{d\chi}= 
\frac{Mp^2}{(p-2-2e\cos\chi)(1+e\cos\chi)^2}
\sqrt{\frac{(p-2)^2-4e^2}{p-6-2e\cos\chi}}.
\label{eq:dt_dchi} 
\end{equation}
Here we have used Eqs.\ (\ref{tdot})--(\ref{rdot}) and then substituted from (\ref{rofchi}) and (\ref{ELep}).

\section{Scatter angle}

An expression $\varphi(\chi)$ along the orbits can be found by integrating 
\begin{equation}
\frac{d\varphi}{d\chi} =\frac{\dot{\varphi}}{\dot{r}}\frac{dr}{d\chi}=
\sqrt{\frac{p}{p-6-2e\cos\chi}},
\label{eq:dphi_dchi}
\end{equation}
where we have used Eqs.\ (\ref{tdot})--(\ref{rdot}) and then substituted from (\ref{rofchi}) and (\ref{ELep}).
This equation has an explicit integral in terms of an Elliptic function: 
\begin{equation}\label{phiofchi}
\varphi(\chi)=\varphi(0)+k\sqrt{p/e}\, \El_1\Big(\frac{\chi}{2};-k^2\Big),
\end{equation}
where 
\begin{equation}\label{k}
k=2\sqrt{\frac{e}{p-6-2e}},
\end{equation}
and $\El_1$ is the incomplete elliptic integral of the first kind:
\begin{equation}
\El_1(\varphi;k)=\int_0^{\varphi} (1-k\sin^2 x)^{-1/2}dx.
\label{eqn:El1}
\end{equation}

From Eq.\ (\ref{phidot}) we see that $\dot\varphi\to 0$ for $r\to\infty$, so $\varphi\to$ const for $\chi\to\pm\chi_\infty$. Let $\varphi_{\rm in}$ and $\varphi_{\rm out}$ be the asymptotic values of $\varphi$ for $\chi\to -\chi_\infty$ and $\chi\to\chi_\infty$, respectively. Eq.\ (\ref{phiofchi}) shows that the difference between them is given by 
\begin{align}
\Delta\varphi:= \varphi_{\rm out}-\varphi_{\rm in} =& \; k\sqrt{p/e}\left[\El_1 \left(\frac{\chi_\infty}{2};-k^2\right)-\El_1\left(-\frac{\chi_\infty}{2};-k^2\right)\right] \nonumber\\
=& \; 2k\sqrt{p/e}\, \El_1\left(\frac{\chi_\infty}{2};-k^2\right).
\label{Deltaphi}
\end{align}
We define the {\it scatter} angle as 
\begin{equation}
\delta\varphi := \Delta\varphi-\pi. 
\label{eqn:ScatterAngleGeodesic}
\end{equation}
A graphical represtentation of the scatter angle is shown in Figure \ref{ImpactParScatterAngle}.

\section{Post-Minkowskian expansion} \label{subsec:PM}

Consider an orbit that lives entirely in the weak-field zone, i.e.\ for which $r_p\sim b\gg M$. We want to obtain $\delta\varphi$ as an expansion in powers of $M/b$ at fixed $E$. Since $E$ and $b$ are ``invariant'' parameters, this should facilitate comparison with other PM calculations (such as Gralla \& Lobo \cite{GrallaLobo2021}).

To this end, we first expand $r_p$. We can use the relation $L=b\uinf$ in Eqs.\ (\ref{r1r2r3}) and expand in $M/b$ to give the result
\begin{equation}\label{rminPM}
r_p=b - \frac{M}{\vinf^2}+\left(\frac{1-4\vinf^2}{2\vinf^4}\right) \frac{M^2}{b}-\left(\frac{4}{\vinf^2}\right) \frac{M^3}{b^2}+ {\cal O}\left(\frac{M^4}{b^3}\right).
\end{equation}
We can use this in Eq.\ (\ref{eofL}) and re-expand in $M/b$ to get
\begin{equation}\label{ePM}
e=\vinf^2\frac{b}{M}-\left(\frac{11E^4-20E^2+8}{2\vinf^2 E^4}\right)\frac{M}{b}+{\cal O}\left(\frac{M}{b}\right)^3,
\end{equation}
which also gives
\begin{equation}\label{chiinfPM}
\chi_\infty =\arccos(-1/e)= \frac{\pi}{2}+\frac{1}{\vinf^2}\, \frac{M}{b}+\left(\frac{17E^4-30 E^2+12}{3\vinf^6 E^4}\right)\frac{M^3}{b^3}+{\cal O}\left(\frac{M}{b}\right)^5.
\end{equation}
Next, we use $p=r_p(1+e)/M$ with the expansions (\ref{rminPM}) and (\ref{ePM}) to obtain
\begin{equation}\label{pPM}
p=\vinf^2\frac{b^2}{M^2}-\left(\frac{4}{E^2}-8\right)+{\cal O}\left(\frac{M}{b}\right)^2.
\end{equation}
Note that $p\propto b^2$ at large $b$ and that $p$ and $e$ have, respectively, even and odd expansions in $M/b$.  

Combining (\ref{ePM}) and (\ref{pPM}) in Eq.\ (\ref{k}) gives
\begin{equation}\label{kPM}
k=\sqrt{\frac{M}{b}}\left[
2+2\frac{M}{b} + \left(\frac{23E^4-28E^2+6}{2\vinf^4 E^4}\right)\frac{M^2}{b^2}+
\left(\frac{61E^4-68E^2+10}{2\vinf^4 E^4}\right)\frac{M^3}{b^3}+{\cal O}\left(\frac{M}{b}\right)^4 \right].
\end{equation}
Finally, we put all these expansions into 
\begin{equation}\label{deltaphi}
\delta\varphi = 2k\sqrt{p/e}\, \El_1\Big(\frac{\chi_\infty}{2};-k^2\Big) -\pi .
\end{equation}
We can choose to expand the Elliptic functions in its index $-k^2$ about $k=0$ or in its argument, $\chi_\infty/2$, about $\pi/4$. Both methods should give the same result.
If we use the general formula 
\begin{align}\label{El1_formula1}
\El_1\Big(x,\kappa\Big)=&
\; x+\frac{1}{8}[2x-\sin(2x)]\kappa +\frac{3}{256}[12x-8\sin(2x)+\sin(4x)]\kappa^2 \nonumber\\
&+\frac{5}{3072}[60x-45\sin(2x)+9\sin(4x)-\sin(6x)]\kappa^3 +{\cal O}(\kappa^4),
\end{align}
to expand about $k=0$ with $x=\chi_\infty/2$ and $\kappa=-k^2$, we obtain 
\begin{align}
\El_1\Big(\frac{\chi_\infty}{2};-k^2\Big)=&
\; \frac{1}{2}\chi_\infty-\frac{1}{8}\left(\chi_\infty-\sin\chi_\infty\right)k^2 +\frac{3}{256}[6\chi_\infty-8\sin\chi_\infty+\sin(2\chi_\infty)]k^4 \nonumber\\
& -\frac{5}{3072}[30\chi_\infty-45\sin\chi_\infty+9\sin(2\chi_\infty)-\sin(3\chi_\infty))]k^6 +{\cal O}(k^8).
\end{align}
Substituting this in (\ref{deltaphi}) and using the expansions for $k$, $p$, $e$ and $\chi_\infty$, gives
\begin{align}
\delta\varphi =& \left(\frac{2(2E^2-1)}{\uinf^2}\right)\frac{M}{b}+ \left(\frac{3\pi(5E^2-1)}{4\uinf^2}\right)\frac{M^2}{b^2} \nonumber \\
&+\left(\frac{2(64E^6-120E^4+60E^2-5)}{3\uinf^6}\right)\frac{M^3}{b^3}+{\cal O}\left(\frac{M}{b}\right)^4,
\end{align}
or
\begin{align}\label{deltaphiPM}
\delta\varphi =& \left(\frac{2(1+\vinf^2)}{\vinf^2}\right)\frac{M}{b}+\left(\frac{3\pi(4+\vinf^2)}{4\vinf^2}\right)\frac{M^2}{b^2}\nonumber \\
&+\left(\frac{2(5\vinf^6+45\vinf^4+15\vinf^2-1)}{3\vinf^6}\right)\frac{M^3}{b^3}+{\cal O}\left(\frac{M}{b}\right)^4.
\end{align}

Alternatively, we can use the expansion about $\chi_\infty/2=\pi/4$ which is given by
\begin{align}\label{El1_formula2}
\El_1\left(\frac{\chi_\infty}{2},-k^2\right)=&\El_1\left(\frac{\pi}{4};-k^2\right)+\frac{(\chi_\infty-\pi/2)}{[2(2+k^2)]^{1/2}}- \frac{k^2(\chi_\infty-\pi/2)^2}{2[2(2+k^2)]^{3/2}}\nonumber \\
&+ \frac{k^4(\chi_\infty-\pi/2)^3}{2[2(2+k^2)]^{5/2}}+{\cal O}(\chi_\infty-\pi/2)^4,
\end{align}
and again we can substitute in (\ref{deltaphi}) and use the expansions for $k$, $p$, $e$ and $\chi_\infty$. When we do this we again arrive at Eq.\ (\ref{deltaphiPM}).

%% file: Correction.tex

In this chapter we calculate the motion of a particle on a scatter orbit around a Schwarzschild black hole using first-order self-force expansions of the orbital parameters, considering only conservative effects. The zeroth-order results are geodesic motion as detailed in Chapter \ref{chapter:geodesics}. We show two separate derivations of formulae for the conservative self-force correction to the scatter angle, both of which involve integrals of the temporal and azimuthal self-force components along the orbit. Post-Minkowsian expansions of the formulae reproduce the results of Gralla \& Lobo for a scalar field \cite{GrallaLobo2021}. 

\section{Self-forced equations of motion}
\label{sec:SFEoM}

Consider endowing the particle with a small charge $q$. The particle experiences a self-force, whose conservative piece we denote $\mu\eta F_{\alpha}$, where $\mu$ is the particle's mass and $F_{\alpha}$ is the corresponding self-acceleration per $\eta:=q^2/(\mu M)$. Note that in the gravitational case $q=\mu$ hence $\eta$ becomes the usual mass ratio $\eta=\mu/M$. If we assume that $\eta\ll 1$ then we can perform a perturbative expansion using $\eta$ for order counting. Our goal is to calculate the resulting change in $\delta\varphi$, for fixed $\{\vinf,b\}$, to order ${\cal O}(\eta)$.

The equation of motion is now
\begin{eqnarray}
\dot{E}&=& - \eta F_t, \label{tdotF}\\
\dot{L} &=& \eta F_\varphi, \label{phidotF}\\
\ddot{r} &=& -\frac{1}{2}\frac{\partial V(r;L)}{\partial r}+\eta F^r,  \label{rdotdotF}
\end{eqnarray}
where $E(\tau):=-u_t$ and $L(\tau):=u_\varphi$ are now functions along the orbit. The conservative self-force components have the symmetry relations
\begin{equation}
F_t(r,\dot{r})= - F_t(r,-\dot{r}),\quad\quad
F_\varphi(r,\dot{r})= - F_\varphi(r,-\dot{r}),\quad\quad
F_r(r,\dot{r})=  F_r(r,-\dot{r}).
\label{eqn:ConsSFr}
\end{equation}

Given the self-force, Eqs.\ (\ref{tdotF}) and (\ref{phidotF}) can be integrated immediately  to give
\begin{equation}
E(\tau) = \Einf + \eta\Delta E(\tau) ,\quad\quad
L(\tau) = \Linf + \eta\Delta L(\tau) ,
\end{equation}
where 
\begin{equation}
\Einf:= E(\tau\to-\infty) = \frac{1}{\sqrt{1-\vinf^2}},\quad\quad
\Linf:= L(\tau\to-\infty) = \frac{b\vinf}{\sqrt{1-\vinf^2}},
\end{equation}
and 
\begin{equation}\label{DeltaEL}
\Delta E(\tau):= - \int_{-\infty}^{\tau} F_t\, d\tau,\quad\quad
\Delta L(\tau):= \int_{-\infty}^{\tau} F_\varphi\, d\tau .
\end{equation}

\section{Self-force correction to $r_p$ }\label{subsec:r_p}

We write 
\begin{equation}
r_p= \rmin + \eta\,r_p^{(1)},
\end{equation}
where $\eta\,r_p^{(1)}$ is the self-force perturbation of $r_p$ defined for fixed $\{\Einf,\Linf\}$ (or, equivalently, fixed $\{\vinf,b\}$). More precisely,
\begin{eqnarray}
\rmin&: =& \lim_{\eta\to 0} r_p(\vinf,b,\eta),\nonumber \\
r_p^{(1)}&: =& \lim_{\eta\to 0}\frac{\partial r_p}{\partial \eta},
\end{eqnarray}
where the limit is taken with fixed $\{\vinf,b\}$. Beware that taking the limit while fixing any other set of parameters (e.g., $\{e,p\}$) will generally give a different $r_p^{(1)}$. From Eq.\ (\ref{r1r2r3}) we have the geodesic limit
\begin{equation}\label{rmin0}
\rmin= \frac{6M}{1-2\zeta \sin\left(\frac{\pi}{6}-\xi\right)},
\end{equation}
with $\zeta$ and $\xi$ given in Eqs.\ (\ref{eqn:zetaxi}).

To obtain $r_p^{(1)}$, we note the normalisation equation $u_\alpha u^\alpha=-1$ still holds for the perturbed orbits, and from it we get
\begin{equation}\label{rdot_sf}
\dot{r} = \pm \sqrt{E(\tau)^2-V(r;L(\tau))},
\end{equation}
where $V$ is the same function as in Eq.\ (\ref{V}). At the periastron we have
\begin{equation}
E(\tau(r_p))^2 = V(r_p,L(\tau(r_p))).
\end{equation}
The linear perturbation of this equation with respect to $\eta$ is 
\begin{equation}\label{rminEq}
2\Einf \Delta E_p = \frac{\partial V(r,L)}{\partial r}\bigg\vert_0 r_p^{(1)}
+\frac{\partial V(r,L)}{\partial L}\bigg\vert_0 \Delta L_p ,
\end{equation}
where the partial derivatives are evaluated at $(r,L)=(\rmin,\Linf)$, and 
\begin{eqnarray} \label{Ep}
\Delta E_p &:=& \Delta E(\tau(\rmin))=- \int_{-\infty}^{\tau(\rmin)} F_t \, d\tau,
\\ \label{Lp}
\Delta L_p &:=& \Delta L(\tau(\rmin))= \int_{-\infty}^{\tau(\rmin)} F_\varphi\, d\tau.
\end{eqnarray}
Solving Eq.\ (\ref{rminEq}) for $r_p^{(1)}$ gives
\begin{equation}\label{rmin1}
r_p^{(1)} = \frac{\rmin(\rmin-2M)\Linf \Delta L_p - \rmin^4\Einf\Delta E_p}{\Linf^2(\rmin-3M)-M\rmin^2}.
\end{equation}

\section{Self-force correction to $p$ and $e$}

We represent the SF-perturbed radial motion again using the form (\ref{rofchi}) such that Eqs.\ (\ref{chiinf}) and (\ref{r1r2pe}) are still valid. By virtue of the normalisation (\ref{rdot_sf}), Eq.\ (\ref{r1r2pe}) implies
\begin{equation}\label{rdot_rmin_sf}
E_p^2 = V\left(\frac{Mp}{1+e},L_p\right),
\end{equation}
which gives one relation between the perturbed $\{p,e\}$ and $\{E_p,L_p\}$, where $E_p := E(\tau(r_p))=\Einf+\eta\Delta E_p$ and similarly for $L_p$. As a second relation we choose
\begin{equation}\label{rdot_rmax_sf}
E_p^2 = V\left(\frac{Mp}{1-e},L_p\right),
\end{equation}
which resembles the choice made in the geodesic case (replacing $E\to E_p$ and $L\to L_p$), and guarantees that the resulting $\{e,p\}$ are ${\cal O}(\eta)$ perturbations of their geodesic counterparts. Since the perturbed $\{e,p\}$ are now related to $\{\Einf,\Linf\}$ exactly as $\{e,p\}$ are related to $\{E,L\}$ in the geodesic case, the former relations are described by Eq.\ (\ref{ELep}) with $E\to E_p$ and $L\to L_p$:
\begin{equation}
E_p^2 =
\frac{(p-2)^2-4e^2}{p(p-3-e^2)}, \qquad
L_p^2 =
\frac{p^2M^2}{p-3-e^2}.
\label{ELinfep}
\end{equation} 

We now write 
\begin{equation}
e=e^{(0)}+\eta e^{(1)}, \qquad\quad p=p^{(0)}+\eta p^{(1)},
\end{equation}
where the perturbation is defined for fixed $\{\Einf,\Linf\}$. The perturbations $e^{(1)}$ and $p^{(1)}$ are determined by varying Eqs.\ (\ref{ELinfep}) with respect to $\eta$ at fixed $\{\Einf,\Linf\}$:
\begin{equation}
2\Einf \Delta E_p = \frac{\partial}{\partial p}\left(\frac{(p-2)^2-4e^2}{p(p-3-e^2)}\right)\bigg\vert_{(p^{(0)},e^{(0)})} p^{(1)}+
\frac{\partial}{\partial e}\left(\frac{(p-2)^2-4e^2}{p(p-3-e^2)}\right)\bigg\vert_{(p^{(0)},e^{(0)})} e^{(1)},
\end{equation}
\begin{equation}
2\Linf \Delta L_p = \frac{\partial}{\partial p}\left(\frac{p^2M^2}{p-3-e^2}\right)\bigg\vert_{(p^{(0)},e^{(0)})} p^{(1)}+
\frac{\partial}{\partial e}\left(\frac{p^2M^2}{p-3-e^2}\right)\bigg\vert_{(p^{(0)},e^{(0)})} e^{(1)}.
\end{equation}
Solving simultaneously for $e^{(1)}$ and $p^{(1)}$ we obtain
\begin{equation}\label{p1}
p^{(1)}=\frac{2(p-3-e^2)}{(p-6)^2-4e^2}\left[
\frac{(p-4)^2}{pM^2}\Linf \Delta L_p -p^2 \Einf \Delta E_p \right],
\end{equation}
\begin{equation}\label{e1}
e^{(1)}=\frac{p-3-e^2}{e[(p-6)^2-4e^2]}\left[
\frac{(e^2-1)[(p-2)(p-6)+4e^2]}{p^2M^2}\Linf \Delta L_p +p(p-6-2e^2) \Einf \Delta E_p \right],
\end{equation}
where here we have dropped the superscripts `$(0)$' off $e$ and $p$ to reduce clutter.

\section{Self-force correction to $\delta\varphi$: Method \RomanNumeralCaps{1}}
\label{sec:SFCorrectionMI}

For the perturbed orbit we have
\begin{equation}
\frac{d\varphi}{d\chi} =\frac{\dot{\varphi}}{\dot{r}}\frac{dr}{d\chi}=
\frac{L(\chi)}{r(\chi)^2\sqrt{E(\chi)^2-V(r(\chi),L(\chi))}}\frac{M p e |\sin\chi|}{(1+e\cos\chi)^2},
\label{dphi_dchi_sf}
\end{equation}
where $r(\chi)=M p/(1+e\cos\chi)$, and we now think of $E$ and $L$ as function of $\chi$ instead of $\tau$. It is convenient to write
\begin{eqnarray}\label{ELchi}
E(\chi)&=&\Einf +\eta \Delta E(\chi) = E_p + \eta(\Delta E(\chi)-\Delta E_p),
\nonumber\\
L(\chi)&=&\Linf +\eta \Delta L(\chi) = L_p + \eta(\Delta L(\chi)-\Delta L_p).
\end{eqnarray}
If we substitute this in (\ref{dphi_dchi_sf}), expand in $\eta$, and then use Eq.\ (\ref{ELinfep}) to replace $\{E_p,L_p\}$ with $\{e,p\}$, we find that the ${\cal O}(\eta^0)$ term has the same form as in the geodesic case, Eq.\ (\ref{eq:dphi_dchi}). But there is now an ${\cal O}(\eta)$ correction coming from the ${\cal O}(\eta)$ term in (\ref{ELchi}). Altogether we have
\begin{align}
\frac{d\varphi}{d\chi} = & \; \sqrt{\frac{p}{p-6-2e\cos\chi}} \nonumber \\
& + \eta f_E(\chi;p,e)\left(\Delta E(\chi)-\Delta E_p\right) + \eta f_L(\chi;p,e)\left(\Delta L(\chi)-\Delta L_p\right)/M,
\label{dphi_dchi_sf2}
\end{align}
where here, and henceforth, $e$ and $p$ take their geodesic values and
\begin{eqnarray}
f_E &=& -\frac{p\sqrt{p-3-e^2}\sqrt{(p-2)^2-4e^2}}{e^2\sin^2\chi\, (p-6-2e\cos\chi)^{3/2}},
\nonumber\\
f_L &=& \frac{\sqrt{p-3-e^2}\left[e^2(p-6)+p-2+2e(p-3-e^2)\cos\chi\right]}{\sqrt{p}\, e^2\sin^2\chi\, (p-6-2e\cos\chi)^{3/2}}.
\end{eqnarray}

The total accumulated phase is 
\begin{align}\label{Deltavarphi_sf}
\Delta\varphi =& \; \varphi_{\rm out}-\varphi_{\rm in}=\int_{-\chiinf}^{\chiinf}\frac{d\varphi}{d\chi}d\chi \nonumber\\
=& \; 2k\sqrt{p/e}\, \El_1\Big(\frac{\chi_\infty}{2};-k^2\Big) \nonumber\\ 
&-\eta \int_{-\chiinf}^{\chiinf} f_E(\chi)\int_{0}^{\chi} F_t(\chi')\tau_{\chi'}d\chi' d\chi+\frac{\eta}{M} \int_{-\chiinf}^{\chiinf} f_L(\chi)\int_{0}^{\chi} F_\varphi(\chi') \tau_{\chi'}d\chi' d\chi
\nonumber\\
=& \; 2k\sqrt{p/e}\, \El_1\Big(\frac{\chi_\infty}{2};-k^2\Big) \nonumber\\
&-2\eta \int_{0}^{\chiinf} f_E(\chi)\int_{0}^{\chi} F_t(\chi')\tau_{\chi'} d\chi' d\chi +2\frac{\eta}{M} \int_{0}^{\chiinf} f_L(\chi)\int_{0}^{\chi} F_\varphi(\chi') \tau_{\chi'}d\chi' d\chi,
\end{align}
where we have recalled Eqs.\ (\ref{Deltaphi}) and (\ref{DeltaEL}), and where the Jacobian $\tau_{\chi}:=d\tau/d\chi$ can be evaluated along the background geodesic:
\begin{equation}
\tau_\chi = \frac{Mp \sqrt{p(p-3-e^2)}}{(1+e\cos\chi)^2\sqrt{p-6-2e\cos\chi}}.
\label{dtaudchi}
\end{equation}
The equality between the 2nd and 3rd lines of (\ref{Deltavarphi_sf}) follows from $F_\alpha(-\chi)=-F_\alpha(\chi)$ for $\alpha=t,\varphi$ [c.f.\ Eqs.\ (\ref{eqn:ConsSFr})]. 

We write the perturbed deflection angle $\delta\varphi=\Delta\varphi-\pi$ as
\begin{equation}\label{phisplit}
\delta\varphi = \delta\varphi^{(0)}+\eta \delta\varphi^{(1)},
\end{equation}
where the split between background and perturbation is, as always, defined with fixed $\{\Einf,\Linf\}$. The background value is 
\begin{equation}\label{deltavarphi0}
\delta\varphi^{(0)} = 2k\sqrt{p/e}\, \El_1\Big(\frac{\chiinf}{2};-k^2\Big)\bigg\vert_{(p^{(0)},e^{(0)})} -\pi, 
\end{equation}
where we recall that $k$ and $\chiinf$ also depend on $p$ and $e$ [recall Eqs.\ (\ref{k}) and (\ref{chiinf})]. The self-force perturbation is obtained by taking the linear perturbation of (\ref{Deltavarphi_sf}) with respect to $\eta$ at fixed $\{\Einf,\Linf\}$:
\begin{align}\label{deltavarphi_sf}
\delta\varphi^{(1)}=&\:
\frac{\partial}{\partial p}\left[2k\sqrt{p/e}\, \El_1\Big(\frac{\chi_\infty}{2};-k^2\Big)\right]p^{(1)}
+
\frac{\partial}{\partial e}\left[2k\sqrt{p/e}\, \El_1\Big(\frac{\chi_\infty}{2};-k^2\Big)\right]e^{(1)}
\nonumber \\
&-2\int_{0}^{\chiinf} f_E(\chi)\int_{0}^{\chi} F_t(\chi')\tau_{\chi'} d\chi' d\chi
+\frac{2}{M}\int_{0}^{\chiinf} f_L(\chi)\int_{0}^{\chi} F_\varphi(\chi') \tau_{\chi'}d\chi' d\chi,
\end{align}
where all $e,p,k,\chiinf$ are to take their geodesic values.

In the first line of Eq.\ (\ref{deltavarphi_sf}) the coefficients of $p^{(1)}$ and $e^{(1)}$ can be written in terms of incomplete elliptic integrals of the 1st and 2nd kinds, using the identities
\begin{equation}
\frac{\partial\El_1(\varphi;k)}{\partial\varphi}=\frac{1}{\sqrt{1-k \sin^2\varphi}},
\end{equation}
\begin{equation}
\frac{\partial\El_1(\varphi;k)}{\partial k}=
\frac{1}{2k(k-1)}\left[\frac{k \cos\varphi\, \sin\varphi}{\sqrt{1-k\sin^2\varphi}}
-(k-1)\El_1(\varphi,k)-\El_2(\varphi,k)
\right],
\end{equation}
where 
\begin{equation}
\El_2(\varphi;k)=\int_0^{\varphi} (1-k\sin^2 x)^{1/2}dx,
\label{eqn:El2}
\end{equation}
is the incomplete elliptic integral of the second kind. Substituting for $p^{(1)}$ and $e^{(1)}$ from Eqs.\ (\ref{p1}) and (\ref{e1}), the first line of (\ref{deltavarphi_sf}) then takes the form
\begin{equation}
\alpha_E(e,p) \Einf \Delta E_p + \alpha_L(e,p) \Linf \Delta L_p/M^2,
\end{equation}
where, we find
\begin{align}\label{alphaE}
\alpha_E = \frac{2(p-3-e^2)p^{3/2}}{e^2(p-6+2e)^2(p-6-2e)^{3/2}}
\Bigg[ &
-(p-6)(p-6+2e)\El_1\Big(\frac{\chi_\infty}{2};-k^2\Big) \nonumber\\
& +(p^2-12p+12e^2+36) \El_2\Big(\frac{\chi_\infty}{2};-k^2\Big)
\nonumber\\
&+\frac{16e^4-(p-6)^2(p-4)+4e^2(p^2-11p+24)}{\sqrt{(e^2-1)(p-4)(p-2e-6)}}
\Bigg],
\end{align}
\begin{align}\label{alphaL}
\alpha_L =& \frac{2(p-3-e^2)}{ e^2 p^{3/2}(p-6+2e)^2(p-6-2e)^{3/2}} \times \nonumber\\
&\Bigg[
(p-6+2e)\left[(p-2)(p-6)+e^2(p^2-8p+24)-4e^4\right]\El_1\Big(\frac{\chi_\infty}{2};-k^2\Big)
\nonumber\\
&\quad+\left[-(p-2)(p-6)^2-e^2(p-2)(p^2-24)+4e^4(p-6)
\right] \El_2\Big(\frac{\chi_\infty}{2};-k^2\Big)
\nonumber\\
&\quad+\sqrt{\frac{(e^2-1)(p-4)}{p-6-2e}}\left[-(p-2)(p-6)^2-2e^2(p-4)(p+6)+8e^4\right]
\Bigg].
\end{align}

The second line of Eq.\ (\ref{deltavarphi_sf}) involves double integrals of the self-force, which would make evaluation inconvenient.  We can do away with this using integration by parts. For example, defining the functions 
\begin{equation}\label{calFEL}
{\cal F}_E(\chi):= \int_{\chiinf}^\chi f_E(\chi') d\chi',
\quad\quad
{\cal F}_L(\chi):= \int_{\chiinf}^\chi f_L(\chi') d\chi',
\end{equation}
we have
\begin{align}
\int_{0}^{\chiinf} f_E(\chi)\int_{0}^{\chi} F_t(\chi')\tau_{\chi'} d\chi' d\chi
= \left({\cal F}_E(\chi) \int_0^\chi  F_t(\chi')\tau_{\chi'} d\chi'\right)\Bigg\vert_0^{\chiinf} -\int_0^{\chiinf} {\cal F}_E(\chi') F_t(\chi')\tau_{\chi'} d\chi' 
\nonumber\\
=-\lim_{\chi\to 0}{\cal F}_E(\chi)\int_0^\chi  F_t(\chi')\tau_{\chi'} d\chi' -\int_0^{\chiinf} {\cal F}_E(\chi') F_t(\chi')\tau_{\chi'} d\chi'.
\end{align} 
Here we note that the limit $\chi\to 0$ evaluates to zero: For $\chi\ll 1$ we have $f_E\sim 1/\chi^2$ and therefore ${\cal F}_E\sim 1/\chi$, but on the other hand $F_t\sim\chi$ (with $\tau_{\chi}$ a smooth function of $\chi$), so $\int_0^\chi  F_t(\chi')\tau_{\chi'} d\chi'$ vanishes at least as $\sim\chi^2$. Similar considerations apply to the second integral in the second line of (\ref{deltavarphi_sf}), which therefore gives
\begin{align}
\int_{0}^{\chiinf} f_L(\chi)\int_{0}^{\chi} F_\varphi(\chi')\tau_{\chi'} d\chi' d\chi
= -\int_0^{\chiinf} {\cal F}_L(\chi') F_\varphi(\chi')\tau_{\chi'} d\chi'.
\end{align} 

Collecting the above results, we can write Eq.\ (\ref{deltavarphi_sf}) in the form  
\begin{align}\label{deltaphi1_almost_final}
\delta\varphi^{(1)}=&\: 
\alpha_E(e,p) \Einf \Delta E_p 
+ \alpha_L(e,p) \Linf/M^2 \Delta L_p \nonumber \\
&+2\int_0^{\chiinf} \left[{\cal F}_E(\chi) F_t(\chi)
-{\cal F}_L(\chi) F_\varphi(\chi)/M\right]\tau_{\chi} d\chi,
\end{align}
or, recalling Eqs.\ (\ref{Ep}) and (\ref{Lp}),
\begin{align}\label{deltaphi1_final}
\delta\varphi^{(1)}=
\int_0^{\chiinf} \left[{\cal G}_E(\chi) F_t(\chi)
-{\cal G}_L(\chi) F_\varphi(\chi)\right]\tau_{\chi} d\chi,
\end{align}
where
\begin{equation}
{\cal G}_E(\chi) = 2{\cal F}_E(\chi) +\alpha_E \Einf ,
\quad\quad
{\cal G}_L(\chi) = 2{\cal F}_L(\chi)/M +\alpha_L \Linf/M^2.
\end{equation}
The functions ${\cal F}_E(\chi)$ and ${\cal F}_L(\chi)$, defined in Eqs.\ (\ref{calFEL}), can be written explicitly in terms of incomplete Elliptic functions of the first and second kind. However, these are cumbersome and we found it more practical to evaluate them numerically. The constants $\alpha_E$ and $\alpha_L$ are given in Eqs.\ (\ref{alphaE}) and (\ref{alphaL}) explicitly in terms of incomplete Elliptic integrals.

\section{Self-force correction to $\delta\varphi$: Method \RomanNumeralCaps{2}}
\label{sec:SFCorrectionMII}

In the second method we avoid $\{e,p\}$ and parametrise directly in terms of $\{\Einf,\Linf\}$ (or, equivalently, $\{\vinf,b\}$). Integrations are done with respect to $r$ instead of $\chi$. The derivation is somewhat simpler but the final result seems to be of comparable complexity.


For the geodesic orbit, the scatter angle is
\begin{equation}\label{methodII_deltaphi}
\delta\varphi^{(0)} = 2 \int_{\rmin}^\infty (\dot\varphi/\dot r) dr -\pi = 2 \int_{\rmin}^\infty \frac{H(r)}{\sqrt{r-\rmin}} dr -\pi,
\end{equation}
with
\begin{equation}
H(r):=H(r;E,L) = \frac{L}{\sqrt{(E^2-1)r(r-r_1)(r-r_3)}},
\end{equation}
where we have used the relation (\ref{eqn:RadialRdot}). Recall that $r_p$, $r_1$, and $r_3$ are roots of $\dot r =0$ given in Eqs.\ (\ref{r1r2r3}). In this section we evaluate all radial integrals on the {\it outbound} leg of the orbit, i.e.\ with $\dot{r}>0$.
For the SF-perturbed orbit, Eq.\ (\ref{methodII_deltaphi}) still applies, but with $E$, $L$, $r_p$, $r_1$ and $r_3$ all becoming slow functions of $r$ along the orbit: $E\to E(r)$, $L\to L(r)$, $r_p\to \tilde r_p(E(r),L(r))$ and $r_{1,3}\to \tilde r_{1,3}(E(r),L(r))$. We use $\tilde r_p$ to distinguish this function-along-the-orbit with the constant SF-perturbed value of the periastron distance calculated in Section \ref{subsec:r_p}, $r_p=\rmin+\eta r_p^{(1)}$. We have the relation $r_p= \tilde r_p(E(r_p),L(r_p))$.

To obtain the perturbation $\delta\varphi^{(1)}$ we need to vary the integral in (\ref{methodII_deltaphi}) with respect to $\eta$ at fixed $\{\Einf,\Linf\}$, and for this we need to evaluate the derivative of the integral with respect to $E(r)$, $L(r)$, $r_p$ and $r_{1,3}$. Varying with respect to $r_p$ is subtle, because of the singularity at the turning point. To overcome this complication, we first integrate by parts:
\begin{equation}\label{ibp}
\delta\varphi = 
4 \sqrt{r-\tilde r_p}\, H(r)\Big|_{\rmin}^\infty
-4 \int_{\rmin}^\infty \sqrt{r-\tilde r_p}\, \frac{dH(r)}{dr}dr 
+2 \int_{\rmin}^\infty \frac{H(r)}{\sqrt{r-\tilde r_p}}\frac{d\tilde r_p}{dr}dr
-\pi.
\end{equation}
The function $H(r)$ is bounded at $r=\rmin$, and falls off like $r^{-3/2}$ at infinity, so the surface terms in Eq.\ (\ref{ibp}) vanish. We are left with
\begin{align}\label{deltavarphi_full}
\delta\varphi =&\:  
-4 \int_{\rmin}^\infty \sqrt{r-\tilde r_p}\left(
\frac{\partial H}{\partial r} - \eta \frac{\partial H}{\partial E} \frac{F_t(r)}{\dot{r}} + \eta \frac{\partial H}{\partial L} \frac{F_\varphi(r)}{\dot{r}}
\right)dr \nonumber \\
&\quad +2\eta \int_{\rmin}^\infty \frac{H}{\sqrt{r-\tilde r_p}} 
\left(-\frac{\partial \tilde r_p}{\partial E}F_t + \frac{\partial \tilde r_p}{\partial L}F_\varphi
\right)\frac{dr}{\dot{r}}
-\pi ,
\end{align}
where $\partial_r$ is taken with fixed $\{E,L\}$, $\partial_E$ is taken with fixed $\{r,L\}$, and $\partial _L$ it taken with fixed $\{r,E\}$.
We have used $dE/dr = \eta d\Delta E/dr = -\eta F_t/\dot{r}$ and $dL/dr = \eta d\Delta L/dr = \eta F_\varphi/\dot{r}$. 
The geodesic limit of this expression is
\begin{equation}
\delta\varphi^{(0)} = -4 \int_{\rmin}^\infty \sqrt{r-\rmin}\, \frac{\partial H_0}{\partial r}\, dr -\pi,
\end{equation}
where $H_0:= H(r,\Einf,\Linf)$ and $\rmin(\Einf,\Linf)$ is the geodesic relation for $r_p$ given in Eq.\ (\ref{r1r2r3}). 
It can be checked that this gives the same geodesic scatter angle as the expression (\ref{deltavarphi0}).

Varying $\delta\varphi$ in Eq.\ (\ref{deltavarphi_full}) with respect to $\eta$ at fixed $\{\Einf,\Linf\}$ we obtain
\begin{align}\label{delta_varphi_v2}
\delta\varphi^{(1)} =& \:
2\int_{\rmin}^\infty \frac{1}{\sqrt{r-\rmin}}\frac{\partial H_0}{\partial r}
\left(\frac{\partial \rmin}{\partial \Einf}\Delta E(r)+ \frac{\partial \rmin}{\partial \Linf}\Delta L(r)
\right)\, dr 
\nonumber\\
&  -4 \int_{\rmin}^\infty \sqrt{r-\rmin}
\left( 
  \frac{\partial^2 H_0}{\partial r\partial\Einf} \Delta E(r) 
  + \frac{\partial^2 H_0}{\partial r\partial\Linf} \Delta L(r) 
  \right) dr 
\nonumber\\
&  +4 \int_{\rmin}^\infty \sqrt{r-\rmin}
\left( 
   \frac{\partial H_0}{\partial\Einf} \frac{F_t(r)}{\dot{r}} - \frac{\partial H_0}{\partial\Linf} \frac{F_\varphi(r)}{\dot{r}}
  \right) dr 
\nonumber\\
& +2\int_{\rmin}^\infty \frac{H_0}{\sqrt{r-\rmin}} 
\left(-\frac{\partial \rmin}{\partial \Einf}F_t + \frac{\partial \rmin}{\partial \Linf}F_\varphi\right)\frac{dr}{\dot{r}}.
\end{align}
The first two lines here involve double integrals of the self-force. These can be turned into single integrals using integration by parts. For instance,
\begin{align}\label{ibp0}
\int_{\rmin}^\infty \frac{1}{\sqrt{r-\rmin}}\frac{\partial H_0}{\partial r}\Delta E(r) dr
=& \left. \left(\int_{\rmin}^\infty \frac{1}{\sqrt{r'-\rmin}}\frac{\partial H_0}{\partial r'} dr' \right)\Delta E \right|_{\rmin}^\infty \nonumber\\
&- \int_{\rmin}^\infty  \left(\int_{\rmin}^r \frac{1}{\sqrt{r'-\rmin}}
  \frac{\partial H_0}{\partial r'}dr' \right)\left(-\frac{F_t}{\dot{r}}\right)dr
\nonumber\\
=& \: \frac{1}{2}\int_{\rmin}^\infty G_r(r) \frac{F_t}{\dot{r}} dr,
\end{align}
where
\begin{equation}
 G_r(r):= 2\int_{\rmin}^{r}\frac{1}{\sqrt{r'-\rmin}}\frac{\partial H_0}{\partial r'}dr'.
\end{equation}
The surface terms in (\ref{ibp0}) both vanish: For $r\to \rmin$, $\Delta E$ is bounded, as is  $\frac{\partial H_0}{\partial r}$, so the term goes like $\sim (r-\rmin)^{3/2}\to 0$. For $r\to\infty$, the conservative SF for {\em bound} motion behaves as $F_t\sim 1/r^2$ hence we expect that {\rm unbound} orbits should fall off as at least $F_t\sim 1/r^2$. This means that for the slowest fall-off $\Delta E(r)\sim rF_t\sim 1/r$, so the surface term vanishes like $1/r$ at least. 

Similarly, for the first term in the second line of (\ref{delta_varphi_v2}) we write
\begin{align}\label{ibp1}
\int_{\rmin}^\infty \sqrt{r-\rmin}
  \frac{\partial^2 H_0}{\partial r\partial\Einf} \Delta E(r) dr=&  
\left. \left(\int_{\rmin}^r \sqrt{r'-\rmin}
  \frac{\partial^2 H_0}{\partial r'\partial\Einf}dr' \right) \Delta E(r) \right|_{\rmin}^\infty \nonumber \\
  &- \int_{\rmin}^\infty  \left(\int_{\rmin}^r \sqrt{r'-\rmin}
  \frac{\partial^2 H_0}{\partial r'\partial\Einf}dr' \right)\left(-\frac{F_t}{\dot{r}}\right)dr.
\end{align}
The surface term again vanishes. In the remaining term we apply another integration by parts:
\begin{equation}
\int_{\rmin}^r \sqrt{r'-\rmin}  \frac{\partial^2 H_0}{\partial r'\partial\Einf}dr' =
\sqrt{r'-\rmin}  \frac{\partial H_0}{\partial\Einf} - \frac{1}{4} G_E(r),
\end{equation} 
where 
\begin{equation}
 G_E(r):= 2\int_{\rmin}^{r}\frac{1}{\sqrt{r'-\rmin}}\frac{\partial H_0}{\partial\Einf}dr'.
\end{equation}
Equation (\ref{ibp1}) then becomes
\begin{equation}\label{ibp2}
\int_{\rmin}^\infty \sqrt{r-\rmin}
  \frac{\partial^2 H_0}{\partial r\partial\Einf} \Delta E(r) dr= \int_{\rmin}^\infty \left(\sqrt{r-\rmin}  \frac{\partial H_0}{\partial\Einf} - \frac{1}{4}G_E(r)\right)
  \frac{F_t}{\dot{r}}dr.
\end{equation}
Similarly, we can write
\begin{equation}\label{ibp3}
\int_{\rmin}^\infty \sqrt{r-\rmin}
  \frac{\partial^2 H_0}{\partial r\partial\Linf} \Delta L(r) dr= -\int_{\rmin}^\infty \left(\sqrt{r-\rmin}  \frac{\partial H_0}{\partial\Linf} - \frac{1}{4}G_L(r)\right)
  \frac{F_\varphi}{\dot{r}}dr,
\end{equation}
with
\begin{equation}
G_L(r):= 2\int_{\rmin}^{r}\frac{1}{\sqrt{r'-\rmin}}\frac{\partial H_0}{\partial\Linf}dr' .
\end{equation}

With these substitutions, Eq.\ (\ref{delta_varphi_v2}) becomes
%
\begin{align}\label{deltaphi1_final_method2}
\delta\varphi^{(1)}=
\int_{\rmin}^{\infty} \left[\tilde{\cal G}_E(r) F_t(r)
-\tilde{\cal G}_L(r) F_\varphi(r)\right] \frac{dr}{\dot{r}},
\end{align}
where
\begin{align}\label{calG_II}
\tilde{\cal G}_E(r) =& G_E(r) +\left(G_r(r) -\frac{2H_0(r)}{\sqrt{r-\rmin}}\right) \frac{\partial \rmin}{\partial\Einf}, \\
\tilde{\cal G}_L(r) =& G_L(r) +\left(G_r(r) -\frac{2H_0(r)}{\sqrt{r-\rmin}}\right) \frac{\partial \rmin}{\partial\Linf}.
\end{align}

It should be possible to confirm that this expression is equivalent to Eq.\ (\ref{deltaphi1_final}). Note, however, that we can {\it not} expect the corresponding $\cal G$ coefficients in the integrand to be in agreement: $\tilde{\cal G}_E(r(\chi)) \ne {\cal G}_E(\chi)$ and 
$\tilde{\cal G}_L(r(\chi)) \ne {\cal G}_L(\chi)$. That's because Eq.\ (\ref{deltaphi1_final_method2}) differs from Eq.\ (\ref{deltaphi1_final}) by some surface terms that are only zero if the self-force satisfies certain vanishing conditions at the integration's boundaries. However, the {\it integrals} should be equal, assuming the self-force satisfies these conditions. In the next section we perform a post-Minkowkian expansion on Eqs.\ (\ref{deltaphi1_final}) and (\ref{deltaphi1_final_method2}) to confirm that the formulae agree in the large-$b$ limit.

\addtocontents{toc}{\protect\newpage}
\section{Post-Minkowskian expansion}

We derive here expressions for $\delta\varphi^{(1)}$ (in terms of the conservative self-force) through the first subleading order in a PM ($M/b$) expansion for both Eqs.\ (\ref{deltaphi1_final}) and (\ref{deltaphi1_final_method2}). This allows us to compare the formulae and confirm that they produce the same results in the large-$b$ limit. In this section we replace $\vinf$ with $v$ for brevity.

\subsection{$\delta\varphi^{(1)}$: Method \RomanNumeralCaps{1}}

First, we copy relevant results from Section \ref{subsec:PM}, restricted to first subleading order:
\begin{eqnarray}
r_p &=& b-\frac{M}{v^2} +{\cal O}(b^{-1}), 
\nonumber\\
e &=& v^2 \frac{b}{M} + {\cal O}(b^{-1}),
\nonumber\\
p &=& v^2 \frac{b^2}{M^2} +{\cal O}(b^0),
\nonumber\\
\chi_{\infty} &=& \frac{\pi}{2}+\frac{M}{v^2 b}+{\cal O}(b^{-3}),
\nonumber\\
k &=& \sqrt{\frac{M}{b}}\left[2+2\frac{M}{b}+{\cal O}(b^{-2})\right].
\end{eqnarray}
Expansions for $\El_1\Big(\frac{\chi_\infty}{2};-k^2\Big)$ and $\El_2\Big(\frac{\chi_\infty}{2};-k^2\Big)$ can now be obtained using the expansion formulae (\ref{El1_formula1}) or (\ref{El1_formula2}), and the equivalent formulae for $\El_2$. We get
\begin{align}
\El_1\Big(\frac{\chi_\infty}{2};-k^2\Big) = &\:
\frac{\pi}{4}+\frac{M}{4v^2b}\left[(2-\pi)v^2+2\right]
+\frac{M^2}{16 v^2 b^2}\left[(\pi-8)v^2-8\right] +{\cal O}(b^{-3}),
\nonumber\\
\El_2\Big(\frac{\chi_\infty}{2};-k^2\Big) = &\:
\frac{\pi}{4}+\frac{M}{4v^2b}\left[(\pi-2)v^2+2\right]
+\frac{M^2}{16 v^2 b^2}\left[(5\pi-8)v^2+8\right] +{\cal O}(b^{-3}).
\end{align}
Here we keep subsubleading terms, as it turns out they are needed for a subleading-order calculation of $\alpha_E$ and $\alpha_L$, due to a cancellation of the leading-order terms.

Using these expansions we obtain 
\begin{eqnarray}
{\cal G}_{E}(\chi) &=& \frac{2}{v^2E}\cot\chi +\frac{6M}{v^2 E b}\csc\chi +{\cal O}(b^{-2}),
\nonumber\\
{\cal G}_{L}(\chi) &=& -\frac{2}{vEb}\cot\chi -\frac{2M(v^2+2)}{v^3 E b^2}\csc\chi +{\cal O}(b^{-3}),
\end{eqnarray}
and 
\begin{equation}
\tau_\chi = \frac{b}{v E}\sec^2\chi + \frac{M}{v^3 E}\sec\chi(v^2-2\sec^2\chi) +{\cal O}(b^{-1}),
\end{equation}
and substituting into (\ref{deltaphi1_final}) then gives
\begin{align}
\delta\varphi^{(1)} =&\int_0^{\chi_{\infty}}\left[
\frac{2b}{v^3 E^2} +\frac{M}{v^5 E^2}(v^2\cos(2\chi)+7v^2-4)\sec\chi +{\cal O}(b^{-1})\right]\csc\chi\sec\chi F_t(\chi)\, d\chi
\nonumber\\
&+ \int_0^{\chi_{\infty}}\left[\frac{2}{v^2 E^2} +\frac{2M}{v^2 E^2 b}\left(\sec\chi+\cos\chi\right)+{\cal O}(b^{-2})\right] \csc\chi\sec\chi  F_\varphi(\chi)\, d\chi.
\end{align} 
The leading PM term is
\begin{equation}\label{1PM_methodI}
\delta\varphi^{(1)}_{\rm 2PM}= \frac{4}{v^2E^2}\int_0^{\pi/2}\left[(b/v)F_t+F_{\varphi}\right] \, d\chi/\sin(2\chi).
\end{equation}
Here care must be taken near the integration boundaries because of the `$0/0$' singularities there. Recall $F_t\sim \chi$ for $\chi\to\ 0$ and $F_t\sim \pi/2-\chi$ for $\chi\to\pi/2$, and similarly for $F_\varphi$.

\subsection{$\delta\varphi^{(1)}$: Method \RomanNumeralCaps{2}}

Using the expansions for $e$ and $p$ and Eqs.\ (\ref{r1r2pe}) and (\ref{r3pe}), we obtain
\begin{eqnarray}
r_p &=& b -\frac{M}{v^2} + {\cal O}(b^{-1}),
\nonumber\\
r_1 &=& -b -\frac{M}{v^2} + {\cal O}(b^{-1}),
\nonumber\\
r_3 &=& 2M +\frac{8M^3}{v^2b^2} +{\cal O}(b^{-4}).
\end{eqnarray}
From $b=L/\sqrt{E^2-1}$ then also follows, at leading order,
\begin{eqnarray}\label{drpdE_PM}
\frac{\partial \rmin}{\partial E}=\frac{\partial b(E,L)}{\partial E} &=& -\frac{b}{v^2 E},
\nonumber\\
\frac{\partial \rmin}{\partial L}=\frac{\partial b(E,L)}{\partial L} &=& \frac{1}{v E}.
\end{eqnarray}

Since for $b\gg M$ we have $r>r_p\sim -r_1\sim b \gg r_3$, the function $H_0(r)$ takes the leading-order form 
\begin{equation}
H_0 = \frac{b}{r\sqrt{r+b}}.
\end{equation}
The integral $G_r(r)$ becomes
\begin{equation}
G_r(r) = 2\int_b^{r} \frac{1}{\sqrt{r'-b}}\frac{\partial H_0}{\partial r'}dr' = 
-\frac{1}{b}\left[\frac{r^2+br-2b^2}{r\sqrt{r^2-b^2}}+\arctan\sqrt{(r/b)^2-1}\right],
\end{equation}
and using (\ref{drpdE_PM}) we also obtain the leading-order integrals
\begin{eqnarray}
G_E(r)&=& 2\int_b^{r} \frac{1}{\sqrt{r'-b}}\frac{\partial H_0}{\partial E}dr'  = 
-\frac{1}{v^2 E}\left[\sqrt{\frac{r-b}{r+b}}+\arctan\sqrt{(r/b)^2-1}\right],
\nonumber\\
G_L(r)&=& 2\int_b^{r} \frac{1}{\sqrt{r'-b}}\frac{\partial H_0}{\partial L}dr'  = 
\frac{1}{v E b}\left[\sqrt{\frac{r-b}{r+b}}+\arctan\sqrt{(r/b)^2-1}\right].
\end{eqnarray}
Finally, at leading order,
\begin{equation}
\dot{r} = vE \sqrt{1-(b/r)^2}.
\end{equation}

Putting everything together in Eq.\ (\ref{calG_II}) we find
\begin{equation}
\tilde {\cal G}_E =\frac{2b}{v^2E \sqrt{r^2-b^2}},
\quad\quad
\tilde {\cal G}_L =-\frac{2}{vE \sqrt{r^2-b^2}},
\end{equation} 
and substituting in Eq.\ (\ref{deltaphi1_final_method2}) finally gives
\begin{equation}\label{1PM_methodII}
\delta\varphi^{(1)}_{\rm 2PM}= \frac{2}{v^2E^2}\int_{b}^{\infty}\left[(b/v)F_t+F_{\varphi}\right] \frac{r\, dr}{r^2-b^2}.
\end{equation}

It can be checked that Eqs.\ (\ref{1PM_methodI}) and (\ref{1PM_methodII}) are in agreement. To see this, note the leading-order relations
\begin{equation}
\cos\chi = \frac{b}{r}, \quad\quad
\sin\chi = \sqrt{1-b^2/r^2}, \quad\quad
\frac{d\chi}{dr} = \frac{b}{r \sqrt{r^2-b^2}},
\end{equation}
leading to
\begin{equation}
\frac{2 d\chi}{\sin(2\chi)} = \frac{r dr}{r^2-b^2}.
\end{equation}

\subsection{Comparison with Gralla \& Lobo \cite{GrallaLobo2021} for a scalar field}

Gralla \& Lobo (GL) \cite{GrallaLobo2021} consider a scalar charge $q$ with mass $\mu$ moving in a straight line in flat space. In Cartesian coordinates $(t,x,y,z)$, the straight line is $x^\mu=(t,b,0,z(t))$, where $z=vt=\sqrt{r^2-b^2}$. The four-velocity is $u^{\mu}=E(1,0,0,v)$, where $E=(1-v^2)^{-1/2}$. The charge now moves in the gravitational field of a mass $M$ located at the centre of coordinates ($x=y=z=0$). There is a ``small charge'' assumption $q^2/(\mu b)\ll 1$ and a ``weak-field" assumption $M/(bv^2)\ll 1$.

To keep with our conventions established in the first paragraph of Section \ref{sec:SFEoM} we write the self-force as $\mu\eta F_{\alpha}$, with $\eta=q^2/(\mu M)$, such that $F_{\alpha}$ is the self-acceleration per $\mu$. GL obtain the leading PM term of the full (dissipative and conservative) scalar-field self-acceleration experienced by the particle:
\begin{eqnarray}
F_x &=& -\frac{2M}{E}A_3, \nonumber\\
F_y &=& 0,\nonumber\\
F_z &=& -\frac{2M}{E}(E^2v A_1+A_2),\nonumber\\
F_t &=& v F_z ,
\end{eqnarray}
where 
\begin{equation}
A_3 = \frac{Mvb\left(2r^2v^3-rz(1+3v^2)-(v^2-3)vz^2\right)}{2r(r-vz)^5}.
\end{equation}
Expressions for $A_1$ and $A_2$ are given in Eqs.\ (39) and (40) of GL, but will not be needed here.
From this we obtain
\begin{equation}
F_t = \frac{2Mv}{E}(E^2 v A_1+A_2),
\qquad 
F_\varphi =\frac{2M}{E}(zA_3-bE^2 vA_1-bA_2),
\end{equation}
(the latter with the help of $\frac{\partial\varphi}{\partial x}=-\frac{z}{r^2}$ and $\frac{\partial\varphi}{\partial z}=\frac{x}{r^2}$), and from this, in turn,
\begin{equation}
F:= (b/v)F_t +F_\varphi = \frac{2M}{E} z A_3,
\end{equation}
[c.f.\ the integrand of Eq.\ (\ref{deltaphi1_final_method2})]. The {\em conservative} piece $F_{\rm cons}(z) = \frac{1}{2}\left[F(z)-F(-z) \right]$ is given by
\begin{equation}
F_{\rm cons} = \frac{2 M^2 b v^2 z \left(r^6 v^2+r^4 \left(10 v^4-8 v^2-1\right) z^2+5 r^2 v^2 \left(v^4-4 v^2+2\right) z^4+v^4 \left(7-4 v^2\right) z^6\right)}{E \left(r^2-v^2 z^2\right)^5}.
\end{equation}

We express $F_{\rm cons}$ as a function of $r$ only using $z=\sqrt{r^2-b^2}$, and then substitute in our leading-PM formula (\ref{1PM_methodII}). The integral can be evaluated analytically, and the result is 
\begin{equation}
\delta\varphi^{(1)}_{\rm 2PM}=-\frac{\pi}{4}\left(\frac{M}{b}\right)^2,
\end{equation}
in complete agreement with Eq.\ (2) of GL.

This agreement between the PM expansions of Eqs.\ (\ref{deltaphi1_final}) and (\ref{deltaphi1_final_method2}) with GL provides a robust check of both formulae. In the rest of the thesis we focus on producing self-force data needed to calculate the correction to the scatter angle for strong-field scatter orbits.

%% file: ScalarVacuum.tex

In the last chapter we determined the components required to calculate the first-order conservative correction to the scatter angle. Here we start to develop the computational infrastructure that would enable the calculation of the self-force components required for the correction to the scatter angle. The feasibility of the proposed 1+1D method is tested with a scalar-field toy model. We consider the vacuum case followed by a point-particle on circular and hyperbolic orbits.

\section{Multipole decomposition}  

The equation for a scalar field $\Psi$ in a vacuum spacetime with metric $\tensor{g}{_\mu_\nu}$ is given by
\begin{equation}
\nabla_\mu\nabla^\mu\Psi = \frac{1}{\sqrt{-g}} \partial_\mu \left(\sqrt{-g} \: \partial^\mu \Psi \right) = 0,
\label{eqn:SourcelessScalarFieldEquation}
\end{equation}
where $g$ is the determinant of $g_{\mu\nu}$. The scalar field can be modally decomposed by expanding in terms of spherical harmonics $Y_{\ell m}$ such that
\begin{equation}
\Psi = \frac{1}{r} \sum_{\ell =0}^{\infty} \sum_{m=-\ell}^{\ell} \psi_{\ell m}(t,r) Y_{\ell m} (\theta, \varphi).
\label{eqn:ScalarSphericalHarmonicExpansion}
\end{equation}
We can show that in Schwarzschild spacetimes the radial and angular parts of the modal decomposition separate. The angular part reduces to $Y_{\ell m}$ hence we can write
\begin{equation}
\left[- \frac{r}{f} \psi_{,tt} + fr \; \psi_{,rr} + \frac{2M}{r} \psi_{,r} - \frac{1}{r} \left( \frac{2M}{r} + \ell(\ell+1) \right) \psi \right] Y_{\ell m}(\theta,\varphi) = 0,
\label{eqn:SourcelessScalarFieldEquationNoAngularDependence}
\end{equation}
where the $\ell m$ subscripts on the field have been dropped for brevity. We can write this in a simpler form using the null coordinates $u=t-r_*$ and $v=t+r_*$ where $r_*= r + 2M\ln[r/(2M)-1]$ is the tortoise coordinate. Rewriting Eq.\ (\ref{eqn:SourcelessScalarFieldEquationNoAngularDependence}) gives a wave equation of the form
\begin{equation}
\psi_{,uv} + V(r) \psi= 0,
\label{eqn:ScalarFieldEquationNull}
\end{equation}
where 
\begin{equation}
V(r) := \frac{f}{4r^2} \left( \frac{2M}{r} + \ell(\ell+1) \right).
\label{eq:ScalarPotential}
\end{equation}
Our convention is that when acting on a function of $u$ and $v$, $\partial_u$ and $\partial_v$ are always taken with fixed $v$ and fixed $u$, respectively.

\section{Vacuum implementation}  
\label{sec:VacuumScalarImplementation}

\begin{figure}[h!]
\centering
\includegraphics[width=0.7\linewidth]{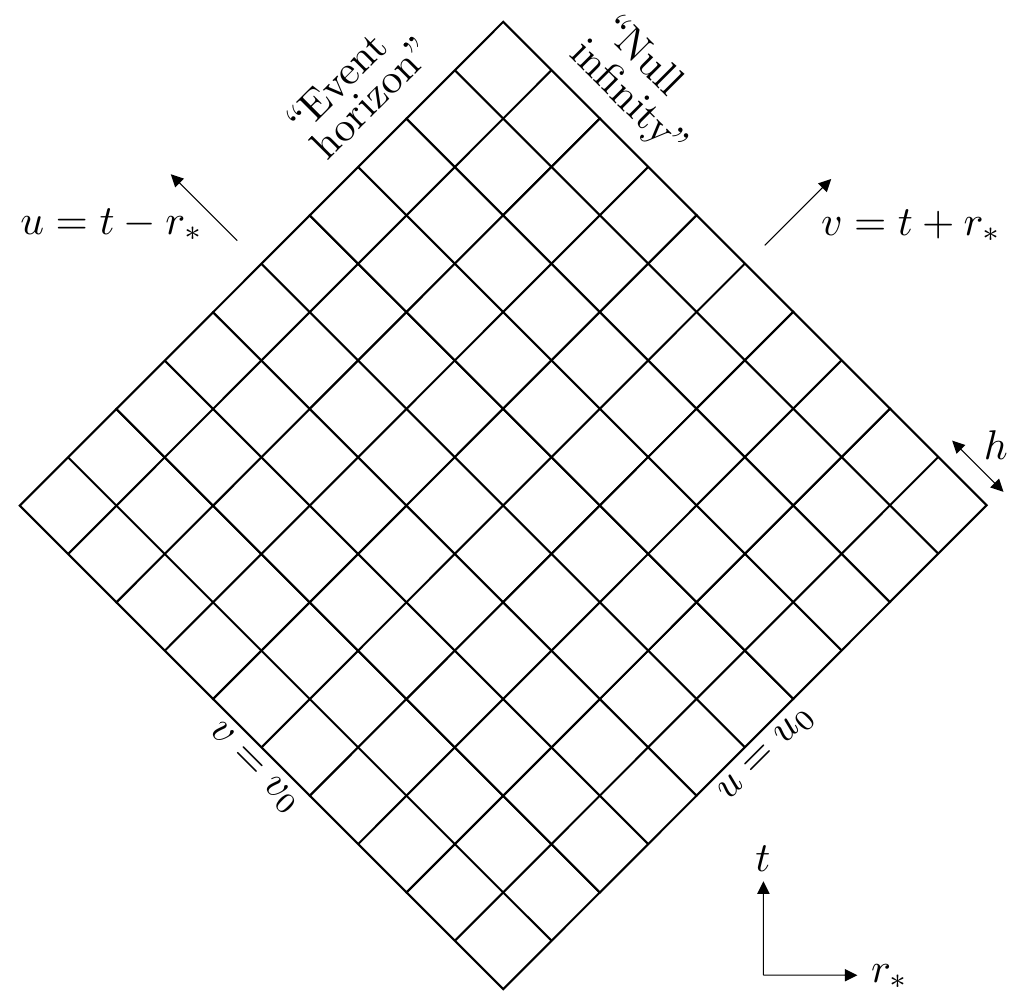}
\caption[Sketch of the 1+1D characteristic grid used in our vacuum numerical evolution]{Sketch of the 1+1D characteristic grid used in our vacuum numerical evolution of the scalar field $\psi_{\ell m}(t,r)$ outside a Schwarzschild black hole. The grid lines are uniformly spaced in Eddington-Finlkelstein coordinates $u,v$. Initial conditions are set on the rays $u=u_0$ and $v=v_0$. The evolution proceeds along successive $u={\rm const}$ rays.}
\label{uvGridVacuum}
\end{figure}

Our numerical implementation is a simple finite-difference solver based on 1+1D characteristic evolution in $u,v$ coordinates. The numerical domain is depicted in Fig.\ \ref{uvGridVacuum}. We use a fixed characteristic mesh, with uniform grid-cell dimensions $h\times h$, where $h$ is a small fraction of $M$ (typically $\sim M/10$ to $\sim M/100$ in our test runs). Characteristic initial data are set on two initial rays $u=u_0$ and $v=v_0$ (see the figure). Specifically, we use an artificial seed of a narrow Gaussian centred on the initial apex $(u_0,v_0)=(0,0)$. The data is evolved using a finite-difference version of Eq.\ (\ref{eqn:ScalarFieldEquationNull}) that has a local discretisation error of $O(h^4)$, leading to a quadratic convergence globally (i.e., the accumulated error scales like $h^2$).  A detailed description of our scheme is provided in Appendix \ref{app:FDSVac}. Our code takes as input the modal numbers $\ell$, $m$, as well as a range of numerical parameters such as $h$ and the coordinate ranges, and returns the scalar field modes $\psi_{\ell m}(t,r)$. 

Since our characteristic numerical domain has no timelike boundaries, there is no need to impose boundary conditions, and no way to actively control violations away from the desired retarded solution. This is not a problem when all ``nonphysical'' vacuum solutions of the field equation decay at late time, but can become a problem when there exist nonphysical solutions that fail to decay, or worse, grow at late time. One purpose of this vacuum calculation is to check there are no nonphysical modes which could contaminate future calculations.

\begin{figure}[H]
\centering
\includegraphics[width=0.8\linewidth]{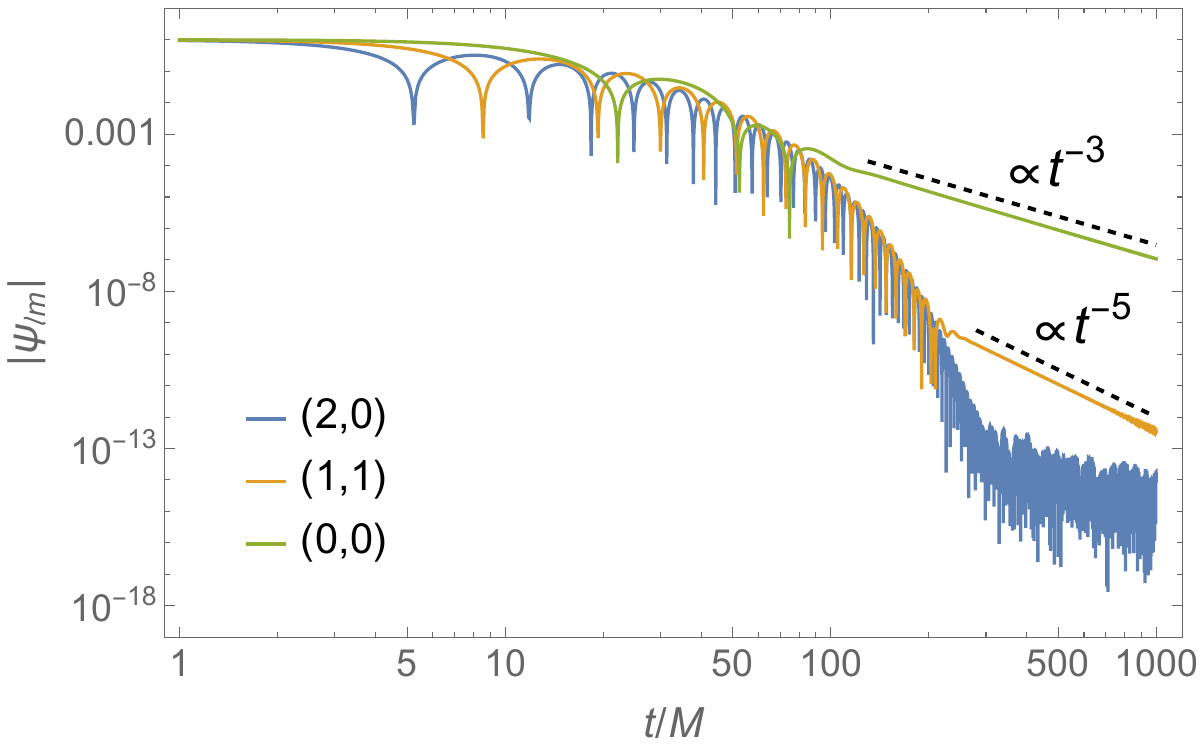}
\caption[A plot of the vacuum scalar field]{A plot of the vacuum scalar field $\psi_{\ell m}$ for various $(\ell,m)$ modes. The data shown is taken along a slice of constant $r_*=0$.  The dashed lines (${\rm cons} \times t^{-3}$ and ${\rm cons} \times t^{-5}$) represent the expected decay rate of the $\ell=0,1$ tails (respectively) and are shown for reference. The $(2,0)$ data becomes dominated by numerical round-off error before the late-time tail appears.}
\label{SourcelessScalarField}
\end{figure}

Figure \ref{SourcelessScalarField} demonstrates the behaviour of the vacuum scalar field $\psi_{\ell m}$ for various $\ell,m$ modes. The early time behaviour is reminiscent of the initial conditions. The next region is the typical quasi normal mode decay of perturbed black hole spacetimes. These quasi normal modes decay exponentially to leave a late time behaviour where the field falls off as $t^{-2 \ell -3}$ \cite{Price1972,Barack:1998bw}. This behaviour is exactly as expected hence we can conclude that future implementations will not contain non-decaying nonphysical vacuum solutions.

%% file: ScalarSource.tex

\section{Point-particle source}
\label{sec:ScalarFieldSource}

If we introduce a point-particle source $S_\Psi$ then the scalar field equation takes the form 
\begin{equation}
\frac{1}{\sqrt{-g}} \partial_\alpha \left(\sqrt{-g} \: \partial^\alpha \Psi \right) = - 4 \pi q \int_{-\infty}^\infty \frac{1}{\sqrt{-g}} \: \delta^4 \left( x^\mu - x_p^\mu(\tau) \right) \: d \tau =:S_\Psi,
\label{eqn:SourcedScalarFieldEquation}
\end{equation}
where $q$ is the scalar charge of the particle, $x_p^\mu(\tau):=(t_p(\tau),R(\tau),\theta_p(\tau),\varphi_p(\tau))$ is the particle's worldline, $\tau$ is proper time and $\delta^4$ is the four dimensional Dirac delta function \cite{BarackBurko2000}. We can expand the source using the relation
\begin{equation}
 \delta \left( \theta - \theta_p \right)  \delta \left( \varphi - \varphi_p \right) = \sum_{\ell ,m} Y_{\ell m}(\theta,\varphi)\bar{Y}_{\ell m}(\theta_p,\varphi_p) \; \sin\theta, 
\end{equation}
and $g=-r^4 \sin^2 \theta$ to give
\begin{equation}
S_\Psi = - 4 \pi q \sum_{\ell ,m} \int_{-\infty}^\infty \delta \left( t - t_p(\tau) \right) \delta \left(r - R(\tau) \right) \frac{1}{r^2} Y_{\ell m}(\theta,\varphi)\bar{Y}_{\ell m}(\theta_p(\tau),\varphi_p(\tau)) \: d \tau.
\end{equation}
We can evaluate the integral using the relation $d \tau = dt \: d\tau/dt = d t f_R(\tau)/E$ where $f_R(\tau) := 1-2M/R(\tau)$. This gives the result
\begin{equation}
S_\Psi(\tau) = - \frac{4 \pi q f_R(\tau)}{E R(\tau)^2} \delta \left(r - R(\tau)\right) \sum_{\ell ,m} Y_{\ell m}(\theta,\varphi) \bar{Y}_{\ell m}(\theta_p(\tau),\varphi_p(\tau)).
\end{equation}

The scalar field can be expanded in terms of spherical harmonics such that
\begin{equation}
\Psi = \frac{2\pi q}{r} \sum_{\ell =0}^{\infty} \sum_{m=-\ell}^{\ell} \psi_{\ell m}(t,r) Y_{\ell m} (\theta, \varphi).
\end{equation}
Note the extra factor of $2\pi q$ relative to Eq.\ (\ref{eqn:ScalarSphericalHarmonicExpansion}). We can utilise the orthogonality of the spherical harmonics to remove the angular dependence of the equation to give the result
\begin{equation}
\psi_{,uv} + V \psi = \frac{f_R^2}{2 E R} \delta \left(r - R\right) \bar{Y}_{\ell m}(\theta_p,\varphi_p).
\label{eqn:SourcedFieldEquationuv}
\end{equation}
This is the form of the scalar field equation that we will implement in the following sections. 

In our point-particle implementations the domain gets split into two vacuum regions ${\cal S}^<$ and ${\cal S}^>$ which are defined by $r<R(t)$ and $r>R(t)$ respectively. The boundary between these regions is the 1D surface $\cal S$ which corresponds to the worldine of the particle (i.e.\ $r=R(t)$). Figures \ref{uvGridCircular} and \ref{uvGridScatter} show the locations of $\cal S$ and $\cal S^\gtrless$ within the numerical grid for circular and scatter orbits respectively. 

%% file: ScalarCircular.tex

\section{Circular-orbit implementation}  
\label{sec:ScalarCirc}

First we consider the simple bound case of a point-particle on an equatorial circular orbit. This simplifies the problem by restricting the orbital parameters such that $x_{\text{p}}^\mu=(t, R, \pi/2, \Omega t)$ where $R$ is the (constant) radius and $\Omega:=\sqrt{M/R^3}$ is the orbital angular velocity. In fact, we can remove all the time dependence when considering only the static ($m=0$) modes using the relation $\bar{Y}_{\ell 0}(\theta_\text{p},\varphi_\text{p}) = \bar{Y}_{\ell 0}(\theta_\text{p},0)$. In this case we can obtain analytic solutions for Eq.\ (\ref{eqn:SourcedFieldEquationuv}) using Legendre polynomials of the first and second kinds, denoted $P_\ell$ and $Q_\ell$ respectively. The solution can be written as
\begin{equation}\label{ScalarAnalyticalSolution}
\psi = A r P_\ell \left( \rho \right) \Theta(r-R) + B r Q_\ell \left(\rho\right)\Theta(R-r),
\end{equation}
where $\Theta$ is the Heaviside step function, $\rho:=(r-M)/M$ and $A$, $B$ are constants to be determined. This form comes from the behaviour of each term in the asymptotic limits. $r P_\ell(\rho)$ diverges at radial infinity but is finite at the horizon. Similarly, $r Q_\ell(\rho)$ diverges at the horizon but is finite at radial infinity. The use of the Heaviside theta functions ensures that we obtain the physically relevant finite solutions. 

We can determine the values of the constants by substituting Eq.\ (\ref{ScalarAnalyticalSolution}) into the field equation (\ref{eqn:SourcedFieldEquationuv}) and comparing coefficients of the Heaviside and Dirac delta functions, which gives
\begin{eqnarray}
A R P_\ell \left( \rho_R \right) - B R Q_\ell \left(\rho_R\right) &=& 0, \\
\left( A r P_\ell \left( \rho \right) - B r Q_\ell \left(\rho\right) \right)_{,r} \Big|_{R} &=& - \frac{2}{E} \bar{Y}_{\ell 0}(\pi/2,0).
\end{eqnarray}
Solving these equations gives the analytical solution as
\begin{equation}
\psi_{\ell 0} = - \frac{2}{Er} \bar{Y}_{\ell 0}(\pi/2,0) \frac{Q_\ell \left( \rho_R \right) P_\ell \left( \rho \right) \Theta(r-R) + P_\ell \left( \rho_R \right) Q_\ell \left(\rho\right)\Theta(R-r)}{ P_{\ell ,r} \left( \rho_R \right) Q_{\ell } \left( \rho_R \right) - P_{\ell } \left( \rho_R \right) Q_{\ell ,r} \left( \rho_R \right)}.
\label{eqn:ScalarCircularAnalytic}
\end{equation}

\begin{figure}[H]
\centering
\includegraphics[width=0.7\linewidth]{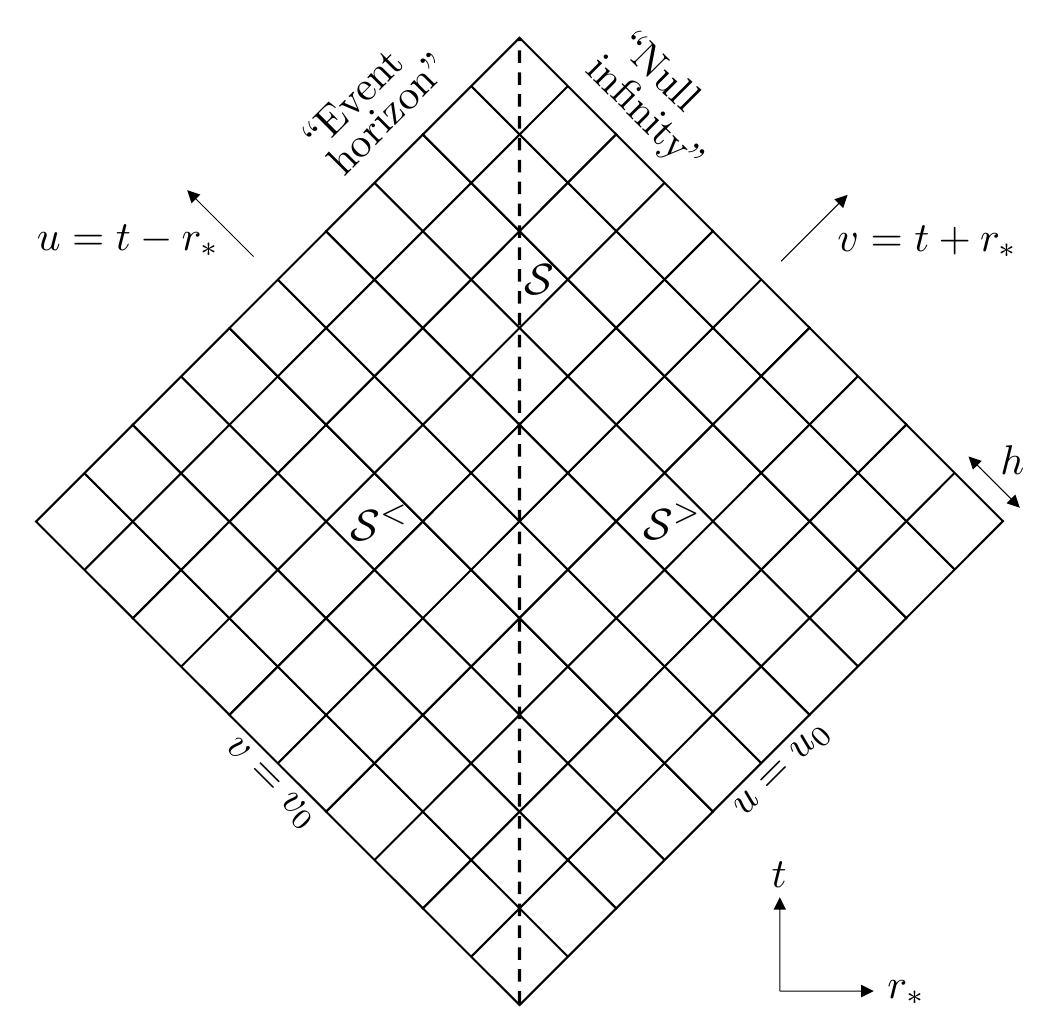}
\caption[Sketch of the 1+1D characteristic grid used in our circular orbit numerical evolution]{Sketch of the 1+1D characteristic grid used in our circular orbit numerical evolution of the scalar field $\psi_{\ell m}(t,r)$ outside a Schwarzschild black hole. The grid is centred along the worldline $\cal S$ corresponding to $r=R$ (dashed). $\cal S^\gtrless$ are the vacuum regions of the numerical domain with $r\gtrless R$.}
\label{uvGridCircular}
\end{figure}

The non-static, $m\neq 0$, modes do not have known analytic solutions. For these, we implement a numerical method to solve the scalar field equation (\ref{eqn:SourcedFieldEquationuv}) sourced by a circular equatorial orbit. The basis of the numerical method is an extension of the vacuum implementation detailed in Section \ref{sec:VacuumScalarImplementation}. Here we choose the initial apex of the grid $(u,v)=(u_0,v_0)$ to correspond to the point $(t,r)=(0,R)$. This ensures that the particle passes directly through the cells centred on $r=R$ and does not pass through any other cells, henceforth known as vacuum cells, as shown in Figure \ref{uvGridCircular}. The characteristic initial conditions are set such that $\psi(u,v_0)=\psi(u_0,v)=0$ (i.e.\ all field values along the initial rays are zero) and the evolution is purely sourced by the source of the field equation. Our finite-difference scheme (detailed in Appendix \ref{app:FDSScalarCircular}) has global quadratic convergence. The code has inputs of the orbital radius $R$, modal numbers $\ell$, $m$ (plus various numerical parameters) and returns the scalar field modes $\psi_{\ell m}(t,r)$. We have produced two identical implementations, one in \texttt{Mathematica} and another in \texttt{C++}, to enable cross-checks.  

\begin{figure}[H]
\centering
\includegraphics[width=0.8\linewidth]{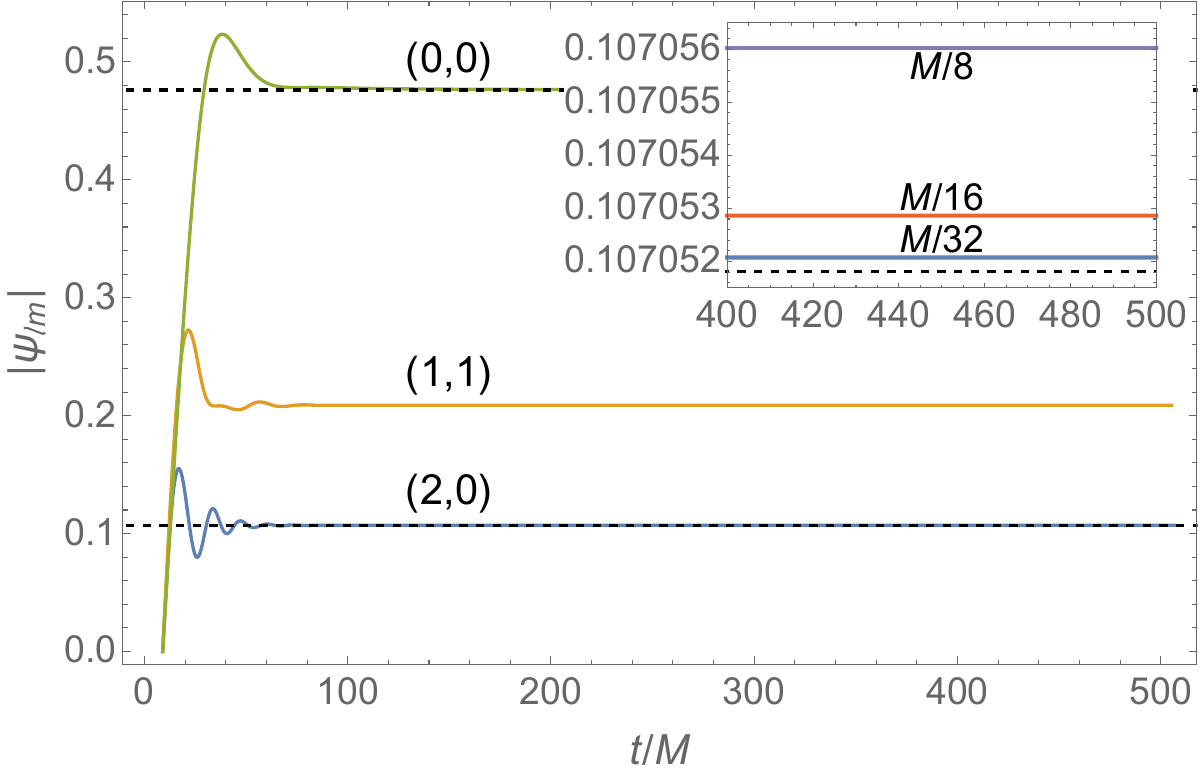}
\caption[A plot of the scalar field for the circular orbit case]{A plot of the scalar field $\psi_{\ell m}$ for the circular orbit case showing various $(\ell,m)$ modes. The data shown is taken along the particle's wordline at constant $r_*=9M$. Dashed lines represent the analytic solutions of the static modes as determined by Eq.\ (\ref{eqn:ScalarCircularAnalytic}). The early part of the data is contaminated by initial junk radiation, and it is to be discarded. The junk radiation decays away quicker for increasing $\ell$ as expected. The inset shows a detail from the $\psi_{20}$ (blue) curve for a sequence of runs with decreasing grid spacing, $h=\left\{\frac{1}{8},\frac{1}{16},\frac{1}{32}\right\}M$. Notice how the runs converge towards the analytic solution with increasing resolution.}
\label{CircularScalarField}
\end{figure}

Figure \ref{CircularScalarField} shows the behaviour of the field $\psi_{\ell m}$ along the particle's worldline, chosen to be at $r_*=9M$. Early time results are dominated by junk radiation from non-physical initial conditions. This radiation behaves in the same way as the compact vacuum perturbations discussed previously hence we expect a $t^{-2 \ell -3}$ decay. The late-time behaviour of the field approaches the true solution. This is explicitly shown for the static modes in Figure \ref{CircularScalarField} where the numerical data tends to the analytic solutions.

%% file: ScalarScatter.tex

\section{Hyperbolic-orbit implementation}
\label{sec:ScalarScatter}

\begin{figure}[H]
\centering
\includegraphics[width=0.65\linewidth]{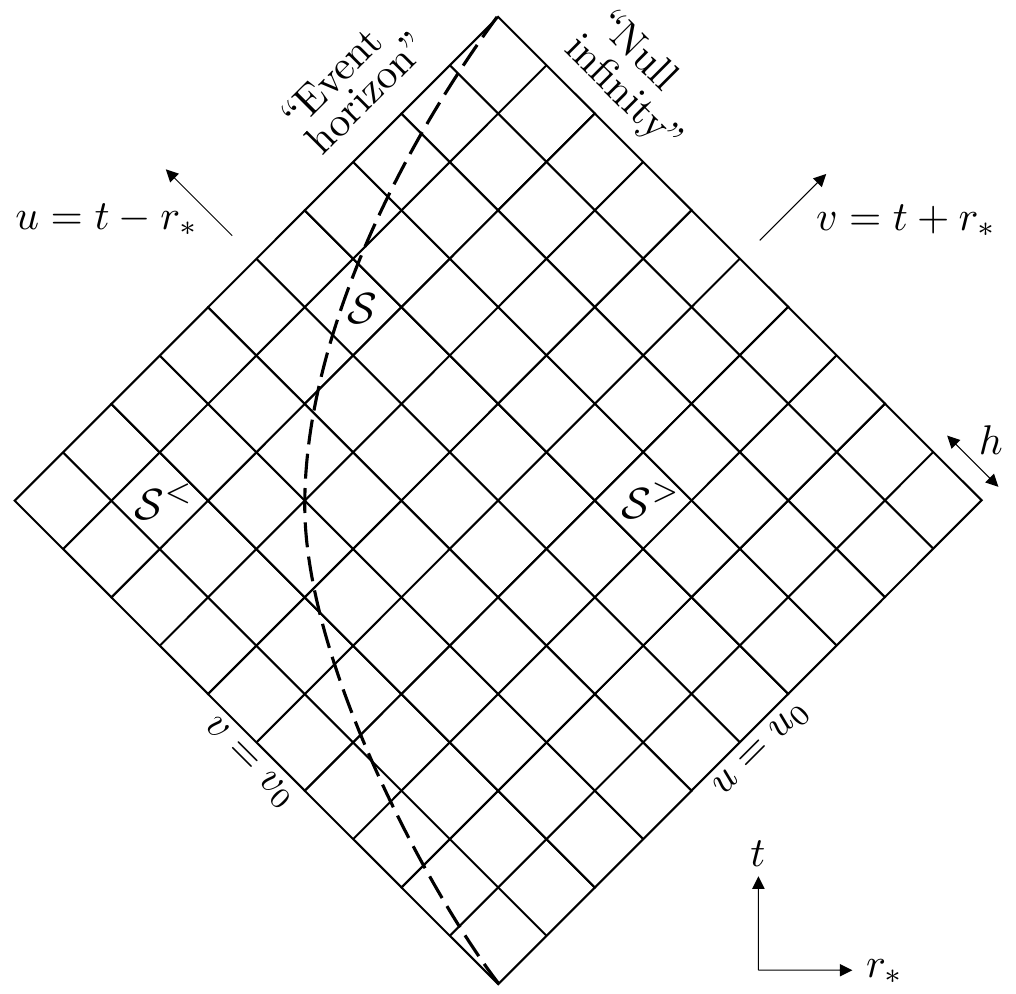}
\caption[Sketch of the 1+1D characteristic grid used in our scatter orbit numerical evolution]{Sketch of the 1+1D characteristic grid used in our scatter orbit numerical evolution of the scalar field $\psi_{\ell m}(t,r)$ outside a Schwarzschild black hole. The grid is centred along the maximal value of the particle's radius within the numerical domain $r=R_{\rm init}$.}
\label{uvGridScatter}
\end{figure}

Here we extend our implementation to be able to calculate the scalar field for unbound orbits. The primary difference between this and the circular orbit case is the trajectory of the particle through the numerical domain. The 1D worldline $\cal S$ no longer passes directly through the vertices of the cells, as shown in Figure \ref{uvGridScatter}. The particle's trajectory is calculated through integration of the geodesic equations of motion (\ref{tdot}) -- (\ref{rdot}) with initial conditions $R(t=0)=R_{\rm min}$ and $\varphi_{\rm p}(t=-\infty)=0$ and is then used to determine which cells the particle enters. This information is fed into the numerical evolution as the entry and exit coordinates of the wordline through each particle cell and is applied through the finite-difference scheme detailed in Appendix \ref{app:FDSScalarGeneric}. The numerical domain is constructed such that the evolutions starts with the initial apex at the point $R=R_{\rm init}$ (a user input) and finishes when the particle returns to $R=R_{\rm init}$ after being scattered. The code takes as inputs the modal numbers $\ell$, $m$, orbital parameters $v_\infty$, $b$, various numerical parameters (including $R_{\rm init}$) and returns the scalar field modes $\psi_{\ell m}(t,r)$.

A more detailed version of the numerical algorithm (as applied to the gravitational case) is presented in Section \ref{sec:NumericalAlgorithmScatter}.

\begin{figure}[h!]
\centering
\includegraphics[width=0.7\linewidth]{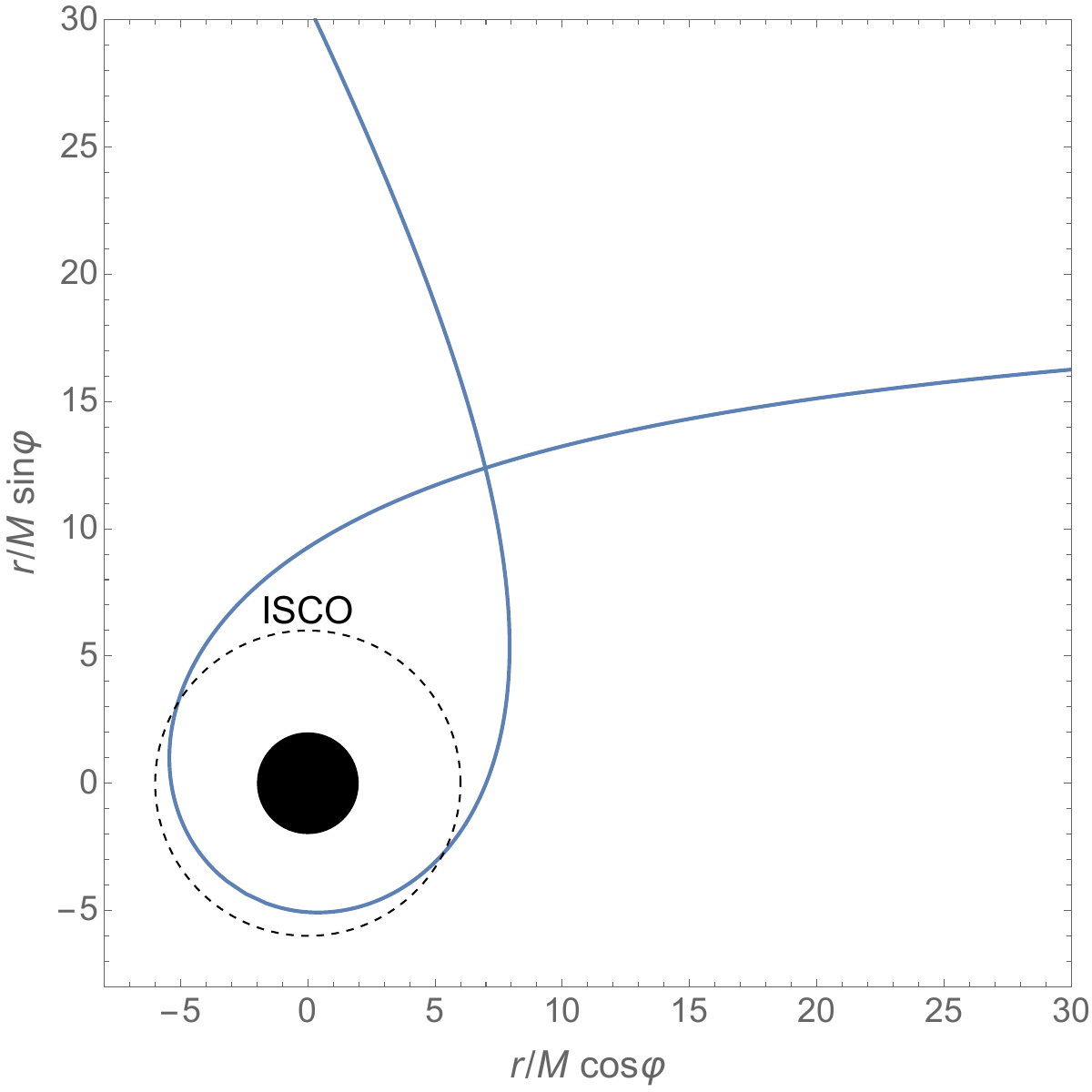}
\caption[The sample scatter geodesic orbit used for our numerical illustration]{The sample scatter geodesic orbit used for our numerical illustration, with parameters given in Eqs.\ (\ref{sampleorbit_vb}) and (\ref{sampleorbit}). The orbit is plotted in the equatorial plane using Cartesian-like coordinates $(x,y)=(r\cos\varphi,r\sin\varphi)$.  The location of the innermost stable circular orbit (ISCO) is shown for reference. The deflection angle of this strong-field orbit is $\delta\varphi \simeq 301^\circ$.}  
\label{orbit}
\end{figure}

For the numerical demonstration to be presented below we have picked a sample strong-field scatter geodesic with 
\begin{equation}\label{sampleorbit_vb}
\vinf = 0.2, \qquad \qquad b=21M,
\end{equation} 
corresponding to 
\begin{align}
R_{\rm min}\simeq 4.98228M,\qquad  E\simeq&\: 1.02062,\qquad L\simeq 4.28661M, \nonumber \\
\quad e\simeq 1.1948, \qquad  p\simeq&\: 10.9351, \qquad \delta\varphi \simeq 301^\circ.
\label{sampleorbit}
\end{align}
The orbit is depicted in Figure \ref{orbit}. 

\begin{figure}[H]
\centering
\includegraphics[width=0.8\linewidth]{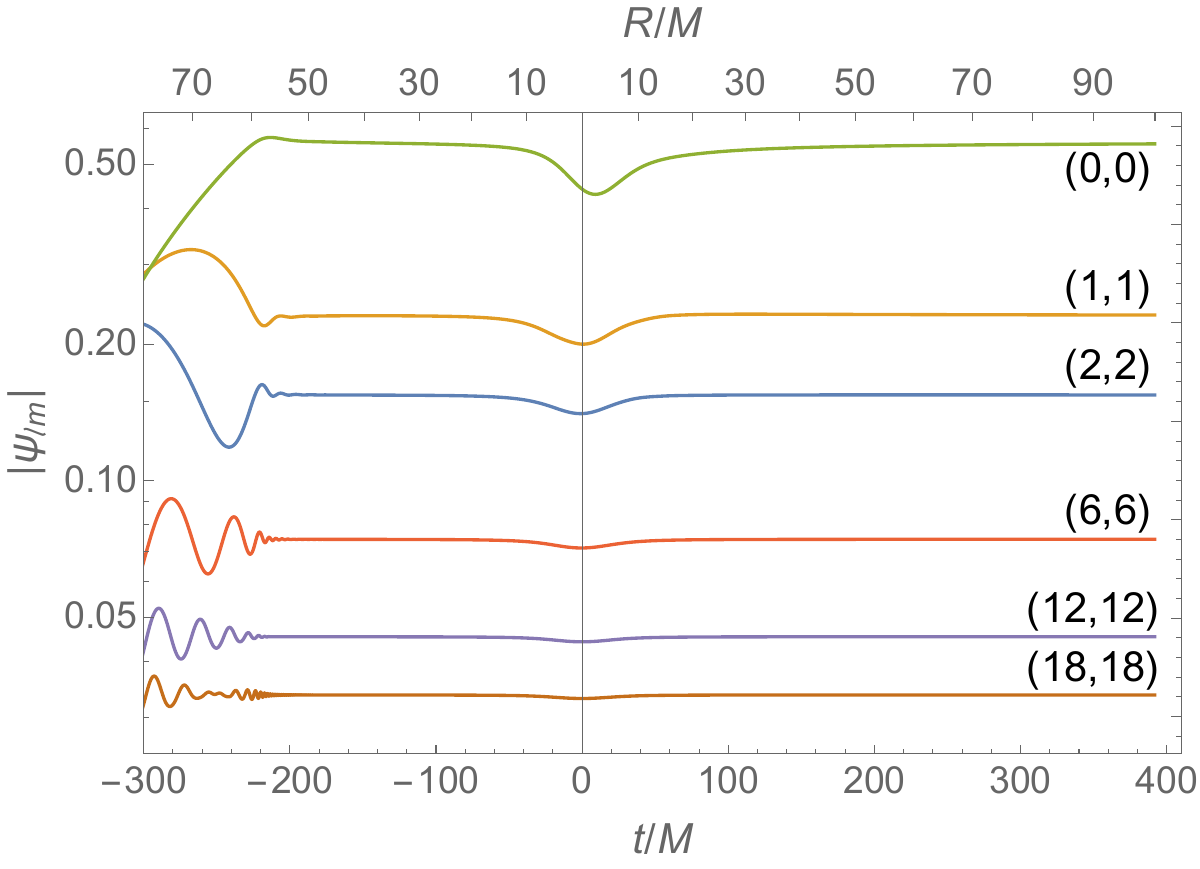}
\caption[The scalar field along the worldine of a hyperbolic orbit]{The scalar field $\psi_{\ell m}$ along the particle’s worldline for the orbit shown in Figure \ref{orbit} for a sample of $(\ell,m)$ modes. Here we show $|\psi_{\ell m}(t,R(t))|$ as a function of time $t$ (lower scale) and orbital radius $R$ (upper scale). The periastron location at $t=0$ is indicated with a vertical line. The early part of the data is contaminated by initial junk radiation, and is to be discarded.}
\label{ScatterScalarField}
\end{figure}

\begin{figure}[h!]
\centering
\includegraphics[width=0.8\linewidth]{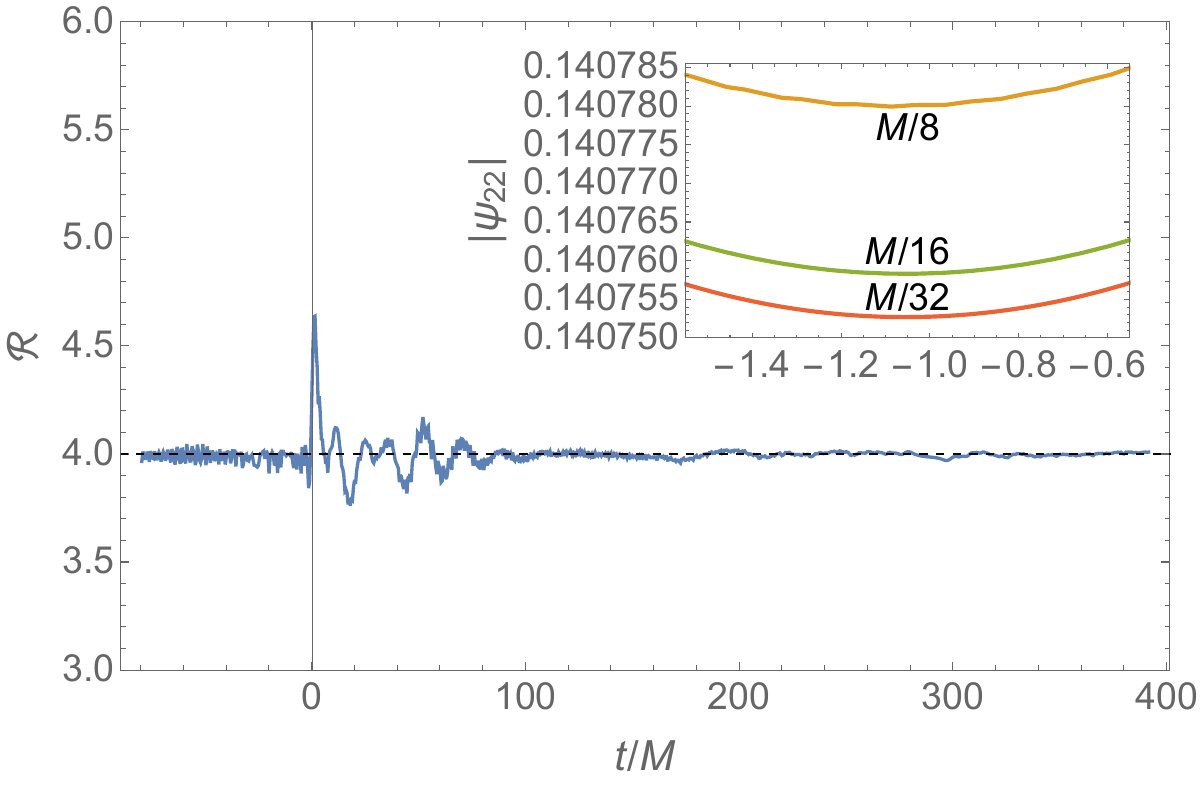}
\caption[Convergence test of the numerical solution of the scalar field]{Convergence test for the $(\ell,m)=(2,2)$ numerical solution. The inset shows a detail from the $|\psi_{22}|$ (blue) curve in Figure \ref{ScatterScalarField}, for a sequence of runs with decreasing grid spacing, 
$h=\left\{\frac{1}{8},\frac{1}{16},\frac{1}{32}\right\}M$. The main plot quantifies the convergence rate: It shows the ratio ${\cal R}:=\left|\psi_8-\psi_{16}\right|/\left|\psi_{16}-\psi_{32}\right|$ as a function of $t$ along the orbit, where a subscript `8' (e.g.)\ denotes a calculation with grid spacing $h=M/8$. A ratio of ${\cal R}=4$ is indicative of quadratic convergence.}
\label{ScalarConv}
\end{figure}

Figure \ref{ScatterScalarField} demonstrates the behaviour of the field $\psi_{\ell m}(t,r)$ along the worldline of the particle, for a sample of $\ell, m$ values. The evolution begins (and ends) when the particle is at $R_{\rm init} = 100M$. We have performed convergence tests to confirm that our code exhibits a quadratic global convergence rate in $h$, as it is designed to do. An example is shown in Figure \ref{ScalarConv}.

\begin{figure}[h!]
\centering
\includegraphics[width=0.8\linewidth]{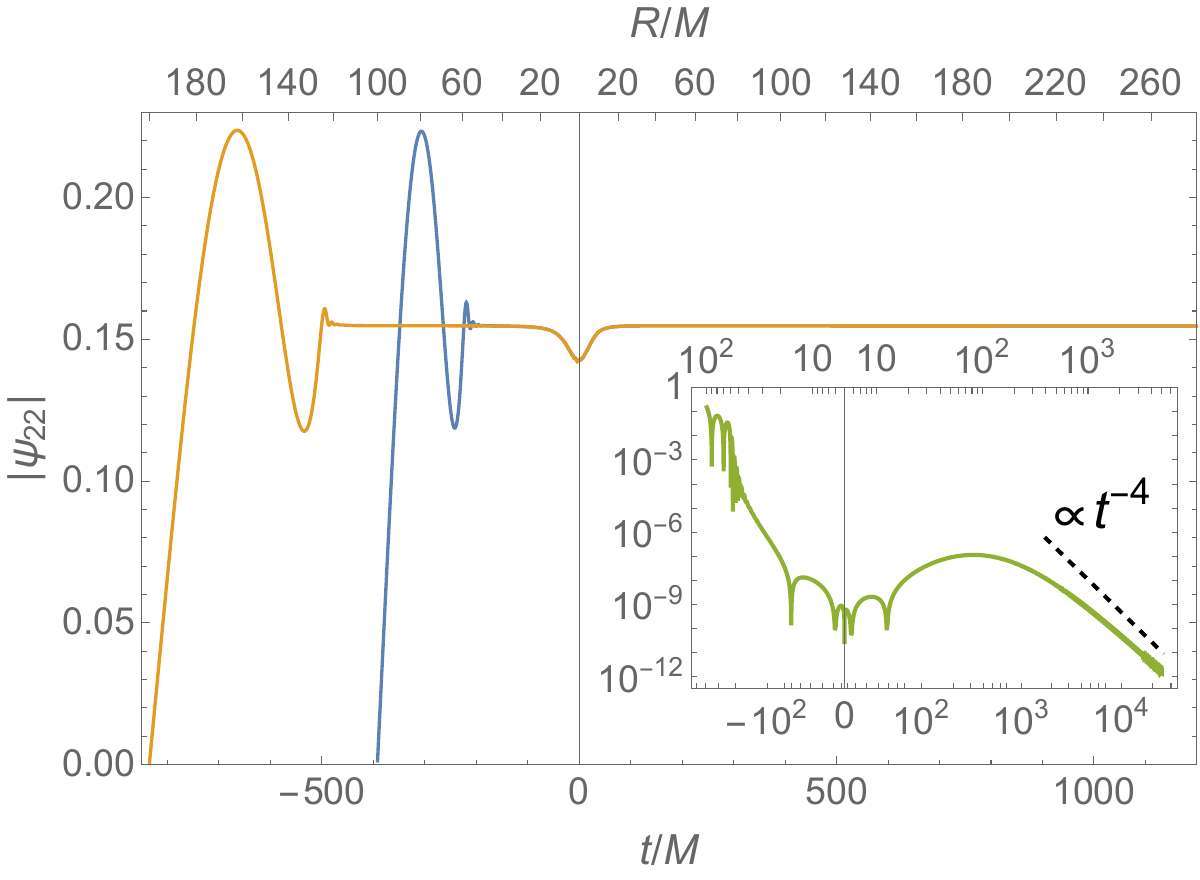}
\caption[A comparison of the scalar field calculated using numerical domains of different sizes]{Numerical results for $|\psi_{22}|$ on the particle's worldline, as calculated with $R_{\rm init}=100M$ (blue) and with $R_{\rm init}=200M$ (orange).  The comparison illustrates how, reassuringly, the ``clean'' portion of the data is insensitive to $R_{\rm init}$, up to a small error that dies off in time. The inset displays the relative difference between the two curves, showing a $t^{-4}$ fall-off at late time, consistent with the theoretically predicted decay rate.}
\label{ScalarRMax}
\end{figure}

Figure \ref{ScalarRMax} illustrates how, reassuringly, the ``clean'' part of the data appears to be insensitive to the value of $R_{\rm init}$, up to a small decaying difference. The decay rate of compact vacuum perturbations at late time $t\gg|r_*|$ is given to leading order in $M/t$ and $M/u$ by
\begin{equation}
\psi \propto \frac{1}{t^{2\ell+3}} r P_\ell \left( \rho \right),
\end{equation}
where we recall $P_\ell$ is the Legendre polynomial of the first kind and $\rho=(r-M)/M$ \cite{Barack:1998bw}. We are interested in the late-time behaviour along the worldline $r=R\propto t$ hence we can expand the Legendre polynomial to give $P_\ell(\rho) \propto r^\ell$ and can write
\begin{equation}
\psi \propto \frac{1}{t^{\ell+2}}.
\end{equation}
As the figure demonstrates, this is the decay rate we see in the difference of the field calculated using different values of $R_{\rm init}$. This means that we can use $R_{\rm init}$ as a control parameter enabling us to evaluate the level of residual contamination from initial junk.

%% file: ScalarSF.tex

In this chapter we calculate first-of-their-kind results for the scalar self-force for a hyperbolic orbit. We then present preliminary results for the calculation of the self-force correction to the scatter angle.

The equation of motion for a particle with mass $\mu$ and scalar charge $q$ is given by
\begin{equation}
\mu\frac{D^2 x_p^\mu}{d\tau^2} = F^\mu,
\end{equation}
where we recall $\tau$ is the proper time, $D/d\tau := u^\mu \nabla_\mu$ is the covariant derivative along the worldline $x_p$, which has an associated four-velocity $u^\mu= dx_p^\mu/d\tau$. We obtain the scalar self-force $F^\mu$ via 
\begin{equation}
F^\mu = q^2 \left( \delta^{\mu}_{\:\:\:\nu} + u^\mu u_\nu \right) {\tilde F}^\nu,
\label{eqn:scalarSF}
\end{equation}
where ${\tilde F}_\nu$ is constructed from the gradient of the scalar field using mode-sum regularisation, as follows.

First, we introduce the $\ell$-mode contribution 
\begin{equation}
{\tilde F}_{\mu\ell}^{\rm full} := 2\pi \sum_{m=-\ell}^\ell \nabla_\mu \left( \frac{1}{r} \psi_{\ell m}(t,r) Y_{\ell m}(\theta, \varphi)\right)\Bigg|_{x_p}.
\label{eqn:ScalarSFModes}
\end{equation}
Note that this is only defined along the worldline of the particle. Ref.\ \cite{BarackOri2000} showed that we can perform mode-sum regularisation at the level of this new function:
\begin{equation}
{\tilde F}_\mu = \sum_{\ell=0}^{\infty} \left[ {\tilde F}^{\rm full}_{\mu\ell} - {\tilde F}_\mu^{[-1]}{\cal L}_{-1} - {\tilde F}_\mu^{[0]} - {\tilde F}_\mu^{[2]} {\cal L}_2 - {\tilde F}_\mu^{[4]}{\cal L}_4 - {\tilde F}_\mu^{[6]}{\cal L}_6 - {\cal O}\left(\ell^{-8}\right) \right],
\label{eqn:ModeSumRegScalar}
\end{equation}
where we recall ${\cal L}_{-1} = 2\ell+1$ and define
\begin{equation}
{\cal L}_2 := \frac{1}{(2\ell-1)(2\ell+3)}, \qquad \qquad {\cal L}_4 := \frac{1}{(2\ell-3)(2\ell-1)(2\ell+3)(2\ell+5)}, \nonumber
\end{equation}
\begin{equation}
{\cal L}_6 := \frac{1}{(2\ell-5)(2\ell-3)(2\ell-1)(2\ell+3)(2\ell+5)(2\ell+7)}.
\label{eqn:ellExpansion}
\end{equation}
The first few regularisation parameters are given by
\begin{equation}
{\tilde F}_t^{[-1]} = \frac{\pm E \dot R }{2f_R(L^2+R^2)}, \qquad {\tilde F}_r^{[-1]} = -\frac{\pm E}{2f_R(L^2+R^2)}, \nonumber
\end{equation}
\begin{equation}
{\tilde F}_t^{[0]} = \frac{E R \dot R}{Lf_R\pi \sqrt{L^2+R^2}} ({\cal K - E}), \qquad {\tilde F}_\varphi^{[0]} = \frac{E R \dot R}{f_R\pi \left(L^2+R^2\right)^{3/2}} ({\cal K} - 2 {\cal E}),\nonumber
\end{equation}
\begin{equation}
{\tilde F}_r^{[0]} = \frac{\left(2E^2R^2-f_R(L^2+R^2)\right){\cal E}-\left(E^2R^2+f_R(L^2+R^2)\right){\cal K}}{f_R R \pi \left(L^2+R^2\right)^{3/2}},
\end{equation}
with
\begin{equation}
{\cal K} := \El_1 \left(\frac{\pi}{2}; \frac{L^2}{L^2+R^2}\right), \qquad {\cal E} := \El_2 \left(\frac{\pi}{2}; \frac{L^2}{L^2+R^2}\right).
\end{equation}
We recall that $E$ and $L$ are the energy and angular momentum of the orbit (respectively), $R(t)$ is the radial coordinate of the orbit, $f_R=1-2M/R$, $\El_1$ and $\El_2$ are the incomplete elliptic integrals of the first and second kind respectively [c.f.\ Eqs.\ (\ref{eqn:El1}) and (\ref{eqn:El2})] and an overdot represents a derivative with respect to $t$. The $\pm$ sign in the expressions for ${\tilde F}_t^{[-1]}$ and ${\tilde F}_r^{[-1]}$ corresponds to whether the derivative of the field at the worldline [as used in Eq.\ (\ref{eqn:ScalarSFModes})] is taken in $\cal S^>$ ($+$) or $\cal S^<$ ($-$). The higher-order regularisation parameters are cumbersome so we refer the reader to Section V.C.\ of Ref.\ \cite{Heffernan2012} or the RegularizationParameters package of the Black Hole Perturbation Toolkit \cite{BHPToolkit}.

\section{Sample results}

\begin{figure}[H]
\centering
\includegraphics[width=0.8\linewidth]{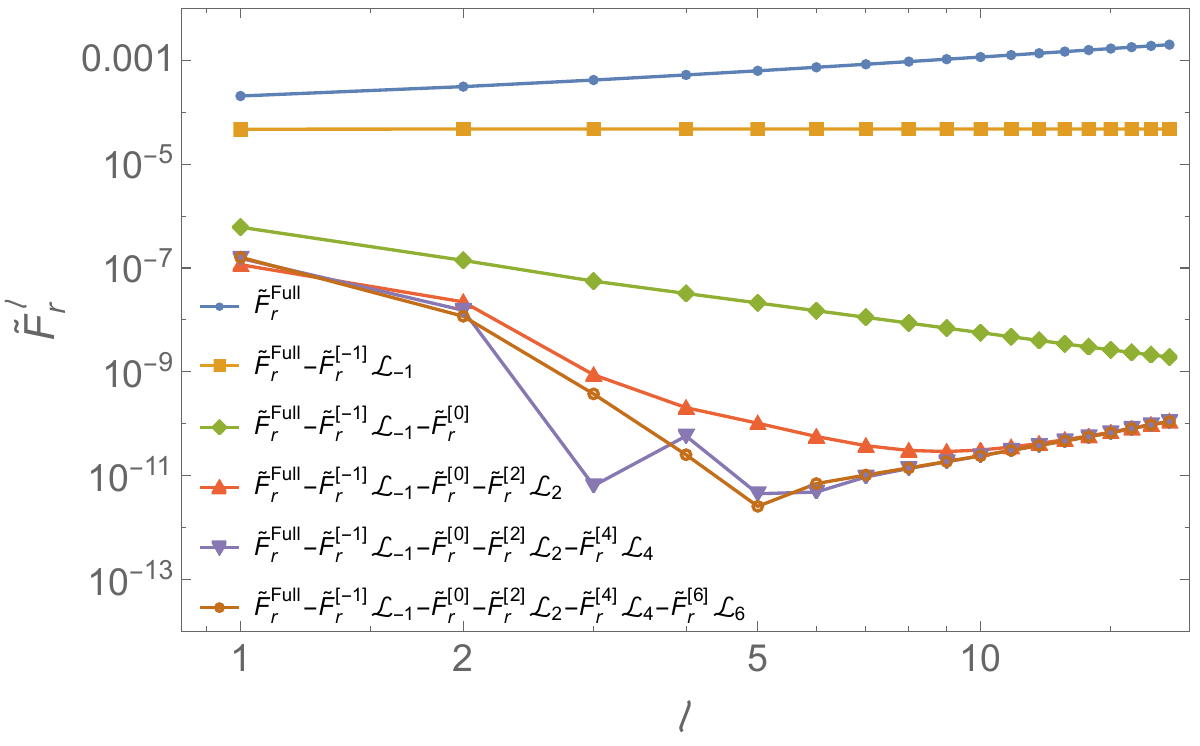}
\caption[The regularised radial derivative of the scalar field for a hyperbolic orbit]{The regularised radial derivative of the scalar field ${\tilde F}_r^\ell $ as a function of $\ell$ for a hyperbolic orbit. The data converges quicker with $\ell$ when additional analytic regularisation parameters are subtracted. The data fails to converge when using $F_r^{[n\geq2]}$ due to other dominant errors in the data.}
\label{FlPlot}
\end{figure}

The results displayed here are constructed from the data presented in Section \ref{sec:ScalarScatter} for the orbit shown in Figure \ref{orbit} with parameters given in Eqs.\ (\ref{sampleorbit_vb}) and (\ref{sampleorbit}). The SF was calculated using Eqs.\ (\ref{eqn:scalarSF}) -- (\ref{eqn:ModeSumRegScalar}) with scalar field modes up to $\ell=18$. We investigated how the solution converged when using high-order regularisation parameters. An example is shown in Figure \ref{FlPlot}. The results initially converge quicker with an increasing number of regularisation parameters until there is a plateau. This suggests that there is another source of dominant error in the data and using regularisation parameters of even higher-order will not improve the accuracy of the results.

\begin{figure}[H]
\centering
\includegraphics[width=0.8\linewidth]{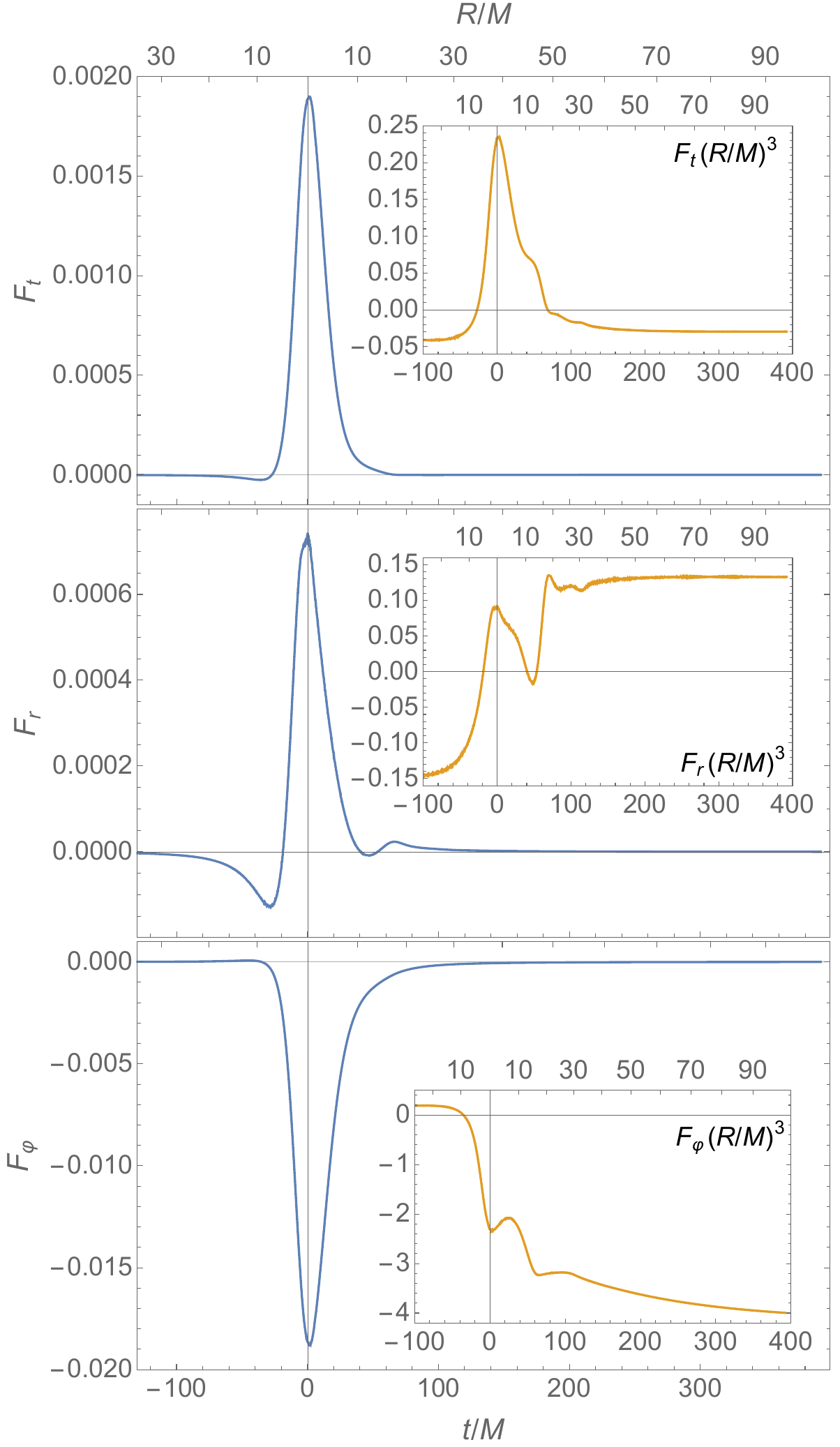}
\caption[Numerical results for the regularised scalar self-force for a hyperbolic orbit]{Numerical results for the regularised scalar self-force $F_t$ ({\it top}), $F_r$ ({\it middle}), and $F_\varphi$ ({\it bottom}) for the orbit shown in Figure \ref{orbit} as a function of time $t$ (lower scale) and orbital radius $R$ (upper scale). The periastron location at $t=0$ is indicated with a vertical line. The insets show the same data rescaled by a factor $(R/M)^3$. The force exhibits the lagging peak and post-periastron undulation features discussed in the text. The multiplication by $(R/M)^3$ makes the undulation features more prominent compared with the main plots.}
\label{FPlots}
\end{figure}

Figure \ref{FPlots} shows the behaviour of the non-zero components of the regularised self-force. Notable features include (i) the small lag between the peak of the field and the periastron passage and (ii) the small undulation in the field amplitude not long after periastron passage. The periastron lag has been observed before in calculations along eccentric orbits (see, e.g., \cite{Haas07}); it is attributed to the effect of ``tail'' contributions to the self-field, which peak in amplitude soon after periastron. The undulation, we suggest, is a weak manifestation of the quasinormal-mode excitation phenomenon observed in self-field calculations for highly eccentric orbits \cite{Nasipak:2019hxh,Thornburg:2019ukt}. Both features are associated with ``tail'' contributions to the self-field, and are less visible at larger $\ell$, where the ``direct'' part of the field is more dominant.  We have not conducted here a more detailed study of the above physical features. 

\section{Correction to the scatter angle}

With the scalar SF results we can now calculate the conservative scatter angle correction $\delta\varphi^{(1)}$ using either of Eqs.\ (\ref{deltaphi1_final}) or (\ref{deltaphi1_final_method2}). For clarity we denote the correction calculated with method I (\ref{deltaphi1_final}) as $\delta\varphi^{(1)}_{\rm I}$ and with method II (\ref{deltaphi1_final_method2}) as $\delta\varphi^{(1)}_{\rm II}$.

In order to perform the calculation we must obtain the conservative part of the SF using
\begin{equation}
F_\alpha^{\rm cons}(t) = \frac{1}{2} \left[ F_\alpha (t) - F_\alpha (-t) \right], \qquad F_r^{\rm cons}(t) = \frac{1}{2} \left[ F_r (t) + F_r (-t) \right],
\label{eqn:ConsSF}
\end{equation}
for $\alpha=t,\varphi$. We now have the necessary inputs to calculate the SF correction to the scatter angle using either of Eqs.\ (\ref{deltaphi1_final}) or (\ref{deltaphi1_final_method2}).

\begin{figure}[h!]
\centering
\includegraphics[width=0.8\linewidth]{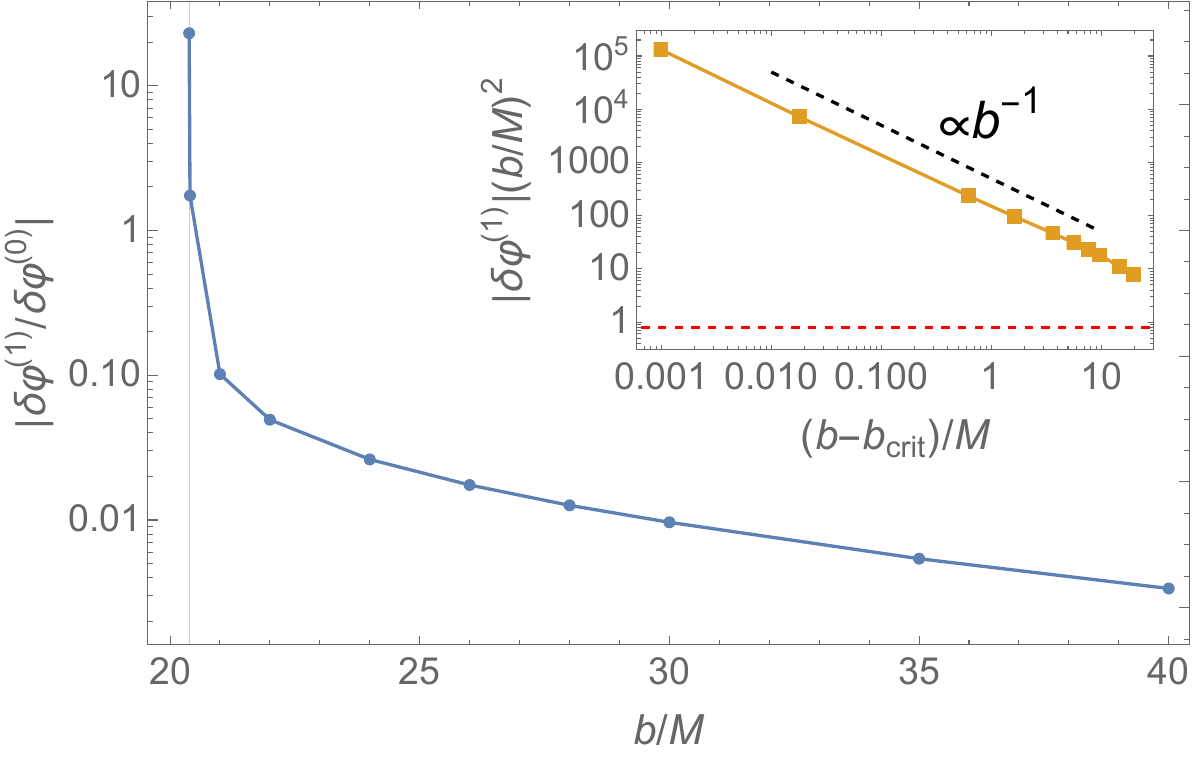}
\caption[The conservative scalar self-force correction to the scatter angle]{The conservative scalar self-force correction to the scatter angle for the orbits shown in Table \ref{tab:Corrv0.2}. Here we show $|\delta\varphi^{(1)}_{\rm II}/\delta\varphi^{(0)}|$ as a function of impact parameter $b$. The vertical line represents the critical value of $b$ at which the orbit is marginally bound $b_{\rm crit}=20.382M$. The inset plot shows the first-order correction $|\delta\varphi^{(1)}_{\rm II}|$ scaled by $(b/M)^2$ as a function of $b-b_{\rm crit}$. The red dashed line represents the 2PM solution $\delta\varphi^{(1)}_{\rm 2PM} = -\pi/(2b)^2$ of Gralla \& Lobo \cite{GrallaLobo2021}. The data decays as a $1/b^3$ power law, as shown by the black dashed reference line.}
\label{ScatAnglePlot}
\end{figure}

Table \ref{tab:Corrv0.2} shows the value of the correction to the scatter angle for a variety of orbits. The results show that there is good agreement between the values as calculated using methods I and II. Figure \ref{ScatAnglePlot} gives a graphical representation of the same data. There is a clear blow up of values (which scales as $\sim 1/b^3$) near the critical value of the impact parameter $b_{\rm crit}$. This is not unexpected as the orbit penetrates deeper into the strong-field for an increased amount of time as $b\rightarrow b_{\rm crit}$, hence we expect a larger correction. The inset plot of Figure \ref{ScatAnglePlot} shows that our results do not match well with the PM results. However, our results do seem to be tending towards the PM value as we approach the large-$b$ regime. 

\begin{table}[h]
\centerline{%
\begin{tabular}{c|c|c|c|c|c}
\toprule
\cmidrule(r){1-2}
$b/M$ & $\rmin/M$ & $\delta\varphi^{(0)}$ & $\delta\varphi^{(1)}_{\rm I}$ & $\delta\varphi^{(1)}_{\rm II}$ & $\left|\left(\delta\varphi^{(1)}_{\rm II}-\delta\varphi^{(1)}_{\rm I}\right)/\delta\varphi^{(1)}_{\rm II}\right|$ \\
\midrule
$20.383$ & $3.89720$ & $13.8325$ & $-319.307$ & $-319.043$ & 0.0827\% \\ 
$20.4$ & $4.01885$ & $9.93644$ & $-17.3432$ & $-17.3474$ & 0.0244\% \\ 
$21$ & $4.98228$ & $5.25737$ & $-0.535591$ & $-0.535503$ & 0.0164\% \\ 
$22$ & $5.95946$ & $4.05494$ & $-0.199893$ & $-0.199869$ & 0.0120\% \\ 
$24$ & $7.64287$ & $3.11464$ & $-0.081656$ & $-0.081661$ & 0.00612\% \\ 
$26$ & $9.25543$ & $2.63589$ & $-0.045992$ & $-0.0460113$ & 0.0419\% \\ 
$28$ & $10.8670$ & $2.32189$ & $-0.0293214$ & $-0.0292618$ & 0.204\% \\ 
$30$ & $12.4959$ & $2.09184$ & $-0.0201692$ & $-0.0201483$ & 0.104\% \\ 
$35$ & $16.6697$ & $1.70280$ & $-0.0091889$ & $-0.0092003$ & 0.124\% \\ 
$40$ & $20.9833$ & $1.44954$ & $-0.0048744$ & $-0.0048748$ & 0.00821\% \\ 
\bottomrule
\end{tabular}}
\caption[Results for the conservative scalar self-force correction to the scatter angle]{Results for the conservative scalar self-force correction to the scatter angle calculated using methods I $(\delta\varphi^{(1)}_{\rm I})$ and II $(\delta\varphi^{(1)}_{\rm II})$ as detailed in Sections \ref{sec:SFCorrectionMI} and \ref{sec:SFCorrectionMII} respectively. All of the orbits shown have $v_\infty=0.2$. We show the geodesic values of the periastron $\rmin$ and scatter angle $\delta\varphi^{(0)}$ as well as the relative difference between the results of the two methods for reference. Recall that the full scatter angle is given by $\delta\varphi=\delta\varphi^{(0)}+\eta \delta\varphi^{(1)}+{\cal O}\left( \eta^2 \right)$ [c.f.\ Eq.\ (\ref{phisplit})].}
\label{tab:Corrv0.2}
\end{table}

To explore the PM regime we need to satisfy the assumption $\vinf^2b\gg1$, however, this is extremely computational expensive. The evolution time of our scalar field code gradually increases with larger $R_{\rm min}$ and/or larger $\vinf$. In large-$R_{\rm min}$ runs we are penalised by the longer evolution time, and in the large-$\vinf$ case the slower decay of initial junk along the orbit requires a larger value of $R_{\rm init}$ (and again a longer run). We estimate that a run with $(\vinf, b)=(0.5,400M)$ [corresponding to $\vinf^2b=100$] would take timescales of months to years with our current capabilities. Our code is far from being optimal so the computational burden could be drastically reduced with appropriate computational techniques. An alternative method would be to use approximate initial conditions to reduce the amplitude of the junk radiation and thus the time they need to decay. This could be achieved by performing an initial low resolution run from which we extract the required initial conditions that feed into a high resolution run.

We performed some tests to determine the dominant source of numerical error. Table \ref{tab:CorrNumParam} shows the values of the corrections as calculated for the orbital parameters given in Eqs.\ (\ref{sampleorbit_vb}) and (\ref{sampleorbit}), whose orbit is depicted in Figure \ref{orbit}. The largest difference, by over a magnitude, is when we increase the size of the numerical domain (i.e.\ increase $R_{\rm init}$), due to the large-$R$ contributions of the SF. One possibility to reduce the error would be to simply increase the value of $R_{\rm init}$ in our numerical runs. However, this would come with a large computational cost as the runtime of the code scales as $\sim R_{\rm init}^2$. An alternative method would be to fit an analytic tail to the SF data such that we can extended our SF results to the limit $R\rightarrow\infty$ and thus perform the entire integrals (\ref{deltaphi1_final}) or (\ref{deltaphi1_final_method2}). Preliminary tests show that adding the analytic tail alters the result of the correction to the scatter angle by $\sim 3\%$. We use this value as a tentative estimate of the error bar on our results for $\delta\varphi^{(1)}$.

\begin{table}[H]
\centerline{%
\begin{tabular}{c|c|c|c|c}
\toprule
\cmidrule(r){1-2}
$l_{\rm max}$ & $M/h$ & $R_{\rm init}/M$ & $\delta\varphi^{(1)}_{\rm II}$ & Rel diff  \\
\midrule
$15$ & $128$ & $100$ & $-0.535503$ & - \\
$18$ & $128$ & $100$ & $-0.535564$ & 0.0114\% \\
$15$ & $256$ & $100$ & $-0.53527$ & 0.0435\% \\
$15$ & $128$ & $150$ & $-0.541908$ & 1.20\% \\
\bottomrule
\end{tabular}}
\caption[The conservative self-force correction to the scatter angle for a variety of numerical parameters]{The conservative self-force correction to the scatter angle using a variety of numerical parameters for the orbit shown in Figure \ref{orbit}. The numerical parameters shown are maximum modal number calculated $l_{\rm max}$, grid spacing $h$, and initial radius of the particle in the numerical domain $R_{\rm init}$. The correction to the scatter angle was calculated using method II as described in Section \ref{sec:SFCorrectionMII}. In the last column we show the relative difference in the values of $\delta\varphi^{(1)}_{\rm II}$ relative to the top row results, for reference.}
\label{tab:CorrNumParam}
\end{table}

These results present a significant step in the calculation of the scatter angle correct to first order in the mass ratio. We have successfully implemented both methods for calculating the SF correction to the scatter angle and shown that they produce the same results. With this numerical framework in place the only requirement for the calculation of the gravitational SF correction to the scatter angle is the SF along a scatter orbit. We continue by extending our 1+1D scalar field solver to be able to calculate gravitational perturbations.

%% file: Reconstruction.tex

We now move on from the scalar toy model and tackle the gravitational case. We start by reviewing vacuum metric reconstruction and then formulate an evolution scheme for the vacuum no-string Hertz potential in $1+1$-dimensions. A numerical implementation shows that there are non-physical modes which quickly dominate the results. We discuss their cause and suggest several ways to mitigate the modes including a transform to a Regge-Wheeler-like equation.

\section{Metric reconstruction in a radiation gauge}
\label{sec:review}

In this section we review essential results concerning the reconstruction of vacuum metric perturbations from curvature scalars first prescribed by Chrzanowski \cite{Chrzanowski1975} and Cohen and Kegeles \cite{CohenKegeles1975}, but we follow here the concise presentation by Wald \cite{Wald1978}. From a certain point we will specialise to a Schwarzschild background, introducing a decomposition of the various fields into multipole modes, but refraining from a further frequency-mode decomposition and instead remaining in the time domain. Our purpose here is to remind readers of the relevant theory, introduce notation, and set up the relevant technical background for the rest of the analysis.

We adopt the Kinnersley null tetrad $e^\alpha_{\bm a}=\{\ell^\alpha,n^\alpha,m^\alpha,\bar m^\alpha\}$ [see Eq.\ (\ref{eq:kerrtetrad})], where boldface Roman indices run over $1,\ldots, 4$ and denote tetrad components such that $A_{\bm a}:=e^\alpha_{\bm a}A_{\alpha}$. The legs $e^\alpha_{\bm a}$ are all mutually orthogonal except for $\ell^\alpha n_\alpha=-1$ and $m^\alpha \bar m_{\alpha}=1$. An overbar denotes complex conjugation. (Covariant) directional derivatives along the tetrad legs are denoted ${\bm D}_\ell = \ell^\alpha\nabla_\alpha$, ${\bm D}_n = n^\alpha\nabla_\alpha$, ${\bm D}_m = m^\alpha\nabla_\alpha$ and ${\bm D}_{\bar m} = \bar m^\alpha\nabla_\alpha$ (corresponding to the more customary but less transparent $\bm D$, $\bm \Delta$, $\bm \delta$ and $\bar {\bm\delta}$, respectively). In what follows hatted sans-serif symbols (${\sf\hat E}, {\sf\hat T}, \ldots$) represent linear differential operators on tensors. 
 
Suppose $h_{\alpha\beta}$ is a solution of the vacuum Einstein equation linearised about the Kerr metric (\ref{eqn:LinearisedEinsteinEquation}) which we recall has the form
\begin{equation}\label{EFEvacuum}
{\sf\hat E} h:= \delta G(h) =0.
\end{equation} 
Here $\delta G_{\mu\nu}$ is the linearised Einstein tensor, thought of as a differential operator ${\sf\hat E}$ acting on $h_{\alpha\beta}$, and we have omitted tensorial indices for brevity. To this perturbation there correspond Weyl curvature scalars $\Psi_0=:\Psi_+$ and $\varrho^{-4}\Psi_4=:\Psi_-$ [see Eq.\ (\ref{eq:psi}); $\varrho=-1/r$ for Schwarzschild]. $\Psi_\pm$ are derived from $h_{\alpha\beta}$ via
\begin{equation}\label{T}
{\sf\hat T}_{\pm} h =\Psi_\pm,
\end{equation} 
where the operators $\sf\hat T_{\pm}$ are given explicitly in Eqs.\ (\ref{hatT}). Let $\sf\hat S_{\pm}$ be the operators that take the linearised Einstein equation into the Teukolsky equations with spins $\pm 2$ such that
\begin{equation}\label{S}
{\sf\hat S}_{\pm} {\sf\hat E}h ={\sf\hat O}_{\pm}\Psi_\pm ,
\end{equation} 
where ${\sf\hat O}_{\pm}$ is the Teukolsky operator given in Eqs.\ (\ref{Ocompact}) and $\sf\hat S_{\pm}$ can be read off the source side of the Teukolsky equation (\ref{eq:kerrteuk}); these operators are given explicitly in Eqs.\ (\ref{eq:kerrsource}) and (\ref{eq:kerrsource-2}). From (\ref{T}) and (\ref{S}) there follows the operator identity
\begin{equation}\label{operator identity}
{\sf\hat S}_{\pm} {\sf\hat E} ={\sf\hat O}_{\pm}{\sf\hat T}_{\pm}  .
\end{equation}

Now let $\Phi_{\pm}$ be (any) solution of the {\em adjoint}\footnote{
For a linear operator $\sf\hat L$ taking an $n$-rank tensor field $\phi$ to an $m$-rank tensor field $\psi$, the adjoint $\sf\hat L^\dagger$ takes $\psi$ to $\phi$ and satisfies 
$({\sf\hat L}^\dagger\psi)\phi=\psi({\sf\hat L}\phi)$
(up to a divergence of an arbitrary vector field). 
} 
vacuum Teukolsky equation,
\begin{equation}\label{Teukolski Hertz}
{\sf\hat O}^\dagger_{\pm}\Phi_{\pm}(={\sf\hat O}_{\mp}\Phi_{\pm}) =0 . 
\end{equation}
Noting $\sf\hat E$ is self-adjoint (${\sf\hat E}={\sf\hat E}^\dagger$), we then have
\begin{equation}
{\sf\hat E} {\sf\hat S}^\dagger_{\pm} \Phi_{\pm} = 
({\sf\hat S}_{\pm}{\sf\hat E})^\dagger \Phi_{\pm}=
({\sf\hat O}_{\pm}{\sf\hat T}_{\pm})^\dagger \Phi_{\pm} =
{\sf\hat T}_{\pm}^\dagger {\sf\hat O}_{\pm}^\dagger \Phi_{\pm} = 0 ,
\end{equation}
where in the second equality we have used Eq.\ (\ref{operator identity}). Thus 
$h^{\pm}:={\sf\hat S}^\dagger_{\pm} \Phi_{\pm}$
are (complex-valued) solutions of the vacuum Einstein equation. 
A real-valued reconstructed solution is given by
\begin{equation}\label{h_rec}
h^{\rm rec}_{\pm}:={\rm Re\, }{\sf\hat S}^\dagger_{\pm} \Phi_{\pm}.
\end{equation}
The explicit form of the reconstruction operator ${\sf\hat S}^\dagger_{\pm}$ is given in Eqs.\ (\ref{eq:kerrh+}). It returns $h^{\rm rec}_{+}$ in an {\em ingoing} radiation gauge (IRG) and $h^{\rm rec}_-$ in an {\em outgoing} radiation gauge (ORG) which obey the relations
\begin{eqnarray}
h^{{\rm rec}+}_{{\bf 1}\beta} = 0\quad\text{(IRG)},\quad\quad
h^{{\rm rec}-}_{{\bf 2}\beta} = 0 \quad\text{(ORG)},
\end{eqnarray}
with both perturbations being traceless.

For $h_{+}^{\rm rec}$ and $h_{-}^{\rm rec}$ to each reproduce the original perturbation $h$, we must have ${\sf\hat T}_{\pm} h_+^{\rm rec} =\Psi_\pm$ and ${\sf\hat T}_{\pm} h_-^{\rm rec} =\Psi_\pm$, leading to
\begin{eqnarray}
{\sf\hat T}_{\pm}{\rm Re\, }{\sf\hat S}^\dagger_{+} \Phi_{+} &=& \Psi_\pm \quad\text{(IRG)}, \label{inversionIRG}\\
{\sf\hat T}_{\pm}{\rm Re\, }{\sf\hat S}^\dagger_{-} \Phi_{-} &=& \Psi_\pm\quad\text{(ORG)}. \label{inversionORG}
\end{eqnarray}
These are the fourth-order ``inversion'' equations. A Hertz potential $\Phi_{+}$ satisfying both the adjoint Teukolsky equation (\ref{Teukolski Hertz}) and either of the two  inversion relations (\ref{inversionIRG}) or (\ref{inversionORG}) will reproduce $h$ up to some perturbation $\Delta h_+$ that is in the Kernel of ${\sf\hat T}_{+}$; and a Hertz potential $\Phi_{-}$ satisfying both (\ref{Teukolski Hertz}) and either of the two  inversion relations (\ref{inversionIRG}) or (\ref{inversionORG}) will reproduce $h$ up to some perturbation $\Delta h_-$ that is in the Kernel of ${\sf\hat T}_{-}$. That is,
\begin{equation}
h= h^{\rm rec}_{\pm} + \Delta h_{\pm},
\end{equation}
where 
\begin{equation}
{\sf\hat T}_{\pm} \Delta h_{\pm} = 0 . 
\end{equation}
Wald \cite{Wald1973} explored the Kernel of ${\sf\hat T}_{\pm}$, and hence the space of $\Delta h_{\pm}$, for vacuum perturbations in Kerr. He found that $\Delta h_{\pm}$ is spanned by pure gauge perturbations (which are also in the Kernel of $\sf\hat E$), in addition to exactly four types of stationary and axially symmetric (algebraically-special) vacuum perturbations: a mass perturbation, an angular-momentum perturbation, and perturbations away from Kerr into the Kerr--NUT or the C-metric solutions.

%% file: HertzFormulation.tex

\section{1+1D evolution scheme for the Hertz potential in vacuum}
\label{Sec:HertzFormulation}

From this point onwards we specialise to a Schwarzschild background. The Schwarzschild reduction of the Teukolsky equation with $s=\pm2$ is sometimes known as the Bardeen--Press equation. Here we refer to it as the Bardeen--Press--Teukolsky (BPT) or Teukolsky equation interchangeably. 

\subsection{Multipole decomposition}

As with the scalar case, we perform a multipole decomposition of the Hertz potential to get the Teukolsky equation in a 1+1D form. We recall the IRG fields $\Phi_{+}$ and ORG fields $\Phi_{-}$ have spin weights $s=-2$ and $s=+2$ respectively. We thus expand 
$\Phi_{\pm}$ in $s=\mp 2$ spin-weighted spherical harmonics such that
\begin{equation} \label{expansionHertz}
\Phi_\pm=
\frac{\Delta^{\pm 2}}{r}
       \sum_{\ell=2}^{\infty}\sum_{m=-\ell}^{\ell}
       \phi_{\pm}^{\ell m}(t,r)\,  {}_{\mp2}\!Y_{\ell m}(\theta,\varphi),
\end{equation}
The normalisation factor $\Delta^{\pm 2}/r$, where $\Delta:=r(r-2M)$, is introduced (following \cite{Barack:2017oir}) to regulate the behaviour of the time-radial fields $\phi_{\pm}^{\ell m}$ at infinity and on the horizon. It is such that the physical (point-particle) solutions (satisfying physical boundary solutions) generally approach constant nonzero values at both ends. A more involved discussion of this choice of normalisation is discussed in Section \ref{sec:Boundaries}. The spherical basis functions ${}_{\pm2}\!Y_{\ell m}$ can be derived from standard spherical harmonics $Y_{\ell m}(\theta,\varphi)$ via 
\begin{equation}\label{Y2}
{}_{\pm 2}\!Y_{\ell m}=\sqrt{\frac{(\ell-2)!}{(\ell+2)!}}\left[\frac{\partial^2 Y_{\ell m}}{\partial\theta^2}
-\left(\frac{\cos\theta\pm 2m}{\sin\theta}\right)\frac{\partial Y_{\ell m}}{\partial\theta} + \left(\frac{m^2\pm 2m\cos\theta}{\sin^2\theta}\right)Y_{\ell m}\right].
\end{equation}
They satisfy the differential equation
\begin{align}\label{YslmEq}
\frac{1}{\sin\theta}\frac{\partial}{\partial\theta}\left(\sin\theta\frac{\partial {}_s\!Y_{\ell m}}{\partial\theta}\right) 
+\left(-\frac{m^2+2ms\cos\theta}{\sin^2\theta} -s^2\cot^2\theta+s+(\ell-s)(\ell+s+1)\right){}_s\!Y_{\ell m}=0 ,
\end{align}
and the symmetry relation
\begin{equation}\label{symmetry}
{}_{\pm 2}\!\bar Y_{\ell m}=(-1)^m {}_{\mp 2}\!Y_{\ell,-m}.
\end{equation}
We also note the symmetry under reflection by the equatorial plane,
\begin{equation}\label{reflection}
{}_{\pm 2}\!Y_{\ell m}(\theta,\varphi)=(-1)^\ell{}_{\pm 2}\!\bar Y_{\ell,-m}(\pi-\theta,\varphi),
\end{equation}
to become useful further below.  

In what follows we frequently drop the labels $\ell,m$ off of $\phi^{\ell m}_{\pm}$ for notational economy. It should be remembered that $\Phi$ is the full 4D fields, while $\phi$ is the corresponding 1+1D reduction. 

\subsection{Bardeen-Press-Teukolsky equation in 1+1D}

With the substitution (\ref{expansionHertz}), the adjoint vacuum BPT equation (\ref{Teukolski Hertz}) separates into $\ell,m$ modes, with each modal function $\phi_{\pm}(t,r)$ satisfying the 1+1D wave equation
\begin{equation} \label{Teukolsky1+1phi}
\phi^{\pm}_{,uv} + U_s(r)\, \phi^{\pm}_{,u} + V_s(r)\, \phi^{\pm}_{,v}  + W_s(r)\phi^{\pm} =0 ,
\end{equation}
with $s=\mp 2$  for $\phi_{\pm}$. Here
\begin{equation} \label{UV}
U_s(r)=-\frac{sM}{r^2},\qquad\qquad
V_s(r)=\frac{sf}{r},
\end{equation}
\begin{equation} \label{W_Sch}
W_s(r)=\frac{f}{4}\left(\frac{(\ell+s +1)(\ell- s)}{r^2}+\frac{2(1+s)M}{r^3}\right),
\end{equation}
where we recall $f=1-2M/r = \Delta/r^2$.

\subsection{Initial/boundary-value formulation}
\label{sec:Boundaries}

Our strategy is to solve the 1+1D vacuum hyperbolic equation (\ref{Teukolsky1+1phi}) directly as a time evolution problem from initial data outside the black hole. The specific form of boundary conditions for $\phi_\pm$ is inherited from that of the reconstructed metric $h_{\pm}^{\rm rec}$. For the applications we have in mind (e.g., a self-force calculation) it is the {\em retarded} (hereafter ``physical'') perturbation that we are after, i.e., the one corresponding to the boundary conditions of having no radiation coming in from $\mathscr{I}^+$ or out of $\mathscr{H}^+$. These requirements can be translated into asymptotic conditions on the behaviour of $\phi_{\pm}$ at infinity and on the horizon. This analysis was carried out in Ref.\ \cite{Barack:2017oir}, and we quote the results here.

For a monochromatic physical perturbation that has the asymptotic form $\sim e^{-i\omega u}/r$ at $\mathscr{I}^+$ (in a suitable Lorentzian frame) and $\sim e^{-i\omega v}$ on $\mathscr{H}^+$ (in a suitable horizon-regular frame), for some frequency $\omega>0$, the corresponding Hertz potential modes admit solutions
\begin{eqnarray}\label{asymptPhys}
\phi_\pm(r\to\infty) &\sim & e^{-i\omega u}\quad \text{(physical)}, \nonumber\\
\phi_\pm (r\to 2M)&\sim & e^{-i\omega v}\quad \text{(physical)}.
\end{eqnarray} 
Note $\phi_\pm$ generically approach constant nonzero values at $\mathscr{I}^+$ ($r\to\infty$ with constant $u$) and on the $\mathscr{H}^+$ ($r\to 2M$ with constant $v$). To achieve this convenient behaviour was the purpose of introducing the radial prefactors in Eq.\ (\ref{expansionHertz}).

For the interpretation of numerical results in Sec.\ \ref{sec:vacuum}, it will be useful to also have at hand the asymptotic behaviour of ``nonphysical'' monochromatic modes, which correspond to waves coming in from $\mathscr{I}^-$, $h_{\pm}^{\rm rec}\sim e^{-i\omega v}/r$, or to waves coming out of $\mathscr{H}^-$, $h_{\pm}^{\rm rec}\sim e^{-i\omega u}$. For such solutions, the asymptotic analysis in Ref.\ \cite{Barack:2017oir} finds
\begin{eqnarray}\label{asymptNonphys}
\phi_\pm(r\to\infty) &\sim & r^{\mp 4} e^{-i\omega v}\quad \text{(nonphysical)}, \nonumber\\
\phi_\pm(r\to 2M) &\sim & \Delta^{\mp 2} e^{-i\omega u} \quad \text{(nonphysical)}.
\end{eqnarray}

\pagebreak

%% file: Vacuum.tex

\section{Time-domain evolution of the Teukolsky equation: problem of growing modes}
\label{sec:vacuum}

\subsection{Numerical method}
\label{TeukNumMethod}

In this subsection, we implement a simple finite-difference Teukolsky solver based on 1+1D characteristic evolution in $u,v$ coordinates. The architecture of the code is virtually identical to that of the vacuum scalar field case detailed in Section \ref{sec:VacuumScalarImplementation}. Here we can evolve both the IRG ($s=-2$) and the ORG ($s=+2$) Hertz potentials. The finite-difference scheme used is precisely identical to the one used in \cite{Barack:2017oir} (as detailed in Appendix B therein) when applied to circular orbits and has global quadratic convergence. Our code takes as input the spin $s=\pm 2$, modal numbers $\ell$, $m$ as well as a range of numerical parameters such as $h$ and the coordinate ranges and returns the Hertz potential modes $\phi^+_{\ell m}(t,r)$ (IRG) or $\phi^-_{\ell m}(t,r)$ (ORG). As with the scalar case, we have produced two identical implementations, one in \texttt{Mathematica} and another in \texttt{C++}, to enable cross-checks.  

The main reason for implementing the vacuum case is to ensure there are no nonphysical vacuum modes which could contaminate future results. As we demonstrate below, the situation with the $|s|=2$ Teukolsky equation is less fortunate than the scalar case. In our evolution, nonphysical modes of the equation, seeded by numerical error, will grow unbounded at late time, spoiling the evolution.  We will discuss the origin of the problem and suggest ways around it but an important question to ask is why has this growth not been seen before, especially in Ref.\ \cite{Barack:2017oir}, which used the same numerical method? Our new code, when run with a circular-orbit source and $s=-2$, does reproduce the numerical results of \cite{Barack:2017oir} in the early stage of the evolution, before the onset of growth as shown later in Figure \ref{HertzCirc}.  It appears that the evolutions performed in that study were simply too short to reveal the problem. The calculation of the Hertz potential along circular orbits did not require very long runs, and evolutions were always terminated before the relatively slowly growing mode ($\sim t^4$ for $s=-2$; see below) had a chance to manifest itself in the data.  Calculations for scatter orbits require much longer evolutions, so here we must deal with the problem. The problem must be dealt with anyway if one is interested in an ORG reconstruction $(s=+2)$, where, as discussed below, the blow-up is exponential. 

In what follows we illustrate the problem of growing modes with numerical examples and describe the range of tests we performed to understand its origin. We then discuss possible remedies.

\subsection{The $s=-2$ case}
\label{sec:Teuks=-2}

\begin{figure}[h!]
\centering
\includegraphics[width=0.8\linewidth]{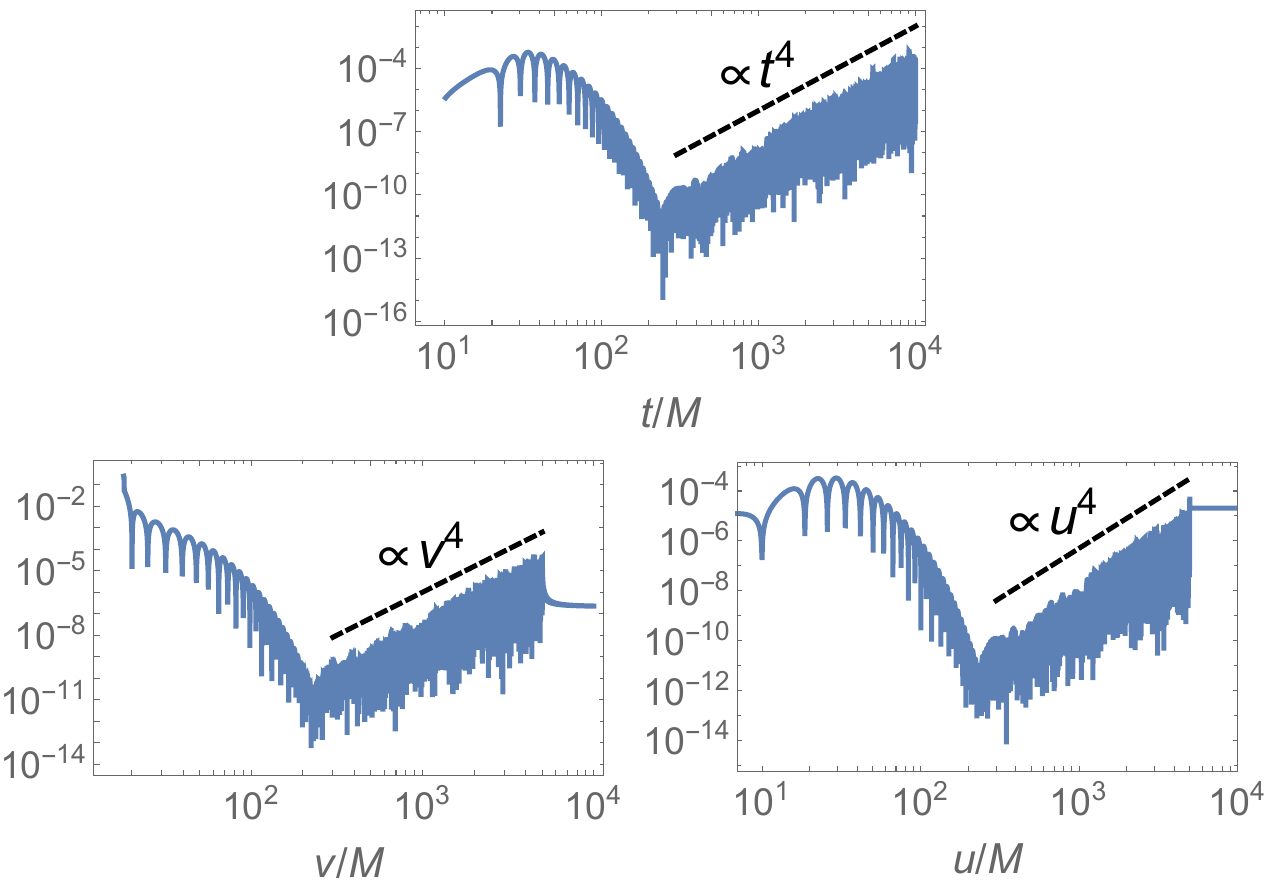}
\caption[Results from evolution of the vacuum 1+1D BPT equation with $s=-2$]{Results from evolution of the $(\ell,m)=(2,0)$ mode of the vacuum 1+1D BPT equation with $s=-2$. The evolution is seeded with a narrow Gaussian near $(t,r_*)\sim (0,9M)$. We show, on a log-log scale, the field amplitude $|\phi^-_{20}|$ sampled along slices of constant $r_*=10M$ ({\it top}), $u=500M$ ({\it lower left}) and $v=500M$ ({\it lower right}). The dashed lines (${\rm const}\times t^4$, ${\rm const}\times v^4$ and ${\rm const}\times u^4$, respectively) are shown for reference.}
\label{TeukVacsm2}
\end{figure}

Figure \ref{TeukVacsm2} shows a typical output from an $s=-2$ numerical evolution in vacuum. After the initial spike of radiation (not shown in the figures), the field decays with characteristic quasinormal ringing. However, at around $t\sim 250M$ the solution becomes dominated by a noisy component, whose amplitude appears to grow approximately like $\sim t^4$. The growth seems to continue indefinitely towards future timelike infinity ($t\to\infty$ with fixed $r>2M$), but the solution settles to a finite value approaching null infinity ($v\to\infty$ with fixed $u$) and also approaching the event horizon ($u\to\infty$ with fixed $v$). A similar behaviour is observed for all values of $\ell$ and $m$ and irrespectively of the choice of compact initial data. The evolution up to the onset of growth is numerically stable and the solution there converges quadratically in grid spacing $h$, as expected. The growing component, however, is not numerically stable. It displays noisy features on grid-size scale, and its amplitude appears to {\em increase} with decreasing $h$ (finer resolution). However, the $\sim t^4$ behaviour seems to be persistent and universal.      

We have performed a series of tests in attempt to understand these results. First, as mentioned, we have tried a variety of initial data, including a point seed at the initial vertex, smooth Gaussians of various configurations and data corresponding to an exact static solution of the BPT equation.  Second, we have tried a range of alternative finite-difference formulas and stepping schemes. Third, we have used our code to solve for the Weyl scalar modes $\psi^-(t,r)$ with jump conditions on $\cal S$ corresponding to a circular geodesic orbits (the necessary jump conditions are derived in Appendix \ref{App:WeylJumps}); we have done so both with ``zero'' initial data and with data corresponding to an exact static solution (for an $m=0$ mode). In all these tests, the $t^4$ growing mode developed just the same. Fourth, we note that the troublesome $t^4$ behaviour is observed \cite{priv_comm_Conor} also in the application to the Teukolsky equation of the recently introduced approach by O'Toole {\it et al.}\ \cite{OToole2021}, in which the Green function (rather than the field itself) is evolved from exact characteristic initial data. Finally, we observe that we are able to successfully suppress the $t^4$ growth using our \texttt{Mathematica} implementation with very high working precision (albeit at considerable computational cost). All this supports the conclusions that (i) the $t^4$ behaviour has a genuine dynamical origin and (ii) the $t^4$ component is seeded by numerical roundoff error. 

We suggest that the $t^4$ mode represents nonphysical incoming radiation sourced by numerical roundoff error near $\mathscr{I}^+$. This can be seen from the following heuristics. First, recall from Eqs.\ (\ref{asymptPhys}) and (\ref{asymptNonphys}) the asymptotic form of monochromatic $s=-2$ solutions in the ``wave zone'' ($r\gg M$ with $v\gg u$): $\phi\sim e^{-i\omega u}$ for physical solutions (purely outgoing waves) and $\phi\sim r^{-4}e^{-i\omega v}$ for nonphysical solutions representing purely incoming waves. More generally, the time-domain solutions are superpositions of such monochromatic modes, and have the forms $\phi\sim F(u)$ (physical) and $\phi\sim r^{-4}G(v)$ (nonphysical) for some functions $F(u)$ and $G(v)$ that depend on the initial data. [These forms can be confirmed more directly by substituting the ans\"{a}tze $\phi=r^{\alpha}F(u)$ and $\phi=r^{\beta}G(v)$ into the BPT equation (\ref{Teukolsky1+1phi}) and solving at leading order in $M/r$ under the wave-zone assumptions $F'(u)\gg F(u)/r$ and $G'(v)\gg G(v)/r$, to obtain $\alpha=0$ and $\beta=-4$.] Consider an outgoing ray $u=\text{const}$ shortly after the start of the evolution. The field on this ray is composed mostly of outgoing radiation $\phi\sim F(u)$, which approaches a constant value at large $v$. However, roundoff error in the numerical data along this ray will inevitably source a small component of nonphysical high-frequency incoming radiation $\phi\sim r^{-4}G(v)$. Since the sourcing field is asymptotically constant at $v\to\infty$, the amplitude of the seeded incoming radiation is also expected to be asymptotically constant on the $u={\rm const}$ ray, i.e.\ $r^{-4}|G(v)|\sim {\rm const}$ for $v\to\infty$. This implies $|G(v)|\sim v^4 \sim(t+r)^4$ at large $v$, and it follows that the incoming-wave component has an amplitude $|\phi|\sim v^4/r^4\sim (t+r)^4/r^4$. At fixed $r$, this will exhibit a $\sim t^4$ growth, at least in the wave zone where our heuristic analysis applies. (To show that this wave-zone behaviour might lead to a $t^4$ growth elsewhere at late time, as evident in the numerical data, would require a more detailed asymptotic matching analysis, which we have not attempted.)

This heuristic description explains the results of our various experiments. The $t^4$ behaviour arises dynamically from roundoff error seeds, so it is persistent, universal and independent of initial data. The amplitude of the $t^4$ component can be suppressed by increasing the precision of the floating-point arithmetic, which reduces the roundoff error. For a fixed floating-point precision, increasing the grid resolution (decreasing $h$) enhances the amplitude of the incoming radiation component, by seeding more of its modes at higher frequencies.     

It is also possible to explain why the $t^4$ growth does not appear to plague other time-domain treatments of the $s=-2$ Teukolsky equation reported in the literature. In the 2+1D Cauchy evolution approach of Khanna {et al.}\ (legacy of \cite{LopezAleman:2003ik,Khanna:2003qv} and many works since), boundary conditions are actively imposed, which presumably suppress the growth of the unphysical component. In the compactified hyperboloidal slicing approaches of Refs.\ \cite{Racz:2011qu,Zenginoglu:2012us,Harms2013}, we suspect it is the {\it compactification} of $\mathscr{I}^+$ that averts the problem, since the wavezone for incoming waves is vastly underresolved on the compactified grid. In contrast, our simple $u,v$-coordinate-based approach resolves the wave-zone equally well for both outgoing and incoming waves. Unfortunately, as we have seen, the resolution of incoming waves near null infinity is harmful in our case.


\subsection{The $s=+2$ case}

\begin{figure}[h!]
\centering
\includegraphics[width=0.8\linewidth]{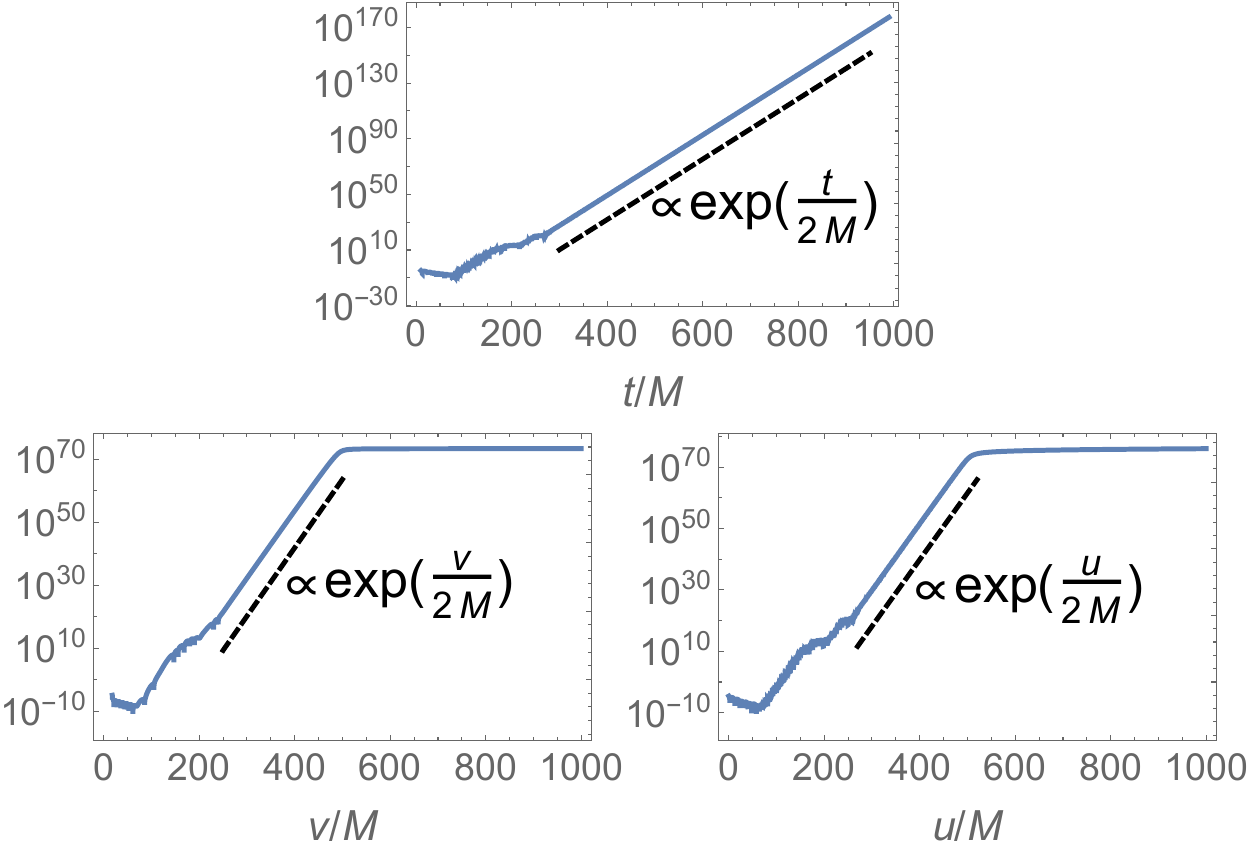}
\caption[Results from evolution of the vacuum 1+1D BPT equation with $s=+2$]{Results for the evolution of the $(\ell,m)=(2,0)$ mode of the vacuum 1+1D BPT equation with $s=+2$. Other details are as in Figure \ref{TeukVacsm2}, except here the scale is semilogarithmic. The dashed lines are for reference.}
\label{TeukVacs2}
\end{figure}

Figure \ref{TeukVacs2} shows a typical output from an $s=+2$ numerical evolution in vacuum. In this case, after a short phase of quasinormal decay (harder to discern on the semilogarthmic scale of Figure \ref{TeukVacs2}), there commences a rapid exponential growth, $\phi\sim\exp[t/(2M)]$. Again, the growth seems to continue indefinitely towards future timelike infinity ($t\to\infty$ with fixed $r>2M$), but the solution settles to finite values towards $\mathscr{I}^+$ and $\mathscr{H}^+$. A similar behaviour is observed for all values of $\ell$ and $m$ and all choices of initial data we have tried and the blow-up exponent $(1/2M)$ seems universal. The growing component is not numerically stable, increasing in amplitude with decreasing $h$ (finer resolution). We have performed similar tests to the ones described above for $s=-2$ and with similar results. The exponential growth is persistent, universal, and can be moderated (in amplitude) only with high-precision arithmetic. 

We again argue that the culprit is a nonphysical growing solution of the BPT equation seeded by roundoff error, this time an exponential mode of the $s=+2$ equation. We can see this most instructively from a simple asymptotic analysis near the horizon, as follows. Working at leading order in $\Delta$, and assuming $\phi$, $\phi_{,u}$ and $\phi_{,v}$ are all of the same order in $\Delta$ near the horizon [this is true of the $r\to 2M$ asymptotic expressions in Eqs.\ (\ref{asymptPhys}) and (\ref{asymptNonphys})], the BPT equation (\ref{Teukolsky1+1phi}) reduces to
\begin{equation}
\phi_{,uv} -k\,\phi_{,u}  =0,
 \label{TeukEqnAsympH+}
\end{equation}
in which $k:=s/(4M)$, and
where we have retained the $s$ dependence to enable us to compare the situation between the two spin values. 
The general solution is
\begin{equation}
\phi = C_1 (u)e^{kv} + C_2 (v),
\label{TeukAsympH+}
\end{equation}
where $C_1(u)$ and $C_2(v)$ are arbitrary functions. Solutions of the form $\phi= C_2(v)$ represent physical perturbations that are purely ingoing at the horizon [compare with Eq.\ (\ref{asymptPhys})], while solutions of the form $\phi = C_1 (u)e^{kv}$ represent nonphysical perturbations coming out of the past horizon [compare with Eq.\ (\ref{asymptNonphys}), noting $\Delta^{s}\sim (2M)^{2s}e^{2kr_*}\sim (2M)^{2s}e^{kv}$ near the horizon]. For $s=+2$ the nonphysical solution blows up exponentially in $v$ along the horizon, while for $s=-2$ it is exponentially suppressed. 

The situation now mirrors what we had near null infinity for the $s=-2$ growth. As the main physical perturbation, of the form $\phi= C_2(v)$, reaches the horizon, roundoff error along the incoming ray seeds a nonphysical component $\sim C_1 (u)e^{kv}$, which, for $s=+2$, blows up exponentially along the horizon. The predicted rate of exponential growth is consistent with that observed in the numerical data: $\sim e^{sv/(4M)}=e^{v/(2M)}$.  To understand the propagation of this exponential growth into other areas of the black hole's exterior would require a detailed asymptotic matching analysis, but it would not be surprising to find a similar exponential growth in time anywhere outside the black hole, as seen in the numerics. We note the fortunate situation in the $s=-2$ case, where all nonphysical modes are exponentially suppressed in a dynamical manner, with no need to actively impose boundary conditions. 

There are in the literature several successful time-domain numerical methods for the $s=+2$ Teukolsky equation (e.g., \cite{Harms2013,Burko2016,Burko2018,Burko2019}), all incorporating horizon-penetrating coordinates in some form. The use of such coordinates (effectively a compactification of our $u,v$ coordinates) under-resolves any outgoing component of the perturbation field near the horizon, thereby avoiding the problem encountered here.  

\subsection{Mitigation}

Although initially surprising to us, it is clear that a standard uni-grid characteristic evolution based on $u,v$ coordinates does not work well for either $s=+2$ or $s=-2$ Teukolsky equations. A remedy based on the use of very high precision arithmetic is clearly impracticable. The preceding discussion and evidence from the literature suggest that compactification of the two asymptotic domains ($\mathscr{I}^+$ and $\mathscr{H}^+$) can offer a solution that is both computationally efficient and practicable. This has already been achieved, e.g., by Harms {\it et al.}\ \cite{Harms2013}, using asymptotically null compactified spacelike slices. It is perceivable that the same could also be achieved within the convenient framework of a fully double-null architecture. This approach is worth exploring. 

Here we choose to apply a different strategy. Instead of tackling the BPT equation directly, we will introduce a transformation of the Hertz potential to a new field variable, which satisfies a field equation free of the above difficulties. From the preceding discussion it is clear that the culprit term in the BPT equation (in both $s=-2$ and $s=+2$ cases) is the one involving $\phi_{,t}$, so we seek a transformation that eliminates that term. The simplest such transformation is a time-domain version of the familiar Chandrasekhar transformation \cite{Chandra1975}, which takes solutions of the BPT equation to solutions of the Regge-Wheeler (RW) equation. Tthe RW equation evolves without a problem on a simple uniform mesh based on $u,v$ coordinates (see, e.g., \cite{Martel:2003jj}), so this approach would require no radical architectural changes to our numerical method.  

In the next section we reformulate the 1+1D evolution problem in terms of a RW-like variable. Then later we demonstrate a full numerical calculation of the Hertz potential for circular and scatter orbits in Sections \ref{sec:RWCirc} and \ref{sec:RWScatter} respectively. 

%% file: RWFormulation.tex

\section{Reformulation in terms of a Regge-Wheeler-like variable} 
\label{sec:RWformulation}

Let the field $X(t,r)$ be a solution of the vacuum Regge-Wheeler equation 
\begin{equation}\label{RWeq}
X_{,uv}+\frac{f}{4}\left(\frac{\lambda_1}{r^2}-\frac{6M}{r^3} \right) X =0 ,
\end{equation}
where $\lambda_1:=\ell(\ell+1)$ and we recall $f=1-2M/r$. Then, as can be easily checked,
\begin{equation}\label{transformationIRG}
\phi_+ = \frac{1}{4r} \tilde{\boldsymbol D}_n^2(r X) 
= f^{-2}\left(X_{,uu}-\frac{r-3M}{r^2}\, X_{,u} \right),
\end{equation}
and 
\begin{equation}\label{transformationORG}
\phi_- = \frac{f^2 r^3}{4}{\boldsymbol D}_\ell^2(r X) 
= r^4\left(X_{,vv}+\frac{r-3M}{r^2}\, X_{,v} \right),
\end{equation}
are solutions of the vacuum BPT equation (\ref{Teukolsky1+1phi}) with $s=-2$ and $s=+2$, respectively. This means that we can use the RW variable $X$ as a generating function for both IRG Hertz potential $\phi_+$ and ORG Hertz potential $\phi_-$. The advantage, of course, is that the RW equation (\ref{RWeq}), unlike the BPT equation (\ref{Teukolsky1+1phi}), does lend itself to a straightforward characteristic evolution in $u,v$ coordinates. The idea now would be to formulate a suitable characteristic initial-value problem for the RW variable $X$, from which the no-string Hertz potential $\phi_+$ [or $\phi_-$] can be obtained by applying the transformation (\ref{transformationIRG}) [or (\ref{transformationORG})] to vacuum RW solutions $X$.

Let us consider boundary conditions. In both asymptotic regions $r\to\infty$ and $r\to 2M$, monochromatic solutions of the RW equation (\ref{RWeq}) are superposition of modes $X\sim e^{-i\omega u}$ and $X\sim e^{-i\omega v}$ (with some generally nonzero constant coefficients at leading order). The ``retarded'' monochromatic solution has the behaviour $X\sim e^{-i\omega u}$ near ${\mathscr{I}}^+$ and $X\sim e^{-i\omega v}$ near ${\mathscr{H}}^+$. It it easily seen that this retarded solution generates the physical IRG Hertz potential $\phi_+\sim e^{-i\omega u}$ near ${\mathscr{I}}^+$ and the physical ORG Hertz potential $\phi_-\sim e^{-i\omega v}$ near ${\mathscr{H}}^+$ [here we recall Eq.\ (\ref{asymptPhys})]. It is harder to show that the retarded RW solution necessarily generates the physical field $\phi_+$ near ${\mathscr{H}}^+$ or the physical field $\phi_-$ near ${\mathscr{I}}^+$ (this would require a higher-order asymptotic analysis), but we can circumvent this with the following observation: From Eq.\ (\ref{asymptNonphys}) we see that nonphysical modes of the the IRG potential $\phi_+$ diverge (like $\Delta^{-2}$) near ${\mathscr{H}}^+$, and that the nonphysical modes of the ORG potential $\phi_-$ diverge (like $r^4$) near ${\mathscr{I}}^+$. Thus, in either case, a nonphysical Hertz potential announces its presence loudly in the form of a strong asymptotic divergence. This is a point made already in Ref.\ \cite{Barack:2017oir}: A solution $\phi_+$ that is regular (bounded) at ${\mathscr{H}}^+$ is automatically the physical one, and so is a solution $\phi_-$ that is regular (bounded) at ${\mathscr{I}}^+$. We can establish a posteriori that our numerical solutions $\phi_+$ or $\phi_-$ satisfy physical boundary conditions simply by checking they are bounded.

We do not present an explicit implementation this new method for the vacuum case due to previous calculations of stable RW evolutions in the literature (see, e.g., \cite{Martel:2003jj}). A full implementation for both circular and scatter orbits will be presented in Chapter \ref{chapter:Implementation}.

%% file: NonVacuumReconstruction.tex

In this chapter we introduce a point-particle into the implementation. First we show how the standard vacuum reconstruction procedure (presented in Section \ref{sec:review}) fails with the introduction of a source and suggest possible remedies including a summary of the new procedure by Green {\em et al.} \cite{GreenHollandsZimmerman2020}. Continuing, we formulate a 1+1D evolution scheme for the no-string Hertz potential and derive closed-form ODEs for the jump conditions of the Hertz potential and the Regge-Wheeler-like field.

\section{Non-vacuum metric reconstruction}
\label{sec:MetricReconSourced}

\subsection{Failure of standard non-vacuum metric reconstruction}  

Let $h_{\alpha\beta}$ be a solution of the non-vacuum equations
\begin{equation}\label{EEFsourced}
{\sf\hat E} h_{\alpha\beta} =T_{\alpha\beta},
\end{equation} 
where $T_{\alpha\beta}$ is the energy-momentum tensor (absorbing an $8\pi$ factor). The corresponding Teukolsky equation now reads
\begin{equation} \label{TeukolskySource}
{\sf\hat O}_{\pm}\Psi_\pm  = {\sf\hat S}_{\pm} {\sf\hat E}h_{\alpha\beta} ={\sf\hat S}_{\pm}T_{\alpha\beta} =: {\cal T}_{\pm},
\end{equation} 
where ${\cal T}_{\pm}$ are the source terms for the $s=\pm 2$ Teukolsky equations. The operator equality (\ref{operator identity}) still applies.

Now let $\Phi_{\pm}$ be a solution of the adjoint sourced Teukolsky equation
\begin{equation}\label{Tsourced}
{\sf\hat O}^\dagger_{\pm}\Phi_{\pm}={\sf\hat O}_{\mp}\Phi_{\pm} =S_{\pm} ,
\end{equation}
where the source $S_{\pm}$ will need to be determined. It can be written that
\begin{equation}\label{Source_rec}
{\sf\hat E} {\sf\hat S}^\dagger_{\pm} \Phi_{\pm} = 
{\sf\hat E}^\dagger {\sf\hat S}^\dagger_{\pm} \Phi_{\pm}=
({\sf\hat S}_{\pm}{\sf\hat E})^\dagger \Phi_{\pm}=
({\sf\hat O}_{\pm}{\sf\hat T}_{\pm})^\dagger \Phi_{\pm} =
{\sf\hat T}_{\pm}^\dagger {\sf\hat O}_{\pm}^\dagger \Phi_{\pm} = {\sf\hat T}_{\pm}^\dagger S_{\pm}.
\end{equation}
Thus $h_{\pm}:={\sf\hat S}^\dagger_{\pm} \Phi_{\pm}+h^{\rm comp}$
are solutions of the sourced Einstein's equation (\ref{EEFsourced}) if $S_{\pm}$ are chosen such that
\begin{equation}\label{SeqnFail}
({\sf\hat T}_{\pm}^\dagger S_{\pm})_{\alpha\beta} + T_{\alpha\beta}^{\rm comp}= T_{\alpha\beta},
\end{equation}
where $T_{\alpha\beta}^{\rm comp}:={\sf\hat E}h^{\rm comp}$ is the energy-momentum associated with the completion piece of the perturbation.

However, it can be shown that (\ref{SeqnFail}) cannot be satisfied in general. The explicit form of the operators ${\sf\hat T}_{\pm}^\dagger$ [Eqs.\ (\ref{Tdagger+Expl}) and (\ref{Tdagger-Expl})] gives
\begin{eqnarray}\label{ldotT=0}
\ell \cdot {\sf\hat T}_{+}^\dagger = &0& = m \cdot {\sf\hat T}_{+}^\dagger,\\
\label{ndotT=0}
n \cdot {\sf\hat T}_{-}^\dagger = &0& = \bar m \cdot {\sf\hat T}_{-}^\dagger.
\end{eqnarray}
This means that (\ref{SeqnFail}) is consistent only if $\ell^\alpha (T_{\alpha\beta}-T_{\alpha\beta}^{\rm comp})=0=m^\alpha (T_{\alpha\beta}-T_{\alpha\beta}^{\rm comp})$ for $s=+2$ or if $n^\alpha (T_{\alpha\beta}-T_{\alpha\beta}^{\rm comp})=0=\bar m^\alpha (T_{\alpha\beta}-T_{\alpha\beta}^{\rm comp})$ for $s=-2$. However, this is generally not the case, and it is never the case for a $T_{\alpha\beta}$ corresponding to a massive point particle. Thus Eq.\ (\ref{SeqnFail}) cannot be solved, and the naïve reconstruction procedure described here fails. 

\subsection{Alternative non-vacuum metric reconstruction procedure}  
\label{sec:NonVacuumMetricReconstruction}

Here we describe our proposed alternative reconstruction procedure designed to circumvent the issues associated with the presence of a source. This work was conducted in the early stages of my research project and while there were promising preliminary results the decision was made to abandon this line of research due to work of Green {\em et al.}\ \cite{GreenHollandsZimmerman2020}. Their method of non-vacuum metric reconstruction solved the issue that we were also tackling. For completeness we include the progress we made on the problem but our work is far from a full formulation. We will give a summary of Green {\em et al.}'s method in the proceeding section.

Consider an alternative choice of $S_{\pm}$ such that
\begin{equation}\label{Seqn}
({\sf\hat T}_{+}^\dagger S_{+})_{\alpha\beta} + ({\sf\hat T}_{-}^\dagger S_{-})_{\alpha\beta} + T_{\alpha\beta}^{\rm comp} = T_{\alpha\beta}.
\end{equation}
Note that this new choice of $S_{\pm}$ is still a solution to the adjoint sourced Teukolsky equation (\ref{Tsourced}).

As $T_{\alpha\beta}$ is real, it is convenient to take the real part of the left-hand side of (\ref{Seqn}). This is because the imaginary parts of $({\sf\hat T}_{\pm}^\dagger S_{\pm})_{\alpha\beta}$ would have to be completely cancelled by $T_{\alpha\beta}^{\rm comp}$ hence no information is added by keeping the imaginary parts. From here onwards, ${\rm Re \:}({\sf\hat T}_{\pm}^\dagger S_{\pm})_{\alpha\beta}$ and ${\rm Re \:}T_{\alpha\beta}^{\rm comp}$ refer to the real part of the tensors.

Consider the explicit form of the operators ${\sf\hat T}_{\pm}^\dagger$ as given in (\ref{Tdagger+Expl}) and (\ref{Tdagger-Expl}). We can see that $({\sf\hat T}_{+}^\dagger S_{+})_{\alpha\beta} \in \{{\bf 11, 13, 33}\}$ and $({\sf\hat T}_{-}^\dagger S_{-})_{\alpha\beta} \in \{{\bf22, 24, 44}\}$, where we recall that boldface Roman indices denote Kinnersley null tetrad components. Taking the real part of $({\sf\hat T}_{\pm}^\dagger S_{\pm})_{\alpha\beta}$ can be done by adding the complex conjugate and halving the result. ${\rm Re \:}({\sf\hat T}_{+}^\dagger S_{+})_{\alpha\beta}$ contains the null vector $m^\alpha$ hence adding the complex conjugate will add $\bar m^\alpha$ terms. This means that ${\rm Re \:}({\sf\hat T}_{+}^\dagger S_{+})_{\alpha\beta} \in \{{\bf 11, 13, 14, 33, 44}\}$. Similarly, ${\rm Re \:}({\sf\hat T}_{-}^\dagger S_{-})_{\alpha\beta} \in \{{\bf22, 23, 24, 33, 44}\}$ as it has $m^\alpha$ terms added by the complex conjugate. This means that ${\rm Re \:}({\sf\hat T}_{\pm}^\dagger S_{\pm})_{\alpha\beta}$ are not in independent spaces.

${\rm Re \:}({\sf\hat T}_{\pm}^\dagger S_{\pm})_{\alpha\beta}$ and $T_{\alpha\beta}$ are both symmetric hence ${\rm Re \:}T_{\alpha\beta}^{\rm comp}$ must also be symmetric. This means that there are a total of 14 unknowns in the problem: the 10 independent components of ${\rm Re \:}T_{\alpha\beta}^{\rm comp}$ and 4 unknowns from the complex $S_{\pm}$. Eq.\ (\ref{Seqn}) contains 10 independent equations which are given by
\begin{eqnarray}
({\sf\hat T}_{+}^\dagger S_{+})_{\bf ab} + T_{\bf ab}^{\rm comp} + c.c. &=& 2T_{\bf ab}, \label{Seqn+} \\
({\sf\hat T}_{-}^\dagger S_{-})_{\bf cd} + T_{\bf cd}^{\rm comp} + c.c. &=& 2T_{\bf cd}, \label{Seqn-}\\
({\sf\hat T}_{+}^\dagger S_{+})_{\bf ef} + ({\sf\hat T}_{-}^\dagger S_{-})_{\bf ef} + T_{\bf ef}^{\rm comp} + c.c. &=& 2T_{\bf ef}, \label{Seqnpm} \\
T_{\bf gh}^{\rm comp} + c.c. &=& 2T_{\bf gh}. \label{Seqnind}
\end{eqnarray}
for ${\bf ab} \in \{{\bf 11, 13, 14}\}$, ${\bf cd} \in \{{\bf 22, 23, 24}\}$, ${\bf ef} \in \{{\bf 33,34}\}$ and ${\bf gh} \in \{{\bf 12, 34}\} $ and where $+c.c.$ represents adding the complex conjugates of the preceding terms.

However, with only $(\ref{Seqn+})$ -- $(\ref{Seqnind})$ the system is underdetermined. The solution to this is to act on (\ref{Seqn}) with ${\sf\hat S}_{\pm}$.  Acting on ${\sf\hat T}_{+}^\dagger S_{+}$ with ${\sf\hat S}_{+}$ gives
\begin{equation}
{\sf\hat S}_{+}{\sf\hat T}_{+}^\dagger S_{+} = {\sf\hat S}_{+}{\sf\hat T}_{+}^\dagger {\sf\hat O}^\dagger_{+} \Phi_+ =  {\sf\hat S}_{+}{\sf \hat E} {\sf\hat S}^\dagger_{+} \Phi_+ = {\sf\hat O}_{+} {\sf\hat T}_{+} {\sf\hat S}^\dagger_{+} \Phi_+ = {\sf\hat O}_{+} {\sf\hat T}_{+} h_+ = {\sf\hat O}_{+} \Phi_+ = {\cal T}_{+}.
\end{equation}
It is also known from (\ref{TeukolskySource}) that ${\sf\hat S}_{+} T_{\alpha\beta} ={\cal T}_{+}$ hence the two terms cancel when acting upon (\ref{Seqn}) with ${\sf\hat S}_{+}$. Similarly, the $({\sf\hat T}_{-}^\dagger S_{-})_{\alpha\beta}$ and $T_{\alpha\beta}$ terms cancel when acted upon (\ref{Seqn}) with ${\sf\hat S}_{-}$. Combining the two results gives the scalar equations
\begin{equation} \label{SeqnSopp}
{\sf\hat S}_{\pm} \left( {\sf\hat T}_{\mp}^\dagger S_{\mp} \right) + {\sf\hat S}_{\pm} \left( T_{\alpha\beta}^{\rm comp}\right) = 0.
\end{equation}
This added set of four equations means the system can be defined uniquely through Eqs.\ (\ref{Seqn+}) -- (\ref{Seqnind}) and (\ref{SeqnSopp}).

There is an alternative method for determining just the source term $S_\pm$ which vastly reduces the number of equations that have to be solved. Consider acting on the IRG inversion relation (\ref{inversionIRG}) with ${\sf\hat O}_{\pm}$ which gives the left-hand side as
\begin{equation}
{\sf\hat O}_{\pm}{\sf\hat T}_{\pm}{\sf\hat S}^\dagger_{+} \Phi_{+} = {\sf\hat S}_{\pm}{\sf\hat E}{\sf\hat S}^\dagger_{+} \Phi_{+} =  {\sf\hat S}_{\pm} ({\sf\hat S}_{+}{\sf\hat E})^\dagger \Phi_{+} = {\sf\hat S}_{\pm} ({\sf\hat O}_{+}{\sf\hat T}_{+})^\dagger \Phi_{+} =  {\sf\hat S}_{\pm} {\sf\hat T}_{+}^\dagger {\sf\hat O}_{+}^\dagger \Phi_{+} ={\sf\hat S}_{\pm} {\sf\hat T}_{+}^\dagger S_+
\end{equation}
hence we can be write
\begin{equation}\label{PointParticleInversionIRG}
{\sf\hat S}_{\pm}{\sf\hat T}_{+}^\dagger S_{+} = {\cal T}_{\pm}, 
\end{equation}
where we have used Eq.\ (\ref{TeukolskySource}). Similarly, acting with ${\sf\hat O}_{\pm}$ on the ORG inversion relation (\ref{inversionORG}) gives
\begin{equation}\label{PointParticleInversionORG}
{\sf\hat S}_{\pm}{\sf\hat T}_{-}^\dagger S_{-} = {\cal T}_{\pm}. 
\end{equation}
Eqs.\ (\ref{PointParticleInversionIRG}) and (\ref{PointParticleInversionORG}) form four equations which can be solved to give the four components of the complex $S_{\pm}$ uniquely. This alternative method is useful if we only require the source term as it vastly reduces the number of equations that need to be solved.

\subsection{GHZ metric reconstruction procedure}  

In recent work \cite{GreenHollandsZimmerman2020}, Green {\em et al.}\ prescribed a modification of the above naïve approach, based on adding a certain ``corrector tensor'' $x$ such that the metric perturbation in the IRG is given by
\begin{equation}
h_{ab} = h^+_{ab} + x_{ab} + (h^{\rm comp}_{+})_{ab},
\end{equation}
where $h^{\rm comp}_{+}$ is the standard completion part in vacuum satisfying ${\sf\hat E} h^{\rm comp}_{+} = 0$. This means to ensure that the left relation of (\ref{ldotT=0}) is satisfied it must be true that
\begin{equation}
(T_{ab}- {\sf\hat E}x_{ab})\ell^b=0.
\label{MRlcomp}
\end{equation}
With the ansatz
\begin{equation}
x_{ab} = 2 m_{(a} \bar m_{b)}x_{m \bar m} - 2 \ell_{(a} \bar m_{b)} x_{nm} - 2 \ell_{(a} m_{b)} x_{n \bar m} + l_a l_b x_{nn},
\end{equation}
$x_{ab}$ has 4 real independent components as $x _{n \bar m}=\bar x_{nm}$ hence we can solve Eq.\ (\ref{MRlcomp}) for the different components by taking different tetrad components. Taking the $l^a$ component gives a second order ODE of the form
\begin{equation}
\hat {\cal A} \: x_{m\bar m} = T_{ll},
\end{equation}
where $\hat{\cal A}$, and all future calligraphic operators, represent a second-order differential operator. Taking the $m^a$ component of Eq.\ (\ref{MRlcomp}) gives another second order ODE of the form
\begin{equation}
\hat {\cal B}_1 \: x_{nm} = T_{lm} + \hat {\cal B}_2 \: x_{m\bar m}.
\end{equation}
Finally, taking the $n^a$ component of Eq.\ (\ref{MRlcomp}) gives a third second order ODE of the form
\begin{equation}
\hat {\cal C}_1 \: x_{nn} = T_{ln} + \hat {\cal C}_2 \: x_{m\bar m} + \hat {\cal C}_3 \: x_{nm}.
\end{equation}
This means that given the source of the Einstein field equations, we can obtain the corrector tensor by integrating a certain hierarchical set of {\it ordinary} differential equations along null directions. There is ongoing work to demonstrate the applicability of this non-vacuum metric reconstruction method in practice \cite{ToomaniZimmerman2021}.

\subsection{Metric reconstruction in a no-string radiation gauge}
\label{sec:PointParticleSource}

A more acute question is whether the standard vacuum reconstruction procedure works in {\em vacuum} regions of spacetime in the presence of sources elsewhere. It has long been known, from analysis of the point-particle source example \cite{BarackOri2001}, that this was not the case: a perturbation $h^{\rm rec}_{\pm}$ reconstructed as in Eq.\ (\ref{h_rec}) (with or without a completion piece $\Delta h_{\pm}$) develops singularities in the vacuum region {\em away} from the particle. This can be appreciated already from the simple example of a static particle in flat space (see Section V.C.\ of \cite{BarackOri2001}, or the more detailed analysis in Section VI of \cite{PoundMerlinBarack2014}). What one finds is that $h^{\rm rec}_{\pm}$ exhibits string-like singularities that emanate from the particle along radial null directions. By adjusting the residual gauge freedom (within the class of radiation gauges) one can arrange to confine the string to either outgoing or ingoing directions, but no choice of a radiation gauge can rid of the strings altogether. The leading-order singular form of the string is described in Table I of \cite{PoundMerlinBarack2014}. The singularly is sufficiently strong that the perturbation field fails to be (absolutely) integrable over a two-dimensional surface intersecting the string, with the result that a multipole decomposition of the field is not even well defined. Thus a mode-by-mode reconstruction procedure cannot work in the entire vacuum part of spacetime containing the string. It should be presumed that an analysis based on the new corrector-tensor method of \cite{GreenHollandsZimmerman2020} would reproduce this basic picture when applied to the point-particle case.

Let us describe the situation more precisely. We are interested in the case of a pointlike particle of mass $\mu$, moving outside a Kerr black hole (to be specialised to Schwarzschild further below) with mass $M\gg\mu$. We assume the particle's stress-energy is given by the distribution
\begin{equation}\label{Tmunu}
T_{\mu\nu}= \mu \int_{-\infty}^{\infty} u_{\mu}u_{\nu}\delta^4(x^\alpha-x_{\rm p}^\alpha(\tau))(-g)^{-1/2}d\tau ,
\end{equation}
where we recall $x_{\rm p}^\alpha(\tau)$ describes the particle's timelike worldline ($\tau$ being proper time), $u^\mu:=dx_{\rm p}^{\mu}/d\tau$ and, as usual, indices are lowered using the background metric $g_{\mu\nu}$ with determinant $g$. In Boyer-Lindquist coordinates we write $x_{\rm p}^\mu(t)=\left(t,R(t),\theta_{\rm p}(t),\varphi_{\rm p}(t)\right)$, so that $R(t)$ is the radial location of the particle at time $t$. We denote by $\cal S$ the 2+1-dimensional closed surface $r=R(t)$; this is a 2-sphere through the particle at each given time.\footnote{Here we use $\cal S$ to represent the 2+1D sphere $r=R(t)$ in spacetime, while in Section \ref{sec:ScalarFieldSource} it was introduced as the curve $r=R(t)$ in the $r,t$ plane. Throughout this work we will continue to use $\cal S$ in both ways; the relevant meaning in each instance should be clear from the context. A similar remark applies to ${\cal S}^<$ and ${\cal S}^>$.} The surface $\cal S$ splits the exterior of the black hole into two disjoint regions, $r>R(t)$ and $r<R(t)$, which we call ${\cal S}^{>}$ and ${\cal S}^{<}$, respectively. 
 
As we have described, a reconstructed radiation-gauge metric $h^{\rm rec}_{\pm}$ generically exhibits a string singularity in both ${\cal S}^{>}$ and ${\cal S}^{<}$: it is a ``full-string'' solution, in the terminology of \cite{PoundMerlinBarack2014}. It is not known how to calculate the physical self-force in such a pathological gauge, so the full-string reconstruction is not useful in the present context. As also described, there is a way to choose a radiation gauge such that the string is confined to ${\cal S}^{>}$ and the reconstructed perturbation, denoted here $h^{\rm <}_{\pm}$, is regular (smooth) anywhere in ${\cal S}^{<}$. Similarly, there is a choice of radiation gauge for which the string is confined to ${\cal S}^{<}$, and the perturbation, denoted $h^{\rm >}_{\pm}$, is regular (smooth) anywhere in ${\cal S}^{>}$. These are the two ``half-string'' solutions. Ref.\ \cite{PoundMerlinBarack2014} showed how the physical self-force may be computed from either of the two half-string solutions using a procedure that involves taking a directional (radial) limit to the particle from its ``regular'' side. This procedure may be suitable for frequency-domain calculations, where one could (in principle) integrate the relevant radial ODE from boundary conditions either on the event horizon or at infinity, towards the particle, working in the regular side of spacetime. However, the half-string reconstructions are not suitable for time-domain calculations, where one evolves the field equations as PDEs on the full exterior of the black hole.

This brings us to the ``no-string'' reconstruction, first advocated in a series of papers by Friedman and collaborators \cite{KeidlFriedmanWiseman2007,Keidl2010}, and later formulated in detail (and received its name) in \cite{PoundMerlinBarack2014}. 
The idea is simple: Take the two regular sides of the two one-string solutions, and glue them together at $\cal S$. The resulting, no-string perturbation is given by
\begin{equation}
h_{\pm}^{\rm nos} = h^{<}_{\pm}\Theta(R(t)-r)+h^{>}_{\pm}\Theta(r-R(t)),
\end{equation}
where $\Theta(\cdot)$ is the Heaviside step function. The perturbation $h_{\pm}^{\rm nos}$ is regular (smooth) in both ${\cal S}^{<}$ and ${\cal S}^{>}$, where it solves the linearised vacuum Einstein's equations. On $\cal S$ itself $h_{\pm}^{\rm nos}$ is not a vacuum solution, even away from the particle, and even when allowing arbitrary completion pieces $\Delta h_{\alpha\beta}$ in and out of $\cal S$ [see Section VI.B.1.\ of \cite{PoundMerlinBarack2014}, where it is shown that, at least in the flat-space example, the completed no-string solution differs from a vacuum solution by a singular perturbation with a distributional support (a delta function) on $\cal S$]. However, this failure of the no-string solution to be regular (or even a valid solution) on $\cal S$ turns out to be inconsequential in practice. Ref.\ \cite{PoundMerlinBarack2014} obtained a formulation of the physical self-force, complete with a practical mode-sum formula, from a no-string metric perturbation. This formulation requires information about the perturbation field (and its derivatives) only in the one-sided radial limits $r\to R(t)^\pm$, which avoid $\cal S$. It is this formulation that forms the basis for Ref.\ \cite{vandeMeent2018}'s calculation of the gravitational self-force for generic orbits in Kerr spacetime, using a frequency-domain method.

Importantly for us here, the no-string reconstruction also, in principle, enables calculations in the time domain. The idea is to solve the relevant evolution equation in each of the two vacuum regions ${\cal S}^{<}$ and ${\cal S}^{>}$, with suitable jump conditions across $\cal S$. In our method, we solve directly for the Hertz potential in the two vacuum regions, $\Phi^{<}_{\pm}$ in ${\cal S}^{<}$ and $\Phi^{>}_{\pm}$ in ${\cal S}^{>}$, with suitable jump conditions that relate between $\Phi^{<}_{\pm}$ and $\Phi^{>}_{\pm}$ on $\cal S$. The key ingredient in this formulation are, indeed, the particular jumps necessary for $\Phi^\gtrless_{\pm}$ to reproduce the no-string perturbation via 
\begin{equation}\label{h_rec<>}
h^{\rm \gtrless}_{\pm}:={\rm Re\, }{\sf\hat S}^\dagger_{\pm} \Phi^\gtrless_{\pm}.
\end{equation}
The derivation of the required jumps, for generic geodesic orbits in a Schwarzschild geometry, will be described in Section \ref{sec:Jumps}.

First, however, we present a formulation of the evolution problem for $\Phi^\gtrless_{\pm}$ via a 1+1D decomposition, henceforth specialising to the Schwarzschild case. 

%% file: HertzNonVacFormulation.tex

\section{1+1D evolution scheme for the no-string Hertz potential}
\label{Sec:HertzNonVacFormulation}

\subsection{Modal decomposition of the Weyl scalars}

For metric reconstruction we will need a decomposition of the Weyl scalars in the same basis. Recalling  $\Psi_{\pm}$ have spin weights $s=\pm 2$, we introduce  
\begin{equation} \label{expansionpsi}
\Psi_{\pm}=
\frac{\Delta^{\mp 2}}{r}
       \sum_{\ell=2}^{\infty}\sum_{m=-\ell}^{\ell}
       \psi^{\ell m}_{\pm}(t,r)\, {}_{\pm2}\!Y_{\ell m}(\theta,\varphi).
\end{equation}
As with the Hertz potential, the modal functions $\psi_{\pm}(t,r)$ of the Weyl scalars satisfy the vacuum 1+1D BPT equations
\begin{align} \label{Teukolsky1+1psi}
\psi^{\pm}_{,uv}+ U_{s}(r)\, \psi^{\pm}_{,u} + V_{s}(r)\, \psi^{\pm}_{,v}  + W_{s}(r)\psi_{\pm} =0 ,
\end{align}
away from the wordline where the explicit forms of the potentials are given in Eqs.\ (\ref{UV}) and (\ref{W_Sch}) with $s=\pm 2$ for $\psi_{\pm}$.

\subsection{Inversion relations in 1+1D}
\label{Sec:InvRel}

In our method we solve for the (modal) Hertz potential $\phi$ directly, making use of neither the BPT equation (\ref{Teukolsky1+1psi}) for $\psi$, nor the inversion relations that link $\psi$ to $\phi$. However, we {\it will} make use of the inversion relations in deriving jump conditions for $\phi$ across $\cal S$ (this will be done in Section \ref{sec:Jumps}), and for that purpose we need these relations in a 1+1D form. 

The inversion relations for the vacuum solutions $\Phi_+$ and $\Phi_-$ were given in Eqs.\ (\ref{inversionIRG}) and (\ref{inversionORG}), respectively. We recall there are two alternative relations for each of the two gauges, one linking (each of) $\Phi_\pm$ to $\Psi_+$, and another linking them to $\Psi_-$. In the Schwarzschild case, the relations read
\begin{subequations}\label{invertion_rad}
\begin{eqnarray}
{\boldsymbol D}_\ell^4\bar\Phi_+ &=& 2\Psi_{+}, 
\label{invertion_rad_IRG}
\\
\Delta^{2}\tilde{\boldsymbol D}_n^4\Delta^2\bar\Phi_- &=& 32\Psi_{-} ,
\label{invertion_rad_ORG}
\end{eqnarray}
\end{subequations}
(``radial'' inversion), and
\begin{subequations}\label{invertion_ang}
\begin{eqnarray}
\bar\eth_{-1}\bar\eth_0\bar\eth_{1}\bar\eth_{2}\bar\Phi_+-12M\partial_t\Phi_{+} &=& 8\Psi_{-}, 
\label{invertion_ang_IRG}
\\
\eth_{1}\eth_0\eth_{-1}\eth_{-2}\bar\Phi_-+12M\partial_t\Phi_- &=& 8\Psi_{+},
\label{invertion_ang_ORG}
\end{eqnarray}
\end{subequations}
(``angular'' inversion).
The differential operators ${\boldsymbol D}_\ell$ and $\tilde{\boldsymbol D}_n$, whose general definition is given just below Eq.\ (\ref{Ocompact}), are, in the Schwarzschild case,
\begin{equation}
{\boldsymbol D}_\ell = (2/f)\partial_v, \quad\quad
\tilde{\boldsymbol D}_n = -(2/f)\partial_u .
\end{equation}
The operators $\eth_{s}$ and $\bar\eth_{s}$ are the ``spin raising'' and ``spin lowering'' angular operators defined in Eq.\ (\ref{eth}), whose action on ${}_{s}\!Y_{\ell m}(\theta,\varphi)$ is described in Eq.\ (\ref{raising&lowering}).

To separate the radial inversion relations (\ref{invertion_rad}) into multipole modes, we first take the complex conjugate of Eq.\ (\ref{expansionHertz}) to obtain 
\begin{eqnarray}\label{expansionbarphi}
\bar\Phi_\pm =
\frac{\Delta^{\pm 2}}{r}
       \sum_{\ell,m}
       \bar\phi_{\pm}^{\ell,-m}(-1)^m {}_{\pm 2}\!Y_{\ell m},
\end{eqnarray}
where use was made of the symmetry relation (\ref{symmetry}). The expansions (\ref{expansionpsi}) and (\ref{expansionbarphi}) then separate Eqs.\ (\ref{invertion_rad}) to give, for each $\ell,m$, the fourth-order ODEs
\begin{subequations}\label{radinversion1}
\begin{equation}\label{radinversion_IRG1}
r\Delta^2 {\boldsymbol D}_\ell^4 \left(\Delta^2 \phi_+^{\ell m}/r\right) = 2 (-1)^m \bar\psi_{+}^{\ell,-m},
\end{equation}
\begin{equation}\label{radinversion_ORG1}
r\tilde{\boldsymbol D}_n^4 \left(\phi_-^{\ell m}/r\right) = 32 (-1)^m \bar\psi_{-}^{\ell,-m}.
\end{equation}
\end{subequations}
These relations can be written in a tidier form when the perturbation possesses a symmetry of refection about the equatorial plane, as in the setup to be considered in this thesis: a particle source moving in the equatorial plane of the Schwarzschild black hole. In this case we have the symmetry relation
\begin{equation}\label{psi_reflection}
\bar\psi_\pm^{\ell,-m}=(-1)^\ell \psi_\pm^{\ell m},
\end{equation}
which follows from the following argument. First, we note that under the reflection transformation $\theta\to \pi-\theta$ (with fixed $t,r,\varphi$) the tetrad legs $\ell^\alpha$ and $n^\alpha$ remain invariant, while $m^\alpha\to -\bar m^\alpha$ and $\bar m^\alpha\to -m^\alpha$ [see Eqs.\ (\ref{eq:kerrtetrad})]. Inspecting Eqs.\ (\ref{eq:psi}), we see this implies $\Psi_{\pm}\to \bar \Psi_{\pm}$, assuming the perturbed Weyl tensor $C_{\alpha\beta\gamma\delta}$ is invariant under such reflection. Thus, using (\ref{reflection}), we have
\begin{align}\label{expansionbarpsi}
\Psi_\pm(\theta)&=\bar\Psi_\pm(\pi-\theta) =
\frac{\Delta^{\mp 2}}{r}
       \sum_{\ell,m}
       \bar\psi_{\pm}^{\ell m}{}_{\pm 2}\!\bar Y_{\ell m}(\pi-\theta)
       \nonumber \\
    &=  \frac{\Delta^{\mp 2}}{r}
       \sum_{\ell,m}
       \bar\psi_{\pm}^{\ell,-m}(-1)^\ell {}_{\pm 2}\!Y_{\ell m}(\theta),
\end{align}
and a comparison with (\ref{expansionpsi}) then leads to (\ref{psi_reflection}). Using Eq.\ (\ref{psi_reflection}) we now write the 1+1D radial inversion relation (\ref{radinversion1}) in their final form,
\begin{subequations}\label{radinversion}
\begin{equation}\label{radinversion_IRG}
r\Delta^2 {\boldsymbol D}_\ell^4 \left(\Delta^2 \phi_+^{\ell m}/r\right) = 2 p \psi_{+}^{\ell m},
\end{equation}
\begin{equation}\label{radinversion_ORG}
r\tilde{\boldsymbol D}_n^4 \left( \phi_-^{\ell m}/r\right) = 32 p \psi_{-}^{\ell m},
\end{equation}
\end{subequations}
where 
\begin{equation}
p:=(-1)^{\ell+m},
\end{equation}
is the ``parity'' factor. We note (\ref{radinversion}) implies the 1+1D Hertz potentials share the same refection symmetry as the 1+1D Weyl scalars:\footnote{More precisely, Eqs.\ (\ref{radinversion}) alone imply (\ref{phi_reflection}) only up to homogeneous solutions of (\ref{radinversion}). However, no homogeneous solution of (\ref{radinversion}) satisfies the BPT equations as required, so such solutions can be excluded.}
\begin{equation}\label{phi_reflection}
\bar\phi_\pm^{\ell,-m}=(-1)^\ell \phi_\pm^{\ell m}.
\end{equation}

Let us next separate the angular inversion formulas (\ref{invertion_ang}). Using (\ref{raising&lowering}) with (\ref{expansionbarphi}) and (\ref{phi_reflection}), we have
\begin{subequations}\label{spinlowering}
\begin{eqnarray}
\bar\eth_{-1}\bar\eth_0\bar\eth_{1}\bar\eth_{2}\bar\Phi_+
&=&\frac{\Delta^{2}}{r}
       \sum_{\ell,m}p\,\lambda_2\,
       \phi_+^{\ell m} {}_{-2}\!Y_{\ell m},\quad\quad
       \\
\eth_{1}\eth_0\eth_{-1}\eth_{-2}\bar\Phi_-
 &=& \frac{\Delta^{-2}}{r}
       \sum_{\ell,m}p\,\lambda_2\,
       \phi_-^{\ell m} {}_{+2}\!Y_{\ell m} ,\quad\quad
\end{eqnarray}
\end{subequations}
where 
\begin{equation}
\lambda_n := \frac{(\ell+n)!}{(\ell-n)!}. 
\end{equation}
With this substitution, Eqs.\ (\ref{invertion_ang}) separate to give, for each $\ell,m$, the first-order ODEs
\begin{equation}\label{anginversion}
\partial_t\phi_{\pm}^{\ell m}  \mp p\alpha \phi_{\pm}^{\ell m} = \mp \frac{2}{3M}\psi_{\mp}^{\ell m},
\end{equation}
where
\begin{equation}\label{alpha}
\alpha:= \frac{\lambda_2}{12M}.
\end{equation}

We note that it is obviously possible to solve (\ref{anginversion}) in closed form in terms of a time integral involving $\psi_{\mp}$ (this was the main result of \cite{LoustoWhiting2002}). That, however, would not serve our purpose here. Recall that the inversion relations (\ref{anginversion}) are only valid in vacuum, and cannot be used (despite temptation) to relate the distributional contents of $\psi_{\mp}$ on $\cal S$ to these of the no-string Hertz potentials $\phi_{\pm}$. The idea, instead, is to use the inversion relations {\em evaluated in the two vacuum domains} ${\cal S}^>$ and ${\cal S}^<$ in order to get information about the jumps in $\phi_{\pm}$ across $\cal S$, given the known jumps in $\psi_{\mp}$. As we show in the next subsection, with some further manipulation [which also involves the radial inversion relations (\ref{radinversion})] this procedure can completely determine the jumps in $\phi_{\pm}$ and all of their derivatives on $\cal S$.

We also note the relation (\ref{anginversion}) means that (given $\psi_{\mp}$) all time derivatives of $\phi_{\pm}$ are determinable {\em algebraically} from $\phi_{\pm}$ itself. For example, taking $\partial_t$ of (\ref{anginversion}) and then substituting for $\partial_t \phi$ back from Eq.\  (\ref{anginversion}), we find
\begin{equation}\label{phitt}
\partial_{tt}\phi_{\pm} = \alpha^2 \phi_{\pm} \mp \frac{2}{3M}\left(\partial_t \psi_{\mp}\pm p\alpha \psi_{\mp}\right).
\end{equation}
Taking $\partial_r$ of (\ref{anginversion}) similarly determines $\partial_{tr}\phi_{\pm}$ algebraically from $\phi_{\pm}$ and $\partial_{r}\phi_{\pm}$.
With the help of the vacuum BPT equation (\ref{Teukolsky1+1phi}), we can then iteratively express $\partial_{rr}\phi_{\pm}$ and all higher derivatives of $\phi_{\pm}$ algebraically in terms of $\phi_{\pm}$ and $\partial_{r}\phi_{\pm}$ alone. The significance of this in the context of this work is as follows: It means we need only determine the jumps across $\cal S$ of $\phi_{\pm}$ and of its first $r$ derivative. The jumps in all $t$, $r$ and mixed derivatives to all orders are obtainable algebraically from these two alone. 

%% file: HertzJumps.tex
\section{Jump conditions for the no-string Hertz potential}
\label{sec:Jumps}

Ref.\ \cite{Barack:2017oir} sketched a method for obtaining the jumps across $\cal S$ for a generic geodesic orbit in Kerr spacetime, but the actual jumps were only calculated for circular orbits in the Schwarzschild case. In the general case (and even in the Schwarzschild limit) the method requires the solution of a complicated set of coupled fourth-order ODEs for the jumps in $\phi$ and in $\phi_{,r}$ along the orbit. There was no attempt to solve these equations (neither analytically nor numerically), except in the circular-orbit case, where they reduce to algebraic equations. 

Here we describe a different method for obtaining the jumps, and apply it to generic orbits in the Schwarzschild case. The method yields a single first-order ODE for the jump in $\phi$ along the orbit, which can be solved in closed form. The jumps in all partial derivatives of $\phi$, at any order, are then obtained algebraically from that solution. There were two key advances that made possible this much simpler and more effective approach: First, we have found a way of utilizing both radial and angular inversion formulas in tandem, in a particular way that simplifies the calculation. Second, we have observed certain algebraic simplifications that were overlooked in Ref.\ \cite{Barack:2017oir}.

We consider here only the IRG Hertz potential $\Phi_+$ (as also in \cite{Barack:2017oir}), but the jumps for the ORG potential $\Phi_-$ can be worked out in just the same way. We henceforth omit the label `$+$' for notational economy, taking $\Phi\equiv \Phi_+$ and $\phi^{\ell m}(t,r)\equiv \phi^{\ell m}_+(t,r)$. We let the interface $\cal S$ be described by the smooth function $r=R(t)$, and denote the jump in $\phi(t,r)$ across $\cal S$ by
\begin{equation}
[\phi]: = \lim_{\epsilon\to 0}\left[\phi^>(t,R(t)+\epsilon) -\phi^<(t,R(t)-\epsilon)\right],
\end{equation}
where $\phi^\gtrless$ are the {\em vacuum} fields in $\cal S^\gtrless$. The jumps in other 1+1D fields are similarly defined: $[\phi_{,r}]$, $[\phi_{,t}]$, $[\psi_{\pm}]$, etc. We think of $[\phi]$ as a function of coordinate time $t$ along the orbit, and note the relation 
\begin{equation}\label{chainrule}
\dot{[\phi]} =  [\phi_{,t}] +\dot{R} [\phi_{,r}],
\end{equation} 
where we recall that an overdot denotes $d/dt$.
 
In what follows we assume that the jumps across $\cal S$ of the modal Weyl scalars $\psi^{\ell m}_\pm(t,r)$ and of their first 3 derivatives are already known and are given. These jumps can be obtained in a straightforward way from the source of the Teukolsky equation. We carry out this calculation in Appendix \ref{App:WeylJumps} for generic (geodesic) orbits, and for both $\psi_-$ and $\psi_+$ (as both will be needed in our approach even if we restrict to the IRG potential $\phi_+$). 

\subsection{Expressions for $[\phi_{,t}]$ and $[\phi_{,r}]$ in terms of $[\phi]$}

Our task is to express each of $[\phi_{,r}]$ and $[\phi_{,t}]$ in terms of [$\phi$] alone (and possibly the known jumps in the Weyl scalars). Substitution into Eq.\ (\ref{chainrule}) would then give a first-order ODE for $[\phi]$. The second half of this task can be accomplished immediately thanks to the angular inversion formula (\ref{anginversion}). By applying Eq.\ (\ref{anginversion}) in both vacuum sides of $\cal S$ in the limit to $\cal S$, we obtain relations between the jumps $[\phi_{,t}]$, $[\phi]$ and $[\psi_{-}]$ (the latter assumed known). These relations are obtained by simply replacing $\phi\mapsto [\phi]$ etc.\ in Eq.\ (\ref{anginversion}) and setting $r\mapsto R(t)$ there to give
\begin{equation}\label{Jphit}
[\phi_{,t}]   = p\alpha [\phi] -\frac{2}{3M}[\psi_{-}].
\end{equation}
The jump $[\psi_{-}]$ is given in Eq.\ (\ref{Jpsipm}) of Appendix \ref{App:WeylJumps}, and recall $\alpha=\lambda_2/(12M)$. To obtain $[\phi_{,r}]$ in terms of $[\phi]$ is harder, and utilizes the fourth-order radial inversion (\ref{radinversion_IRG}), using a procedure we now describe.

First, we write (\ref{radinversion_IRG}) more explicitly in terms of coordinate derivatives. Using ${\boldsymbol D}_\ell = 2(r^2/\Delta)\partial_v$ (taken with fixed $u$), a calculation yields 
\begin{align}\label{inversion_explicit}
\partial_v^4 \phi
+\frac{2}{r^2}(3r-5M)\partial_v^3 \phi
+\frac{1}{r^4}(9r^2-26Mr+15M^2)\partial_v^2 \phi& \nonumber \\
+ \frac{1}{2r^5}(6r^2-21 Mr+16 M^2)\partial_v \phi
&= \frac{p}{8r^8}\psi_{+}.
\end{align} 
We now act with $\partial_u$ (fixed $v$) on both sides of (\ref{inversion_explicit}), and use the vacuum BPT equation (\ref{Teukolsky1+1phi}) to substitute for each mixed derivative $\phi_{,uv}$ in terms of $\phi_{,u}$, $\phi_{,v}$ and $\phi$. In the resulting expression we then substitute for $\partial_v^4 \phi$ from Eq.\ (\ref{inversion_explicit}). We arrive at a third-order ODE of the form 
\begin{equation}\label{inversion_3rd}
\sum_{n=0}^3 \hat A_n(r) \partial_v^n \phi
= \sum_{n=0}^1 \hat B_n(r)\partial_u^n \psi_{+},
\end{equation} 
where $\hat A_n(r)$ and $\hat B_n(r)$ are certain (rational) functions. Notably, no $u$ derivatives occur on the left-hand side. Repeating this procedure with a second application of $\partial_u$, this time replacing $\partial_v^3 \phi$ from Eq.\ (\ref{inversion_3rd}), yields a second-order ODE of the form 
\begin{equation}\label{inversion_2nd}
\sum_{n=0}^2 \tilde A_n(r) \partial_v^n \phi
= \sum_{n=0}^2 \tilde B_n(r)\partial_u^n \psi_{+},
\end{equation} 
with some other (rational) functions $\tilde A_n(r)$ and $\tilde B_n(r)$. Again, we find that no $u$ derivatives occur on the left-hand side. One last application of $\partial_u$ reduces the inversion relation to a first-order differential equation, which, however, is now a PDE, since it features both $\phi_{,u}$ and $\phi_{,v}$. We can, however, reduce this to an ODE by first converting to $r_*$ and $t$ derivatives using
$\phi_{,v}=\frac{1}{2}(\phi_{,t}+\phi_{,r_*})$ and $\phi_{,u}=\frac{1}{2}(\phi_{,t}-\phi_{,r_*})$, and then eliminating $\phi_{,t}$ using the angular inversion relation (\ref{anginversion}). This leads to a first-order ODE for $\phi$, which has the form
\begin{equation}\label{1stradial}
\phi_{,r_*} + A(r)\phi = p \sum_{n=0}^3  B_n(r)\partial_u^n \psi_{+} + B(r)\psi_{-}.
\end{equation}
An explicit calculation gives 
\begin{equation}
A(r)=-\alpha+\frac{2M\left[(2\lambda_1-3)r-6M\right]}{r^2(\lambda_1 r-6M)},
\end{equation}
for odd-parity modes ($p=-1$), and 
\begin{align}
A(r)=\alpha+
\frac{M\left[
4r^3\alpha^2(2\lambda_1+3)+2r^2\alpha \lambda_1(\lambda_1+4)+\lambda_1^2(3r-2M)\right]}
{r^2\left[2\alpha^2 r^2(\lambda_1 r+6M)+M\lambda_1(6\alpha r+\lambda_1)\right]} ,
\end{align}
for even-parity modes ($p=+1$).
The other radial coefficients in Eq.\ (\ref{1stradial}) are found to be given by
\begin{eqnarray}
B_0 &=&\lambda_1 f^3 r(\lambda^2 r^2+3M\lambda r+6M^2)/C(r) \ ,
\nonumber\\
B_1 &=& \big[18\alpha\lambda M r^4 + 8M^2 r^3\alpha(9-7\lambda) +M^2 r^2(4\lambda^3-9\lambda^2-31\lambda+24)  \nonumber \\ 
&& \quad +2M^3 r(\lambda+3)(7\lambda-13) +12M^4(\lambda+5)\big]/C(r)\ ,
\nonumber\\
B_2 &=& 12M r^2 \big[\alpha\lambda r^3-\alpha(\lambda-5)Mr^2  +2(1-2M\alpha)Mr -4M^2 \big]/C(r)\ ,
\nonumber\\
B_3 &=& 4Mr^3\left[\alpha r^2(\lambda r+6M)+3Mr f\right]/C(r) \ ,
\nonumber\\
B &=&  -4f^3 M^2 r^2\big[6\alpha r^4(\alpha^2 r^2+\lambda)-r^3\lambda(\lambda^2-4) +9M r^2(1-6M\alpha)-36M^2(r-M) \big]/C(r) , \nonumber\\
\end{eqnarray}
with
\begin{equation}
C=18 M^3 f^3 r^4\alpha \left[-2\alpha^2 r^4-\lambda r^2+2Mr(\lambda-1)+6M^2\right].
\end{equation}
Here we have introduced 
\begin{equation}
\lambda:=\lambda_2/\lambda_1 = (\ell+2)(\ell-1),
\end{equation}
and we remind $\lambda_1 = \ell(\ell+1)$ and $\alpha= \lambda_2/(12M)$.

Using Eq.\ (\ref{1stradial}) (imposed in the limit to $S$ on both sides of $\cal S$) we can finally express the jump $[\phi_{,r_*}]=f(R)[\phi_{,r}]$ in terms of the jump $[\phi]$ (and the known jumps in the Weyl scalars $\psi_\pm$):
\begin{equation}\label{Jphir}
[\phi_{,r_*}]   = - A(R) [\phi] 
+p\sum_{n=0}^3  B_n(R)[\partial_u^n \psi_{+}] + B(R)[\psi_{-}].
\end{equation}

\subsection{First-order ODE for $[\phi]$ and its solution}

Substituting Eqs.\ (\ref{Jphit}) and (\ref{Jphir}) in (\ref{chainrule}) now gives a simple first-order ODE for $[\phi]$ as a function along the orbit:
\begin{equation}\label{ODE}
\dot{[\phi]} + \Big(A(R)\dot{R}_*-p \alpha\Big)[\phi] = {\cal F}
\end{equation}
where $\dot{R}_*=\dot{R}/f(R)$. The source term here is 
\begin{equation}\label{calF}
{\cal F} =  p \dot{R}_*\sum_{n=0}^3  B_n(R)[\partial_u^n \psi_{+}] + \left(\dot{R}_* B(R)-\frac{2}{3M}\right)[\psi_{-}].
\end{equation}

Equation (\ref{ODE}) admits simple homogeneous solutions, given by (any constant multiple of)
\begin{equation}\label{[phi]AsympEven}
[\phi]_{\rm h}= \left(\frac{R(\lambda_1 R-6M)}{(R-2M)^2}\right)\times e^{-\alpha(t-R_*)}
\end{equation}
for odd-parity modes, or
\begin{equation}\label{[phi]AsympOdd}
[\phi]_{\rm h}= \left(\frac{R^3\lambda^2\lambda_1+6MR^2\lambda^2+36M^2 R\lambda+72M^3}{R(R-2M)^2} \right) \times e^{\alpha(t-R_*)}
\end{equation}
for even-parity modes.
The general inhomogeneous solution of (\ref{ODE}) reads
\begin{equation}
[\phi] = [\phi]_{\rm h}\int_{t_0}^t \frac{{\cal F}(t')}{[\phi]_{\rm h}}\, dt',
\end{equation} 
where $t_0$ is an a-priori arbitrary integration constant. We determine $t_0$ from the physical requirement that $[\phi]$ remains bounded for $t\to\pm\infty$. Observing that $[\phi]_{\rm h}$ blows up like $e^{\pm\alpha t}$ at  $t\to\pm \infty$ ($+$ for even parity modes, $-$ for odd-parity modes), while ${\cal F}(t)$ is at worst polynomial in $t$, it is easy to see that the requirement of boundedness necessitates $t_0=\pm\infty$ for $p=\pm 1$. 
Hence, the unique physical solution of (\ref{ODE}) is  
\begin{equation}\label{HertzJump}
[\phi] = [\phi]_{\rm h}\int_{\pm\infty}^t \frac{{\cal F}(t')}{[\phi]_{\rm h}} dt' \quad \text{(for $p=\pm 1$)}.
\end{equation} 

Equation (\ref{HertzJump}) gives the jumps across $\cal S$ that the no-string Hertz potential modes must satisfy, for an arbitrary orbit in Schwarzschild spacetime. (It requires as input the jumps in the modes of the Weyl scalars, which in Appendix \ref{App:WeylJumps} we give explicitly specialised to {\em geodesic} orbits; but given the Weyl scalar jumps, there is no further assumption on whether the orbit is geodesic.) This is one of the main results of presented in this thesis.

We recall that the jumps in the field's derivatives, $[\phi_{,t}]$ and $[\phi_{,r}]$ (or $[\phi_{r_*}]$), can be obtained algebraically from $[\phi]$, using Eqs.\ (\ref{Jphit}) and (\ref{Jphir}), respectively. In principle, knowledge of the jumps in the field and its first derivatives should suffice in our formulation. However, in practice it is also useful to have at hand the jumps in higher derivative, which eases the formulation of finite-difference schemes that have high-order convergence properties. Once the jumps in the field and its first derivatives are known, it is straightforward to obtain the jumps in higher derivatives in an iterative manner using the procedure described in the last paragraph of Section\ \ref{Sec:InvRel} [the paragraph containing Eq.\ (\ref{phitt})]. The application of this procedure up to third derivatives is illustrated in Appendix \ref{app:WeylJumpHighDerivatives} (as applied to modes of the Weyl scalars). 



\subsection{Large-$R$ asymptotics for scatter orbits}
\label{subsec:asymptotics}

We were not able to evaluate the integral in (\ref{HertzJump}) analytically for a generic orbit, but it is straightforward to compute $[\phi](t)$ numerically for any given geodesic orbit. In practice we find it easier to obtain $[\phi]$ by (numerically) solving the first-order ODE (\ref{ODE}). For the class of scatter orbits of interest to us in this paper, we need to integrate the equation over $-\infty < t< \infty$. We choose to do so forward in time for for odd-parity modes but backward in time for even-parity modes, in each case going ``against'' the direction of exponential growth of the homogeneous solutions (\ref{[phi]AsympEven}) and (\ref{[phi]AsympOdd}). This prevents the growth of nonphysical modes from numerical error. We now derive the leading-order asymptotic form of $[\phi]$ at $t\to \pm\infty$. One of these two asymptotic values will be used as an initial value for the ODE solver, and the other will be used to check the result of the numerical integration.

We consider a timelike scatter geodesic orbit in Schwarzschild spacetime, parametrised by specific energy $E>1$ (``gamma factor'') and angular momentum $L$. We let
\begin{equation}
\Rdotinf:= \pm\left|\dot{R}(t\to\pm\infty)\right| = \pm\frac{\sqrt{E^2-1}}{E}
\end{equation}
be the ``velocity at infinity'' (with respect to coordinate time $t$), so that $\Rdotinf$ is negative (positive) for the inbound (outbound) asymptotic states (Note that $|\dot R_\infty|=v_\infty$). We formally expand $[\phi]$ as a power series in $1/R$ at large $R(t)$, and seek to obtain the leading term of that expansion. 

To this end, we first obtain the large-$R$ asymptotic form of $\cal F$ in Eq.\ (\ref{calF}). Using as input the asymptotic expressions derived in Appendix \ref{WeylJumpsAsympt} for $[\psi_-]$ and $[\partial^n_u\psi_+]$ ($n=0,\ldots,3$), a direct calculation leads to
\begin{equation}\label{calFasy} 
{\cal F} = c_0 R^{-3} +O(R^{-4}),
\end{equation}
where
\begin{align}\label{c0_initial}
c_0 = \frac{4\pi\mu(1+\Rdotinf)}{3\lambda_2}
\Big[
i(L/M)\lambda_2 \Big(\partial_\theta - m \Big){}_{-2}\! \bar{Y}_{\ell m} + 6E\dot{R}_{\infty}
\Big(\partial_{\theta\theta} -2m\partial_{\theta} +(m^2-2)  \Big) {}_{-2}\! \bar{Y}_{\ell m}
\Big].
\end{align}
Here all angular functions are evaluated at $\theta=\pi/2$ and $\varphi=\varphi_{\rm in}$ (or $\varphi=\varphi_{\rm out})$, with $\varphi_{\rm in}$ ($\varphi_{\rm out}$) being the asymptotic value of the particle's azimuthal phase at $t\to -\infty$ ($t\to +\infty$). 
Equation (\ref{c0_initial}) takes a neater form when written in terms of spin-$0$ spherical harmonics. With the aid of (\ref{Y2}), we find
\begin{align}
c_0 = \frac{4\pi\mu (1+\Rdotinf)}{3M\sqrt{\lambda_2}}
\left[\Big(6ME\Rdotinf -im\lambda L \Big)\bar Y -i\lambda L \bar Y_{\theta}\right], 
\end{align}
where $\bar Y:= \bar Y_{\ell m}\!\left(\frac{\pi}{2},\varphi_{\rm in/out}\right)$ and $\bar Y_\theta:= \partial_\theta\bar Y_{\ell m}\!\left(\frac{\pi}{2},\varphi_{\rm in/out}\right)$.

The asymptotic form of $[\phi]$ can now be obtained either by evaluating (\ref{HertzJump}) with the asymptotic form (\ref{calFasy}), or directly from the ODE (\ref{ODE}) using a power-law ansatz. Either way, we arrive at 
\begin{align}
\label{JphiAsymp}
[\phi]_{R\to\infty} =
-\frac{16\pi\mu}{\lambda_2^{3/2}}\left(\frac{1+\dot{R}_{\infty}}{1-\dot{R}_{\infty}}\right)
\Big[6M \dot{R}_{\infty}E \bar Y + i \lambda L \Big(\bar Y_{\theta}- m \bar Y\Big)
 \Big] R^{-3} + O(R^{-4}).
\end{align}
We note $\bar Y_{\theta}=0$ for even-parity modes, and $\bar Y=0$ for odd-parity modes. 
Our result (\ref{JphiAsymp}) can be checked against the $m=0$, circular-orbit expression given in Eq.\ (87) of Ref.\ \cite{Barack:2017oir}, by setting $\dot{R}_{\infty}=0$, $r_0=R$, $\Omega=\sqrt{M/R^3}$ and ${\cal Y}_\theta = -\lambda\bar Y_{\theta}/\sqrt{\lambda_2} $. We find an agreement.

In the next section, for reasons that will become clear there, we will require also the asymptotic forms of the jumps $[\phi_{,v}]$ and $[\phi_{,vv}]$. The jump $[\phi_{,v}]=\frac{1}{2}([\phi_{,t}]+[\phi_{,r_*}])$ is obtained using Eqs.\ (\ref{Jphit}) and (\ref{Jphir}) with the known asymptotic expressions for $[\phi]$, $[\psi_-]$ and $[\partial^n_u\psi_+]$. The result is
\begin{equation}\label{JphivAsymp}
[\phi_{,v}]_{R\to\infty} = \frac{4\mu\pi E (1+\Rdotinf)}{\sqrt{\lambda_2}}\bar Y R^{-3} + O(R^{-4}).
\end{equation}
The asymptotic form of $[\phi_{,u}]$ is obtained in a similar way.
The jump $[\phi_{,vv}]$, in turn, can be written in terms of lower-derivative jumps as explained in the last paragraph of Section\ \ref{Sec:InvRel}, and, substituting the asymptotic expressions already obtained for these, one finds
\begin{equation}\label{JphivvAsymp}
[\phi_{,vv}]_{R\to\infty} = -\frac{4\mu\pi E (2+\Rdotinf)}{\sqrt{\lambda_2}}\bar Y R^{-4} + O(R^{-5}).
\end{equation}

%% file: RWJumps.tex

\section{Jump conditions for the Regge-Wheeler-like variable}
\label{sec:RWJumps}

It remains to translate the jumps in $\phi$ and its derivatives across $\cal S$, obtained in Section\ \ref{sec:Jumps}, to jumps in $X$ and its derivatives there. For brevity we only discuss here the IRG case, but jumps for the ORG case can be obtained in a similar manner. 

We could not find an explicit inverse of the transformation (\ref{transformationIRG}), but (given $\phi_+$) it is easy to obtain two independent algebraic relations between $X$, $X_{,u}$ and $X_{,v}$, which will suffice for our purpose. First, taking $\partial_v$ of (\ref{transformationIRG}), and using $(\ref{RWeq})$ and (\ref{transformationIRG}) to substitute for $X_{,uv}$ and $X_{,uu}$, respectively, leads to 
\begin{align}\label{chiu}
X_{,u}=\frac{f}{r(\lambda r+6M)}\left(3M X-8Mr^2 \phi_+
-4r^4 \phi_{+,v}\right),
\end{align}
where, recall, $\lambda=(\ell+2)(\ell-1)$.
Second, taking $\partial_{vv}$ of (\ref{transformationIRG}), then using the $u$ and $v$ derivatives of the RW equation (\ref{RWeq}) to replace for $X_{,uvu}$ and $X_{,uvv}$, and finally using $(\ref{RWeq})$ and (\ref{transformationIRG}) again to replace for $X_{,uv}$ and $X_{,uu}$, we obtain  
\begin{equation}\label{chiv}
X_{,v}=-\left(\alpha+\frac{3Mf}{r(\lambda r+6M)}\right)X + \frac{8M r (\lambda+3)}{3(\lambda r+6M)}\,\phi_+ +\frac{4r^3[\lambda r+(\lambda+9)M]}{3M(\lambda r+6M)}\, \phi_{+,v}
+\frac{4r^4}{3M}\,\phi_{+,vv} .
\end{equation}
By applying Eqs.\ (\ref{chiu}) and (\ref{chiv}) in both vacuum sides of $\cal S$ in the limit to $\cal S$, we obtain relations between the jumps $[X_{,u}]$ and $[X_{,v}]$ on the one hand, and the jumps $[X]$, $[\phi_+]$, $[\phi_{+,v}]$ and $[\phi_{+,vv}]$ (the latter 3 assumed known) on the other hand; these relations are obtained by simply replacing $X\mapsto [X]$ etc.\ in Eqs.\ (\ref{chiu}) and (\ref{chiv}), and setting $r\mapsto R(t)$ there. 

Now, along the orbit we also have the relation 
\begin{equation}\label{chainruleX}
\dot{[X]} =  [X_{,t}] +\dot{R}_* [X_{,r_{*}}],
\end{equation} 
where, recall, an overdot denotes $d/dt$. Using $[X_{,t}]=[X_{,v}]+[X_{,u}]$ and $[X_{,r_*}]=[X_{,v}]-[X_{,u}]$ and substituting $[X_{,u}]$ and $[X_{,v}]$ from Eqs.\ (\ref{chiu}) and (\ref{chiv}), we thus obtain a simple first-order ODE for the jump $[X]$ along the orbit, of the form [compare with (\ref{ODE})]
\begin{equation}\label{RWODE}
\dot{[X]} + \Big(A_X(R)\dot{R}_*+\alpha\Big)[X] = {\cal F}_X .
\end{equation}
The coefficient $A_X$ here is given by 
\begin{equation}
A_X= \alpha+\frac{6M(R-2M)}{R^2 (\lambda  R+6M)},
\end{equation}
and the sourcing function ${\cal F}_X$ involves the known jumps in the Hertz potential and its derivatives:
\begin{align}
{\cal F}_X =& \frac{8 M \left[6 M \left(f_R - \dot R \right)+ R \left(\lambda  (f_R+\dot{R})+6 \dot{R}\right)\right]}{3 f_R (\lambda  R+6M)} \J{\phi_+} +\frac{4 R^4 (f_R+\dot{R})}{3 M f_R} \J{\phi_{+,vv}}\nonumber \\
&+ \frac{4 R^2}{3 M f_R (\lambda  R+6M)} \Big[6 M^2 (f_R-\dot{R}) +\lambda  R (M+R) (f_R+\dot{R})+6 M R (f_R+2\dot{R})\Big] \J{\phi_{+,v}}, 
\label{RWODESource}
\end{align}
recall that $f_R=1-2M/R$.

Equation (\ref{RWODE}) admits simple homogeneous solutions, given by (any constant multiple of) 
\begin{equation}\label{X_hom}
[X]_{\rm h}= \left(\lambda+\frac{6M}{R}\right)\, e^{-\alpha(t+R_*)}.
\end{equation}
Note that all these homogeneous solutions (except the trivial zero one) blow up exponentially at $t\to -\infty$. There is a unique particular solution of the full inhomogeneous equation (\ref{RWODE}) that remains bounded; it is given by
\begin{equation}\label{XJump}
[X] = [X]_{\rm h}\int_{-\infty}^t \frac{{\cal F}_X(t')}{[X]_{\rm h}} dt' .
\end{equation} 
This describes the jump in the RW variable needed for it to reproduce the no-string Hertz potential. 

\begin{figure}[h!]
	\begin{center}
        \includegraphics[width=0.8\linewidth]{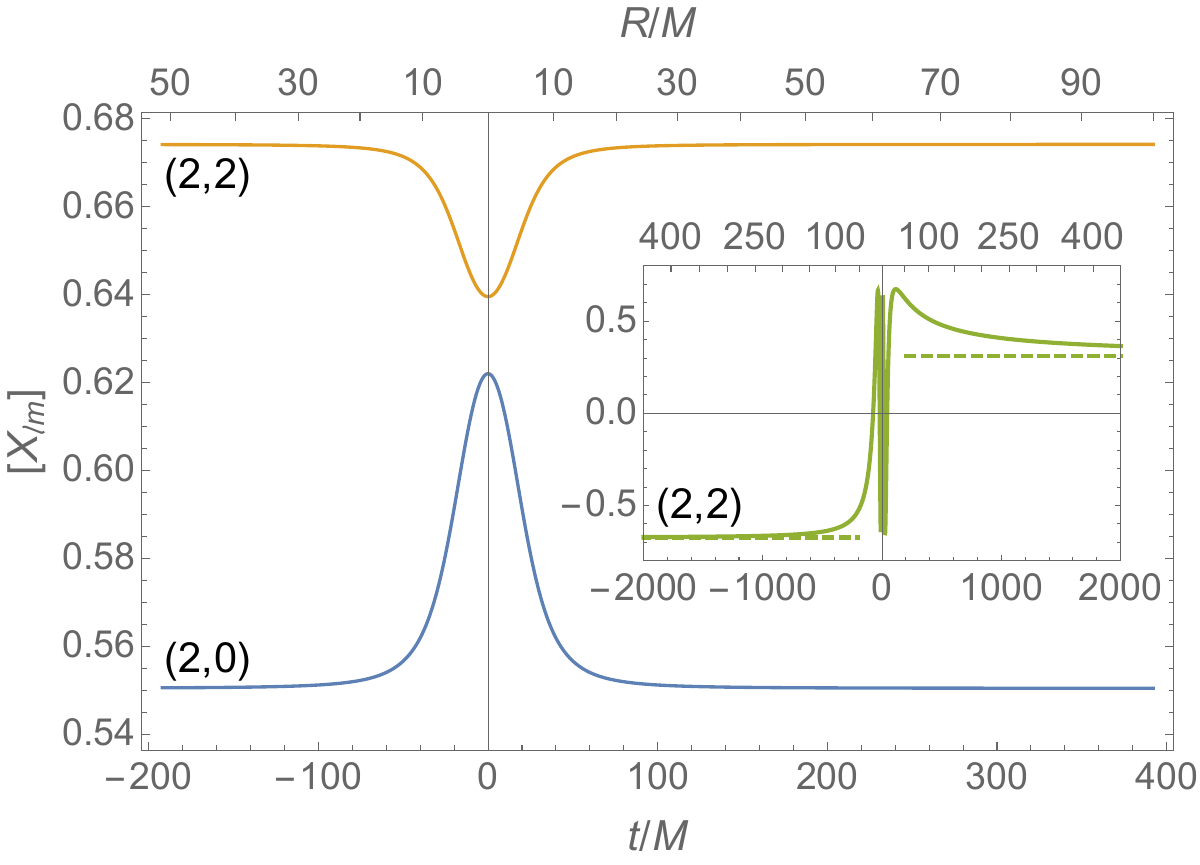} 
\caption[The jump in the ``no-string'' Regge-Wheeler field along a hyperbolic orbit]{The modulus of the jump $[X_{\ell m}]$ in the ``no-string'' field $X_{\ell m}$ along the geodesic scatter orbit depicted in Figure \ref{orbit}, for the $(\ell,m)=(2,0)$ and $(2,2)$ modes. This is obtained by numerically integrating the first-order ODE (\ref{RWODE}) forwards in time with the initial condition (\ref{IC}) at large negative $t$ and $\varphi_{\rm in}=0$. As a check, the solution approaches the asymptotic value given in (\ref{IC}) with $\varphi_{\rm out}\simeq481^\circ$ (obtained by integrating the geodesic equation). The inset plot demonstrates this for the real part of $[X_{22}]$, with dashed lines indicating the analytical asymptotic values. This jump function inputs into our 1+1D characteristic evolution scheme, whose application is illustrated in Sections \ref{sec:RWCirc} and \ref{sec:RWScatter}.
}  
        \label{Fig:X_jump}
	\end{center}
\end{figure}

In practice we find it easier to calculate $[X]$ not from Eq.\ (\ref{XJump}) but via a numerical integration of the first-order ODE (\ref{RWODE}). It is best to integrate forward in time from $t\to -\infty$, going ``against'' the direction of exponential growth of the homogeneous solutions (\ref{X_hom}), in order to prevent the growth of such nonphysical modes from numerical error. For this integration we need an initial condition at $t\to -\infty$, which in the case of a scatter orbit corresponds to $R\to\infty$ (with $\dot{R}_\infty<0$). The condition is obtained from a simple asymptotic analysis of the ODE (\ref{RWODE}): Assuming $[X]$ has a power-law behavior at infinity, we have $\dot{[X]}\ll [X]$ at large $R$, so the derivative term in (\ref{RWODE}) may be dropped at leading order. Then, using the $R\to\infty$ limits $A_X \to \alpha$ and 
\begin{equation}
{\cal F}_X \to -\frac{16\pi\mu E}{3M\sqrt{\lambda_2}}(1+\dot{R}_\infty) \bar Y,
\end{equation}
[obtained with the help of Eqs.\ (\ref{JphiAsymp})--(\ref{JphivvAsymp})], we arrive at 
\begin{equation}\label{IC}
[X]_{R\to\infty} = -\frac{64\pi\mu E}{\lambda_2^{3/2}}\, \bar Y_{\ell m}\!\left(\frac{\pi}{2},\varphi_{\rm in/out}\right) ,
\end{equation}
which applies with $\varphi_{\rm in}$ for $t\to -\infty$, and with $\varphi_{\rm out}$ for $t\to +\infty$.
In our implementation we use (\ref{IC}) to set the initial value of $[X]$ at $t\to -\infty$, integrate the ODE (\ref{RWODE}) forward in time, and then use (\ref{IC}) again to check the result of integration at $t\to +\infty$. Figure \ref{Fig:X_jump} illustrates the result of applying this procedure along a particular strong-field scatter orbit (the one depicted in Figure \ref{orbit}).

Once we have $[X]$, it is straightforward to get jumps in derivatives of the field, also needed for our 1+1D evolution scheme. This can be done algebraically. From Eqs.\ (\ref{chiu}) and (\ref{chiv}) one immediately gets $[X_{,u}]$ and $[X_{,v}]$, and using the RW equation (\ref{RWeq}) one gets $[X_{,uv}]$. The jump $[X_{,uu}]$ is subsequently obtained from the transformation equation (\ref{transformationIRG}), and $[X_{,vv}]$ can be found from the $v$ derivative of Eq.\ (\ref{chiv}). The jumps in third and higher derivative can be found recursively in a similar manner.  

In the chapter we showed how the vacuum metric reconstruction procedure fails with the addition of a source and how this can be overcome for a point-particle using a ``no-string'' radiation gauge by only solving in vacuum regions. We developed a 1+1D evolution scheme capable of calculating the no-string Hertz potential. The crucial ingredient for this implementation is the jump conditions across the worldine of the particle. We derive closed-form equations for jumps in the Hertz potential and the RW variable, as well as their derivatives. This gives us all the required inputs to perform a calculation of the Hertz potential for a point-particle source, which will be undertaken in the next chapter.

%% file: RWCircular.tex

In this chapter we implement our 1+1D method formulated in the last chapter. In our circular-orbit implementation we show that the new method produces the correct solutions for the IRG Hertz potential by comparing to static analytic solutions and the results of an implementation which directly solves the Teukolsky equation (but contains the $t^4$ IRG non-physical mode). We extend our implementation to the scatter orbit case and present first-of-their-kind calculations of gravitational self-force quantities for unbound orbits.

\section{Circular-orbit implementation} 
\label{sec:RWCirc}

Here we present an implementation of our new method for a circular equatorial orbit. This allowed us to compare our new method against known analytic solutions for static modes and the previous work of Barack \& Giudice \cite{Barack:2017oir}.

\subsection{Jumps across $\cal S$}

In order to numerically evolve the Regge-Wheeler equation (\ref{RWeq}) we first need to obtain the jump conditions $[X]$ along $\cal S$. For this we need jumps in the Hertz potential, and its derivatives, which in turn require the jumps in the Weyl scalars, and their derivatives. $[\psi]$, and jumps in the derivatives, can be obtained from the results in Appendix \ref{App:WeylJumps} by using the simplifications $\dot R=0$ and the fact that any time derivative is taken using $\partial_t \to - i m \Omega$, due to the fact that the only time dependence comes from $\varphi_\text{p}=\Omega t$ in the spherical harmonics. $[\phi]$ can be obtained by using the same simplifications when solving Eq.\ (\ref{ODE}), which reduces the ODE to an analytic equation.  Explicit expressions for the jumps in the Weyl scalar (for $s=-2$) and IRG Hertz potential for the circular orbit case are given in Appendix A and Section IV of \cite{Barack:2017oir} respectively.

Similarly, we can apply the circular orbit simplifications to Eq.\ (\ref{RWODE}) to obtain an analytic equation for the jumps in the RW field:
\begin{equation}
[X] = \frac{16 }{\lambda_2-12 i m M \Omega} \left( 2M^2 \J{\phi_+}+ R^2(M+R) \J{\phi_{+,v}} R^4 \J{\phi_{+,vv}}\right).
\end{equation}
With this result we can easily obtain jumps in the fields derivatives as detailed in the last paragraph of Section \ref{sec:RWJumps}. Using this method gives the higher-order jump conditions needed for the finite-difference scheme (see Appendix \ref{app:FDSJumpsCircular}) as
\begin{align}
\J{X_{,u}} =& \frac{f_R}{R(6M+R\lambda)} \left( 3M \J{X} - 8M R^2 \J{\phi_+} -4R^4 \J{\phi_{+,v}} \right), \\
\J{X_{,uv}} =& -\frac{f_R}{4}\left(\frac{\lambda_1}{R^2}-\frac{6M}{R^3} \right) \J{X},\\
\J{X_{,uu}} =& -i m \Omega \J{X_{,u}} - \J{X_{,uv}}.
\end{align}

\subsection{Analytic solutions for static modes}  

As with the scalar field on a circular orbit case, we can obtain analytic solutions for the fields $\phi$ and $X$ in the static case where $m=0$. The analytic solution for both fields consists of a superposition of two independent vacuum solutions, one valid for the regime $r < R$ and the other valid for $r > R$. The independent solutions for the Hertz potential are given by
\begin{equation}
\phi^<_{\ell0} = \frac{A P^2_\ell(\rho)}{fr}, \qquad\qquad \phi^>_{\ell0} = \frac{B Q^2_\ell(\rho)}{fr},
\end{equation}
where $P^m_\ell$ and $Q^m_\ell$ are the associated Legendre polynomials of the first and second kinds respectively, $A$ and $B$ are constants and we recall $\rho = (r-M)/M$. The solutions have this split form across ${\cal S}$ due to their asymptotic behaviour. $\phi^<_{\ell0}$ is finite at $r\rightarrow 2M$ but is divergent in the limit $r\rightarrow \infty$. Conversely,  $\phi^>_{\ell0}$ diverges in the limit $r\rightarrow 2M$ but is finite at $r\rightarrow \infty$.

The independent solutions for the RW field are
\begin{eqnarray}
X^<_{\ell0} &=& C z^{-\ell-1} \: {}_2  F_1 (-\ell-2,- \ell+2;-2\ell;z), \\
X^>_{\ell0} &=& D z^{\ell} \: {}_2  F_1 (\ell-1,\ell+3;2\ell+2;z),
\end{eqnarray}
where $z:= 2M/r$, $C$ and $D$ are constants and ${}_2 F_1$ are hypergeometric functions defined by
\begin{equation}
x(1-x) \partial_x^2 \: y(x) + \left( c-(a+b+1)x \right)\partial_x \: y(x) - ab \: y(x) = 0,
\end{equation}
with $y(x) = {}_2 F_1 (a,b;c;x)$. Again, the split across $\cal S$ is chosen due to their asymptotic behaviour as $X^<_{\ell0}$ ($X^>_{\ell0}$) is finite (divergent) as $r\rightarrow 2M$ and divergent (finite) as $r\rightarrow \infty$.

We can determine the constants of the solutions by taking the difference of the solutions (and one of their derivatives) at the wordline $r=R$ and then use the appropriate jump conditions to form a set of simultaneous equation. The solutions for the constants are cumbersome hence we are not writing them here. For the explicit form of the IRG Hertz potential solutions we refer the reader to Section IV A of \cite{Barack:2017oir}.

\subsection{Numerical method}
\label{sec:NumericalMethodCircularRW}

Our numerical implementation has the same setup to that of the scalar field with a point-particle source as described in Section \ref{sec:ScalarCirc}, but here we evolve the RW equation (\ref{RWeq}). We centre the grid on the orbital radius of the particle $r=R$ such that $\cal S$ passes directly through a set of field values. However, this creates an issue as the field values are not defined on the worldline. To overcome this we shift $\cal S$ to an infinitesimally larger $r>R$ such that we calculate $X^<$ along $r=R$. Given $X^<$ we can trivially obtain the field value corresponding to the $\cal S^>$ limit to the worldline using the definition of the jump conditions $X^> = X^< + [X]$. We set the field values on the initial rays as $X^<(u,v_0)=X^>(u_0,v)=0$ and let the evolution be sourced by the jump conditions across the surface $\cal S$.

This is done at the level of the finite-difference scheme as detailed in Appendix \ref{app:FDSJumps}. Our code takes as input the radius of the orbit $R$, multipolar numbers $\ell,m$ as well as various numerical parameters, and returns the generating-function fields $X^\gtrless_{\ell m}(t,r)$ and the IRG no-string modal Hertz-potential fields $\psi^\gtrless_{\ell m}(t,r)$. A more detailed description of our numerical algorithm, as applied to the hyperbolic-orbit case, is given in Section \ref{sec:NumericalAlgorithmScatter}.

\subsection{Sample results}

\begin{figure}[h!]
\centering
\includegraphics[width=0.8\linewidth]{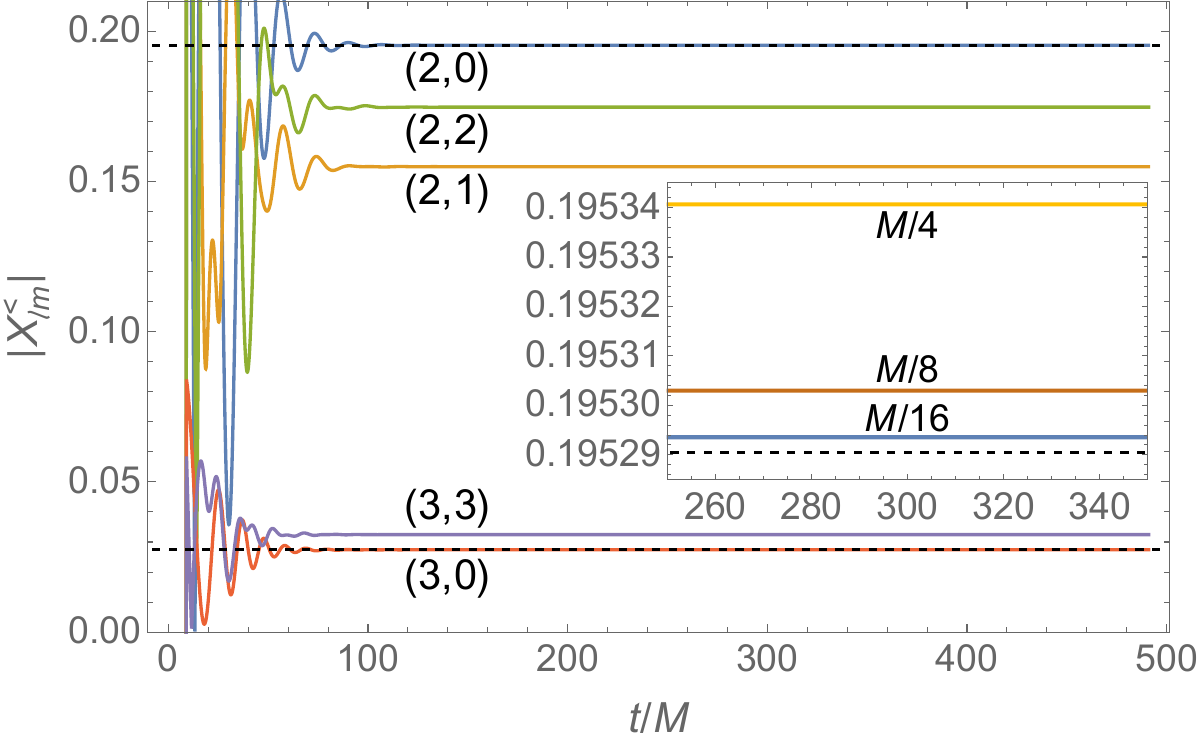}
\caption[The Regge-Wheeler-like field along the worldline of a circular orbit]{The behaviour of the Regge-Wheeler-like field $|X^<_{\ell m}|$ along the particle's (chosen to be at $r_*=9M$) for various $(\ell,m)$ modes. The early part of the data is contaminated with initial junk radiation, which is to be discarded. We show the analytic solutions for static modes (dashed) for comparison. The inset shows how the $|X^<_{20}|$ results converge to the analytic solution (dashed) with decreasing grid spacings $h=\left\{\frac{1}{8},\frac{1}{16},\frac{1}{32}\right\}M$.}
\label{RWCirc}
\end{figure}

Figure \ref{RWCirc} shows the behaviour of the field $X^<_{\ell m}$ along the worldline of the particle, chosen to be at $r_*=9M$, for a selection of modes. $X^>_{\ell m}$ shows similar behaviour. The early-time behaviour is dominated by junk radiation which decays in time (as an inverse $\ell$-dependent power law). The late time solution approaches the true analytic solution, as shown explicitly by comparing the numerical and analytic solutions of the static modes.

\begin{figure}[H]
\centering
\includegraphics[width=0.8\linewidth]{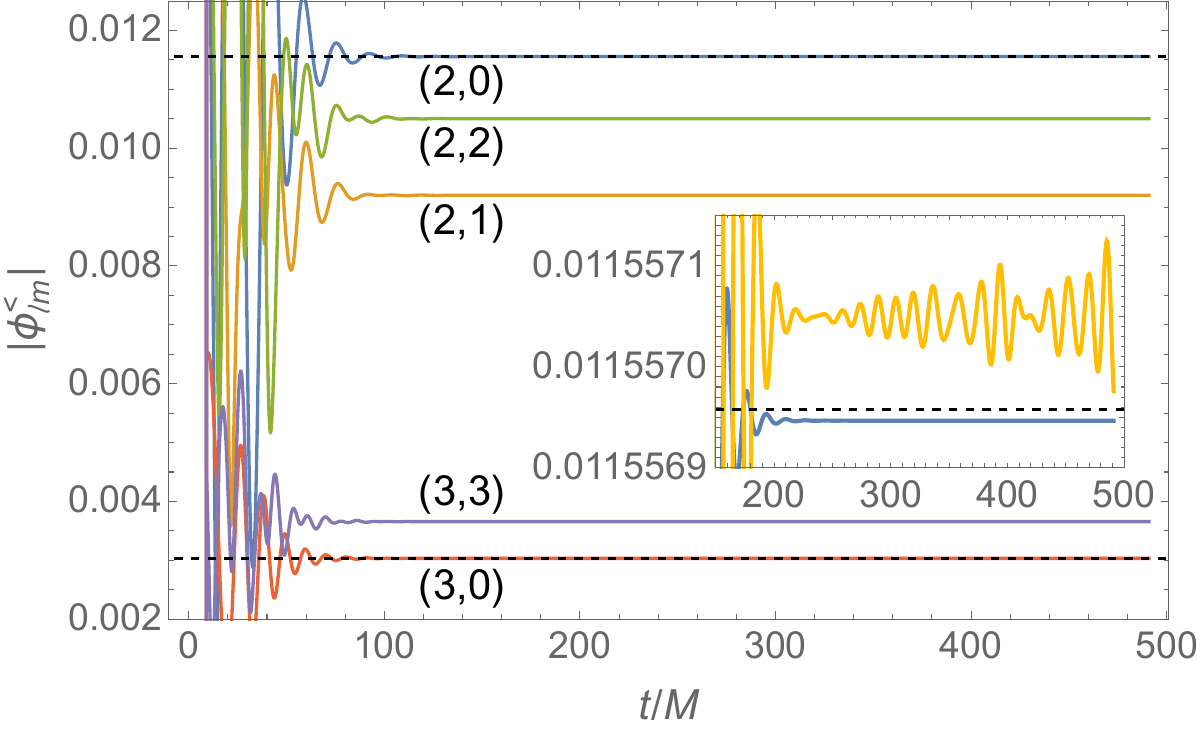}
\caption[The IRG Hertz potential along the worldline of a circular orbit]{The IRG Hertz potential $|\phi^<_{\ell m}|$ along the particle's worldline for an orbit of $r_*=9M$ for a selection of $(\ell,m)$ modes. Dashed lines represent analytic solutions (obtained from Sec.\ IV A of \cite{Barack:2017oir}) and are shown for reference. The inset shows a subset of the $|\phi^<_{20}|$ data as calculated by our method (blue) and directly from the Teukolsky equation (yellow) using the method of \cite{Barack:2017oir}, which shows evidence of the $t^4$ growth discussed in Section \ref{sec:Teuks=-2}.}
\label{HertzCirc}
\end{figure}

We can form a solution for the IRG Hertz potential modes $\phi_{\ell m}$ by applying the Chandrasekhar transformation (\ref{transformationIRG}) to the generating function $X_{\ell m}$. The behaviour of the Hertz potential along the worldline is shown in Figure \ref{HertzCirc}. The late-time results produced by our new method were compared with those from an implementation that directly solves the Teukolsky equation (using the method detailed in Ref.\ \cite{Barack:2017oir}) and were found to be in good agreement before the comparison data becomes dominated by the $t^4$ growth. From this we can conclude that our new method can successfully obtain the no-string IRG Hertz potential without the presence of non-physical modes of the Teukolsky equation.

%% file: RWScatter.tex

\section{Hyperbolic-orbit implementation} 
\label{sec:RWScatter}

We now present a full implementation of our method for a strong-field geodesic scatter orbit. Our code takes as input the parameters of the geodesic orbit, along with multipolar numbers $\ell,m$, and returns the generating-function fields $X^\gtrless_{\ell m}(t,r)$ and the IRG no-string modal Hertz-potential fields $\psi^\gtrless_{\ell m}(t,r)$.  

\subsection{Numerical algorithm}
\label{sec:NumericalAlgorithmScatter}

Our method is based on a characteristic numerical evolution in $u,v$ coordinates, as described in Section \ref{sec:ScalarScatter}---only here we are evolving the Regge-Wheeler-like equation (\ref{RWeq}) instead of the scalar field equation, and we impose suitable jump conditions along $\cal S$ (see Figure \ref{uvGridScatter}) compatible with the no-string IRG solution for our sample scatter orbit. A detailed description of our finite-difference scheme is given in Appendix \ref{app:FDSJumps}, where we also explain how the jump conditions are incorporated into the scheme so as to achieve a (global) quadratic rate of numerical convergence. Here we lay out the main steps of the numerical algorithm.

{\it Input.} The code takes as input the two orbital parameters $\vinf$ and $b$, the orbital radius $r=R_{\rm init}$ at the start (and end) of the numerical evolution, the field's multipole numbers $\ell,m$, and the finite-difference interval $h:=\Delta u=\Delta v$. 

{\it Step 1: Calculate geodesic orbit.} Given $\vinf$ and $b$, the code calculates $E$ and $L$ and from these $e$ and $p$, as well as $R_{\rm min}$. The functions $R(t)$ and $\varphi_{\rm p}(t)$ are then derived in the range $R_{\rm min}\leq R\leq R_{\rm init}$ by numerically integrating $\dot{R}$ and $\dot{\varphi}_{\rm p}$ [as obtained from Eqs.\ (\ref{tdot})--(\ref{rdot})], with initial conditions $R(0)=R_{\rm min}$ and $\varphi_{\rm p}(-\infty)=0$. The code also calculates $t_{\rm tot}$, the time it takes the particle to get from $R_{\rm init}$ back to $R_{\rm init}$ after being scattered.  

{\it Step 2: Set characteristic grid.} The code then prepares a $2\times 2$ array of $u,v$ coordinate values representing the nodes of the characteristic mesh shown in Figure\ \ref{uvGridScatter}. For the initial rays we take $v_0=-t_{\rm tot}/2+R^*_{\rm init}$ and $u_0=-t_{\rm tot}/2-R^*_{\rm init}$ with $R^*_{\rm init}:=r_*(R_{\rm init})$. This is so that the initial vertex $(u,v)=(u_0,v_0)$ is crossed by the particle at $(t,r)=(-t_{\rm tot}/2,R_{\rm init})$. The stepping interval is set at $h$, and the grid's dimensions are taken such that the final characteristic rays are at $u=t_{\rm tot}/2-R^*_{\rm init}$ and $v=t_{\rm tot}/2+R^*_{\rm init}$ such that the particle crosses the upper vertex at $(t,r)=(t_{\rm tot}/2,R_{\rm init})$ on its way out. Finally, the coordinate values of all intersubsections of the orbit with grid lines are calculated and stored. 

{\it Step 3: Obtain $[\phi]$ along the orbit.} Using the analytical expressions in Appendix \ref{App:WeylJumps}, we calculate the jumps in the Weyl scalars $\psi_{\pm}$ and their derivatives along the orbit, for the $\ell,m$ mode in question. Specifically, we compute $[\psi_-]$ and $[\partial_u^n\psi_+]$ for $n=0,\ldots,3$, and from these, using (\ref{calF}), we analytically construct the source function ${\cal F}(t)$ in Eq.\ (\ref{ODE}). We next numerically integrate the first-order ODE (\ref{ODE}) with the initial condition (\ref{JphiAsymp}), to obtain the jump $[\phi_{+}]$ in the Hertz potential along the orbit. From $[\phi_{+}]$, we algebraically obtain $[\phi_{+,v}]$ and $[\phi_{+,vv}]$ using the procedure described in the last paragraph of Section \ref{Sec:InvRel}.

{\it Step 4: Obtain $[X]$ along the orbit.} We now construct the source function ${\cal F}_X$ using Eq.\ (\ref{RWODESource}), and then numerically solve the first-order ODE (\ref{RWODE}) for $[X]$ with the initial condition (\ref{IC}). From $[X]$ we algebraically obtain also $[X_{,v}]$, $[X_{,u}]$, $[X_{,vv}]$, $[X_{,uv}]$ and $[X_{,uu}]$, using the procedure described in the last paragraph of Section \ref{Sec:InvRel}. The jump values are computed at all intersubsections of the particle's worldline with grid lines, and stored as vector datasets.

{\it Step 5: Obtain the generating function $X^{\gtrless}_{\ell m}$.} We evolve the RW equation (\ref{RWeq}) using the second-order-convergent finite-difference scheme described in Appendix \ref{app:FDSJumps}. The scheme requires as input the field jumps calculated in the previous step at intersubsections of the worldline with grid lines. The evolution starts with zero initial data along $v=v_0$ and $u=u_0$ and proceeds along successive lines of $u=$ const. The outcome is a finite-difference approximation to the generating field $X$ in each of the vacuum regions ${\cal S}^>$ and ${\cal S}^<$.

{\it Step 6: Derive the Hertz potential $\phi^{\gtrless}_{\ell m}$.} Given $X$, the Hertz potential mode $\phi$ is calculated in each of the two vacuum regions using Eq.\ (\ref{transformationIRG}), where derivatives are taken numerically. 

{\it Output.} In principle, the code can make available the Hertz potential $\phi$ anywhere in the computational domain. For our initial tests and for the purpose of illustration in this thesis, we output both $X$ and $\phi$ as functions of $t$ along the orbit (on either of its sides) and as functions of $u$ along the final $v=$ const ray (approximating $\mathscr{I}^+$).  

\subsection{Sample results}

For the numerical demonstration to be presented below we use the same sample strong-field scatter geodesic as in Section \ref{sec:ScalarScatter} whose parameters are given in Eqs.\ (\ref{sampleorbit_vb}) and (\ref{sampleorbit}). The orbit is depicted in Figure \ref{orbit}.
 
\begin{figure}[h!]
\centering
\includegraphics[width=0.8\linewidth]{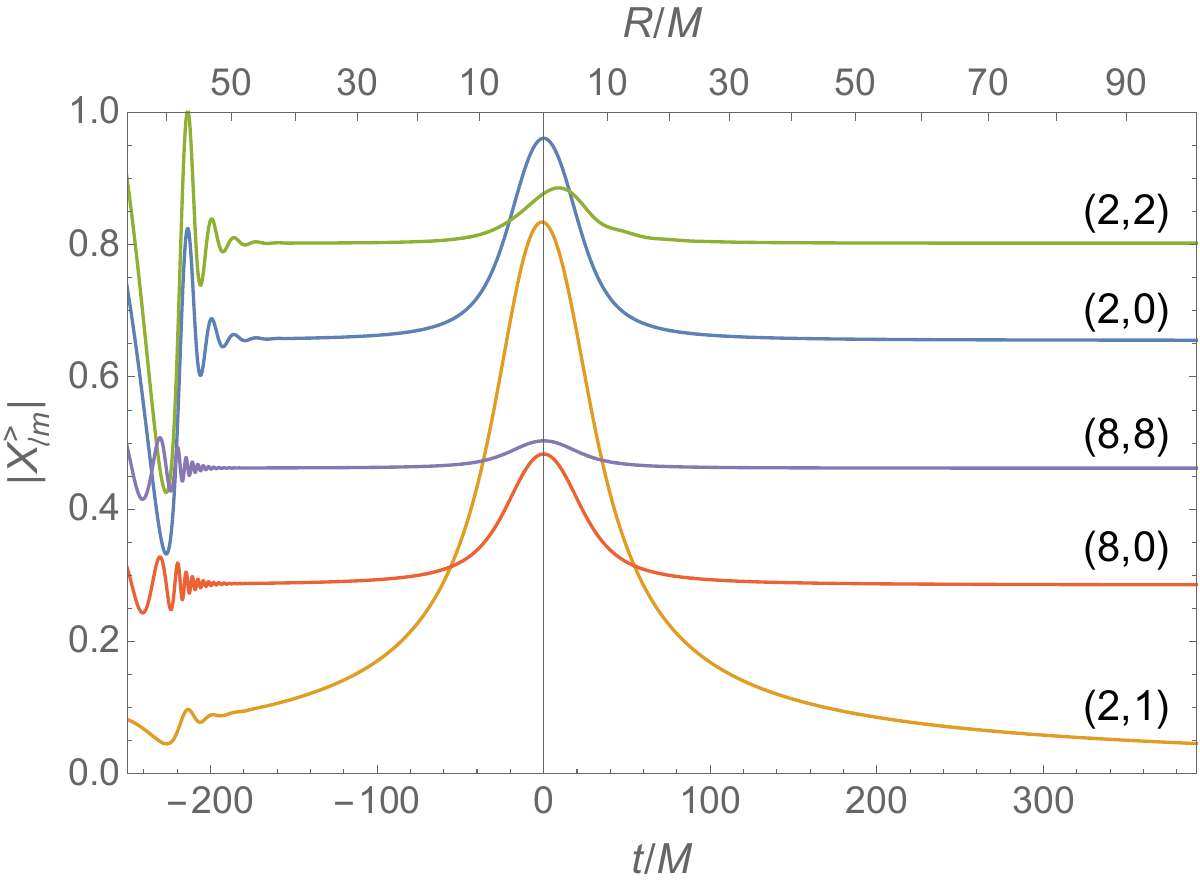}
\caption[The Regge-Wheeler-like field along the worldine of a hyperbolic orbit]{The RW field $X_{\ell m}$ along the particle's worldline for the orbit shown in Figure \ref{orbit} for a sample of $(\ell,m)$ values. Here we show $|X^>_{\ell m}(t,R(t))|$ as a function of time $t$ (lower scale) and orbital radius $R$ (upper scale). Curves are labelled with their $(\ell,m)$ values, with $\ell=8$ data shown amplified by a factor $\times 600$. The periastron location at $t=0$ is indicated with a vertical line. The early part of the data is contaminated by initial junk radiation, and it is to be discarded.
}
\label{RWField}
\end{figure}

Figure \ref{RWField} demonstrates the behaviour of the field $X^>_{\ell m}$ along the worldline of the particle, for a sample of $\ell,m$ values (the field $X^<_{\ell m}$ has a similar behaviour). The evolution begins when the incoming particle is at $R_{\rm init}=100M$, and ends when the outflying particle is back at $100M$. 

\begin{figure}[h!]
\centering
\includegraphics[width=0.8\linewidth]{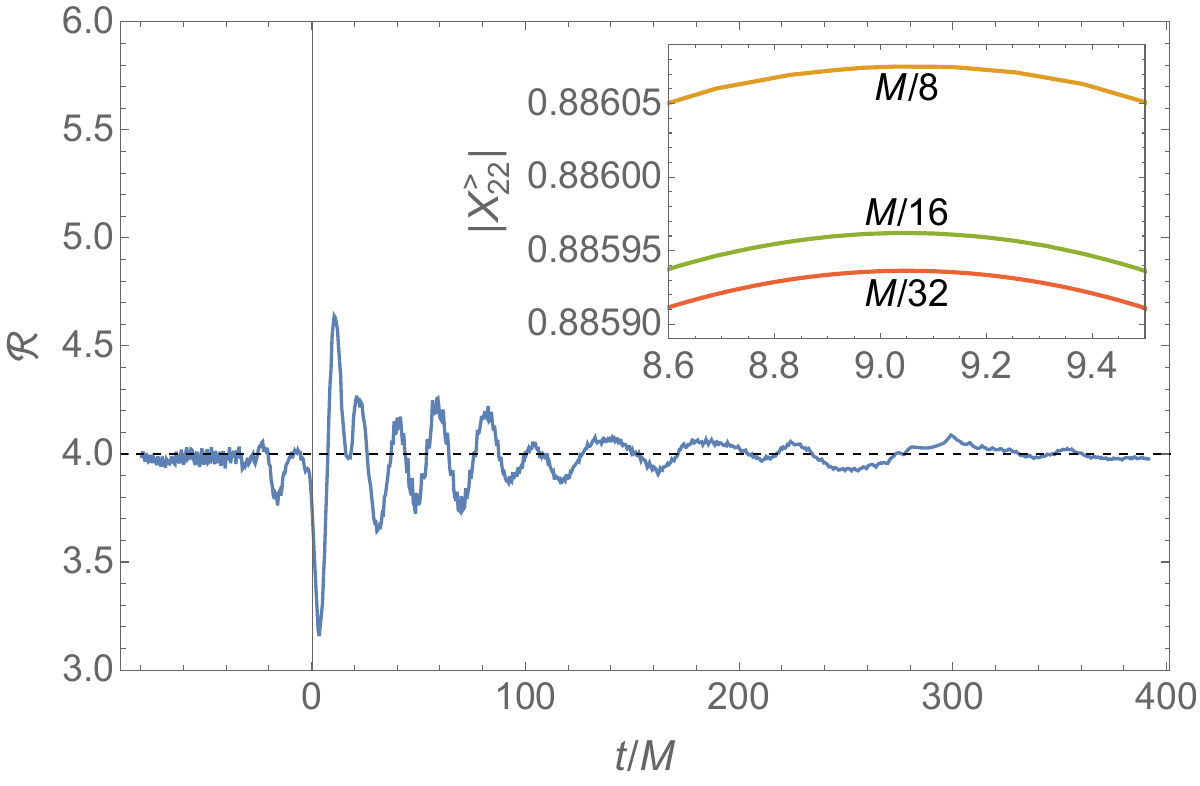}
\caption[Convergence test of the numerical solution of the Regge-Wheeler-like field]{Convergence test for the $(\ell,m)=(2,2)$ numerical solution. The inset shows a detail from the $|X^>_{22}|$ (green) curve in Figure \ref{RWField}, for a sequence of runs with decreasing grid spacing, 
$h=\left\{\frac{1}{8},\frac{1}{16},\frac{1}{32}\right\}M$. The main plot quantifies the convergence rate: It shows the ratio ${\cal R}:=\left|X^>_8-X^>_{16}\right|/\left|X^>_{16}-X^>_{32}\right|$ as a function of $t$ along the orbit, where a subscript `8' (e.g.)\ denotes a calculation with grid spacing $h=M/8$. A ratio of ${\cal R}=4$ is indicative of quadratic convergence.
}
\label{RWConv}
\end{figure}

We have performed convergence tests to confirm that our code exhibits a quadratic global convergence rate in $h$, as it is designed to do. An example is shown in Figure \ref{RWConv}. The global rate of convergence is very sensitive to the implementation details of the jump conditions in the finite-difference scheme (see Appendix \ref{app:FDSJumps}), so the observed quadratic convergence provides important reassurance that these jumps are implemented correctly. 

\begin{figure}[h!]
\centering
\includegraphics[width=0.8\linewidth]{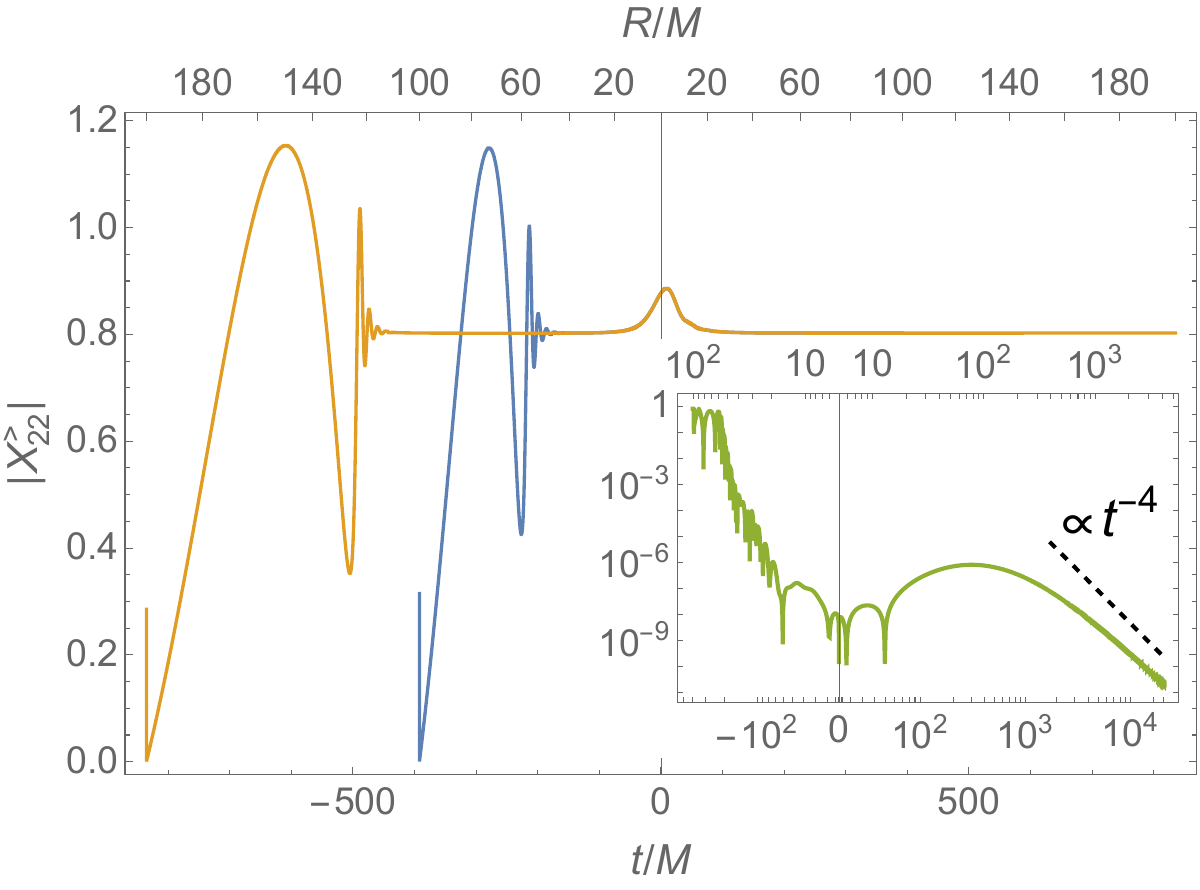}
\caption[A comparison of the Regge-Wheeler-like field calculated using numerical domains of different sizes]{Numerical results for $|X^>_{22}|$ on the particle's worldline, as calculated with $R_{\rm init}=100M$ (blue) and with $R_{\rm init}=200M$ (orange).  The comparison illustrates how, reassuringly, the ``clean'' portion of the data is insensitive to $R_{\rm init}$, up to a small error that dies off in time. The inset displays the relative difference between the two curves, showing a $t^{-4}$ fall-off at late time, consistent with the theoretically predicted $t^{-\ell-2}$ decay rate for compact vacuum perturbation along a curve $r=R\propto t$ [see, for instance, Eq.\ (89) of Ref.\ \cite{Barack:1998bw}].
}
\label{RWRMax}
\end{figure}

As can be seen in Figure \ref{RWField}, initially the data is contaminated by junk radiation which decays over time to reveal the true, physical solution.  The decay appears faster for higher values of $\ell$, as expected from theory. Figure \ref{RWRMax} illustrates how the ``clean'' part of the data appears to be insensitive to the value of $R_{\rm init}$, up to a small decaying difference. As the figure demonstrates, using $R_{\rm init}$ as a control parameter enables us in practice to evaluate the level of residual contamination from initial junk. 
 
\begin{figure}[h!]
\centering
\includegraphics[width=0.8\linewidth]{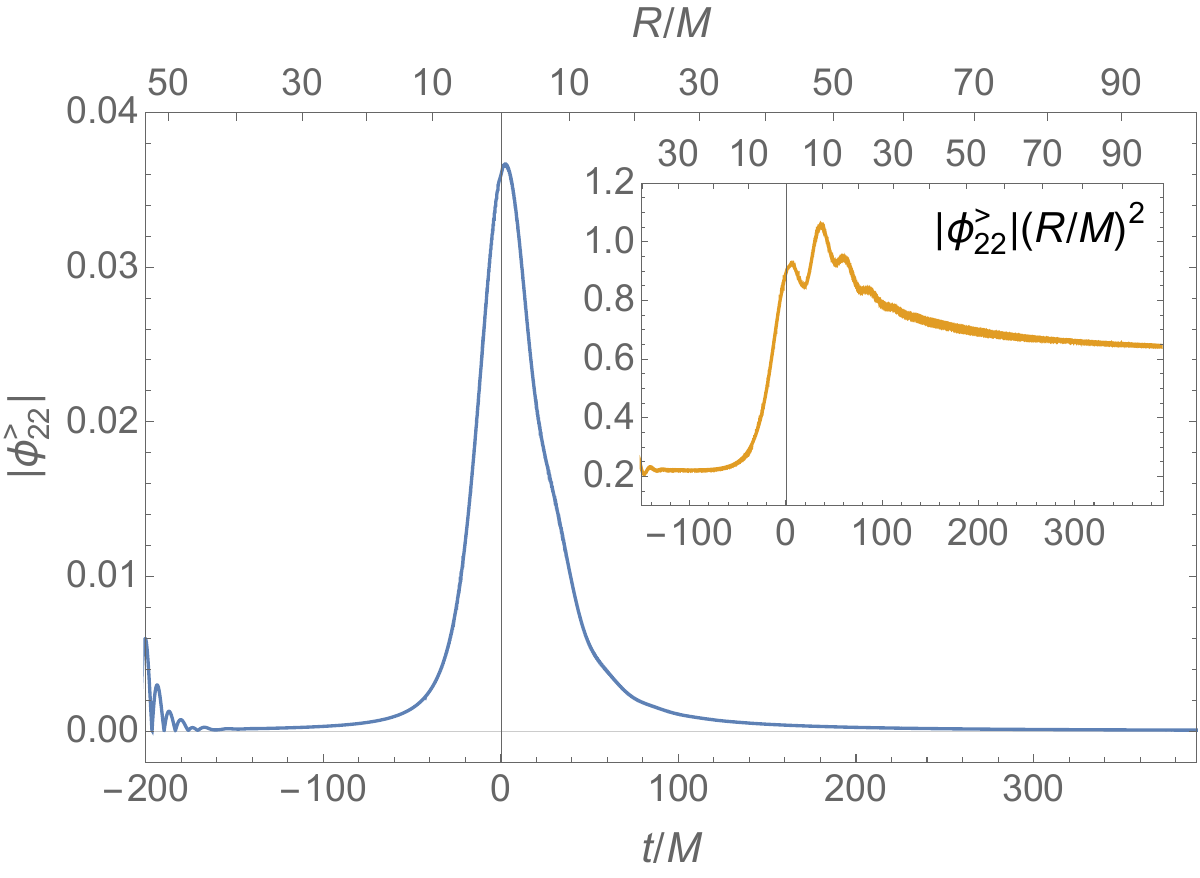}
\caption[The IRG Hertz potential along the worldine of a hyperbolic orbit]{The modulus of the ``no-string'' IRG Hertz potential $\phi^>_{22}$ along the particle's worldline. The field falls off as $X\sim t^{-2}$ at large $R$. The inset shows the same data rescaled by a factor $(R/M)^2$. The field exhibits the lagging peak and post-periastron undulation features discussed in the text. (The multiplication by $(R/M)^2$ makes more distinct the undulation feature, only barely visible in the main plot.)
}
\label{HertzField}
\end{figure}

Figure \ref{HertzField} shows the no-string IRG Hertz potential $\phi^>_{22}$ derived from $X^>_{22}$, as a function along the orbit. The gravitational case has similar notable physical to the scalar case including: (i) the small lag between the peak of the field and the periastron passage, and (ii) the small undulation in the field amplitude not long after periastron passage. (Both features are visible already at the level of the generating function $X$, and are numerically stable.) As with the scalar case, we suggest these are both features associated with ``tail'' contributions to the self-field, and are less visible at larger $\ell$, where the ``direct'' part of the field is more dominant.

\begin{figure}[H]
\centering
\includegraphics[width=0.8\linewidth]{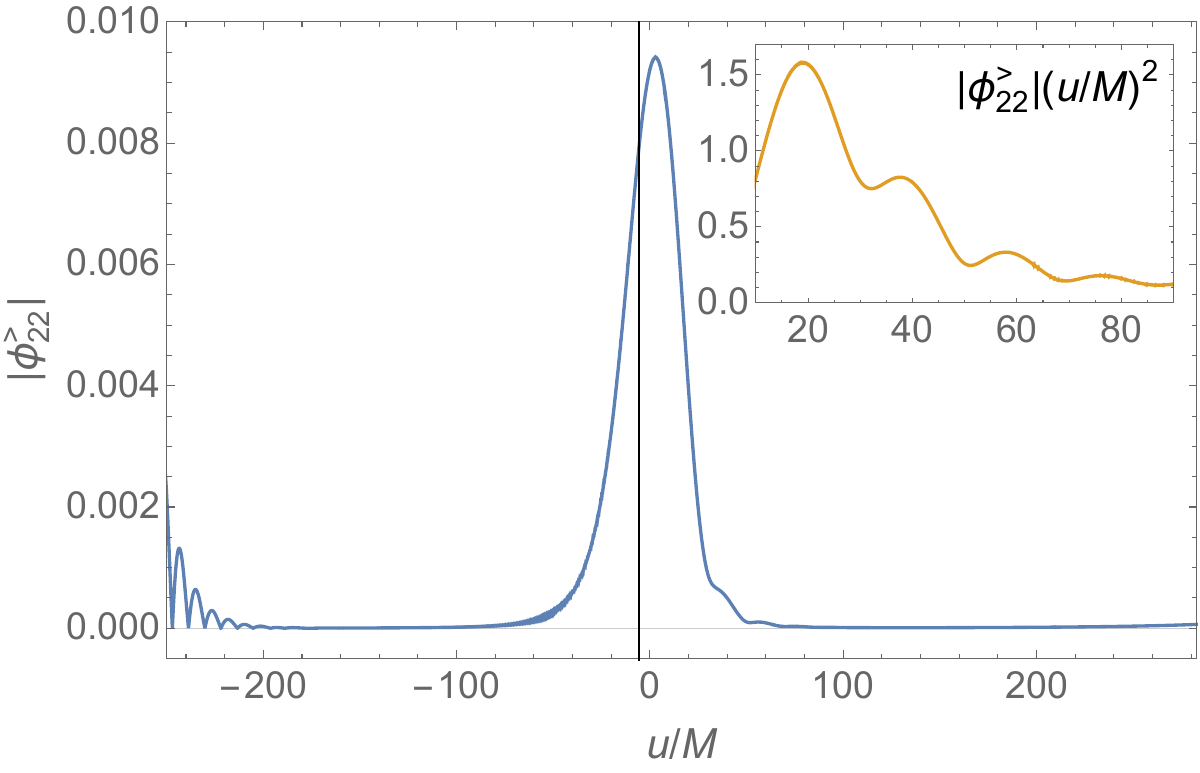}
\caption[The IRG Hertz potential along $\mathscr{I}^+$ for a hyperbolic orbit]{The modulus of the Hertz potential $\phi_{22}^>$ as a function of $u$ at $v={\rm const}=499M$ (approximating $\mathscr{I}^+$). The vertical line represents $u$ at periastron. The inset shows the same data rescaled by a factor $(u/M)^2$ to again highlight the post-periastron undulation feature.
}
\label{Hertzscri}
\end{figure}

Finally, Figure\ \ref{Hertzscri} shows the behaviour of $\phi^>_{22}$ near $\mathscr{I}^+$, as a function of retarded time $u$. The periastron lag and post-periastron undulation are also visible in the radiation field in this domain.

\pagebreak

In this chapter we implemented our new method to calculate the mode decomposed 1+1D IRG Hertz potential. The comparison of our circular-orbit results with known analytic solutions and other implementations shows that our method calculates the Hertz potential without the nonphysical modes which are present in other similar methods. We then extended our implementation to be able to produce the first calculations of certain self-force quantities for a hyperbolic encounter. Our results showed evidence of new physics including a post-periastron peak and undulation in the field amplitude.

%% file: Conclusions.tex
\chapter{Concluding remarks}
\label{chapter:conclusions}

The main results of this work are fivefold. First, we derived two separate formulae to calculate the conservative self-force correction to the scatter angle (at fixed initial velocity $\vinf$ and impact parameter $b$) as certain integrals of the self-force along the orbit. Second, we calculated the {\em scalar} self-force correction to the scatter angle for a strong-field orbit, the first calculation of its kind. Third, we have provided the details of a practical method for a time-domain calculation of the Hertz potential for point-particle metric perturbations in Schwarzschild spacetime. The main ingredients were jump conditions that the Hertz potential must satisfy along the particle's worldline (in a 1+1D multipolar reduction of the problem), which we derived in explicit form for generic geodesic orbits. Fourth, considering the numerical implementation strategy, we have demonstrated that a straightforward approach based on evolution of the Teukolsky equation in $(u,v)$ coordinates does not work (even for vacuum problems), and explained the reason for that failure. Fifth, we have proposed a way around the problem and demonstrated its applicability with an end-to-end numerical calculation of the Hertz potential for a scatter orbit. 

Since this work primarily concentrates on method development, we have not explored in detail the performance of our code near the extremes of the parameter space for scatter orbits. Relevant asymptotic domains of interest are that of large $R_{\rm min}$ (weak-field regime) and that of large $\vinf$ (ultrarelativistic regime), where useful comparisons can be made with analytical approximations. Preliminary experiments suggest that, as expected, the performance of our code gradually deteriorates with larger $R_{\rm min}$ and/or larger $\vinf$. In large-$R_{\rm min}$ runs we are penalised by the longer evolution time required, and in the large-$\vinf$ case the slower decay of initial junk along the orbit requires a larger value of $R_{\rm init}$ (and again a longer run). We estimate, nonetheless, that our current (admittedly suboptimal) method and code can comfortably handle $R_{\rm min}\lesssim 50M$ and $\vinf\lesssim 0.6$. Note that we virtually have no limit on how large the impact parameter $b$ can be taken to be (indeed, in the marginally-bound case studied in \cite{Baracketal2019} via a similar time-domain method one has $b\to\infty$).

\section{Outlook}

Here we discuss possible followups to the work presented in this thesis. This includes completing the calculation of the gravitational self-force for an unbound orbit in a Schwarzschild background and the possible interfaces with other two-body GR models. We continue by discussing extending our calculation to a Kerr background and adding in second-order self-force effects.

\subsection{Gravitational self-force}

Let us review here the additional steps necessary towards a calculation of the gravitational self-force correction to the scatter angle, starting from the baseline of the computational method and code developed here.
\begin{itemize}
\item Given the Hertz potential, the no-string radiation-gauge metric perturbation is reconstructed (mode by mode) via Eq.\ (\ref{h_rec}). This involves taking two derivatives of the numerical variables $\phi_{\ell m}$ (and hence four derivatives of $X_{\ell m}$) along the orbit, on either side of it. For the eventual self-force calculation one requires the gradient of the metric perturbation, which therefore requires {\em three} derivatives of $\phi_{\ell m}$ (and hence {\em five} derivatives of $X_{\ell m}$). The computational implications are discussed further below. 
\item One has to separately compute the ``completion'' piece of the metric perturbation, which is not accounted for by the Hertz potential \cite{Merlin2016,vandeMeent:2017fqk}. In the Schwarzschild problem this corresponds precisely to the determination of the $\ell=0,1$ perturbation modes. Of these, the axially-symmetric modes $(l,m)=(0,0)$ and $(1,0)$, which describe mass and angular-momentum perturbations, are easily determinable using the results of \cite{Merlin2016}. The modes $(l,m)=(1,\pm 1)$, which regulate the centre-of-mass location, require a more careful analysis, similar to the one performed in \cite{Baracketal2019} for marginally-bound orbits. 
\item Once all the modes of the metric perturbation and its gradient are available, the self-force along the orbit is straightforwardly obtained via the no-string radiation-gauge version of the mode-sum formula, prescribed in \cite{PoundMerlinBarack2014}. It is also easy to separately extract the dissipative and conservative components of the self-force, utilising the symmetries of the geodesic scatter orbit about the periastron point [e.g.\ using Eqs.\ (\ref{eqn:ConsSF}) to determine the conservative SF].
\item One can then calculate the self-force correction to the scatter angle by using Eq.\ (\ref{deltaphi1_final}) or Eq.\ (\ref{deltaphi1_final_method2}). Additional physical quantities, such as the time delay induced by the self-force, or the integrated particle's spin precession and tidal-field invariants, may also be calculated, though the latter two would require evaluating higher derivatives of the metric perturbation. The self-force information allows calculation of all these effects with or without dissipation. 
\end{itemize}

We have noted above that a calculation of the metric perturbation and self-force involves taking high-order derivatives (fourth and fifth, respectively) of the numerical evolution field $X_{\ell m}$. This is an obvious computational disadvantage of our approach. It can be mitigated if a method is employed that allows a direct evolution of the Teukolsky equation for the Hertz potential $\phi_{\ell m}$, which would reduce the required number of derivatives to only two for the metric perturbation, and three for the self-force. As mentioned in Section \ref{sec:vacuum}, there already exist such methods, based on compactification of $\mathscr{I}^+$ and the use of horizon-penetrating coordinates---which appear to automatically eliminate the problematic non-physical growing solutions of the Teukolsky equation. Existing codes employ asymptotically null (hyperboloidal) Cauchy slicing of the numerical domain. We propose that, in our context, it might be advantageous to retain the convenience and simplicity of a fully double-null treatment, taking advantage of the domain split across $\cal S$. What we have in mind is a scheme where on ${\cal S}^+$ we use the original Eddington-Finkelstein coordinate $u$ with a compactified $v$ coordinate, while on ${\cal S}^-$ we use the original $v$ coordinate with a compactified $u$. The coordinate discrepancy along $\cal S$ is then incorporated into the jump conditions.

\subsection{Interfaces with other two-body GR models}

With full results of the gravitational self-force for unbound orbits we can inform and compare with a variety of other methods of modelling two-body systems within GR:
\begin{itemize}
\item One of the most exciting prospects utilises the results that the self-force correction to the scatter angle can be used to inform {\em bound} two-body models. As discussed earlier, information of scattering dynamics at first-order (second-order) in the mass ratio determines the complete two-body Hamiltonian through 4PM (6PM) order to all orders in the mass ratio \cite{Damour2020}. This opportunity to extract highly accurate information about BBH collisions will allow us to form models capable of meeting the requirements for the next generation of GW detectors.
\item The highly penetrating nature of scatter orbits, as shown in Figure \ref{periastron}, makes them uniquely positioned to provide information about the ultra-strong-field. We can use this data to callibrate EOB calculations without the weak-field approximation that the current benchmarks rely on. This is another way that scatter calculations can inform {\em bound} waveform models for GW observations.
\item Comparisons with other two-body scattering calculations with GR, such as PM, PN, and NR, gives us the opportunity to test the relative merits of each model in different regions of the parameter space shown in Figure \ref{GWModelling}. 
\item Methods from high-energy physics are being used to calculate quantities with gravitational physics by mapping information about scatter observables into some form of two-body gravitational potential \cite{ChoKalin2021}. We can compare our classical GR models with the results from these QFT/EFT methods. Additionally, the use of scattering data from GR to calculate quantities within QFT (see Ref.\ \cite{BautistaGuevara2021}) suggests that it may be possible to extract high-energy information from gravitational scattering. This connection between high-energy and gravitational physics provides unique opportunities for progress in both fields.
\end{itemize}

\subsection{Extension to a Kerr background}

It is natural to ask about the prospect of an extension to orbits in Kerr geometry. This has been discussed in some detail in Ref.\ \cite{Barack:2017oir}. A 1+1D treatment of the Teukolsky equation in Kerr is still possible, albeit with the additional complication of coupling between $\ell$ modes. The field equation, together with jump conditions on $\cal S$, can be recast in a narrow band-diagonal matrix form, and solved for all $\ell$ modes simultaneously (with a cutoff at a sufficiently high $\ell_{\rm max}$). The application of this mode-coupling approach has been demonstrated in vacuum problems \cite{Dolan_2013,Barack:2017oir}, but it is yet to be applied with a particle source, and the appropriate no-string jump conditions are yet to be derived. In the Kerr case there is no known way of transforming to a RW-like variable in the time domain (the Sasaki-Nakamura formulation achieves that in the frequency domain only \cite{SasakiNakamura1982, Hughes2000}), which further motivates an approach based on a direct evolution of the Hertz potential with a suitable form of domain compactification.

\subsection{Extension to second-order self-force}

Another extension would add corrections that are second-order in the mass ratio. While second-order SF theory is now well developed, the case of hyperbolic motion presents unique theoretical issues that have not been considered so far. The current formulation of second-order SF calculations is motivated by EMRI observations and thus is highly specialised to bound orbits. These calculations are based on a two-timescale expansion which separates quantities into those that evolve on the timescale of the radiation-reaction time and those that evolve on much shorter timescales, such as the orbital frequencies \cite{KevorkianCole1996, PoundWardell2020}. This expansion is not valid in the scatter problem. There is also a great deal of subtlety in the behaviour of the spacetime metric at large distances in a scatter scenario; the spacetime does not possess the usual asymptotically flat structure and thus there are potential repercussions regarding the source of the second order Einstein field equation (\ref{eqn:2GSFEFE}), which will need to be explored. 

\pagebreak

The outlook for black-hole scattering is promising. The revelations that unbound encounters can inform bound waveforms has launched the ``scattering revolution". The scientific community are now tackling the problem from multiple angles and developing new techniques to produce groundbreaking results that will shape the future of gravitational wave astronomy.

%% file: BPTReconstruction.tex

\chapter{Bardeen--Press--Teukolsky equation and metric reconstruction}
\label{App:convention}

We give here a more detailed technical account of the background material presented 
in Section \ref{sec:review}, and in particular we give explicit expressions for the operators ${\sf\hat T}_{\pm}$, ${\sf\hat O}_{\pm}$ and ${\sf\hat S}_{\pm}$, and their adjoints. Our sign conventions for the Newman--Penrose formalism are adopted from Ref.\ \cite{Merlin2016};  Appendix A therein gives a useful summary.   

In this paper we use Kinnersley's null tetrad basis on a Schwarzschild background with metric $g_{\alpha\beta}$ and mass parameter $M$. In Schwarzschild coordinates $(t,r,\theta,\varphi)$, the tetrad legs are given by

\begin{align}
\label{eq:kerrtetrad}
e_{\bm 1}^{\alpha}= \ell^\alpha &=\SP{\frac{r^2}{\Delta},1,0,0},\nonumber \\
e_{\bm 2}^\alpha= n^\alpha &=\frac{1}{2}\SP{1,-\frac{\Delta}{r^2},0,0 },\nonumber \\
e_{\bm 3}^\alpha = m^\alpha &=\frac{1}{\sqrt{2}\, r}\bP{0,0,1,\frac{i}{\sin\theta}}, \nonumber \\
e_{\bm 4}^\alpha = \bar m^\alpha &=\frac{1}{\sqrt{2}\,r}\bP{0,0,1,-\frac{i}{\sin\theta}},
\end{align}
where $\Delta:=r(r-2M)$, and overbars denote complex conjugation. We have $g_{\alpha\beta}e_{\bm a}^\alpha e_{\bm b}^{\beta}=0$ for all ${\bm a}$ and ${\bm b}$, except $\ell^\alpha n_\alpha=-1$ and  $m^\alpha\bar{m}_\alpha=1$. 
The corresponding spin coefficients are 
$
\gamma_{\bm abc} := g_{\mu\lambda}\tet{a}{\mu}\tet{c}{\nu}\nabla_{\nu}\tet{b}{\lambda}.
$
Up to trivial index permutations, the only nonzero coefficients in the Schwarzschild case are 
\begin{align}
\varrho &:= -\gamma_{314}= -\frac{1}{r},\nonumber\\
\mu &:= -\gamma_{243} = -\frac{\Delta}{2r^3},\nonumber\\
\gamma &:=-\frac{1}{2}(\gamma_{212}+\gamma_{342}) =  \frac{M}{2r^2},\nonumber\\
\beta &:= -\frac{1}{2}(\gamma_{213}+\gamma_{343})= \frac{\cot\theta}{2\sqrt{2}\, r},\nonumber\\
\alpha &:= -\frac{1}{2}(\gamma_{214}+\gamma_{344})=-\frac{\cot\theta}{2\sqrt{2}\, r}.
\end{align}
The Weyl curvature scalars $\Psi_0$ and $\Psi_4$ are defined in terms of the Weyl tensor $C_{\alpha\beta\gamma\delta}$  as
\begin{align}
\Psi_0=&\: C_{\alpha\beta\gamma\delta}\,\ell^\alpha m^\beta \ell^\gamma m^\delta , \nonumber\\
\Psi_4=& \: C_{\alpha\beta\gamma\delta}\, n^\alpha \bar m^\beta n^\gamma \bar m^\delta .\label{eq:psi}
\end{align}
Both $\Psi_0$ and $\Psi_4$ vanish in the Schwarzschild background, and so, for the sake of economy but in a slight abuse of notation, we use these symbols to represent the linear perturbations in these quantities. 
We define $\Psi_{+}:=\Psi_0$ and $\Psi_{-}:=\varrho^{-4}\Psi_4$ for notational ease. In terms of the metric perturbation $h_{\alpha\beta}$, we have
${\sf\hat T}_{\pm} h_{\alpha\beta} =\Psi_\pm$ [Eq.\ (\ref{T})], where the second-order differential operators ${\sf\hat T}_{\pm}$ are given by
\begin{align}\label{hatT}
(\sf\hat T_{+})^{\alpha\beta} =& \frac{1}{2}\left(
 \ell^{(\alpha} m^{\beta)} m^\gamma\ell^\delta
-m^\alpha m^\beta \ell^\gamma\ell^\delta
-\ell^\alpha \ell^\beta m^\gamma m^\delta
+m^{(\alpha} \ell^{\beta)} \ell^\gamma m^\delta
\right)\nabla_\delta\nabla_\gamma , \nonumber \\
(\sf\hat T_{-})^{\alpha\beta} =& \frac{1}{2}\left(
 n^{(\alpha} \bar m^{\beta)} \bar m^\gamma n^\delta
-\bar m^\alpha \bar m^\beta n^\gamma n^\delta
-n^\alpha n^\beta \bar m^\gamma \bar m^\delta
+\bar m^{(\alpha} n^{\beta)} n^\gamma \bar m^\delta
\right)\nabla_\delta\nabla_\gamma .
\end{align}
Here $\nabla_\alpha$ is the covariant derivative compatible with the Schwarzschild background metric $g_{\alpha\beta}$, and parenthetical indices are symmetrised, as in $A_{(\alpha\beta)}=\frac{1}{2}(A_{\alpha\beta}+A_{\beta\alpha})$.

The perturbation fields $\Psi_{\pm}$ satisfy the Teukolsky equation with spin parameter $s=\pm 2$, whose Schwarzschild reduction is sometimes referred to as the Bardeen--Press equation. Here we refer to it as the Bardeen--Press--Teukolsky (BPT) equation. It has the form
\begin{equation}\label{eq:kerrteuk} 
{\sf\hat O}_{\pm}\Psi_{\pm}= {\cal T}_\pm,
\end{equation}
where the differential operators on the left are
\begin{align}\label{Ocompact}
{\sf\hat O}_{+}=&\: \Delta\left(\boldsymbol{D}_\ell+2\frac{r-M}{\Delta}\right)
\left(\tilde{\boldsymbol{D}}_n+4\frac{r-M}{\Delta}\right) +\eth_1 \bar\eth_2 -6r\partial_t ,
\nonumber\\
{\sf\hat O}_{-}=&\: \Delta\left(\tilde{\boldsymbol{D}}_n-2\frac{r-M}{\Delta}\right)
\boldsymbol{D}_\ell
+\bar\eth_{-1} \eth_{-2} +6r\partial_t.
\end{align}
Here 
$\boldsymbol{D}_\ell:=\ell^\alpha\nabla_\alpha$, 
$\boldsymbol{D}_n:=n^\alpha\nabla_\alpha$,
$\tilde{\boldsymbol D}_n:=-(2r^2/\Delta){\boldsymbol D}_n$, 
and we have introduced the ``spin raising and lowering'' operators, respectively
\begin{align} \label{eth}
\eth_s&:=-\partial_\theta-{i\csc\theta}\partial_\vf+s \cot\theta, 
\nonumber\\
\bar\eth_s&:=-\partial_\theta+{i\csc\theta}\partial_\vf-s \cot\theta,
\end{align}
whose action on spin-weighted spherical harmonics ${}_{s}\!Y_{\ell m}(\theta,\varphi)$ is described by  
\begin{align}\label{raising&lowering}
\eth_s\,{_s\!Y_{\ell m}} &= +\sqrt{(\ell-s)(\ell+s+1)} \, {_{s+1}\!Y_{\ell m}},
\nonumber \\
\bar\eth_s\,{_s\!Y_{\ell m}} &= -\sqrt{(\ell+s)(\ell-s+1)} \, {_{s-1}\!Y_{\ell m}}.
\end{align}
The source terms ${\cal T}_\pm$ in (\ref{eq:kerrteuk}) are obtained from the energy-momentum tensor $T_{\alpha\beta}$ using 
\begin{align} \label{eq:kerrsource}
{\cal T}_{+}&= {\sf\hat S}_+ T_{\alpha\beta}=-8\pi r^2
\Big[(\boldsymbol{D}_m  -2\beta )\boldsymbol{D}_m   T_{\boldsymbol{11}}
 - (\boldsymbol{D}_\ell -5\varrho)(\boldsymbol{D}_m -2\beta ) T_{\boldsymbol{13}}
 \nonumber\\  &
\qquad \qquad \qquad \qquad -(\boldsymbol{D}_m  -2\beta )(\boldsymbol{D}_\ell -2\varrho ) T_{\boldsymbol{13}}
 +(\boldsymbol{D}_\ell -5\varrho)(\boldsymbol{D}_\ell -\varrho ) T_{\boldsymbol{33}} 
 \Big], \nonumber\\
 \end{align}
\begin{align}
{\cal T}_{-}&= {\sf\hat S}_- T_{\alpha\beta}=8\pi r^6 
\Big[
(\boldsymbol{D}_{\bar m}-2\beta )\boldsymbol{D}_{\bar m}  T_{\boldsymbol{22}} 
-(\boldsymbol{D}_n +2\gamma +5\mu )(\boldsymbol{D}_{\bar m} -2\beta )T_{\boldsymbol{24}}
\nonumber\\  &
\qquad \qquad \qquad \qquad -(\boldsymbol{D}_{\bar m} -2\beta )(\boldsymbol{D}_n +2\gamma +2\mu ) T_{\boldsymbol{24}}
+(\boldsymbol{D}_n +2\gamma +5\mu)(\boldsymbol{D}_n +\mu )T_{\boldsymbol{44}} \Big],
\label{eq:kerrsource-2}
\end{align}
where $T_{\boldsymbol{11}}=T_{\alpha\beta}e_{1}^{\alpha}e_1^{\beta}$, etc., and we have also introduced 
$\boldsymbol{D}_m:=m^\alpha\nabla_\alpha$ and $\boldsymbol{D}_{\bar m}:=\bar m^\alpha\nabla_\alpha$.


As described in Section \ref{sec:review}, the metric reconstruction procedure involves the operators adjoint to ${\sf\hat O}_{+}$, ${\sf\hat T}_{+}$ and ${\sf\hat S}_{+}$.\footnote{
Recall that for a linear operator $\sf\hat L$ taking an $n$-rank tensor field $\phi$ to an $m$-rank tensor field $\psi$, the adjoint $\sf\hat L^\dagger$ takes $\psi$ to $\phi$ and satisfies 
$({\sf\hat L}^\dagger\psi)\phi=\psi({\sf\hat L}\phi)$ (up to a divergence of an arbitrary vector field).} These adjoint operators can be obtained by integrating each operator against a suitable test function and manipulating using integrations by parts. In this fashion it is straightforward to show that 
\begin{equation}
{\sf\hat O}^\dagger_{\pm}={\sf\hat O}_{\mp},
\end{equation}
i.e., solutions $\Phi_{\pm}$ to the adjoint BPT equation with spin $s=\pm 2$ are also solutions to the standard BPT equation with spin $s=\mp 2$.  
For the metric reconstruction operators [see Eq.\ (\ref{h_rec})] a calculation gives 
\begin{eqnarray} \label{eq:kerrh+}
{\sf\hat S}_+^\dagger &=&  -2\ell_\alpha \ell_\beta
\SP{\boldsymbol{D}_m +2\beta}\SP{\boldsymbol{D}_m+4\beta } - 2m_\alpha m_\beta \SP{\boldsymbol{D}_\ell-\varrho}\SP{\boldsymbol{D}_\ell +3\varrho}, 
\nonumber\\
 && +2\ell_{(\alpha} m_{\beta)} \BB{
 \SP{\boldsymbol{D}_m +4\beta}\SP{\boldsymbol{D}_\ell+3\varrho}
+\boldsymbol{D}_\ell \SP{\boldsymbol{D}_m +4\beta}
}
\end{eqnarray}
\begin{eqnarray}\label{eq:kerrh-}
{\sf\hat S}_-^\dagger &= &  -2r^4\Big[
n_\alpha n_\beta
\SP{\boldsymbol{D}_{\bar m}+2\beta}\SP{\boldsymbol{D}_{\bar m}+4\beta} +\bar m_\alpha \bar m_\beta \SP{\boldsymbol{D}_n+5\mu-2\gamma}\SP{\boldsymbol{D}_n +\mu-4\gamma}
\nonumber\\
 && \qquad \qquad -n_{(\alpha} \bar m_{\beta)} \Big(
 \SP{\boldsymbol{D}_{\bar m} +4\beta}\SP{\boldsymbol{D}_n+\mu-4\gamma} +\SP{\boldsymbol{D}_n +4\mu-4\gamma}\SP{\boldsymbol{D}_{\bar m} -2\beta}
\Big) \Big].
\nonumber\\
\end{eqnarray}
Finally, for the ``source reconstruction'' operators [see Eq.\ (\ref{Source_rec})]
one finds
\begin{eqnarray}\label{Tdagger+Expl}
({\sf\hat T}^\dagger_{+})^{\alpha\beta} &=&
-\frac{1}{2}\ell^\alpha \ell^\beta (\boldsymbol{D}_m+2\beta)(\boldsymbol{D}_m+4\beta)- \frac{1}{2}m^\alpha m^\beta \left(\boldsymbol{D}_\ell-\varrho\right)^2 
\nonumber\\
&&+\frac{1}{2}\ell^{(\alpha} m^{\beta)}\Big[\boldsymbol{D}_\ell\left(\boldsymbol{D}_m+4\beta\right) 
+\left(\boldsymbol{D}_m +4\beta\right)\left(\boldsymbol{D}_\ell-\varrho\right)\Big],
\end{eqnarray}
\begin{eqnarray}\label{Tdagger-Expl}
({\sf\hat T}^\dagger_{-})^{\alpha\beta} &=&
-\frac{1}{2}n^\alpha n^\beta(\boldsymbol{D}_{\bar m}+2\beta)(\boldsymbol{D}_{\bar m}+4\beta) -\frac{1}{2}\bar m^\alpha \bar m^\beta(\boldsymbol{D}_{n}+\mu -2\gamma)(\boldsymbol{D}_{n}+\mu-4\gamma)
\nonumber\\
&&+\frac{1}{2}n^{(\alpha} \bar m^{\beta)}\Big[\left(\boldsymbol{D}_{n}-4\gamma\right)\left(\boldsymbol{D}_{\bar m}+4\beta \right)
+\left(\boldsymbol{D}_{\bar m}+4\beta\right)\left(\boldsymbol{D}_{n}+\mu-4\gamma\right)\Big].
\end{eqnarray}

%% file: WeylJumps.tex

\chapter{Jumps in the Weyl-scalar modes $\psi^{\pm}_{\ell m}$}
\label{App:WeylJumps}

In this appendix we derive the jumps across $\cal S$ in the Weyl-scalar modal fields $\psi^{\pm}_{\ell m}(t,r)$ and their first 3 derivatives, for a generic geodesic orbit in a Schwarzschild background. We do so analytically, and for both spins $s=\pm 2$. These jumps are necessary input for the calculation of the no-string Hertz-potential jumps $[\phi_{\pm}]$ in Section \ref{sec:Jumps}. In our method we require both $[\psi_+]$ and $[\psi_-]$ for either $[\phi_+]$ (IRG potential) or $[\phi_-]$ (ORG potential); cf.\ Eq.\ (\ref{ODE}) with (\ref{calF}). At the end of this appendix we derive asymptotic expressions for $[\psi_{\pm}]$ at large radii in the case of scatter orbits which are used in the asymptotic analysis of Section \ref{subsec:asymptotics}.

\section{BPT equation with a point-particle source}

Let $\Psi_{+}\equiv \Psi_{0}$ and $\Psi_{-}\equiv r^4\Psi_{4}$ be the Weyl scalars associated with the physical metric perturbation sourced by a geodesic point particle with stress-energy as in Eq.\ (\ref{Tmunu}).
We recall our notation: $\mu$ is the particle's mass, and $x^\mu=x_{\rm p}^\mu(\tau)$ describes its geodesic worldline, with proper time $\tau$ and four-velocity $u^\alpha:= dx_{\rm p}/d\tau$. For convenience, we set the Schwarzschild coordinates so that the orbit lies in the equatorial plane ($\theta_{\rm p}\equiv \pi/2$), and write $R(\tau)\equiv r_{\rm p}(\tau)$. The conserved (specific) energy and angular momentum of the orbit are $E=f_Ru^t$ and $L=r^2 u^\varphi$, respectively, where $f_R:=1-2M/R$. The Schwarzschild components of the four-velocity are
\begin{equation}
u^\mu_{\rm p} = \left(E/f_R,(E/f_R)\dot{R},0, L/r^2\right),
\end{equation}   
where an overdot denotes $d/dt$.

The Weyl scalars $\Psi_{\pm}$ satisfy the $s=\pm 2$ BPT equations (\ref{eq:kerrteuk}), where the source $\cal T_\pm$ is derivable from $T^{\alpha\beta}$ by means of Eqs.\ (\ref{eq:kerrsource}) and (\ref{eq:kerrsource-2}).
Expanding both $\Psi_{\pm}$ and $\cal T_\pm$ in $s=\pm 2$ spherical harmonics, as in Eq.\ (\ref{expansionpsi}), separates  the BPT equation into modal equations for each of the time-radial fields $\psi^{\pm}_{\ell m}(t,r)$. The modal equations are (dropping the indices $\ell m$ for brevity)
\begin{align} \label{Teukolsky1+1psi_take2}
\psi^\pm_{,uv} + U_s(r)\psi^\pm_{,u} + V_s(r)\psi^\pm_{,v}  + W_s(r)\psi^\pm =T_\pm(t,r) ,
\end{align}
where the radial functions on the left are those given in Eqs.\ (\ref{UV}) and (\ref{W_Sch}), with 
$s=\pm 2$ for $\psi_{\pm}$. The modal source term $T_\pm$ can be written in the form 
\begin{align}\label{Tsource}
T_\pm(t,r)=s^\pm_0(t)\delta[r-R(t)]+s^\pm_1(t)\delta'[r-R(t)] +s^\pm_2(t)\delta''[r-R(t)],
\end{align}
where a prime denotes a derivative with respect to the argument, and the source functions $s^\pm_n(t)$ are certain functions along the orbit. The explicit expressions for $s^\pm_n(t)$ are rather unwieldy, unfortunately, but they are essential within our method, so we give them here. They are 

\begin{align} \label{s0}
s^+_0=&
\frac{\pi\mu}{E}\bigg\{
-L^2 m f_R^2 R \left(i \ddot\varphi_{\rm p}+m \dot\varphi_{\rm p}^2\right) +Lf_R\left[L(1+2y)-2imER\right]\ddot R \nonumber\\
&+2Lf_R\Big[iL m\left((2+y)f_R-(1+2y)\Rdot\right) + m^2ER(f_R-\Rdot) \Big]\dot\varphi_{\rm p} \nonumber\\
& +\left[-2iLEm+8L^2y^2/R-(m^2-2)E^2 R\right]\Rdot^2
\nonumber\\
&
+2f_R \Big[-(L^2/R)y(6y+7)+2iLmE(1+3y) +(m^2-2)E^2 R\Big]\Rdot
\nonumber\\
&
-f_R^2\left[- 12 (L^2/R)y + 2iLmE(1+4y)+(m^2-2)E^2 R\right]
\bigg\} {\cal Y}^+(t) \nonumber\\
&+\pi\mu\bigg\{ 2  R(f_R-\Rdot)m\left[Lf_R \dot\varphi_{\rm p}-E(f_R-\Rdot)\right] 
\nonumber\\
&
+2i L \Big[-f_R R \ddot R-\Rdot^2 +2f_R(1+3y)\Rdot -f_R^2(1+4y)\Big]
\bigg\} {\cal Y}^+_\theta(t)\nonumber\\
&-\pi\mu E R (f_R-\Rdot)^2{\cal Y}^+_{\theta\theta}(t),
\end{align}
\begin{align}\label{s1}
s^+_1=&
2i\pi\mu L f_R R(f_R-\Rdot)\Big[(f_R-\Rdot)\left[{\cal Y}^+_\theta(t)+m{\cal Y}^+(t)\right] 
-mf_R(L/E)\dot\varphi_{\rm p}{\cal Y}^+(t)\Big]
\nonumber\\
&
+\pi\mu (L^2/E) f_R \Big[-f_R R \ddot R-2(1+2y)\Rdot^2 +2f_R(3+5y)\Rdot-4f_R^2(1+y)\Big] {\cal Y}^+(t),
\end{align}
\begin{equation} \label{s2}
s^+_2= \pi\mu(L^2/E) f_R^2 R (f_R-\Rdot)^2 {\cal Y}^+(t),
\end{equation}
and
\begin{align}
s^-_0 =& -\frac{\pi \mu }{4 E f_R^2 R^9} \bigg\{ L^2 m f_R^2 R^6 \left(m \dot{\varphi }_{\rm p}^2+i \ddot{\varphi }_{\rm p}\right) + L f_R R^5 \left(3 L f_R-2 i E m R\right) \ddot{R} \nonumber\\
&-2 L m R^5 f_R \Big[E m R \dot{R}+f_R \Big(E m R +i L \left(3 \dot{R}+2-5y\right)\Big)\Big] \dot{\varphi }_{\rm p} \nonumber\\
& + R^4 \Big[-12 L^2 f_R^2+E R \Big(E \left(m^2-2\right) R  +2 i L m (3-4 y)\Big)\Big]\dot{R}^2 \nonumber\\
&+ 2 R^4 \Big[-L^2 (8-25 y) f_R^2 +E R \Big(E \left(m^2-2\right) R f_R +2 i L m (3-7 y) f_R\Big)\Big]\dot{R} \nonumber\\
&+ R^4 \Big[-4 L^2 (1-5y) f_R^3+E R \Big(E \left(m^2-2\right) R f_R^2 +2 i L m (3-8 y) f_R^2\Big)\Big] \bigg\} {\cal Y}^-(t) \nonumber\\
&+ \frac{\pi  \mu }{2 f_R^2 R^7} \bigg\{  m R^4 (f_R+\dot{R}) \left(E (f_R+\dot{R})-L f_R \dot{\varphi }_{\rm p}\right) \nonumber\\
& -i L R^3 \left[f_R R \ddot{R} - (3-4y)\dot{R}^2 -2f_R (3-7 y)\dot{R} - (3-8y) f_R^2 \right]\bigg\} {\cal Y}^-_\theta(t)\nonumber\\
&-\frac{\pi  E \mu  (f_R+\dot{R})^2}{4 f_R^2 R^3} {\cal Y}^-_{\theta\theta}(t),
\end{align}
\begin{align}\label{s1psi4leor}
s_1^- =& \: 
\frac{i\pi\mu L (f_R+\Rdot)}{2f_RR^3}\Big[(f_R+\Rdot)\left[{\cal Y}^-_\theta(t)-m{\cal Y}^-(t)\right] +mf_R(L/E)\dot\varphi_{\rm p}{\cal Y}^-(t)\Big]\nonumber\\
&+\frac{\pi\mu L^2}{4E R^4} \Big[-R \ddot R+6\Rdot^2 +2(5-13y)\Rdot+4f_R(1-3y)\Big] {\cal Y}^-(t),
\end{align}
\begin{equation} \label{s2psi4}
s_2^- = \frac{\pi  \mu  L^2 \left(f_R+\Rdot\right){}^2}{4 E R^3} \mathcal{Y}^-(t) .
\end{equation}
Here we have introduced 
\begin{equation}\label{calY}
y:=\frac{M}{R},\quad\quad
{\cal Y}^\pm(t):={}_{\pm2}\!\bar Y_{\ell m}\left(\frac{\pi}{2},\varphi_{\rm p}(t)\right),
\end{equation}
with ${\cal Y}^\pm_\theta$ and ${\cal Y}^\pm_{\theta\theta}$ being the first and second derivatives of ${}_{\pm2}\!\bar Y_{\ell m}\left(\theta,\varphi_{\rm p}(t)\right)$ with respect to $\theta$, evaluated at $\theta=\pi/2$.

Note that in Eqs.\ (\ref{s0})--(\ref{s2psi4}) we have not yet specialised to a timelike geodesic. With such specification, the time derivatives featuring in these expressions can be expressed in terms of $R(t)$ alone (as well as $E$ and $L$), as follows: 
\begin{align}\label{dotR}
\Rdot &= \pm (f_R/E)\left[E^2-f_R(1+L^2/R^2)\right]^{1/2}, \\
\ddot R&=\frac{f_R^2(1-5y)L^2+yR^2f_R(2E^2-3f_R)}{R^3 E^2},
\end{align}
\begin{equation}\label{dotvarphi}
\dot\varphi_{\rm p}= \frac{f_RL}{R^2 E},\quad\quad
\ddot\varphi_{\rm p}=-\frac{2L(1-3y)\Rdot}{ER^3}.
\end{equation}
The sign in (\ref{dotR}) is $(-)$ for the incoming leg of the orbit and $(+)$ for the outgoing leg.

\section{The jumps in $\psi_\pm$ and their first derivatives}

The jumps in the 1+1D Weyl-scalar fields $\psi_\pm$ are determined by requiring that Eq.\ (\ref{Teukolsky1+1psi_take2}) is satisfied as a distributional equation, with the ansatz
\begin{align}
\psi_\pm=&\psi_\pm^>(t,r)\Theta[r-R(t)]+\psi_\pm^<(t,r)\Theta[R(t)-r]+\psi_\delta^\pm(t)\delta[r-R(t)].
\end{align}
Here $\Theta[\cdot]$ is the Heaviside step function, and $\psi^\pm_\delta(t)$ is to be determined.
Balancing the coefficients of $\Theta[r-R(t)]$ and $\Theta[R(t)-r]$ imply that $\psi_\pm^>(t,r)$ and $\psi_\pm^<(t,r)$ are homogeneous solutions of Eq.\ (\ref{Teukolsky1+1psi_take2}). The remaining terms are supported on $r=R(t)$ only, and are each proportional to either $\delta$, $\delta'$ or $\delta''$. We use the distributional identities 
\begin{eqnarray}
F(r)\delta(r-R) &=& F(R)\delta(r-R),\nonumber\\
F(r)\delta'(r-R) &=&F(R)\delta'(r-R)-F'(R)\delta(r-R),\nonumber\\
F(r)\delta''(r-R)&=&F(R)\delta''(r-R)-2F'(R)\delta'(r-R) +F''(R)\delta(r-R),
\end{eqnarray} 
[valid for any smooth function $F(r)$] to eliminate the $r$ dependence of the coefficients of each of these terms, and then compare the coefficient values across the two sides of Eq.\ (\ref{Teukolsky1+1psi_take2}), recalling the form of $T_\pm$ in Eq.\ (\ref{Tsource}).

From the coefficient of $\delta''$ we immediately obtain
\begin{equation}\label{psidelta}
\psi^\pm_\delta(t)=-\frac{4s^\pm_2(t)}{f^2_R-\Rdot^2} . 
\end{equation}
The coefficient of $\delta'$ then determines the jump:
\begin{align}\label{Jpsipm}
\J{\psi_\pm}= &
-\frac{1}{f_R^2-\Rdot^2} \left(4s_1(t)+2 \dot{\psi}^\pm_{\delta}\Rdot + \left[-2f_R\big(s(1-y)+3y\big)+2s(1-3y)\Rdot+R\ddot R\right]\frac{\psi^\pm_\delta}{R} \right),
\end{align}
with $s=\pm 2$ for $\psi_\pm$.
Finally, comparing the coefficients of $\delta$ gives a relation between the jumps $\J{\psi^\pm_{,t}}$ and $\J{\psi^\pm_{,r}}$ we get
\begin{equation}\label{Jeq1}
s^\pm_0(t)= 
-\frac{1}{2}\Rdot\J{\psi^\pm_{,t}}
-\frac{1}{4}(f_R^2+\Rdot^2)\J{\psi^\pm_{,r}} +P_\pm(t),
\end{equation}
where
\begin{align}
P_\pm=
& \frac{\J{\psi_\pm}}{4R}\left[2f_R\big(s(1-y)+y\big)-2s(1-3y)\Rdot-R \ddot R\right] +\frac{1}{4} \ddot{\psi}^\pm_\delta +\frac{s}{2R}(1-3y)\dot{\psi}^\pm_\delta \nonumber \\
&+\frac{\psi^\pm_\delta}{4R^2}\Big[\lambda f_R +2+s-s^2+2y(1+s^2-4s) +8y^2(s-2) -2s(1-6y)\Rdot\Big],
\end{align}
with $s=\pm 2$ for $P_\pm$. Recall $\lambda=(l+2)(l-1)$.

A second relation between $\J{\psi^\pm_{,t}}$ and $\J{\psi^\pm_{,r}}$ is obtained by writing 
\begin{equation}\label{Jeq2}
\dot{[\psi_\pm]}=\J{\psi^\pm_{,t}}+\Rdot\J{\psi^\pm_{,r}},
\end{equation}
where we recall an overdot represents a time derivative $d/dt$. Solving (\ref{Jeq1}) and (\ref{Jeq2}) as a simultaneous set then gives
\begin{equation}
{
\J{\psi^\pm_{,r}}=\frac{4(P_\pm-s^\pm_0)-2\Rdot \dot{[\psi_\pm]}}{f_R^2-\Rdot^2},
}
\end{equation}
and
\begin{equation}
{
\J{\psi^\pm_{,t}}=\frac{-4(P_\pm-s^\pm_0)\Rdot+ (f_R^2+\Rdot^2)\dot{[\psi_\pm]} }{f_R^2-\Rdot^2}.
}
\end{equation}
The corresponding jumps in the $u$ and $v$ derivatives are
\begin{equation}\label{Jpsiv}
{
\J{\psi^\pm_{,v}}=\frac{4(P_\pm-s^\pm_0)+ (f_R-\Rdot)\dot{[\psi_\pm]} }{2(f_R+\Rdot)},
}
\end{equation}
\begin{equation}\label{Jpsiu}
{
\J{\psi^\pm_{,u}}=\frac{-4(P_\pm-s^\pm_0)+ (f_R+\Rdot)\dot{[\psi_\pm]} }{2(f_R-\Rdot)}.
}
\end{equation}

Equations (\ref{Jpsipm}), (\ref{Jpsiv}) and (\ref{Jpsiu}) give the jumps in $\psi_\pm$ and its first derivatives for a generic geodesic orbit.

\section{Jumps in the second and third derivatives of $\psi_\pm$}
\label{app:WeylJumpHighDerivatives}

We can get $\J{\psi^\pm_{,uv}}$ directly from the vacuum BPT equation:
\begin{align} 
\J{\psi^\pm_{,uv}}= - U_{\pm 2}(R)\J{\psi^\pm_{,u}} - V_{\pm 2}(R)\J{\psi^\pm_{,v}}  - W_{\pm 2}(R)\J{\psi_\pm} ,
\end{align}
where the jumps $\J{\psi^\pm_{,v}}$, $\J{\psi^\pm_{,u}}$ and $\J{\psi^\pm}$ can be substituted for from Eqs.\ (\ref{Jpsiv}), (\ref{Jpsiu}) and (\ref{Jpsipm}).
Then $\J{\psi^\pm_{,uu}}$ and $\J{\psi^\pm_{,vv}}$ can be obtained from the chain rules
\begin{eqnarray}
\dot{\J{\psi^\pm_{,u}}} &=& \dot v\J{\psi^\pm_{,uv}} + \dot u\J{\psi^\pm_{,uu}}, \nonumber\\
\dot{\J{\psi^\pm_{,v}}} &=& \dot v\J{\psi^\pm_{,vv}} + \dot u\J{\psi^\pm_{,uv}},
\end{eqnarray} 
where 
\begin{equation}
\dot v= 1 +\Rdot/f_R,\quad\quad
\dot u= 1 -\Rdot/f_R.
\end{equation}

The jumps in the third derivatives are obtained in a similar fashion: First, we obtain $\J{\psi^\pm_{,uvu}}$ and $\J{\psi^\pm_{,uvv}}$ from the $u$ and $v$ derivatives of the vacuum BPT equations. Then, $\J{\psi^\pm_{,uuu}}$ and $\J{\psi^\pm_{,vvv}}$ are determined from the appropriate chain rule; e.g.,
\begin{equation}
\J{\psi^\pm_{,uuu}} = \frac{\dot{\J{\psi^\pm_{uu}}}-\dot v\J{\psi^\pm_{,uuv}}}{\dot u},
\end{equation} 
where the jumps on the right-hand side are known from previous steps. We may proceed in this recursive manner to determine the jumps in all higher derivatives.  In the calculation performed in this paper we require jump information only up to the third derivatives. 

\section{Large-$R$ asymptotics for scatter orbits}\label{WeylJumpsAsympt}
 
For our asymptotic analysis in Section \ref{subsec:asymptotics} (where we derive initial conditions for the Hertz potential's jump equations), it is useful to have at hand the large-$R$ asymptotic form of the Weyl-scalar jumps calculated above, in the case of a scatter orbit coming from infinity (i.e., the class of geodesic scatter orbits described in Chapter \ref{chapter:geodesics}). Specifically, we need the asymptotic forms of $\J{\psi_{\pm}}$ as well as $\J{\psi^+_{,u}}$, $\J{\psi^+_{,uu}}$ and $\J{\psi^+_{,uuu}}$.
 
As input for this calculation, we need the asymptotic form of the source coefficients $s_n^\pm$ in Eqs.\ (\ref{s0})--(\ref{s2psi4}). Specializing to scatter geodesics and working at leading order in $y=M/R$ (at fixed $E,L$), we find
\begin{eqnarray}
s_n^+ &=& \sigma_n^+ R + O(R^0), \nonumber\\
s_n^- &=& \sigma_n^- R^{-3} + O(R^{-4}),
\end{eqnarray}
for $n=0,1,2$. The coefficients needed for our purpose are given explicitly by 
\begin{eqnarray}\label{s^+_nAsymp}
\sigma_0^+ &=& -\mu E\pi(1-\Rdotinf)^2\left[(m^2-2){\cal Y}^+
+2m{\cal Y}^+_{\theta} +{\cal Y}^+_{\theta\theta} \right]_{\infty},
\nonumber\\
\sigma_1^+ &=&\mp 2i \mu L\pi(1-\Rdotinf)^2\left(m{\cal Y}^++{\cal Y}^+_{\theta} \right)_{\infty},
\end{eqnarray}
and 
\begin{eqnarray}\label{s^-_nAsymp}
\sigma_1^- &=& \pm \frac{1}{2} i \mu L (1+\Rdotinf)^2\left(m \mathcal{Y}^- -\mathcal{Y}^-_{\theta}\right)_\infty ,
\end{eqnarray}
where subscripts `$\infty$' imply an evaluation at $t\to \pm\infty$, 
depending on whether it is the ``in'' or `'out'' states being considered. In the expressions for $\sigma_1^+$ and $\sigma_1^-$, the upper sign is for the out state ($\Rdotinf>0$) and the lower sign is for the in state ($\Rdotinf<0$).

A straightforward leading-order calculation now gives 
\begin{eqnarray}
\J{\psi_{+}} &=& -4 \sigma_{1}^{+}E^2 R+ O(R^0), \\
\J{\psi_{-}} &=& -4 \sigma_{1}^{-}E^2 R^{-3}+ O(R^{-4}),
\end{eqnarray}
as well as 
\begin{equation}
\J{\psi^{+}_{,t}} = -\Rdotinf \J{\psi^{+}_{,r}} = 4 \sigma_{0}^{+}E^2 \Rdotinf R+ O(R^0),
\end{equation}
\begin{equation}\label{Ju}
\J{\psi^{+}_{,u}} = \frac{2 \sigma_{0}^{+}}{1-\Rdotinf}R+ {\cal O}(R^0),
\end{equation}
\begin{equation}\label{Juu}
\J{\psi^{+}_{,uu}} = -\frac{2 \sigma_{0}^{+}(2-3\Rdotinf)}{(1-\Rdotinf)^2}+ {\cal O}(R^{-1}),
\end{equation}
\begin{equation}\label{Juuu}
\J{\psi^{+}_{,uuu}} = \frac{\sigma_{0}^{+} \left[8(1-2\Rdotinf) + \lambda(1+\Rdotinf)\right]}{2 (1-\Rdotinf)^2}\, R^{-1} + O(R^{-2}).
\end{equation}

%% file: FDS.tex

\chapter{Finite-difference scheme}
\label{app:FDS}

In this appendix we detail the finite-difference (FD) schemes employed for solving the 1+1D field equations associated with the scalar field (\ref{eqn:SourcedFieldEquationuv}) and the Regge-Wheeler-like field (\ref{RWeq}). Here we focus on the FD scheme itself, at the grid-cell level. We detail schemes for field equations of the form
\begin{equation}
X_{,uv} + V(r) X = S_X,
\label{eqn:FDSFieldEquation}
\end{equation}
where $W(r)$ is a generic radial potential. The source $S_X$ can be zero (vacuum), an explicitly known point-particle source which is localised to the particle's 1D worldline $\cal S$ (scalar field) or have information on the source available through known jumps conditions across a 1D surface $\cal S$ associated with the particle's worldline (RW-like field). In deriving the schemes we follow the method of Ref.\ \cite{BarackSago2010} (which itself builds on a long history of time-domain work in the self-force literature, e.g.\ \cite{Lousto05, Haas07}).

\begin{figure}[h!]
\centering
\includegraphics[width=0.5\linewidth]{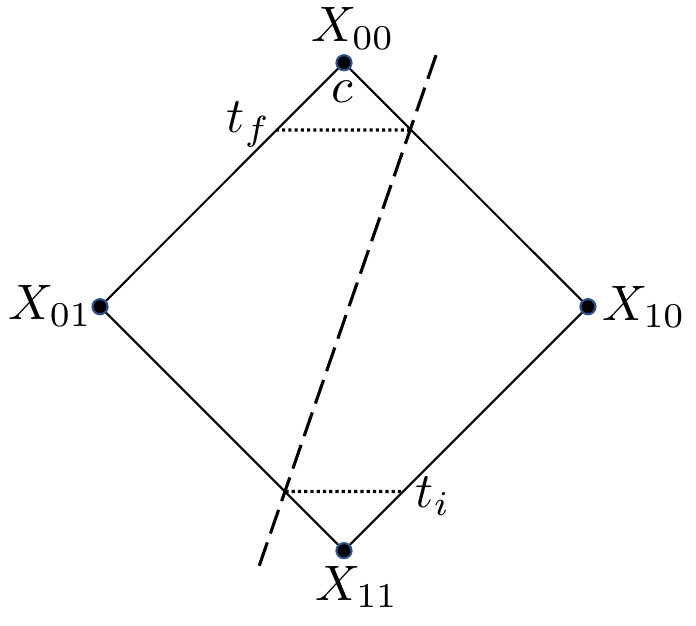}
\caption[Particle cells for a generic orbit with a known source]{A particle cell is traversed by the particle's worldline (dashed curve). The apex of the cell is the point $c$ at $(u,v) = (u_c,v_c)$ and the field point $X_{nk}$ is at the grid point with coordinates $(u,v)=(u_c- nh,v_c-kh)$. The particle enters (exits) the cell at time $t=t_i$ ($t=t_f$).}
\label{ScalarGenericCell}
\end{figure}

Recall Figure \ref{uvGridVacuum}, which shows the 1+1D numerical grid for the vacuum case, based on $u,v$ coordinates with uniform cell dimensions $h\times h$. Consider an arbitrary grid point $c$ with coordinates $(u,v)=(u_c,v_c)$, and in reference to that point denote by $X_{nk}$ the value of the numerical field variable $X$ at the grid point with coordinates $(u,v)=(u_c- nh,v_c-kh)$, as shown in Figure \ref{ScalarGenericCell} for a field sourced by a point-particle. Our goal is to prescribe a FD expression for $X_{00}$ (the field at $c$), given the values $X_{nk}$ for all $n,k>0$, assumed to have been obtained in previous steps of the characteristic evolution. We wish to achieve a global quadratic convergence, i.e.\ an accumulated error in $X$ that scales as $h^2$. Since the total number of grid points over which the error accumulates is $\propto h^{-2}$, this demands a local (single-point) FD error not larger than $O(h^4)$ in general. 

In the non-vacuum cases, our grid is traversed by the curve $\cal S$ representing the timelike geodesic trajectory of the particle, recall Figure \ref{uvGridCircular} for circular orbits and Figure \ref{uvGridScatter} for hyperbolic orbits. The curve is fixed and known in advance, and the coordinates of all of its intersections with grid lines are calculated in advance of the numerical evolution. In reference to the grid cell $C$ with top vertex $c$, we distinguish between two scenarios: Either $C$ is traversed by $\cal S$ (``particle cell'') or it is not (``vacuum cell''). We deal with each of these two scenarios separately below. 

\section{Vacuum cells}
\label{app:FDSVac}

First we consider the simplest case where the field equation is homogeneous or the particle's worldline does not cross the integration cell. Then a sufficiently accurate FD approximation for $X_{00}$ can be written based on the three value $X_{01}$, $X_{10}$ and $X_{11}$ alone. Integrating each of the two terms on the left-hand side of Eq.\ (\ref{eqn:FDSFieldEquation}) over the grid cell $C$, we have 
\begin{equation}
\int_C X_{,uv} \: du dv = X_{00} - X_{01} - X_{10} + X_{11}
\end{equation}
(exactly), and 
\begin{equation}
\int_C V(r)X \: du dv = \frac{1}{2} h^2V(r_c) \left( X_{01} + X_{10} \right) + O(h^4),
\end{equation}
where 
$r_c$ is the value of $r$ at point $c$. The homogeneous version of Eq.\ (\ref{eqn:FDSFieldEquation}) then gives
\begin{equation}
X_{00} = - X_{11} + (X_{01} + X_{10}) \left( 1 - \frac{h^2}{2} V(r_c) \right)+O(h^4),
\label{eqn:FDSVac}
\end{equation}
which we use as our FD formula for vacuum points.

\section{Particle cells from a source}
\label{app:FDSSource}

The vacuum formula (\ref{eqn:FDSVac}) does not work for cells that are traversed by the worldline as there is also a contribution from the source of the field equation. We calculate this contribution by integrating the source over the cell such that the FD scheme is given by
\begin{equation}
X_{00} = - X_{11} + (X_{01} + X_{10}) \left( 1 - \frac{h^2}{2} V(r_c) \right) + \int_C S_X \: dudv +O(h^4).
\label{eqn:FDSSourceGeneric}
\end{equation}

\subsection{Scalar point-particle on a circular orbit}
\label{app:FDSScalarCircular}

In the scalar point-particle case we know the source of the field equation (\ref{eqn:SourcedFieldEquationuv}). For a circular equatorial orbit at $r=R$ it has the form
\begin{equation}
S_\psi = \frac{f_R^2}{2 E R} \delta \left(r - R\right) \bar{Y}_{\ell m}(\pi/2, \Omega t),
\end{equation}
where $R$ is the (constant) radius and $\Omega:=\sqrt{M/R^3}$ is the orbital angular velocity. Due to the simplicity of the orbit we can integrate this source analytically over a single $h\times h$ cell to give the contribution to the FD scheme. In our implementation the particle passes directly through the cells centred on $r=R$ and does not pass through any other (vacuum) cells. The contribution from the source to the particle cells is obtained by integrating the source over the cell to give
\begin{equation}
\int_C \frac{f^2_R}{2 E R} \delta \left(r - R\right) \bar{Y}_{\ell m}(\pi/2,\Omega t) \: d u d v = \frac{i f_R}{E m \Omega R} \left(-1 + e^{-im\Omega h}  \right)\bar{Y}_{\ell m}(\pi/2,\Omega t_c),
\end{equation}
where $t_c$ is $t$ evaluated at point $c$ and we have used the relations $d u d v = 2 f^{-1} \: d r d t$ and $\bar{Y}_{\ell m}(\pi/2,\Omega t)= \bar{Y}_{\ell m}(\pi/2,0) e^{-im\Omega t}$. Including this contribution gives the FD for a scalar particle on a circular orbit as
\begin{equation}
\psi_{00} = - \psi_{11} + (\psi_{01} + \psi_{10}) \left( 1 - \frac{1}{2} h^2 V \right) + Z,
\label{eqn:FDSScalarCircular}
\end{equation}
where
\begin{equation}
Z = 
\begin{cases}
0 & \text{if particle does not enter cell} \\
\frac{i f_R}{E m \Omega R} \left(-1 + e^{-im\Omega h}  \right)\bar{Y}_{\ell m}(\pi/2,\Omega t_c) & \text{if particle does enter cell}
\end{cases}
\end{equation}

\subsection{Scalar point-particle on a generic orbit}
\label{app:FDSScalarGeneric}

Recall the form of the source of the scalar field equation (\ref{eqn:SourcedFieldEquationuv})
\begin{equation}
S_\psi = \frac{f_R^2}{2 E R(t)} \delta \left(r - R(t)\right) \bar{Y}_{\ell m}(\pi/2, \varphi_p(t)),
\end{equation}
where now we use $f_R:=1-2M/R(t)$. Integrating this source over the FD cell for a generic orbit is more complicated as in general we do not have an analytic expression for the worldline. This does not cause any issues for integrating over the radial derivative as we can directly integrate the delta function of the source of the field equation (\ref{eqn:SourcedFieldEquationuv}) to give
\begin{equation}
\int_C S_\psi \: dudv = \int_{t_i}^{t_f} \frac{f_R}{E R(t)} \bar{Y}_{\ell m}(\pi/2, \varphi_p(t)) \: dt,
\label{eqn:FDSSourceInt}
\end{equation}
where $t_i$ ($t_f$) are where the particle enters (exits) the cell as shown in Figure \ref{ScalarGenericCell}. 

For a generic orbit we cannot evaluate this integral analytically so we must use other methods. One possibility would be to numerically integrate the source term for all of the particle terms. We take the alternative approach of expanding the integrand of Eq.\ (\ref{eqn:FDSSourceInt}) about the point $c$ which gives
\begin{equation}
\frac{f_c}{E R_c} \bar{Y}_{\ell m}(\pi/2,\varphi_c) - \frac{(R_c-4M) \dot R_c + i m f_c R_c^2 \dot \varphi_c}{E R_c^3} \bar{Y}_{\ell m}(\pi/2,\varphi_c) (t-t_c) + {\cal O}(t^2),
\end{equation}
where subscript $c$ represents the quantity evaluated at point $c$ and we recall that an overdot represents a time derivative $d/dt$. We can integrate this result analytically to give the source's contribution to the FD scheme as
\begin{equation}
Z = \frac{f_c}{E R_c} \bar{Y}_{\ell m}(\pi/2,\varphi_c) (t_f-t_i) + \frac{(R_c-4M) \dot R_c + i m f_c R_c^2 \dot \varphi_c}{E R_c^3} \bar{Y}_{\ell m}(\pi/2,\varphi_c) (2t_c-t_f-t_i).
\label{eqn:FDSZScalarGeneric}
\end{equation}
We can show that this contribution is ${\cal O}(h^3)$ as $t_f-t_i\sim h$ and $2t_c-t_f-t_i \sim h^2$. As the worldine is a 1D surface the total number of particle cells scales as $h$ hence the global convergence of $Z$ is quadratic as required.

The second-order convergent FD scheme for a scalar particle on a generic orbit is given by Eq.\ (\ref{eqn:FDSScalarCircular}) with $Z=0$ for vacuum cells or $Z$ given by Eq.\ (\ref{eqn:FDSZScalarGeneric}) for particle cells.

\section{Particle cells from field jumps}
\label{app:FDSJumps}

In this section we address the case where we do not know the source but we have information about the source through jumps in the field and its derivatives. Here we have a discontinuity in the field across the 1+1D worldline $\cal S$ which corresponds to $r=R(t)$. The worldline splits $C$ into two disjoint vacuum regions, $C^>$ and $C^<$. Since $X$ is smooth on each of the two vacuum regions, we can expand it piecewise in a Taylor series about point $c$, in the form 
\begin{equation}
X^{\gtrless} = \sum_{i+j=0}^N \frac{c_{ij}^{\gtrless}}{i! j!} \tilde u^i \tilde v^j + O(h^{N+1}),
\label{2DTaylorX}
\end{equation}
where $\tilde u := u-u_c$, $\tilde v := v-v_c$, and different expansion coefficients apply on each side of $\cal S$: $c_{ij}^{<}$ are used in $C^<$, and $c_{ij}^{>}$ are used in $C^>$. The idea now is to derive the values of $c_{ij}^{\gtrless}$ based on a sufficient number of data points $X^{\gtrless}_{nk}$, plus the analytically known jumps in $X$ and its derivatives on $\cal S$. We note that, since the total number of particle cells scales as $h^{-1}$, it is acceptable for our local FD scheme to have an error as great as $O(h^3)$ (but not greater) at each particle cell. 

To achieve such accuracy we take $N=2$ in Eq.\ (\ref{2DTaylorX}), leaving us with 12 coefficients $c_{ij}^{\gtrless}$ to determine. We use the 6 data points $\{X_{00},X_{01},X_{10},X_{11},X_{02},X_{20}\}$ to supply 6 constraints, and 6 additional constraints are obtained from the known jumps $\{[X],[X_{,u}],[X_{,v}],[X_{,uu}],[X_{,uv}],[X_{,vv}]\}$, imposed at the point where the worldline exits the cell $C$. Solving the 12 simultaneous equations for $c_{ij}^{\gtrless}$ and then substituting these coefficients back in (\ref{2DTaylorX}), gives an expression for $X^{\gtrless}$, accurate through $O(h^2)$ in the vicinity of point $c$, in terms of the above 6 field points (which include the unknown $X_{00}$) and above 6 jumps.

Considering first the principal part of the Eq.\ (\ref{eqn:FDSFieldEquation}), we thus obtain, 
\begin{align}\label{FD scheme:Xvuterm}
X^{\gtrless}_{,uv} &= c^{\gtrless}_{11} +O(h)
\nonumber\\
&= h^{-2}\left(X_{00}-X_{01}-X_{10}+X_{11}+J_1^A\right) +O(h),
\end{align}
where the explicit form of $J_1^A$ will be discussed for the circular and hyperbolic cases below. For the potential term of Eq.\ (\ref{eqn:FDSFieldEquation}) we wish to obtain a FD approximation that does not involve $X_{00}$. The form of (\ref{FD scheme:Xvuterm}) implies that we only require a leading-order, $O(h^0)$ approximation for this term. We choose to achieve this by taking $N=1$ in Eq.\ (\ref{2DTaylorX}), and then solving for the six coefficients $c^{\gtrless}_{ij}$ ($i+j\leq 1$) using the 3 data points $\{X_{01},X_{10},X_{11}\}$ and 3 jumps $\{[X],[X_{,u}],[X_{,v}]\}$, again evaluated when the particle leaves the cell. This gives 
\begin{align}
X^{\gtrless} &= c^{\gtrless}_{00} +O(h)
\nonumber\\
& = X_{01} + X_{10}- X_{11} + J_2^A +O(h),
\end{align}
where again we will consider the form of $J_2^A$ in later sections. Hence we can write
\begin{equation}\label{FD scheme:WXterm}
VX^{\gtrless} = V(r_c) (X_{01} + X_{10}- X_{11} + J_2^A) +O(h).
\end{equation}

Imposing finally the vacuum field equation $X_{,uv}+VX=0$, we obtain, using (\ref{FD scheme:Xvuterm}) and (\ref{FD scheme:WXterm}),
\begin{align}
X_{00} =& (X_{01} + X_{10}- X_{11}) \left( 1 - h^2V(r_c) \right) - J_1^A - h^2V(r_c)J_2^A + O(h^3),
\label{FD schemeparticle}
\end{align}
which is our FD formula for particle cells.

Note that our second-order-convergent FD scheme, consisting of Eq.\ (\ref{eqn:FDSVac}) for vacuum cells with Eq.\ (\ref{FD schemeparticle}) for particle cells, requires as input only the three field data points $X_{01}$, $X_{10}$ and $X_{11}$ (as well as the known jumps). This is convenient, as it means that at each characteristic evolution step we require data on a single previously calculated characteristic ray.  

\subsection{Circular orbit}
\label{app:FDSJumpsCircular}

\begin{figure}[H]
\centering
\includegraphics[width=0.7\linewidth]{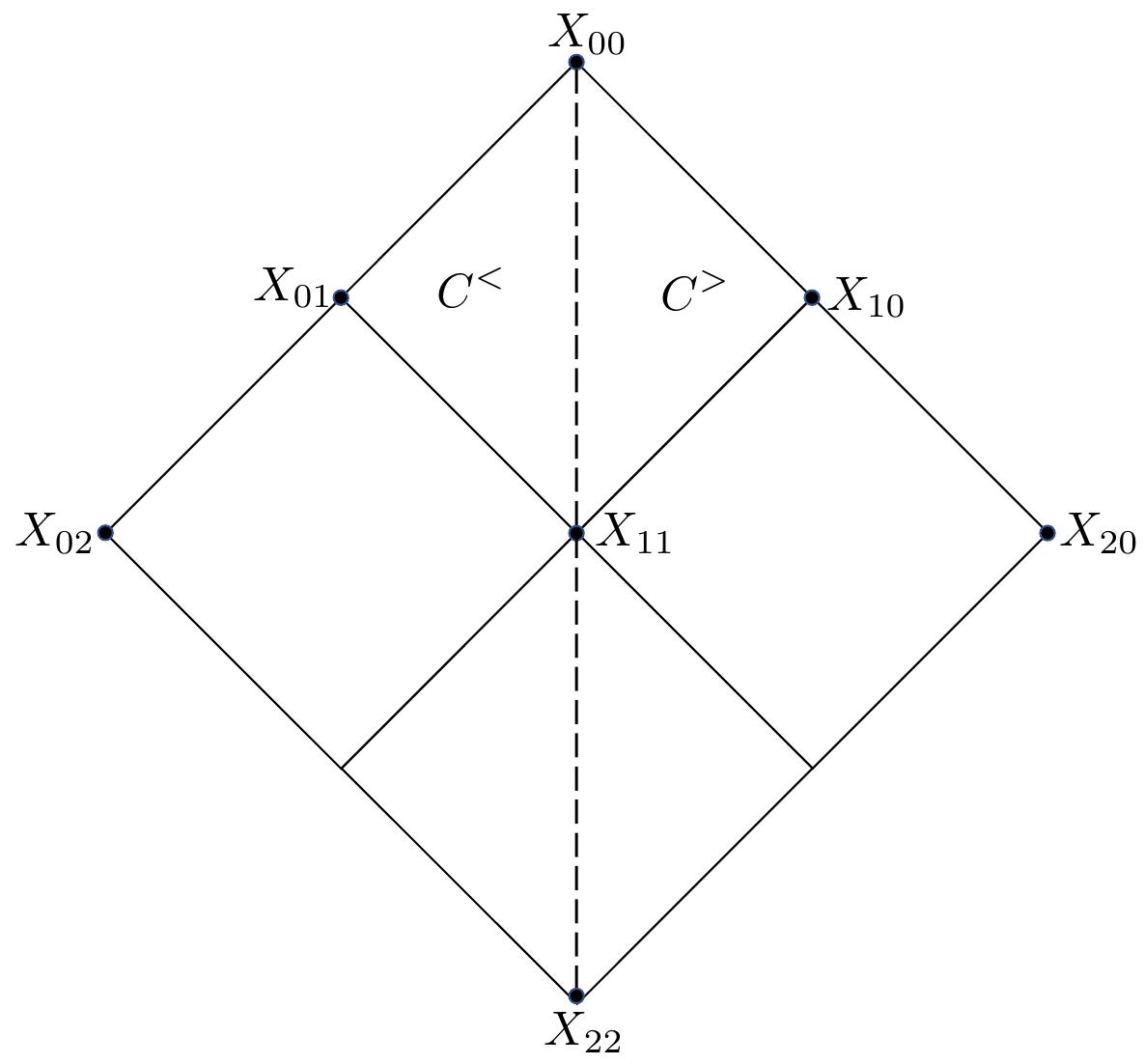}
\caption[Particle cells for a circular orbit with known jump conditions]{Particle cells are traversed by the particle's worldline (dashed curve) either directly through the centre of the cell (e.g.\ cell with $X_{00}$ at the apex) or through the leftmost corner of the cell (e.g.\ cell with $X_{10}$ at the apex) which are known as `P'  and `R' cells respectively. A different variant of the FD formula applies in each case, as described in the text. The vacuum portions of the cell left and right of the worldline are $C^<$ and $C^>$ respectively, as shown for the uppermost P cell.}
\label{CircularNonVacCells}
\end{figure}

As described in Section \ref{sec:NumericalMethodCircularRW} we construct the grid such that the worldline passes directly through cells centred on $r=R$ as shown in Figure \ref{CircularNonVacCells}. However, it is not possible to calculate the field directly on the worldine due to the discontinuity. Instead we shift the worldline to an infinitesimally larger $r>R$ such that we calculate the field points $X^<$ along $r=R$. This setup results in two kinds of cells which cannot use the vacuum FD formula (\ref{eqn:FDSVac}). These cells correspond to cells centred on $r=R$ (`P' cells) and $r=R+h/2$ (`R' cells). The additional terms from these cells to the FD scheme (\ref{FD schemeparticle}) are given by
\begin{align}
J_1^{\rm P} &= -[X] + h [X_{,u}] - \frac{h^2}{2} \left( [X_{,uu}] + [X_{,uv}] \right) \nonumber \\
&= -[X] + \frac{h}{2} [X_{,u}] (2+i h m \Omega), \\
J_2^{\rm P} &= 0,  \\
J_1^{\rm R} &= [X] = -J_2^{\rm R}.
\end{align}
Recall that all the jumps are evaluated where the particle exits the cell (i.e.\ $(u_{\rm c},v_{\rm c})$ for P cells and $(u_{\rm c},v_{\rm c}-h)$ for R cells). The simplification of the $J_1^{\rm P}$ expression arises from using the chain rule to rewrite jumps in higher derivatives in terms of lower derivatives (see Appendix \ref{app:WeylJumpHighDerivatives}) and the relation $\dot{[X]} = -i m \Omega [X]$ for a circular orbit with orbital frequency $\Omega = \sqrt{M/R^3}$.

\subsection{Generic orbit}
\label{app:FDSJumpsGeneric}

\begin{figure}[h!]
\centering
\includegraphics[width=0.8\linewidth]{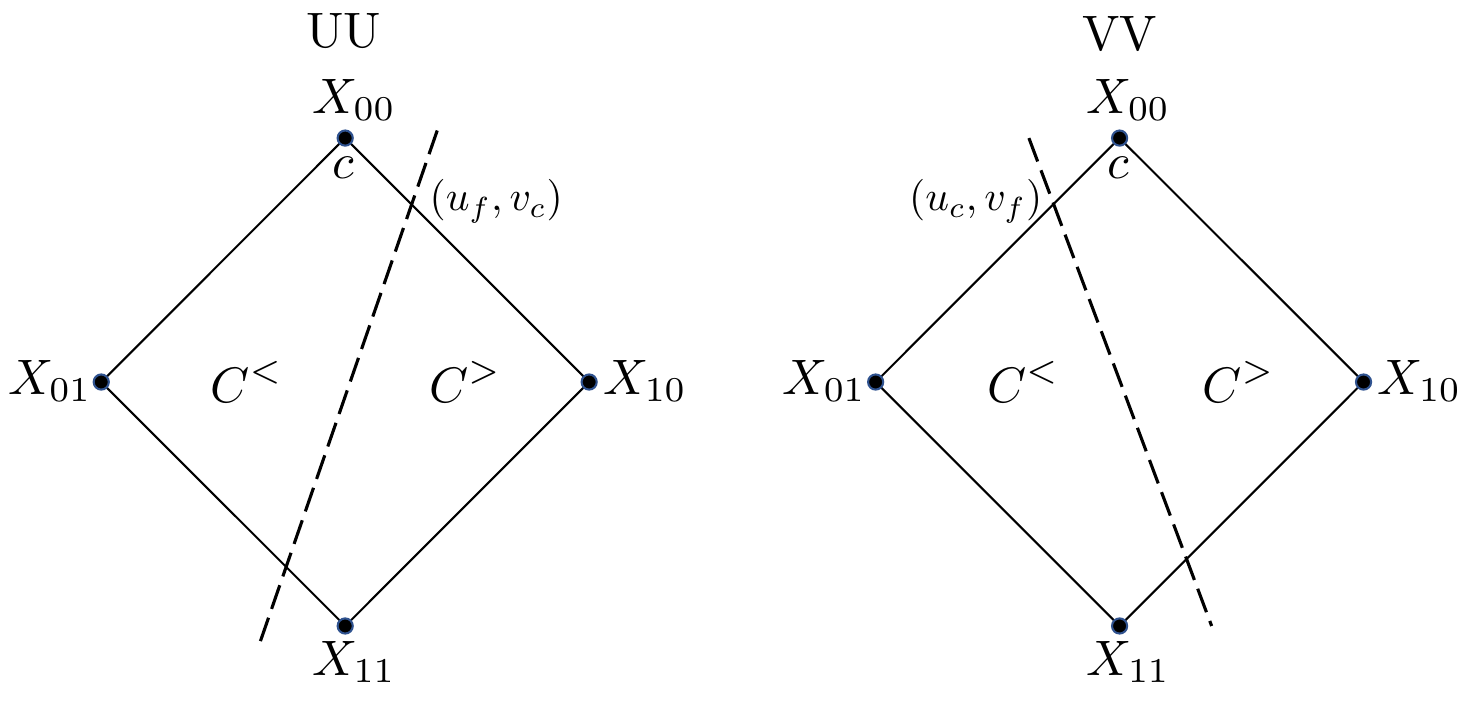} \\
\vspace{3mm}
\includegraphics[width=0.8\linewidth]{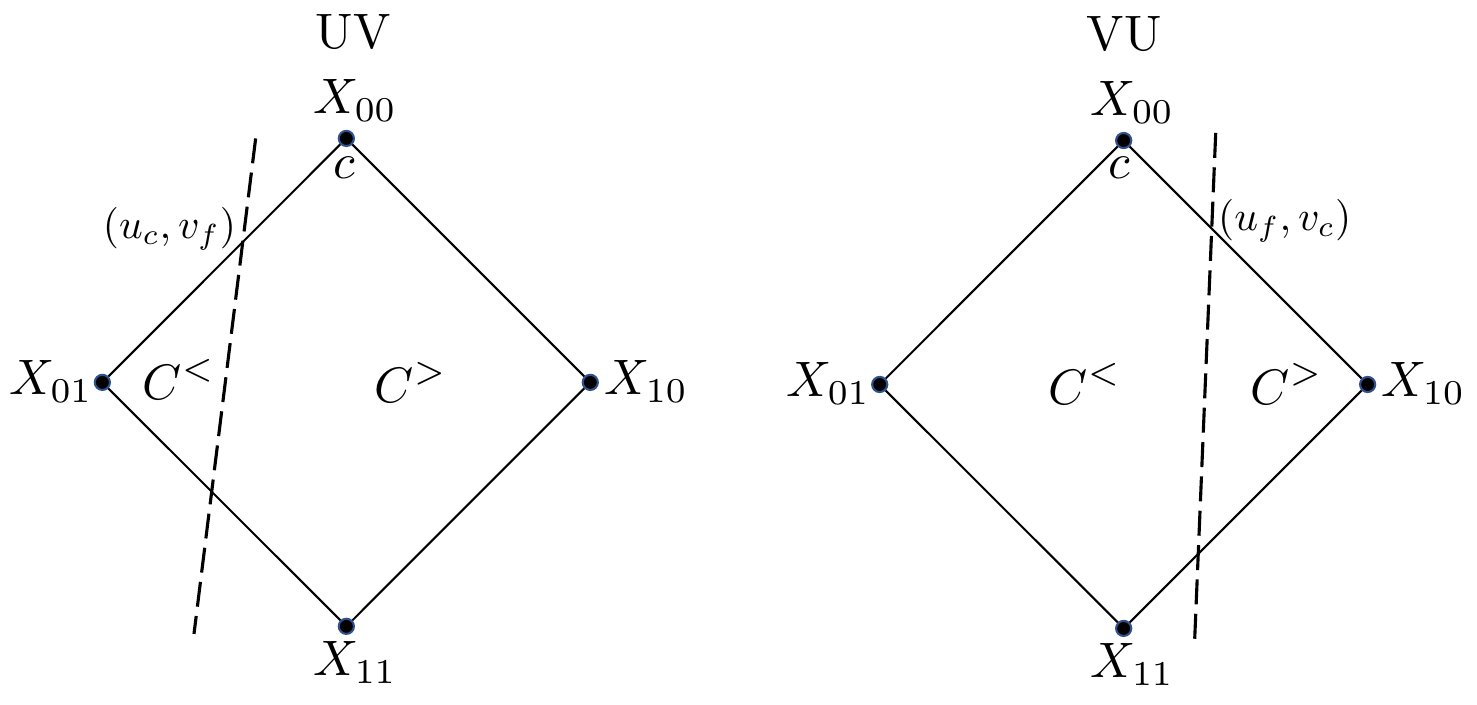}
\caption[Particle cells for a generic orbit with known jump conditions]{A particle cell is traversed by the particle's worldline (dashed curve) in one of four possible ways: cases `UU', `VV', `UV' and `VU', illustrated here. A different variant of the FD formula applies in each case, as described in the text.   The apex of the cell is the point $c$ at $(u,v) = (u_c,v_c)$, and the particle exits the cell at $(u_f,v_c)$ or $(u_c,v_f)$, depending on the case.
The vacuum portions of the cell left and right of the worldline are $C^<$ and $C^>$ respectively.}
\label{ScatterNonVacCells}
\end{figure}

For a generic timelike geodesic the worldline splits $C$ into two distinct vacuum regions, $C^<$ and $C^>$, as shown in Figure \ref{ScatterNonVacCells}, which shows the four possible scenarios. We impose the jump conditions where the particle exits the cell [i.e.\ referring to Figure \ref{ScatterNonVacCells}, either the point $(u_f,v_c)$ or the point
$(u_c, v_f)$, depending on the case]. The additional contributions to Eq.\ (\ref{FD schemeparticle}) are given by
\begin{align}
J_1^{UU}&=h[X_{,v}]+h(u_c-u_f)[X_{,uv}]-\frac{h^2}{2}\left(2[X_{,uv}]+[X_{,vv}]\right),
\\
J_1^{VV}&=-h[X_{,u}]-h(v_c-v_f)[X_{,uv}]+\frac{h^2}{2}\left(2[X_{,uv}]+[X_{,uu}]\right),
\\
J_1^{UV}&=-[X]+(h-v_c+v_f)[X_{,v}]-\frac{1}{2}(h-v_c+v_f)^2[X_{,vv}],
\\
J_1^{VU}&=[X]-(h-u_c+u_f)[X_{,u}]+\frac{1}{2}(h-u_c+u_f)^2[X_{,uu}].
\end{align}
and
\begin{align}
J_2^{UU}=0=J_2^{VV},
\quad\quad
J_2^{UV}=[X]=- J_2^{VU}.
\end{align}

